\documentclass[12pt,floatfix,preprintnumbers,amsmath,amssymb,superscriptaddress,tightenlines]{report}
\pdfoutput=1 
\usepackage{fancyheadings}
\usepackage{url}
\usepackage{amsmath,amssymb,amsfonts}
\usepackage{graphicx}
\usepackage{enumerate} 
\usepackage{colordvi} 
\usepackage{bm}
\usepackage{epsfig}
\usepackage{float}
\usepackage{verbatim}
\usepackage{subfig}
\usepackage[toc,page]{appendix}

\makeatletter
\newenvironment{chapquote}[2][2em]
  {\setlength{\@tempdima}{#1}%
   \def\chapquote@author{#2}%
   \parshape 1 \@tempdima \dimexpr\textwidth-2\@tempdima\relax%
   \itshape}
  {\par\normalfont\hfill--\ \chapquote@author\hspace*{\@tempdima}\par\bigskip}
\makeatother

\def\de{\delta}

\newcommand{\bfv}{\mbox{\boldmath$v$}}
\newcommand{\bfx}{\mbox{\boldmath$x$}}
\newcommand{\bfw}{\mbox{\boldmath$w$}}
\newcommand{\bfk}{\mbox{\boldmath$k$}}
\newcommand{\bfp}{\mbox{\boldmath$p$}}
\newcommand{\bfq}{\mbox{\boldmath$q$}}

\newcommand{\bfr}{\mbox{\boldmath$r$}}
\newcommand{\bfu}{\mbox{\boldmath$u$}}

\newcommand{\ad}[1]{\langle\de^{#1}\rangle}

\begin{document}

\thispagestyle{empty}

\begin{titlepage}
\begin{center}
{\LARGE{\bf Institute of Cosmology and Gravitation}}
\end{center}
\begin{center}
{\LARGE University of Portsmouth}
\end{center}
\begin{center}
\includegraphics[width=4cm,height=4cm]{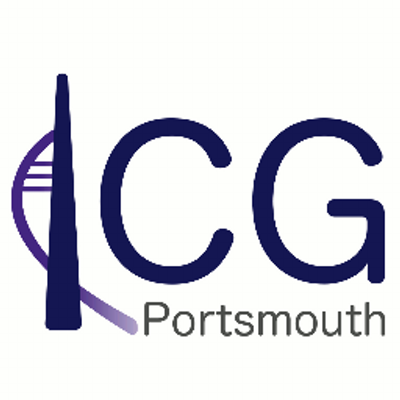}
\end{center}
\vspace{25px}
\begin{center}
\rule{\linewidth}{1pt} \newline \newline {\LARGE{\bf Cosmological Tests of Gravity }} \newline \newline   \rule{\linewidth}{1pt}
\end{center}
\begin{center}
{\LARGE Ph.D. Thesis}
\end{center}
\vspace{40px}
\begin{minipage}{0.4 \textwidth}
\begin{flushleft}
\emph{Candidate} \\
{\bf Benjamin Bose} 
\end{flushleft}
\end{minipage}
 \begin{minipage}{0.6\textwidth}
\begin{flushright} 
\emph{Supervisors}  \\ 
{\bf Prof. Kazuya Koyama} \\
{\bf Prof. Gong-Bo Zhao}      
\end{flushright}
\end{minipage}  
\vspace{50px}
\begin{center}
{\footnotesize THIS THESIS IS SUBMITTED IN PARTIAL FULFILMENT OF THE  REQUIREMENTS FOR THE AWARD OF THE DEGREE OF DOCTOR OF  PHILOSOPHY BY THE UNIVERSITY OF PORTSMOUTH}
\end{center}
\begin{center}
\today
\end{center}
\end{titlepage}
\newpage 

{\LARGE Declaration and Dissemination} 
\newline
\newline
\noindent Whilst registered as a candidate for the above degree, I have not been registered for any other research award. The results and conclusions embodied in this thesis are the work of the named candidate and have not been submitted for any other academic award. This thesis is based on the published works:
\begin{enumerate}
\item
B. Bose and K. Koyama. A Perturbative Approach to the Redshift Space Power Spectrum:  Beyond the
Standard Model. JCAP, 1608, 032 (2016), arXiv:1606.02520 [astro-ph.CO].
\item
B. Bose and K. Koyama. A Perturbative Approach to the Redshift Space Correlation Function:  Beyond the
Standard Model. JCAP, 1608, 032 (2017), arXiv:1705.09181 [astro-ph.CO].
\item
 B. Bose, K. Koyama, W. A. Hellwing, G.-B. Zhao and H. A. Winther. Theoretical Accuracy in Cosmological Growth Estimation. Phys.Rev. D96 no.2, 023519 (2017), arXiv:1702.02348 [astro-ph.CO].
\end{enumerate}
\begin{center}
Approximate Word Count : 31,644
\end{center} 
\vfill
\copyright \hspace{10px}  2017 by Benjamin Bose. All rights reserved.\newline
The copyright of this thesis rests with the Author. Copies (by any means) either in full, or in extracts, may not be made without the prior written consent of the Author. 
 \newpage
{\LARGE {\bf  Abstract}} 
\newline
\newline 
\noindent General relativity has proved itself to be an incredibly robust theory having passed many high precision solar system tests as well as being able to describe the evolution of the universe and its constituents. Its cosmological application necessarily requires large and mysterious energy components to match observations. In particular, the observation of cosmic acceleration forces general relativity to adopt a dominant dark energy component. This has prompted a flurry of investigation into alternatives to the concordance model of gravity which can offer self-acceleration. These alternatives must deal with strong priors coming from the local tests of gravity. This has lead to the development of theories that exhibit so called screening mechanisms which allow them to be observationally equivalent to general relativity at small scales. On the other hand, these theories generically predict distinguishing phenomena at larger scales such as an enhanced gravitational force. This makes galaxy clusters and the LSS of the universe a great testbed for modifications to gravity. 
\newline
\newline
In particular, the anisotropy of galaxy clustering in redshift space, which arises due to the peculiar velocities of galaxies, offers a promising means of testing gravity. These so called redshift space distortions involve the velocity components of galaxies which in turn strongly involves the gravitational force and the growth rate of structure. This links us up with the fundamental laws of nature at large scales. As astronomical surveys become more and more precise, our measurements of the growth of structure become ever more powerful to detect departures from general relativity. This is true only if our theoretical modelling is up to the challenge these surveys propose. Specifically, with upcoming large volume, spectroscopic surveys such as Euclid and DESI, statistical errors will become tiny leaving room for various theoretical biases to enter the game. One such bias is that from using general relativity as a standard in data comparisons. Put in other words, the next era of astronomy will put us in a position to move beyond consistency tests of general relativity. 
\newline
\newline
In this thesis we develop a {\tt c++} code that numerically and consistently constructs LSS observables, accounting for the redshift space distortion phenomenon. By consistently we mean that this can be done for a large class of alternative theories of gravity and dark energy models. This is done using perturbation theory which treats over-densities and velocities as small perturbations upon a homogeneous and isotropic expanding background spacetime. The construction provides the first order contribution in non-linear dynamics which gives a more accurate description of the observables, an ever growing necessity when looking to extract the most information in data comparisons. We focus on the redshift space power spectrum and correlation function, two commonly used statistics that are measured from galaxy surveys. Specifically, we adopt the Taruya, Nishimichi and Saito redshift space power spectrum model and Gaussian streaming model redshift space correlation function, which both employ beyond-linear treatments of the redshift space distortions. The perturbative approach makes the pipeline ideal for statistical inference analyses that require very quick model computations. 
\newline
\newline
We test this framework against suites of numerical simulations that by and large describe the full dynamics of structure growth. Specifically, we make use of simulations within three models of gravity : general relativity, the 5-dimensional DGP brane world model and an $f(R)$ model. The two latter models both exhibit screening which makes them viable contenders to the former. Comparing the perturbative approach to these simulations gives us a handle on its accuracy in modelling the dynamics of the perturbations. We also validate the pipelines' numerical predictions against well established analytic forms for DGP and general relativity. We find very good agreement in all comparisons and a significant improvement in accuracy over the linear treatment.
\newline
\newline
 This is followed by a comparison of different approaches to the redshift space correlation function, specifically the Gaussian streaming model and the Fourier transform of the Taruya, Nishimichi and Saito spectrum. We find that these two approaches are consistent to within a few percent at scales around the baryon acoustic oscillation. 
\newline
\newline
Finally, we use dark matter simulation data to perform a test for bias introduced by incorrectly modelling gravity. This is done by using DGP simulation data in a Markov Chain Monte Carlo analysis for two different templates for the redshift space power spectrum. The first template employs general relativity to model the perturbations while the second models them within the DGP model. We fit next generation like survey error bars on the data to put the analysis in that context. It is then found that for upcoming surveys, using general relativity as a benchmark model may only safely be used in consistency tests of general relativity but not to infer constraints on alternative models. 
\newpage 
\tableofcontents
\listoftables
\listoffigures
\newpage
\begin{center}
{\Large{\bf Conventions, Constants and Abbreviations}}
\end{center}
\noindent The following notation and conventions are used throughout this work:
\begin{itemize}
\item 
Greek indices will take values from the set $(0,1,2,3)$ which will refer to general four-dimensional spacetime coordinates, $0$ being the time coordinate and $(1,2,3)$ being the spatial coordinates.
\item
Roman indices will take values from the set $(1,2,3)$. These will be used both to indicate spatial coordinates as well as order for the kernels and perturbations. This distinction will be clear in the context. 
\item
The Einstein Summation convention will be used, in which if an index appears as both an upper and lower index, it indicates a summation over all coordinates. For example  $g_{\mu \nu}u^{\mu}= \sum_{i=0}^3 g_{i \nu}u^{i}$.
\item
The signature ($-$,+,+,+) is used for all spacetime metrics. 
\item 
The partial derivative with respect to the variable $x^\mu$ will be written $\partial_\mu$, i.e.$\frac{\partial}{\partial x^{\mu}} = \partial_\mu$. 
\item
The Christoffel symbols are given by $ \Gamma^{\rho}{}_{\mu \nu}= \frac{1}{2}g^{\rho \lambda}(\partial_{\mu}g_{\nu \lambda} + \partial_{\nu}g_{\mu \lambda} - \partial_\lambda g_{\mu \nu}) $.
\item
A subscript of $0$ on any time dependent parameter (not spacetime coordinate) value will generally denote a present day value. 
\item
Unless otherwise stated, an overdot $\dot{}$ will denote a derivative with respect to the FLRW metric time coordinate $t$ and a prime $'$ will denote a derivative with respect to the scale factor $a$. 
\item
Unless otherwise stated, an overbar $\bar{}$ will denote a background quantity.   
\item
We will use the summation convention $\bfk_{1 \dots n} = \bfk_1 + \dots + \bfk_n$.
\item
Unless otherwise stated, we work in the system of units in which $c = 1$. 
\end{itemize}
In addition to the above, a number of abbreviations are made use of throughout this work. These are listed in Table.\ref{abbrevtab}.
\begin{table}[htbp]
\centering
\caption[Physical Constants]{Physical Constants }
 \begin{tabular}{| l | c | r | }
 \hline
  \textbf{Physical Constant} & \textbf{Symbol} & \textbf{Value (S.I)} \\ \hline
 Gravitational constant & $G$ & $6.674 28 \times 10^{{-11}} $ m${}^3 $kg${}^{-1} $s${}^{-2}$ \\ \hline
 Speed of light & $c$ & $299,792,458 $ m$ $ s${}^{-1}$ \\ \hline
 Solar Mass & $\mbox{M}_\odot $&$ 1.988 \times 10^{30}$ kg \\ \hline
  \end{tabular}
 \end{table}
  \begin{table}[htbp]
 \centering
\caption[Abbreviations]{Abbreviations}
 \begin{tabular}{ l | r  }
 \textbf{Abbreviation} & \textbf{Expression}  \\ \hline
  BAO & Baryon Acoustic Oscillations \\
  CDM & Cold Dark Matter \\ 
  CMB & Cosmic Microwave Background \\
  COLA & COmoving Lagrangian Acceleration \\ 
  DGP & Dvali-Gabadadze-Porratti \\ 
  EDS & Einstein-de Sitter \\ 
 EFToLSS & Effective Field Theory of Large Scale Structure \\ 
  FoG & Fingers of God\\ 
  FLRW & Friedman-Lema$\hat{\mbox{i}}$tre-Robertson-Walker \\ 
  FT & Fourier Transform \\ 
 GR & General Relativity \\
 GSM & Gaussian Streaming Model  \\ 
  IR & Infra-Red  \\ 
 JBD & Jordan-Brans-Dicke \\
 KGE & Klein-Gordon Equation \\
  LCDM & $\Lambda$-CDM model of Cosmology  \\ 
 LOS & Line of Sight  \\ 
 LPT & Lagrangian Perturbation Theory  \\
 LSM & Linear Streaming Model  \\ 
  LSS & Large Scale Structure  \\ 
 MCMC & Markov Chain Monte Carlo \\
 MG & Modified Gravity \\
 PT & Perturbation Theory \\ 
 RegPT & Regularised Perturbation Theory  \\
 RSD &  Redshift Space Distortions \\
 SPT & Standard Perturbation Theory  \\
 STT & Scalar Tensor Theory \\
 TNS & Taruya-Nishimichi-Saito  \\ 
 \end{tabular} 
 \label{abbrevtab}
 \end{table}
 \newpage
 

 
\chapter{Introduction}
\begin{chapquote}{J.D}
``The man stood there, on a small island of grass, at the foot of a great  Mountain.''
\end{chapquote}
I often envisage mankind's attempt at understanding the natural world as a man climbing the highest of mountains in order to survey the surroundings. We come from that which we wish to observe - nature or the base of the allegorical mountain. At some point in our past, we became conscious of our environment, and at each stage since we've deepened our understanding of it - a height scaled and a distance further seen upon and around the mountain. The peak in this metaphor would be the summit of knowledge, and all that can be seen from that peak, absolute understanding of the observable universe. 
\newline
\newline 
It was the ancient Greeks who arguably first took breath and deeply surveyed the natural world and our place in it, and importantly birthed disciplines devoted to extracting fundamental truths. 2000 years before the Copernican revolution, the Pythagoreans proposed the notion of a non-geocentric configuration for the universe \cite{aristotle1} with a fire lying at the centre. This fire was even identified with the Sun by Aristarchus of Samos (c.310-230 B.C), the father of heliocentrism according to Copernicus himself. Record of this is by Archimedes \cite{archquote}
\newline
\newline
\emph{`` ...  the universe is many times greater than the ``universe" \footnote{This refers to the Earth itself.} just mentioned. His hypotheses are that the fixed stars and the sun remain unmoved, that the earth revolves about the sun on the circumference of a circle, the sun lying in the middle of the orbit ..." }
\newline
\newline
Jumping forward two millennia, the Polish astronomer Nicolaus Copernicus pulled heliocentrism back from the obscure annals of history, where it lay as pure proposition, and pushed it into the realm of a maturing scientific sphere using his own observations of celestial bodies and geometry. This theory came into strong contention with the long standing geocentric model. One important prediction of a heliocentric configuration is that of the apparent relative motion of far away stars, an effect of Earth's own motion around the Sun causing a shift in perceived positions \footnote{This is known as stellar parallax.}. Tycho Brahe, a Danish contemporary of Copernicus whose astronomical observations set unprecedented accuracy on celestial and stellar positions and motions, revealed no such stellar parallax. The constraints from this non-detection on stellar distances were almost 3 orders of magnitude larger than that postulated by Copernicans. It was only in 1838 (some 300 years later!),  that this effect was observed by Friedrich Bessel. The Heliocentric theory was proved 100 years earlier from the effect of stellar aberrations by James Bradley, but this example of non-observation until technology and theory is at a capacity to facilitate will serve as a strong basis for this work's motivation. Sometimes old proposed paths must wait for technological accuracy to bring our tread to them, while current ones are proven obsolete. 
\newline
\newline
But is not just technological power that steers our way within modern science. We must always keep in mind that a theoretical paradigm sets the interpretation of observations. Indeed, the famous perihelion shift of Mercury, observed 20 years after Bessel's observation of stellar parallax, was first interpreted in terms of Newtonian gravity by the French mathematician Urbain Le Verrier through the postulation of a new planet called Vulcan. Backtracking is hard work and with a claimed discovery of Vulcan not a year later by amateur astronomer Edmon Lescarbault, the two received the highest French honorific award, the L\'{e}gion d'honneur for the discovery. Predictions for Vulcan's transits repeatedly failed, each time theoretical parameters being retuned. Numerous claimed sightings were also reported by both amateurs and professional astronomers over the next 2 decades and despite many failed sightings  Verrier died in 1877 believing he'd discovered a new planet. Modern day astronomical accuracy would have shown no such planet existed (at least none larger than 5.7km in diameter \cite{Steffl:2013qk}!). It was 50 years later that Einstein proposed corrections to Mercury's orbit using his theory of general relativity (GR) \cite{Einstein:1916vd} giving the currently accepted paradigm of the gravitational interaction. 
\newline
\newline
This moves us to the 21st century and the age of extremely high precision astronomical observations. Under Moore's law, technology has reached an almost runaway evolution and it has fallen back into the hands of theorists to devise ways of solving modern physical problems. The biggest problem in physics today comes from an observation made almost 20 years ago, in 1998. Measurements of distant stellar explosions indicated that the universe's rate of expansion was smaller in the distant past than it is today - a so called accelerated expansion. Using the currently accepted theory of gravity, as Verrier did 150 years ago, we require an unobservable energy component making up $70\%$ of the entire observable universe's energy density - a problem not just slightly larger than an unseen Vulcan. This issue has led a flurry of research into alternative theories of gravity, revising old proposed paths as Copernicus had, and proposing new ones as Einstein had. These competitors all hope to move upon the surgeon's table of precision observations.  
\newline
\newline
But once again, back-tracking is hard and the currently accepted theory dominates the data-comparisons literature. For current astronomical surveys seeking to test the gravitational interaction at the largest scales, the level of observational precision has been insufficient to warrant significant improvement in theoretical accuracy. This is not true for the next generation of surveys \cite{Weinberg:2012es} such as the European Space Agency's EUCLID\footnote{\url{www.euclid-ec.org}} \cite{Laureijs:2011gra} mission, WFIRST\footnote{\url{https://wfirst.gsfc.nasa.gov/}} \cite{Spergel:2013tha}, the Dark Energy Spectroscopic Instrument (DESI)\footnote{\url{http://desi.lbl.gov/}} \cite{Levi:2013gra}, the Large Synoptic Survey Telescope (LSST)\footnote{\url{https://www.lsst.org/}} (see \cite{Chang:2013xja} for example) or the Square Kilometre Array (SKA)\footnote{\url{https://skatelescope.org/}}. These electromagnetic surveys will observe extremely large volumes of spacetime with unparalleled precision giving them the power to place exquisite consistency tests on GR, the Big Bang paradigm, cosmic acceleration and our current understanding of nature. To boot, the neonatal observational field of gravitational waves, birthed by the  `Advanced Laser Interferometer Gravitational-Wave ObservatoryÕ (aLIGO) \cite{Abbott:2016blz,TheLIGOScientific:2016pea} offers a completely new dimension to test these long held ideas at smaller, astrophysical scales. 
\newline
\newline
This thesis is concerned with testing gravity at the largest observable scales in the universe through measurements of structure formation at late cosmological times. We present a framework in which one can produce theoretically consistent predictions for structure growth for a large class of competing gravitational theories to GR. The {\tt c++} code {\tt MG-Copter}, developed by the author and based on Martin White's Copter code \cite{Copter}, hosts this framework and has been used to produce all original results \footnote{Not counting simulation measurements.} in the works described in the dissemination on pg.1 and herein. The motivation of this consistent modelling is to move towards unbiased constraints on gravity at the dawn of the so called stage-IV electromagnetic surveys described above. 
\newline
\newline
This thesis is organised as follows. The rest of this chapter is dedicated to introducing the currently accepted model of the cosmos. This involves the expanding universe picture along with Einstein's gravity. In brief we discuss key observational evidence of this paradigm as well as that of acceleration. Finally we look at standard modelling of cosmological matter and energy content. In {\bf Chapter 2} we describe the modelling of density and velocity perturbations upon the homogenous and isotropic expanding background, introduced in Chapter 1, and how they evolve under expansion and the gravitational interaction to form the large scale structure (LSS) observable today. We look at common approximations that can be made in explaining structure formation and introduce the power spectrum and correlation function - two key statistics in survey measurements. Finally, we look at the successes, shortcomings and possible improvements of perturbation theory. {\bf Chapter 3} is concerned with alternatives to GR, how they can offer different explanations for the observation of cosmic acceleration and how they can compete with 100 years of successful tests of GR. We describe a general class of gravitational theories that are able to do this and accommodate them into the modelling of the correlation function and power spectrum described in Chapter 2. {\bf Chapter 4} discusses the so called redshift space distortion (RSD) phenomenon that is a necessary part of LSS modelling and provides a direct probe of the growth of structure, and consequently of gravity. {\bf Chapters 5} provides comparisons of theoretical predictions for the LSS statistical observable, the redshift space power spectrum, generated by {\tt MG-Copter} for different gravity models against fully non-linear, non-perturbative treatments in the form of N-body simulations. {\bf Chapter 6} moves in the same vein as Chapter 5 but is concerned with Fourier dual of the power spectrum, the correlation function. Here we compare different models for the correlation function within 3 models of gravity, quantifying their consistency and individual merits. {\bf Chapter 7} provides a test validation of gravitational modelling consistency in the context of stage-IV surveys. We compare different theoretical templates for the dark matter redshift space power spectrum against a suite of N-body simulations and provide a measure of the bias introduced when inconsistently modelling this statistic. Finally, {\bf Chapter 8} provides a summary of the results presented, ongoing work and relevant improvements to the presented framework before it is applied to real survey data. The reader should note that the entirety of chapters 5, 6 and 7 are original work, each taken from the papers listed in the dissemination on pg.1 respectively. Further original work has been made in earlier chapters and we will clearly indicate where. For the reader's convenience the list of equations which have been derived as part of this work is; Eq.\ref{hornmu}-Eq.\ref{horng3}, Eq.\ref{frg2}, Eq.\ref{frg3}, Eq.\ref{g2dgp}, Eq.\ref{g3dgp}, Eq.\ref{bispectrum}, Eq.\ref{tnsfterm}, Eq.\ref{cterm}, Eq.\ref{gsmt1}, Eq.\ref{gsmt2}, Eq.\ref{sigmaiso} and Eq.\ref{gsmt3} - Eq.\ref{gsmt4}. From Eq.\ref{bispectrum} onwards, these are generalised forms of equations already derived in the literature for GR. 
\newline
\newline
We begin 100 years ago, when Albert Einstein abolished Vulcan and made a giant leap up the mountainside $\dots$ 

\section{General Relativity} 
The natural beginning for a thesis on testing gravity should begin with the currently accepted model of gravity - Einstein's GR. This theory is described by the Einstein-Hilbert action 
\begin{equation}
S_{\rm EH} = \int d^4x \frac{\sqrt{-g}}{2\kappa}(R-2\Lambda)+S_m[g_{\mu \nu};\Psi_M],
\label{ehaction}
\end{equation}
where $\kappa = 8\pi G$,  $R = g^{\mu \nu}R_{\mu \nu}$ is the Ricci scalar, $R_{\mu \nu}$ is the Ricci tensor, $g_{\mu \nu}$ is the metric tensor and $g$ is its determinant. $S_m$ is the action governing the matter distribution and $\Psi_M$ represents the matter fields. We have also included the constant $\Lambda$ which, as we will see, has a special significance for cosmology, discussed in the next section. By applying the principle of least action and varying the action with respect to the metric tensor we obtain the Einstein field equations \footnote{We impose the variation of the metric at the boundary is zero.}
\begin{equation}
G_{\mu\nu} = \kappa T_{\mu \nu} - \Lambda g_{\mu \nu},
\label{EFES}
\end{equation}
where $G_{\mu \nu} = R_{\mu \nu} - Rg_{\mu \nu}/2$ is the Einstein tensor and $T_{\mu \nu}$ is the energy-momentum tensor defined as 
\begin{equation}
\frac{\delta S_m}{\delta g_{\mu \nu}} = -\frac{1}{2}\sqrt{-g} T^{\mu \nu}.
\end{equation}
Energy and momentum are conserved within this theory, $\nabla_\mu T^{\mu \nu}=0$, which is a direct consequence of the Bianchi identity $\nabla_\mu G^{\mu \nu} = 0$ \footnote{$\nabla_\mu$ is the covariant derivative compatible with $g_{\mu \nu}$.}.
\newline
\newline
We can test this description of gravity through its predictions for particle motions. The motion of particles in GR are dictated by the \emph{geodesic equation} 
\begin{equation}
u^\mu \nabla_\mu u^\nu = 0,
\label{geoeqn}
\end{equation}
where $u_\mu = dx_\mu/d\lambda$ is the 4-velocity of the particle, $\lambda$ being an affine parameter that parametrises the particle trajectory in terms of the spacetime coordinates, $x^\mu(\lambda)$. This equation can be derived by applying the principle of least action to the action for a point particle of mass $m$ 
\begin{equation}
S_p = -\int m \sqrt{-g_{\mu \nu} u^\mu u^\nu} d\lambda.
\end{equation}  
This tells us that particles follow paths that minimise the spacetime interval - a generalisation of straight lines for general backgrounds. It is this equation that provides correct predictions for the perihelion shift of Mercury's orbit discussed in the introduction. 
\newline
\newline
Any viable succeeding theory should be able to reproduce previous results, in this case GR should be able to reproduce Newtonian gravity in the appropriate limit. This limit is the non-relativistic, low curvature regime. If we consider a non-relativistic particle ($dx^i/dx^0 \ll 1$) in a background close to flat, that can be described by perturbing the Minkowski metric in the \emph{Newtonian gauge}
\begin{equation}
ds^2 = -(1+2\Phi(x,y,z)) dt^2 + (1-2\Psi(x,y,z))(dx^2 + dy^2 + dz^2),
\label{perturbedmink}
\end{equation}
we should arrive at Newton's description of particle motion. The geodesic equation reads 
\begin{equation}
 \frac{d^2 \bfx}{dt^2}= -\nabla \Phi,
\end{equation}
where $\bfx = (x,y,z)$. This is identical to Newton's second law with the time coordinate perturbation being identified with the Newtonian potential. Further if we consider a non-relativistic density distribution $T^{\mu \nu} = diag(\rho,p,p,p)$, $\rho$ and $p$ being the density and pressure respectively with $p/\rho \ll c^2$, in Eq.\ref{EFES}, we arrive at the following relation
\begin{equation} 
\nabla^2 \Phi = \frac{\kappa}{2} \rho  - \Lambda,
\label{poissoneqn1}
\end{equation}
which in the limit $\Lambda \rightarrow 0 $ is simply the Newtonian Poisson equation. We note from Eq.\ref{poissoneqn1} that a positive cosmological constant works in the opposite direction of matter in that it reduces the gravitational force. This will become important in the next section but let's investigate its importance within the local universe. Consider ourselves to be in vacuum ($\rho=0$) outside a spherically symmetric mass distribution 
\begin{align}
\nabla^2 \Phi(r) &=  \frac{1}{r^2}\frac{d}{dr}\left(r^2\frac{d\Phi}{dr}\right), \nonumber \\
 \Rightarrow \frac{d\Phi}{dr}  &= g_N - \frac{\Lambda}{3}r,
 \label{newtonianc}
\end{align}
where we have swapped to spherical coordinates and 
\begin{equation}
g_N = \frac{GM}{r^2},
\label{newtoniang}
\end{equation} 
is the Newtonian acceleration due to gravity with 
\begin{equation}
M = 4\pi \int r^2 \rho(r) dr.
\end{equation}
We know that Newtonian gravitation is a very good first approximation within the solar system and so the value of the cosmological constant must be tiny and further by Eq.\ref{newtonianc} it will only be relevant at very large distances. If we consider the solar system, which has a radius of $r_s\sim 4.5 \times 10^{12}$ m, we have 
\begin{equation}
\frac{\Lambda}{G} \ll \frac{3M_\odot}{r_s^3} = 1.38 \times 10^{-6} \mbox{kg}/\mbox{m}^3.
\end{equation}
If we do a similar calculation for conservative values of the Milky Way we get $\Lambda/G \ll \mathcal{O}(10^{-22}) \mbox{kg}/\mbox{m}^3$. The latest cosmic microwave background (CMB) measurements \cite{Planck:2015xua} put  $\Lambda/G = (1.501\pm 0.042) \times 10^{-25} \mbox{kg}/\mbox{m}^3$. This tells us that a cosmological constant would play a role on scales larger than our own galaxy, so called cosmological scales.
\newline
\newline
Note that when considering motions within the solar system and further, the motions of galaxies, the Newtonian approximation becomes valid. This can be understood in terms of velocities which satisfy $v \ll c$. This implies motion within a very weak gravitational field. Consequently, the metric perturbation $\Phi_N$, quantifying deviations from flat Minkowski spacetime, is very small. For example, typical velocities in the Virgo cluster are observed to be $v < 3000$ km/s \cite{Karachentsev:2010nw}. Applying the virial theorem in the Newtonian limit
\begin{equation}
v^2 = \Phi_N,
\label{virialtheorem}
\end{equation}
 to these objects yields $\Phi_N \sim 10^{-4}$ \footnote{Recall we are working in a system of units where $c=1$.}. This implies the Newtonian approximation should work well for this galaxy cluster, with relativistic corrections beyond our current abilities to probe. Further, the presence of more dominant observational systematics set limits on measurement accuracy. The average separation of galaxies in the Virgo cluster is around $17$Mpc. At separations larger than this we can expect even lower values for $\Phi_N$. The LSS is that of super clusters and hosts of clusters, so the intra separation of galaxies within a cluster can serve as a lower bound for cosmological scales. 



\section{Standard Model of Cosmology}
Moving beyond our galaxy, we reach the domain of cosmology. This is the sphere of physics concerned with modelling the universe - its evolution and \emph{large scale} constituents. It is on these scales that isotropy is observed, both in the very early universe through CMB measurements \cite{Planck:2015xua,Hinshaw:2012aka,Mather:1993ij} and in the late universe through the observed galaxy distribution \cite{Marinoni:2012ba,Abazajian:2008wr}. Further, upon applying the assumption that isotropy holds at every point in the universe, usually called the \emph{Copernican principle}, \footnote{In absence of hubris, this is a fair assumption!} we arrive at a universe that is both spatially homogeneous and isotropic. This is one of the pillars of modern cosmology and is aptly called the \emph{cosmological principle}. Adopting this principle and assuming GR on all scales we end up with the \emph{standard model of cosmology}. In summary, the two core assumptions of the standard model are 
\begin{enumerate} 
\item 
{\bf Copernican principle}: The standard model implicitly assumes Earth occupies no special place in the universe. Of course to directly test this one would need to make observations at other positions in the universe, although consistency tests can be done by testing the models that assume it. Novel tests of homogeneity have shown that this assumption is consistent on scales $r \geq 100$ Mpc \cite{Ntelis:2017nrj,Laurent:2016eqo,Sarkar:2009iga}. On the other hand, it is worth noting that the observed accelerated expansion of the universe can be explained by placing Earth in a privileged position, such as at the centre of a large under-density region \cite{February:2009pv}.
\item
{\bf GR holds on all scales}: The standard model also assumes that GR holds on all scales. As mentioned in the previous section GR has been exquisitely tested within the solar system, a scale of $\sim 10^{-4}$pc, whereas the Laniakea supercluster, a galaxy cluster of which the Milky Way is part of, is of the order $\sim10^8$pc. By using GR on cosmological scales we are making more than a 12 order of magnitude extrapolation in scale! Essentially this work is concerned with testing this assumption. To do this we must break it, but we will assume the description of gravity to be geometrical, i.e. that spacetime is described by some metric tensor. The reader should realise that non-geometric descriptions of gravity have been and are still being developed \cite{Milgrom:2014usa,Verlinde:2010hp}, but the geometric description does both very well in explaining a host of phenomena as well as being very theoretically rich. 
\end{enumerate}
We now move on to develop all this mathematically.  
\subsection{The Metric, Hubble Parameter, Redshift and Distance}
\begin{figure}[H]
  \captionsetup[subfigure]{labelformat=empty}
  \centering
  \subfloat[]{\includegraphics[width=15.5cm, height=7.2cm]{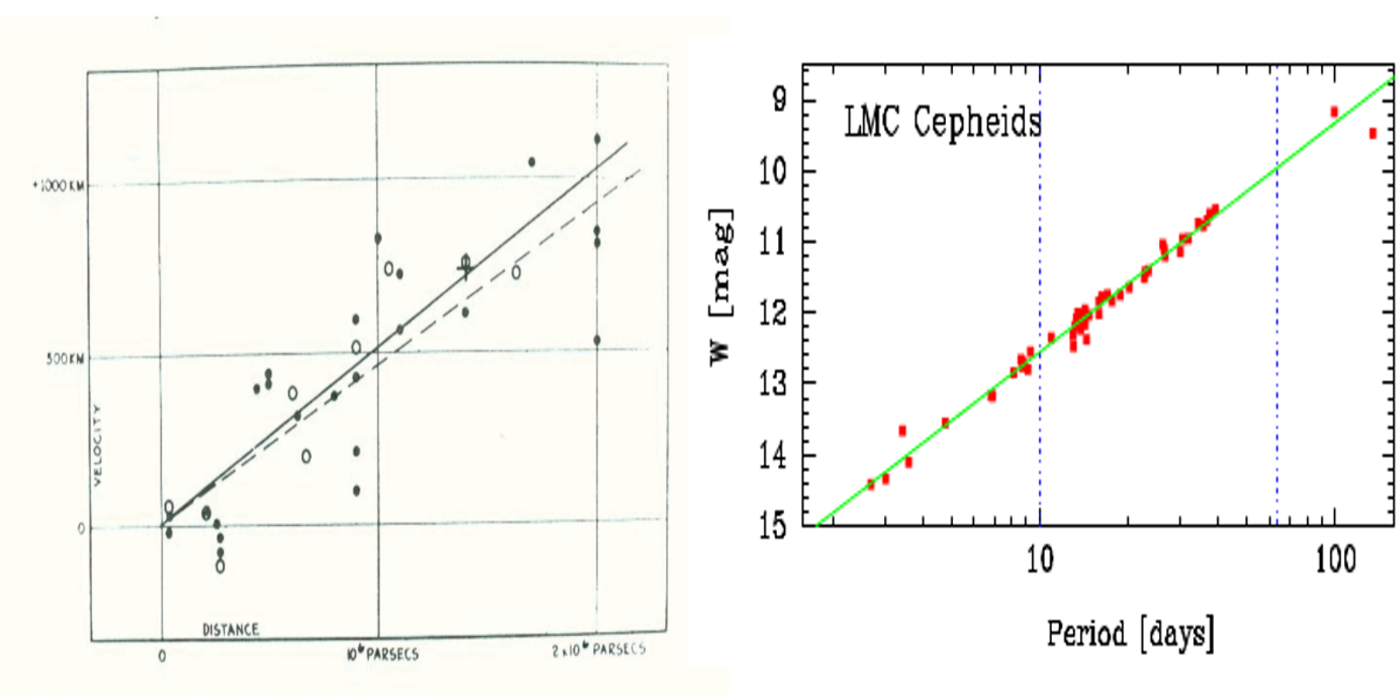}} \quad
  \caption{Hubble's original plot (left) of the recessional velocities (redshift of spectra measurements) of 24 nebulae and their physical distances inferred from Leavitt's cepheid period-luminosity relation, a recent version of which is shown on the right \cite{Efstathiou:2013via}.}
\label{hubble_ceph}
\end{figure}

Our starting point is the Friedman-Lema$\hat{\mbox{i}}$tre-Robertson-Walker (FLRW) metric - the most general spacetime that respects the cosmological principle. The FLRW line element can be written as  
\begin{equation}
ds^2 = -dt^2 + a^2(t)\left[\frac{dx^2}{1-kx^2}+ x^2d\Omega^2\right],
\label{FLRWmetric}
\end{equation}
where $d\Omega$ is the line element on the unit 2-sphere and $x$ is the \emph{comoving} radial coordinate. $k \in \{-1,0,1\}$ represents the spatial curvature with values representing a spatially open (hyperbolic), flat (Euclidean) and closed (spherical) universe respectively. We will take $k=0$ for the rest of this work which is consistent with the latest CMB measurements \cite{Planck:2015xua}. 
\newline
\newline
Lastly, $a(t)$ is the dimensionless scale factor that gives the size of the spatial slices within the spacetime as well as whether they are expanding or contracting. We can change to \emph{physical} coordinates by the scaling $\bfr = a(t)\bfx$. Throughout this work we will normalise $a(t)$ such that $a(t=0) \equiv a_0 = 1$ where $t=0$ is the present time. In general, a $0$ subscript will indicate a present day value. From this quantity we can define the \emph{Hubble parameter}  
\begin{equation}
H(t) \equiv \frac{\dot{a}}{a},
\label{hubbledef}
\end{equation}
where the over-dot means a derivative with respect to the metric coordinate $t$. Its present day value, $H_0$, is usually called \emph{Hubble's constant}. This is usually parametrised in terms of the dimensionless quantity $h=H_0/(100 \mbox{km/s/Mpc})$. The first measure of this constant was performed in 1929, credited to Edwin Hubble after whom the constant is named. By using redshift measurements by Vesto Slipher and Milton Humason as well as distances from the period-luminosity relationship of cepheids developed by Henrietta Leavitt (a recent relation shown on right of Fig.\ref{hubble_ceph}), Hubble used 24 nebulae to get the value of $\sim 500$ km/s/Mpc (see left of Fig.\ref{hubble_ceph}). Assuming the universe is described by GR, current measurements of the CMB put the value at $H_0=67.74 \pm 0.46$km/s/Mpc  \cite{Planck:2015xua} while recent measurements using gravitational lensing put this value at $H_0=71.9^{+2.4}_{-3.0}$km/s/Mpc \cite{Bonvin:2016crt} which is consistent with other late universe measurements using supernovae and cepheids \cite{Zhang:2017aqn,Riess:2016jrr,Efstathiou:2013via}. This tension is the subject of ongoing investigation and may be relieved with better understanding of observational systematics. It may also be relieved through modifications to LCDM.
\newline
\newline
The redshift measured by Slipher and Humason comes from the loss of energy that photons experience as they travel through an expanding universe. Say we have a galaxy at a comoving distance of $d$ that emits an electromagnetic signal at $\tau_1$ over \emph{conformal time} interval $\delta \tau$, related to the metric time interval as $\delta t = a(\tau) \delta \tau$. Say further we receive this signal at $\tau_0$. The value of $\delta \tau$ at observation is identical to that at emission but the observed metric time interval is not. We have 
\begin{align}
\delta t_1 & = a(\tau_1) \delta \tau, \nonumber \\
\delta t_0 & = a(\tau_0) \delta \tau.
\end{align}
We can take $\delta t$ to be the period of the light wave, in which case we get the following relationship 
\begin{equation}
\frac{\lambda_0}{\lambda_1} = \frac{a(\tau_0)}{a(\tau_1)},
\end{equation}
$\lambda_0$ being the wavelength of light observed at present and $\lambda_1$ being the wavelength of light emitted by the source. Using our normalisation, $a(\tau_0)=1$, we arrive at the following relationship 
\begin{equation}
a(t) = \frac{1}{1+z},
\label{azrelation}
\end{equation}
where $z = (\lambda_0-\lambda_1)/\lambda_1$ is the cosmological redshift. If we now Taylor expand the scale factor about the present time $t_0$ up to first order, making this only valid for local sources, we get 
\begin{equation}
a(t_1) = 1 + H_0(t_1-t_0),
\end{equation}
where we have used $a(t_0)=1$ and Eq.\ref{hubbledef}. Light follows null geodesics with $ds^2=0$ giving $dt^2 = a(t)^2dx^2$. The physical distance between source and observer is then $d = a(t)\delta x = (t_0-t_1)$. Using this we arrive at Hubble's relation 
\begin{equation} 
z \approx H_0d,
\end{equation}
where we have used the binomial expansion of $(1+z)^{-1}$ up to first order in $z$ (valid for $z\ll1$). Indeed, Hubble made his first measurement of $H_0$ using nebulae at redshifts $z\leq 0.03$. The distances were measured using the relation on the right hand side of Fig.\ref{hubble_ceph}\footnote{ This is the 2nd rung of the so called `distance' ladder if you omit main sequence star fitting which is only valid in the Milky Way.}, which is calibrated using nearby cepheids whose distance can be measured through direct means such as parallax (1st rung of the distance ladder).
 \begin{figure}[H]
  \captionsetup[subfigure]{labelformat=empty}
  \centering
  \subfloat[]{\includegraphics[width=15.5cm, height=7.2cm]{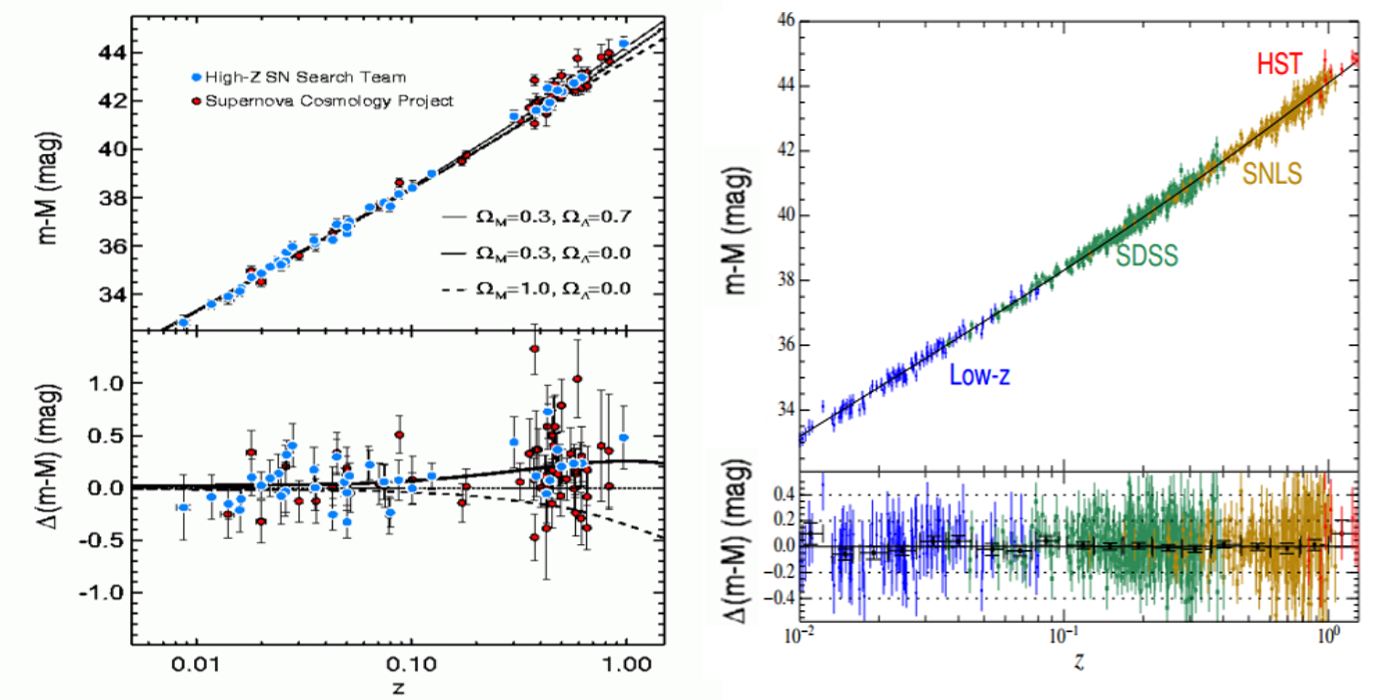}} \quad
  \caption{The highest rung on the distance ladder - type 1a supernovae. On the left there is the original plot of Riess et al \cite{Riess:1998cb} which showed the need for a non-zero cosmological constant in 1998. On the right there is a more recent plot with new measurements from various surveys \cite{Conley:2011ku}.}
\label{sn1afig}
\end{figure}
\noindent The highest employed rung on the distance ladder are supernovae type 1a. It is observed that nearby supernovae, of distance measured using the lower rungs on the ladder, have the same \emph{absolute magnitude} (M) \footnote{Or can be standardised through the Phillips relationship \cite{Phillips:1993ng} which relates the peak luminosity with the decay of the luminosity from its peak value.}, which is a measure of the intrinsic brightness of a celestial object. We can infer this by having a measure of the distance to these nearby supernovae (see Eq.\ref{magnitude}). The quality of standard brightness can be used as a new distance-redshift relation reaching far further than the cepheid method because of the large brightness of these transients \footnote{We note one must assume the measured absolute magnitude holds at higher redshift. This was shown not to be exactly true for the cepheid relation and amendments were made to Leavitt's original relation to account for this (see for example \cite{Kennicutt:1997dm}).}.
Specifically we can use the relation
\begin{equation}
d = 10^{\frac{m-M -10}{5}},
\label{magnitude}
\end{equation}
where $m$ is the \emph{apparent magnitude}, a measure of the observed brightness. Fig.\ref{sn1afig} shows the distance-redshift plot of type 1a supernovae, the left from the Nobel prize winning work by the High-Z supernova search team \cite{Schmidt:1998ys} and the Supernova Cosmology Project \cite{Perlmutter:1998np} released in 1998 \cite{Riess:1998cb} and on the right a plot showing data from recent surveys such as the Sloan Digital Sky Survey (SDSS) \footnote{\url{www.sdss.org}} and the Hubble Space Telescope (HST) \footnote{\url{https://asd.gsfc.nasa.gov/archive/hubble/}} \cite{Conley:2011ku}. The lines show the prediction of the FLRW metric in the context of the standard model of cosmology using the best fit parameters which we discuss next. These measurements constitute the key evidence for a non-zero cosmological constant. 


\subsection{Matter Content and Dynamics}
\begin{figure}[H]
  \captionsetup[subfigure]{labelformat=empty}
  \centering
  \subfloat[]{\includegraphics[width=15.5cm, height=6.6cm]{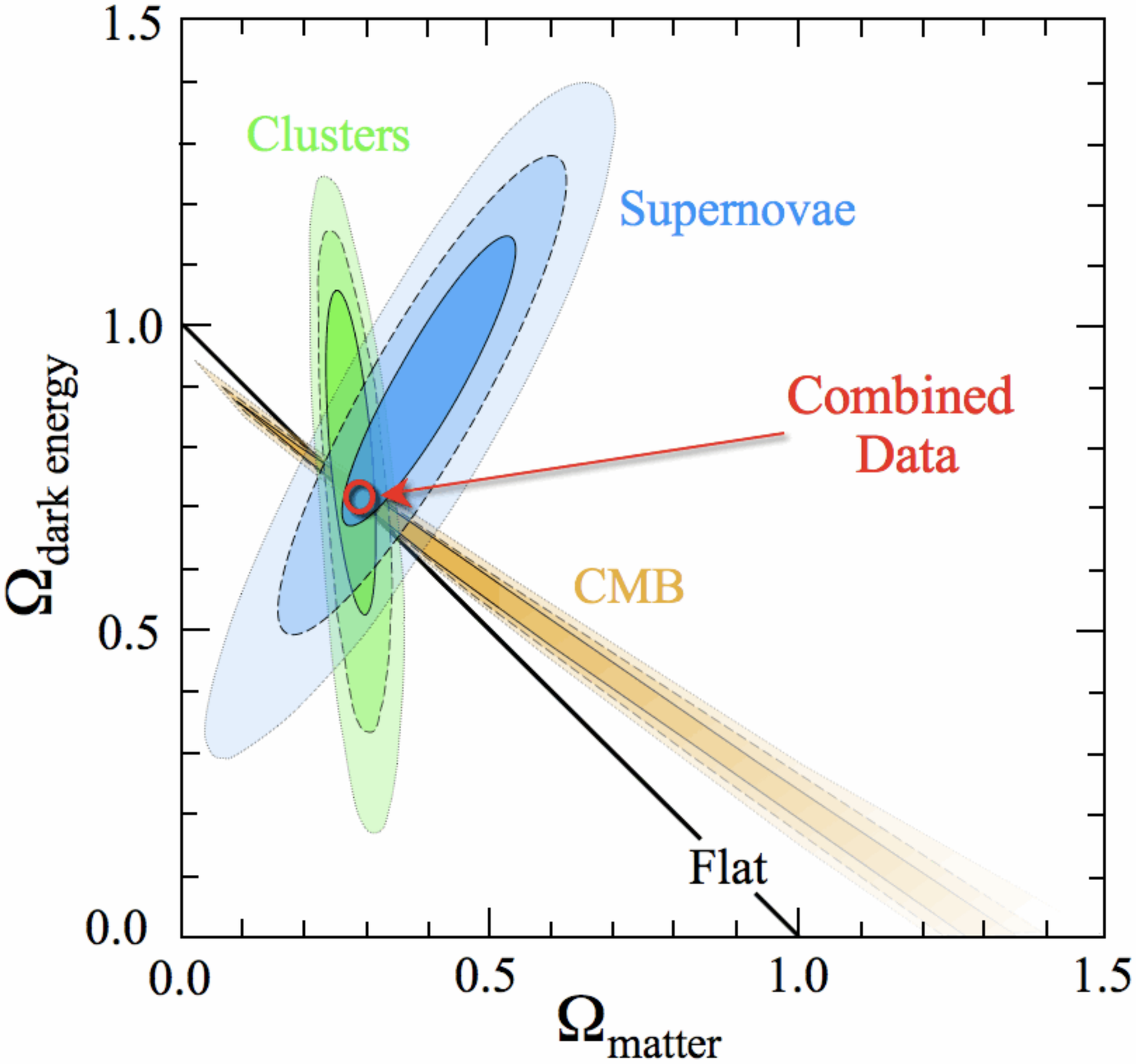}} 
  \caption{The 2D marginalised constraints of $\Omega_{\Lambda,0}$ and $\Omega_{m,0}$ from a variety of cosmological probes (BAO, galaxy clustering, supernovae and the CMB) with the theoretical prediction of a flat LCDM universe shown by a black line. These constraints were compiled in \cite{Amanullah:2010vv} and the image was retrieved from \cite{combconst}.}
\label{combconst}
\end{figure}

Having discussed the metric and how we can use it to obtain predictions for small distances and redshifts, we now turn to a fuller description by discussing the matter content within the standard model. Under the assumptions of homogeneity and isotropy, matter within the universe is forced to behave as a perfect fluid with energy-momentum tensor 
\begin{equation}
T_{\mu \nu} = (\rho + p)u_\mu u_\nu + pg_{\mu \nu},
\label{perfectfluid}
\end{equation} 
where $u_\mu$ is the 4-velocity of the fluid, $\rho$ is the energy density and $p$ is the pressure of the fluid. We work in the fluid's rest frame where we have $u_\mu = (-1,0,0,0)$. Plugging in Eq.\ref{FLRWmetric} and Eq.\ref{perfectfluid} into Eq.\ref{EFES} we arrive at the dynamical equations for the scale factor and matter content, known as the \emph{Friedmann equations}
\begin{align}
\frac{H^2}{H_0^2} &= \frac{\kappa \rho}{3H_0^2} - \frac{\Lambda}{3H_0^2} - \frac{k^2}{H_0^2a^2}\label{friedmann1},\\
\frac{\dot{H}}{H_0^2} + \frac{H^2}{H_0^2} &= -\frac{\kappa}{6H_0^2}\left(\rho + 3p\right) + \frac{\Lambda}{3H_0^2},
\label{friedmann2}
\end{align}
where the over-dot again refers to a derivative with respect to metric time $t$ and we have temporarily re-introduced curvature. Historically $\Lambda$ was first proposed by Einstein who envisaged a closed ($k=1$) static universe with $\dot{a}=0$. This was compatible with Mach's principle, which states that the global matter distribution affects local inertial frames. This was also compatible with the astronomical observations at the time, prior to Edwin Hubble's paper. Upon observation of Eq.\ref{friedmann1} we see $\dot{a}=0$ does not necessarily require $|\Lambda|>0$, but Eq.\ref{friedmann2} tells us without this requirement $\ddot{a}$ will never be zero for normal energy densities and pressures. The need for a static universe was removed upon the discovery of the large scale recession of celestial objects. Some eighty years later, $\Lambda$ was reinstated to enhance the dynamics of the spatial section of the metric rather than suppress it. Again, we will set $k=0$ for the rest of this section. 
\newline
\newline 
The density's dynamics are given by the conservation equation $\nabla_\mu T^{\mu \nu} = 0$, that reduces to
\begin{equation}
\dot{\rho} + 3H(\rho +p) = 0,
\label{conteqn1}
\end{equation}
which holds for each energy component (e.g. baryons or radiation) independently if they do not interact\footnote{True for GR.}. For a complete description we also need an equation of state (EOS) for the energy components
\begin{equation}
 p=\omega \rho.
 \end{equation} 
 $\omega$ is used to parametrise the relation and its value for common energy types is given in Table.\ref{energytable}. 
\begin{table}[b]
\centering
\caption[Common EOS Parameter Values]{Common EOS Parameter Values}
 \begin{tabular}{| c | c | }
 \hline
  \textbf{Energy Species} & \textbf{$\omega$}  \\ \hline
 Relativistic fluid & 1/3 \\ \hline
 Non-relativistic fluid & 0 \\ \hline
 Cosmological Constant & -1 \\ \hline
  \end{tabular}
  \label{energytable}
 \end{table}
Solving Eq.\ref{conteqn1} for constant $\omega$ yields the following solution for $\rho$
\begin{equation}
\rho = \rho_0a^{-3(1+\omega)},
\label{rhoeqn}
\end{equation}
where $\rho_0$ is the present day density. For completeness, if we let $\omega$ vary with time, Eq.\ref{conteqn1} becomes 
\begin{equation}
\frac{d\rho}{da} = -\frac{3(1+\omega(a))}{a}.
\end{equation} 
Performing the integral we get the following equation for the density 
\begin{equation} 
\rho(a) = \rho_0 \exp{\left[3\int^1_a [1+\omega(a')]d\ln{a'}\right]}.
\label{vareos}
\end{equation}
If we consider a matter dominated universe with $\Lambda=0$ in Eq.\ref{friedmann1} we find that 
\begin{equation}
\rho = \rho_c \equiv \frac{3H^2}{\kappa},
\end{equation}
which is called the \emph{critical density} for the fluid. This term comes from the fact that if we don't assume a Euclidean (flat) geometry (see Eq.\ref{FLRWmetric}), a value for the matter density just below critical leaves us with an open universe which expands eternally. On the other hand, a value just above critical would leave us with a closed universe which would eventually start a contraction phase. Eq.\ref{friedmann1} is usually cast in terms of the ratio of the density to the critical density, called the \emph{density parameter}
\begin{equation}
\Omega \equiv \frac{\rho}{\rho_c} = \frac{\kappa \rho}{3H^2}.
\label{omegam}
\end{equation}
We can define the critical density for the cosmological constant by considering Eq.\ref{EFES} together with Eq.\ref{perfectfluid} and then defining the total energy momentum tensor 
\begin{equation}
\tilde{T}_{\mu \nu} = (\tilde{\rho} + \tilde{p})u_\mu u_\nu - \tilde{p}g_{\mu \nu},
\end{equation}
where 
\begin{align}
\tilde{\rho} &= \rho + \frac{\Lambda}{\kappa} = \rho + \rho_\Lambda,  \\
\tilde{p} & = p - \frac{\Lambda}{\kappa},
\end{align}
where we have defined $\rho_\Lambda$. This leads to the following definition 
\begin{equation}
\Omega_\Lambda \equiv \frac{\rho_\Lambda}{\rho_c} = \frac{\Lambda}{3H^2}.
\label{omegal}
\end{equation}
Substituting Eq.\ref{omegam} and Eq.\ref{omegal} into Eq.\ref{friedmann1} gives
\begin{equation}
1 = \Omega_m + \Omega_r +  \Omega_\Lambda,
\label{friedmann1b}
\end{equation}
where we have included a radiation component $\Omega_r$. Lastly, using Eq.\ref{rhoeqn}, we can write the density parameters in terms of their current day values which we will often choose to do 
\begin{align}
\Omega_m &= \frac{\kappa \rho_{m,0}}{3H^2}a^{-3} = \frac{H_0^2}{H^2}\Omega_{m,0} a^{-3}, \label{omega0} \\
\Omega_r &= \frac{\kappa \rho_{r,0}}{3H^2}a^{-4} = \frac{H_0^2}{H^2}\Omega_{r,0} a^{-4},  \\
\Omega_\Lambda &= \frac{\Lambda}{3H^2} = \frac{H_0^2}{H^2}\Omega_{\Lambda,0}. 
\end{align}
Substituting these expressions in Eq.\ref{friedmann1b} gives the following form of Eq.\ref{friedmann1}
\begin{equation}
\frac{H^2}{H_0^2} = \Omega_{m,0}a^{-3} +  \Omega_{r,0}a^{-4} + \Omega_{\Lambda,0}.
\label{friedmann1c}
\end{equation}
A number of independent observations strongly suggest the universe is described by a flat FLRW metric with energy components of matter, radiation and a cosmological constant. Fig.\ref{combconst} shows the constraints from different cosmological probes leading to the best fit $\Lambda$-CDM (LCDM) model, where CDM stands for cold dark matter. CDM is an indirectly observed matter component of the universe fives times more abundant than the observable matter component (baryonic matter). Evidence for this component comes in many forms such as galaxy rotation curves, the bullet cluster, and the CMB temperature correlations (see \cite{Roos:2012cc} for a nice review). We note that a non-zero curvature component is compatible with many of the observations in isolation, but together are consistent with a flat universe.  
\newline
\newline
The $\Lambda$ component, or vacuum energy component (as it remains constant as space expands) was found to be non-zero through supernovae measurements published in 1998 (see Fig.\ref{sn1afig}). Its presence means we live in a universe whose spatial expansion is actually accelerating ($\ddot{a}>0$). This led to the 2011 Nobel prize and has become one of the biggest problems in physics, primarily in relation to the {\it `old cosmological constant problem'} \cite{Weinberg:1988cp,Martin:2012bt}. This is an issue of fine tuning, with a tiny observed value (cosmology) and a very large predicted vacuum energy (quantum theory) for $\rho_\Lambda$. In order to reconcile them, a tuning of the bare vacuum energy value would be needed to an accuracy of 120 decimal places. This would of course be very unstable against higher energy corrections. We do not focus on this problem but rather look for alternative explanations than a vacuum energy for the accelerated expansion. In Chapter 3 we discuss theories which do exactly this. The reader should keep in mind that in these cases, the non-observation of quantum vacuum energy still needs to be explained, but is not the scope of this thesis. 
\newline
\newline
A second problem comes from the relative value of $\Omega_{\Lambda,0}$ and $\Omega_{m,0}$. Since $a(t)$ is a monotonically increasing function, with $a(t)\ll 1$ at the time of recombination and arriving at 1 today, Eq.\ref{friedmann1c} tells us that different energy components came to dominate the universe at different epochs. At some early time radiation would have been dominant over both the cosmological constant term and matter. We can put bounds on the era when radiation and matter densities were equal at $z>3321$ \cite{Planck:2015xua}. Today radiation is completely diluted, being orders of magnitude less than matter or the $\Lambda$ term. On the other hand, $\Lambda$ dominates with matter being a mere factor of 2 less. If we consider that radiation domination happened billions of years ago, it seems a strange coincidence that human observations occur at a time when matter and the $\Lambda$ term are almost equal. This has been aptly named the \emph{coincidence problem} and can be remedied by considering interactions between these two energy sectors \cite{Simpson:2010vh,Lesgourgues:2015wza,Pourtsidou:2016ico,Baldi:2016zom,Buen-Abad:2017gxg}. These models are also able to resolve the tension in $H_0$ between CMB measurements and late time measurements briefly mentioned in the previous section. 
\newline
\newline
The value of the present day cosmological parameters which best fits the latest CMB-only measurement \cite{Planck:2015xua} are given in Table.\ref{planckparams}. The subscript $c$ indicates CDM while $b$ indicates baryons with $\Omega_m = \Omega_c+\Omega_b$. $\Omega_{\nu,0}$ is the present day neutrino density and radiation is two orders of magnitude less that the neutrino density upper bound. $\Omega_{tot}$ gives the constraint on the curvature within the universe with a value of $1$ being flat. By including a weak lensing measurement and a baryon acoustic oscillation (BAO) measurement, this value is $\Omega_{tot} =  1.000\pm0.005$ \cite{Planck:2015xua} which indicates no spatial curvature. Note that this constraint on curvature is derived by fitting GR theoretical templates to the CMB data. In general, for a consistent constraint of theories beyond GR one would need to refit to the data, but as the theories considered in this thesis all behave like GR at the epoch around recombination, as well as light paths being unaffected by these modifications, we can safely assume a zero spatial curvature for the rest of this work.
\newline
\newline
This concludes our discussion on the background metric and its dynamics. We will assume this background for the rest of this thesis unless stated otherwise. The next chapter will focus on tiny deviations from isotropy and homogeneity by looking at perturbations in the matter density and velocity fields. These will be used to construct important observables which will be the focus of later chapters. 
\begin{table}[b]
\centering
\caption[Present Cosmological Parameters]{Present Cosmological Parameters}
 \begin{tabular}{| c | c | }
 \hline
  \textbf{Cosmological Parameter} & \textbf{Value}  \\ \hline
 $H_0$ & 67.31 \\ \hline
$ \Omega_{\Lambda,0}$ & 0.685 \\ \hline
 $\Omega_{c,0} $& 0.264  \\ \hline
 $\Omega_{b,0} $& 0.049 \\ \hline
 $\Omega_{\nu,0}$ & $<0.008$ \\ \hline
 $\Omega_{tot} $& $1.052_{-0.055}^{+0.049}$ \\ \hline
 \end{tabular}
  \label{planckparams}
 \end{table}

 
\chapter{Perturbation Theory} 
\begin{chapquote}{J.D}
``Brick upon brick, stone upon stone, tiny perturbations that form the whole.''
\end{chapquote}
In the standard model of cosmology, the largest structures in the universe today began with primordial fluctuations that were enhanced through the gravitational interaction on a spatially expanding background geometry. These structures are primarily comprised of CDM which only interacts gravitationally and is treated as a non-relativistic perfect fluid. Further, far inside the Hubble horizon \footnote{This is the boundary at which objects recede from us at the speed of light due to metric expansion.} where recessional velocities are non-relativistic, we can use the Newtonian limit of the full geometric gravitational theory. We will assume this CDM description of the matter content of the universe in all gravitational theories considered in this work, although modifications to gravity have been also motivated to remove the CDM component of the standard model (for example \cite{Milgrom:2014usa,Verlinde:2010hp}). Our concern will be general modifications, groups of which have been shown to provide \emph{self-accelerating solutions}, meaning they can produce an accelerated expansion without the need for a dark energy component. A full discussion of these modifications and deviations from the standard model is given in the next section but we will begin to characterise our ignorance of the true theory of gravity in this section.
\newline
\newline
The framework we will look at for treating the evolution of primordial fluctuations and their statistical description is known as standard perturbation theory (SPT). The first three sections of this chapter heavily follow the comprehensive review by Bernardeau et al. \cite{Bernardeau:2001qr} and we refer the interested reader to this for more details, a larger scope and extensions. We begin with looking at the evolution of the {\it perturbations} upon the FLRW background. 


\section{The Evolution of Perturbations}
\subsection{Conservation: The Vlasov Equation} 
We begin with a set of CDM particles distributed in an expanding space with some density $\rho(\bfr)$, where $\bfr$ is a position vector in spherical physical coordinates. Say these particles induce a Newtonian gravitational potential $\varphi(\bfr)$, then the acceleration of a particle at position $\bfr$ is given by
\begin{equation}
\frac{d\bfv}{dt} = -\frac{\partial \varphi}{\partial \bfr},
\label{acceleqn}
\end{equation}
where $\bfv(\bfx,t)$ is the velocity field. Again, we can make the change to comoving coordinates by the transformation $\bfr = a(t) \bfx$. Since we want to study tiny deviations from the background density it is useful to define the \emph{density contrast} or \emph{over density}
\begin{equation}
\delta(\bfx,t) \equiv \frac{\rho(\bfx,t) - \bar{\rho}(t)}{\bar{\rho}(t)},
\label{densitycontrast}
\end{equation}
where $\bar{\rho}$ is the average background density. We will also study the \emph{peculiar velocity} field, which is just the difference between the total velocity $\bfv(\bfx,t)$ and the velocity due to the expansion of space
\begin{equation}
\bfu(\bfx,t) \equiv \bfv(\bfx,t) - Ha\bfx.
\label{peculiarvelocity}
\end{equation}
Finally, we also want the gravitational potential that the density perturbations induce. We can define the gravitational potential perturbation as 
\begin{equation}
\Phi(\bfx,t) \equiv \varphi(\bfx,t) + \frac{a x^2}{2}\frac{\partial (Ha)}{\partial t}.
\end{equation} 
Applying the Laplacian with respect to $\bfx$ to the above equation we have
\begin{align}
\nabla^2 \Phi_{{GR}}(\bfx,t) &= a^2 \nabla_r^2 \varphi(\bfr,t) + \frac{a^2}{2}\left(-\frac{\Omega_m H^2}{2} + \frac{\Lambda}{3}\right) \nabla^2 x^2 \nonumber \\
 &= \frac{3}{2}\Omega_mH^2a^2 \delta(\bfx,t), 
 \label{poisson2}
\end{align}
where we have used the 2nd Friedmann equation (Eq.\ref{friedmann2}) with $p=0$, the Newtonian Poisson equation for $\varphi$ (Eq.\ref{poissoneqn1}) and Eq.\ref{densitycontrast}. $\nabla_r$ indicates derivatives with respect to spherical physical coordinates whereas no subscript indicates with respect to comoving coordinates $\bfx$. Eq.\ref{poisson2} is the Poisson equation for the potential perturbation in GR in the Newtonian limit, which as seen is induced by the density contrast. At this level we can naively introduce a quantification of our ignorance of the theory of gravity through a general isotropic function $\mu(x;a)$ as 
\begin{equation}
\nabla^2 \Phi(\bfx,t) = \frac{3}{2}\Omega_mH^2a^2 \delta(\bfx,t) \mu(x;a).
\label{poissonmglin}
\end{equation} 
We assume $\mu(x;a)=1$ for the entirety of this section but in general it can assume different values as we will see in the next chapter.
\newline
\newline
Eq.\ref{acceleqn} can then be used to obtain the peculiar acceleration in terms of the potential perturbation  
\begin{equation}
\frac{d\bfp}{dt} = -a^2m\nabla\Phi(\bfx,t),
\end{equation}
where $\bfp = am\bfu$ is the peculiar momentum. We can now consider the collisionless Boltzmann equation that expresses the conservation of the phase space distribution function $f(\bfx,\bfp,t)$ \footnote{$f(\bfx,\bfp,t)$ is the time dependent probability distribution function of the CDM particles in phase space $(x,y,z,p_x,p_y,p_z)$.}. This conservation relation is commonly called the \emph{Vlasov equation} 
\begin{equation}
\frac{df}{dt} =  \frac{\partial f}{\partial t} + \frac{\bfp}{ma^2} \cdot \nabla f - m\nabla \Phi \cdot \frac{\partial f}{\partial \bfp} = 0.
\label{vlasov}
\end{equation}
This equation will provide us with the evolution of the perturbations from their initial conditions to a given future time as we will see next. 


\subsection{Evolution: The Continuity and Euler Equations} 
\subsubsection{Evolution in Configuration Space}
Observations in cosmology are generally concerned with the spatial distribution of the velocity and density fields. These come from the momentum moments of $f(\bfx,\bfp,t)$. In particular we look at the first three moments 
\begin{align}
\int d^3\bfp f(\bfx,\bfp,t) & \equiv n(\bfx,t),  \\
\int d^3\bfp \frac{\bfp}{am} f(\bfx,\bfp,t) &\equiv \rho(\bfx,t)\bfu(\bfx,t), \\
\int d^3\bfp \frac{p_ip_j}{a^2m^2}f(\bfx,\bfp,t) &\equiv \rho(\bfx,t)\bfu_i(\bfx,t)\bfu_j(\bfx,t) + \sigma_{ij}(\bfx,t),
\end{align}
where $\sigma_{ij}(\bfx,t)$ is the \emph{stress tensor} and $n(\bfx,t)$ is the number density of particles. This quantity gives a measure of the deviation of particle motions from a single stream, which at the very early stages of gravitational collapse of a pressureless perfect fluid is very small. After collapse this term becomes non-negligible, but taking $\sigma_{ij}\approx 0$ is still a useful simplifying assumption at large enough scales where velocity dispersions are small and the single stream approximation can still be used. Once multiple streams are present, commonly known as \emph{shell crossing}, we have the generation of velocity dispersions and anisotropic stresses which must be accounted for. For the scales relevant in this work we can assume the single stream approximation and so set $\sigma_{ij}=0$. The zeroth moment of Eq.\ref{vlasov} gives the conservation of mass equation, or \emph{continuity equation}
\begin{equation}
\frac{\partial \delta(\bfx,t)}{\partial t} + \frac{1}{a(t)}\nabla \cdot \{[1+\delta(\bfx,t)]\bfu(\bfx,t)\} = 0.
\label{cont1}
\end{equation}
The first moment of Eq.\ref{vlasov} minus [Eq.\ref{cont1} $\times$  $\bfu(\bfx,t)$] gives the conservation of momentum equation, or \emph{Euler equation}
\begin{equation}
\frac{\partial \bfu(\bfx,t)}{\partial t} + H(t)\bfu(\bfx,t) + \frac{1}{a(t)} \bfu(\bfx,t) \cdot \nabla \bfu(\bfx,t) = - \frac{1}{a(t)} \nabla \Phi(\bfx,t),
\label{euler1}
\end{equation}
where $\nabla \bfu(\bfx,t)$ is the directional derivative of $\bfu$ in the $\bfx$ direction. Eq.\ref{cont1} and Eq.\ref{euler1} completely describe the evolution of the zero pressure CDM fluid perturbations in an expanding background. At first order in the perturbations $\delta(\bfx,t)$ and $\bfu(\bfx,t)$ we have the following linear relations 
\begin{align}
\frac{\partial \delta(\bfx,t)}{\partial t} + \frac{1}{a(t)}\nabla\cdot \bfu(\bfx,t) & = 0, \label{contlin} \\ 
\frac{\partial \bfu(\bfx,t)}{\partial t} + H(t)\bfu(\bfx,t) & = - \frac{1}{a(t)} \nabla \Phi(\bfx,t).
\label{eulerlin}
\end{align}
It is useful at this stage to Helmholtz decompose the peculiar velocity field 
\begin{equation}
\bfu(\bfx,t) = - \nabla \Theta(\bfx,t) + \nabla \times \bfu(\bfx,t),
\end{equation}
where we have defined the curl-free part and curl of $\bfu(\bfx,t)$ as $\nabla \Theta(\bfx,t)$ and $\bfw(\bfx,t)$ respectively. We will also define $\bar{\theta}(\bfx,t) = -\nabla \cdot \nabla \Theta(\bfx,t) = \nabla \cdot \bfu(\bfx,t)$. Taking the divergence and curl of Eq.\ref{eulerlin} gives the following linear relations for the vector's components 
\begin{align}
\frac{\partial \bar{\theta}(\bfx,t)}{\partial t} + H(t)\bar{\theta}(\bfx,t) + \frac{3}{2}\Omega_mH^2a(t) \delta(\bfx,t) &= 0, \\
\frac{\partial \bfw(\bfx,t)}{\partial t} + H(t)\bfw(\bfx,t) & = 0, \label{curleqn}
\end{align}
where we have used the Poisson equation (Eq.\ref{poisson2}). The solution to Eq.\ref{curleqn} gives $\bfw \propto a^{-1}$, and so the initial linear vorticity decays as the universe expands. So we can safely describe the peculiar velocity field as being irrotational, especially in the linear regime - so at early times or at large scales.
\newline
\newline
Before discussing the evolution equations further, we first make a comment on the curl of the velocity field. We found that this decays in the linear regime. The non-linear Euler equation for the curl of the peculiar velocity field is
\begin{equation}
\frac{\partial \bfw(\bfx,t)}{\partial t} + H(t)\bfw(\bfx,t) - \nabla\times[\bfu(\bfx,t) \times \bfw(\bfx,t)]  =  0,
\label{vorteqn}
\end{equation}
where we have kept the assumption $\sigma_{ij}=0$. Now, if $\bfw(\bfx,0) = 0$ then the vorticity remains zero throughout the evolution of the universe. If this is not true and there is an initial non-zero vorticity, the vorticity can grow via the last term of Eq.\ref{vorteqn}. In this work we assume zero initial vorticity and so the peculiar velocity field can be completely characterised by its curl-free part $\bar{\theta}(\bfx,t)$. 
\newline
\newline
The fields have so far been described in what we will call \emph{configuration space}, with the first argument of $\delta$ and $\bar{\theta}$ being the physical position. The Fourier description of the fields is described next. 


\subsubsection{Evolution in Fourier Space} 
We begin by defining the Fourier transform (FT) for a field $f(\bfx)$ 
\begin{align}
f(\bfx) &= \int \frac{d^3\bfk}{(2\pi)^3} e^{i\bfk\cdot\bfx} \tilde{f}(\bfk), \\ 
\tilde{f}(\bfk) & = \int d^3\bfx e^{-i\bfk\cdot\bfx} f(\bfx),
\end{align}
and by introducing a useful normalisation for $\theta(\bfx,t)$ 
\begin{equation}
\theta(\bfx,t) = \frac{\bar{\theta}(\bfx,t)}{a(t)H(t)}.
\label{thetadef}
\end{equation}
Fourier transforming Eq.\ref{cont1} and the divergence of Eq.\ref{euler1} we get 
\begin{align}
a \frac{\partial \delta(\bfk,a)}{\partial a} + & \theta(\bfk,a) \nonumber \\
  &= -\int\frac{d^3\bfk_1d^3\bfk_2}{(2\pi)^3}\delta_{\rm D}(\bfk-\bfk_{12})
\alpha(\bfk_1,\bfk_2)\,\theta(\bfk_1,a)\delta(\bfk_2,a),
 \label{cont2} 
 \end{align} 
 \begin{align}
a \frac{\partial \theta(\bfk,a)}{\partial a} + &\left(2+\frac{aH(a)'}{H(a)}\right) \theta(\bfk,a) -\left( \frac{k}{aH(a)}\right)^2 \Phi  \nonumber \\
&= -\frac{1}{2}\int\frac{d^3\bfk_1d^3\bfk_2}{(2\pi)^3}
\delta_{\rm D}(\bfk-\bfk_{12})
\beta(\bfk_1,\bfk_2)\,\theta(\bfk_1,a)\theta(\bfk_2,a),
\label{euler2}
\end{align}
where $\delta_D$ is the Dirac delta function in 3-dimensions which ensures conservation of momentum. We use the summation convention $\bfk_{12} = \bfk_1+\bfk_2$ and we have switched to the scale factor $a$ as our time variable using the chain rule
\begin{equation}
\frac{\partial}{\partial t} = Ha\frac{\partial}{\partial a}.
\end{equation}
Henceforth a prime $'$ will denote a scale factor derivative. The mode mixing functions are given by
\begin{eqnarray}
\alpha(\bfk_1,\bfk_2)=1+\frac{\bfk_1\cdot\bfk_2}{|\bfk_1|^2},
\quad\quad
\beta(\bfk_1,\bfk_2)=
\frac{(\bfk_1\cdot\bfk_2)\left|\bfk_1+\bfk_2\right|^2}{|\bfk_1|^2|\bfk_2|^2}.
\label{alphabeta}
\end{eqnarray}
Lastly, the Poisson term in GR is given by Eq.\ref{poisson2} 
\begin{equation}
-\left(\frac{k}{a H(a)}\right)^2\Phi_{GR}=
\frac{3 \Omega_m(a)}{2} \delta(\bfk,a),
\label{eq:poisson0}
\end{equation}
where $\Omega_m(a) = \kappa \bar{\rho}_m/3 H^2$. So far we have only invoked conservation of the probability distribution in phase space together with the assumptions of zero vorticity, zero anisotropic stress and that GR is the true theory of gravity. The first two of these assumptions are justified at the scales we consider. This third assumption enters our equations only through the Poisson term which can be relaxed by the general parametrisation \cite{Koyama:2009me} 
\begin{equation}
-\left(\frac{k}{a H(a)}\right)^2\Phi=
\frac{3 \Omega_m(a)}{2} \mu(k;a)\,\delta(\bfk,a) + S(\bfk,a),
\label{eq:poisson1}
\end{equation}
where the function $S(\bfk)$ will be called the \emph{non-linear source term} and characterises new mode coupling terms. Up to third order in the perturbations we can write this generically as 
\begin{align}
S(\bfk,a) =&
\int\frac{d^3\bfk_1d^3\bfk_2}{(2\pi)^3}\,
\delta_{\rm D}(\bfk-\bfk_{12}) \gamma_2(\bfk, \bfk_1, \bfk_2;a)
\delta_0(\bfk_1)\,\delta_0(\bfk_2)
\nonumber\\
+ &
\int\frac{d^3\bfk_1d^3\bfk_2d^3\bfk_3}{(2\pi)^6}
\delta_{\rm D}(\bfk-\bfk_{123})
\gamma_3( \bfk, \bfk_1, \bfk_2, \bfk_3;a)
\delta_0(\bfk_1)\,\delta_0(\bfk_2)\,\delta_0(\bfk_3),
\label{eq:Perturb3}
\end{align}
where $\gamma_2( \bfk, \bfk_1, \bfk_2; a)$  and $\gamma_3(\bfk, \bfk_1, \bfk_2, \bfk_3;a)$ are symmetric under the exchange of $\bfk_i$. To recover the equations as described in GR we must set $\gamma_2( \bfk, \bfk_1, \bfk_2; a) = \gamma_3(\bfk, \bfk_1, \bfk_2, \bfk_3;a) = 0$ and $\mu(x;a) = 1$. We will provide motivation for these terms in the next chapter where we review modifications to the standard model. Also $\delta_0(\bfk) = \delta_1(\bfk,1)$ is the linear density contrast today. The general evolution equations at 1st order in the perturbations are then given by 
\begin{align}
a\frac{\partial \delta(\bfk,a)}{\partial a} + \theta(\bfk,a) & = 0, \label{contlin1} \\ 
a\frac{\partial \theta(\bfk,a)}{\partial a} + \left(2+\frac{aH(a)'}{H(a)}\right) \theta(\bfk,a)-\left(\frac{k}{a H}\right)^2\Phi &= 0. \label{eulerlin1}
\end{align}
We note that at linear order, valid at large scales where the perturbations are small, the Fourier modes evolve independently, i.e. there is no mode mixing. When we move to smaller scales and begin to include non-linearities, the Fourier modes start mixing as shown by the right hand sides of Eq.\ref{cont2} and Eq.\ref{euler2}.  


\section{Separability and the EdS Approximation} 
We now look at solutions to Eq.\ref{cont2} and Eq.\ref{euler2} for GR. Beginning with linear order, Eq.\ref{contlin1} and Eq.\ref{eulerlin1}, can be solved analytically by making the assumption that the growth of the linear perturbations is scale independent. In GR, this is true on large scales as different Fourier modes are uncoupled in their evolution. 
This amounts to the separability ansatz 
\begin{align}
\delta(\bfk,a) &= F_1(a) \delta_0(\bfk), \\ 
\theta(\bfk,a) &= G_1(a) \delta_0(\bfk), 
\end{align}
where $F_1(a)$ is called the \emph{linear growth factor}. Taking the scale factor derivative of Eq.\ref{contlin1} and substituting the 2nd term with Eq.\ref{eulerlin1} we get 
 \begin{equation}
 aF_1(a)'' + \left(3+\frac{aH(a)'}{H(a)}\right) F_1(a)' + \frac{3}{2}\Omega_m(a) \delta(\bfk,a) = 0.
 \label{linearevol}
 \end{equation}
 This second order equation has a general solution given by 
 \begin{equation}
 \delta(\bfk,a) = F_1^+(a)A(\bfk) + F_1^-(a)B(\bfk), 
 \end{equation}
 with $A(\bfk)$ and $B(\bfk)$ being arbitrary functions describing the initial density spatial distribution and $F_1^+(a)$ being the fastest growing of the two solutions. Using the linear order continuity equation (Eq.\ref{contlin1}) we obtain a similar solution for $\theta$
 \begin{equation}
 \theta(\bfk,a) = G_1^+(a)A(\bfk) +G_1^-(a)B(\bfk),
 \end{equation}
 with $G_1^+(a)$ again being the fastest growing of the velocity divergence solutions. Again using Eq.\ref{contlin1}, we can write $G_1^{+/-}$ in terms of the \emph{logarithmic growth factors}
 \begin{align}
 f(a) &\equiv \frac{d\ln{F_1^+(a)}}{d\ln a} = -\frac{G_1^+(a)}{F_1^+(a)}, \\ 
 g(a) & \equiv  \frac{d\ln{F_1^-(a)}}{d \ln a}  = -\frac{G_1^-(a)}{F_1^-(a)}.
 \end{align}
In an Einstein-de Sitter (EdS) universe ($\Omega_{m,0} =1,\Omega_\Lambda=0$), one obtains the following solutions for the growth factors
\begin{equation}
F_1^+ =a, \quad F_1^-=a^{-3/2}, \quad f=1, \quad g = -\frac{3}{2}.
\end{equation}
From this we see that the decaying mode, $B(\bfk)$, would have needed to be very close to zero in the early, matter-dominated universe to have any reasonable solution to the primordial density contrast. With this in mind, we take the regular solution $B(\bfk)=0$ and so only consider the growing mode. Further, we will normalise the linear growth such that $F_1(a=1) = 1$ so that $A(\bfk) = \delta_0(\bfk)$. We can now look at the non-linear regime. 
\newline
\newline
The core assumption of perturbation theory is that the fields $\delta$ and $\theta$ can be expressed as a power series expansion about the first order solutions \cite{Goroff:1986ep} , whose spatial component we have just seen is the primordial density field $\delta_0$
\begin{align} 
\delta_{NL}(\bfk,a) &= \sum^{\infty}_{n=1}\delta_n(\bfk,a), \label {genexp0} \\  
\theta_{NL}(\bfk,a) &= \sum^{\infty}_{n=1}\theta_n(\bfk,a). \label{genexp}
\end{align}
In this way, higher order contributions are treated as small corrections to the linear solution with $\delta_n \sim \delta_0^n$ and similarly by Eq.\ref{contlin1}, $\theta_n \sim \delta_0^n$.  We note that this is only a valid description as long as the perturbations remain small ($\delta \ll 1$ or $\theta \ll 1$), which generally breaks down towards the centre of virialised objects. These scales are very small and highly non-linear, making our only current probe into their inner workings so called \emph{N-body simulations}. We discuss these computer based experiments in Appendix D.
\newline
\newline
To start, we will again consider the EdS solution. The linear order solutions are simply $\delta_1 = F_1(a) \delta_0(k)$ and $\theta_1 = G_1(a)\delta_0(k)$ with $F_1(a) = -G_1(a) = a$. By substituting these into the right hand side of Eq.\ref{cont2} and Eq.\ref{euler2} we can verify that the time dependence of the 2nd order solutions goes as $F_2(a) =a^2 $ and $G_2(a) =-a^2 $. Extending this gives the perturbative expansion 
\begin{align} 
\delta^{EdS}_{NL}(\bfk,a) &= \sum^{\infty}_{n=1}a^n\delta_n(\bfk), \\  
\theta^{EdS}_{NL}(\bfk,a) &= -\sum^{\infty}_{n=1}a^n\theta_n(\bfk),
\label{edsexp}
\end{align}
where each order's time dependence can be verified by recursively using the Euler and continuity equations. The scale dependent parts can be obtained similarly. By substituting the above expansions up to 2nd order into the Euler and continuity equations, and using the first Friedmann equation (Eq.\ref{friedmann1c} with only $\Omega_m=1$) we obtain \cite{Jain:1993jh}
\begin{align}
2 \delta_2(\bfk) + & \theta_2(\bfk) \nonumber \\
  &= -\int\frac{d^3\bfk_1d^3\bfk_2}{(2\pi)^3}\delta_{\rm D}(\bfk-\bfk_1-\bfk_2)
\alpha(\bfk_1,\bfk_2)\,\delta_0(\bfk_1)\delta_0(\bfk_2),\\
\frac{5}{2}\theta_2(\bfk) + & \frac{3}{2}\delta_2(\bfk)  \nonumber \\
&= -\frac{1}{2}\int\frac{d^3\bfk_1d^3\bfk_2}{(2\pi)^3}
\delta_{\rm D}(\bfk-\bfk_1-\bfk_2)
\beta(\bfk_1,\bfk_2)\,\delta_0(\bfk_1)\delta_0(\bfk_2),
\end{align}
where the time scaling $a^2$ has dropped out of the equations. We can solve these simultaneously to obtain the scale dependence. The 2nd order solutions are then given by 
\begin{align} 
\delta_2(\boldsymbol{k}, a) &= \frac{1}{(2\pi)^{3}} \int d^3\boldsymbol{k}_1d^3 \boldsymbol{k}_2 \delta_D(\boldsymbol{k}-\boldsymbol{k}_{12}) F_2(\boldsymbol{k}_1,\bfk_2;a) \delta_0(\boldsymbol{k}_1)\delta_0(\boldsymbol{k}_2), \label{nth1} \\ 
\theta_2(\boldsymbol{k},a) &=  \frac{1}{(2\pi)^{3}}\int d^3\boldsymbol{k}_1d^3 \boldsymbol{k}_2 \delta_D(\boldsymbol{k}-\boldsymbol{k}_{12}) G_2(\boldsymbol{k}_1,\bfk_2;a) \delta_0(\boldsymbol{k}_1)\delta_0(\boldsymbol{k}_2). \label{nth2}
\end{align}
The second order kernels $F_2$ and $G_2$ are given by 
\begin{align}
F^{EdS}_2(\bfk_1,\bfk_2;a) &=a^2\left( \frac{5}{14} \{ \alpha(\bfk_1,\bfk_2) + \alpha(\bfk_2,\bfk_1)\} + \frac{1}{7}\beta(\bfk_1,\bfk_2)\right),
\label{eds2ndordera}  \\
G^{EdS}_2(\bfk_1,\bfk_2;a) &= -a^2\left( \frac{3}{14} \{ \alpha(\bfk_1,\bfk_2) + \alpha(\bfk_2,\bfk_1)\} + \frac{2}{7}\beta(\bfk_1,\bfk_2)\right),
\label{eds2ndorderb} 
\end{align}
where $\alpha(\bfk_1,\bfk_2)$ and $\beta(\bfk_1,\bfk_2)$ are given in Eq.(\ref{alphabeta}). These kernels have been symmetrised by summing over all permutations in their spatial arguments and dividing by the number of permutations. At 2nd order there are only 2 such permutations. We note that symmetrising will not change the field solution because the kernel arguments are integration variables.  Higher order solutions can be obtained in the same way and in general the solutions are of the form 
\begin{align} 
\delta_n(\boldsymbol{k} ; a) &= \frac{1}{(2\pi)^{3(n-1)}}\int d^3\boldsymbol{k}_1...d^3 \boldsymbol{k}_n \delta_D(\boldsymbol{k}-\boldsymbol{k}_{1...n}) F_n(\boldsymbol{k}_1,...,\boldsymbol{k}_n ; a) \delta_0(\boldsymbol{k}_1)...\delta_0(\boldsymbol{k}_n), \label{nth1} \\ 
\theta_n(\boldsymbol{k}; a) &=  \frac{1}{(2\pi)^{3(n-1)}}\int d^3\boldsymbol{k}_1...d^3 \boldsymbol{k}_n \delta_D(\boldsymbol{k}-\boldsymbol{k}_{1...n}) G_n(\boldsymbol{k}_1,...,\boldsymbol{k}_n; a) \delta_0(\boldsymbol{k}_1)...\delta_0(\boldsymbol{k}_n). \label{nth2}
\end{align}
In this work we will only be interested in solutions up to 3rd order and so for completeness we here quote the 3rd order EdS solutions 
\begin{align}
F^{EdS}_3(\bfk_1,\bfk_2,\bfk_3;a) = \frac{a^3}{3} &\left[ \frac{2}{63} \beta(\bfk_1,\bfk_{23}) \Big\{\beta(\bfk_2,\bfk_3)+\frac{3}{4}\left(\alpha(\bfk_2,\bfk_3)+\alpha(\bfk_3,\bfk_2)\right) \Big\} \right.  \nonumber \\
& + \frac{1}{18} \alpha(\bfk_1,\bfk_{23}) \Big\{\beta(\bfk_2,\bfk_3)+\frac{5}{2}\left(\alpha(\bfk_2,\bfk_3)+\alpha(\bfk_3,\bfk_2)\right)\Big\}  \nonumber \\
& \left. + \frac{1}{9} \alpha(\bfk_{23},\bfk_1) \Big\{\beta(\bfk_2,\bfk_3)+\frac{3}{4}\left(\alpha(\bfk_2,\bfk_3)+\alpha(\bfk_3,\bfk_2)\right) \Big\} \right. \nonumber \\ 
& + (\mbox{cyclic perm.}) \Big],
\label{eds3rdordera} 
\end{align}
\begin{align}
G^{EdS}_3(\bfk_1,\bfk_2,\bfk_3;a) = -\frac{a^3}{3}& \left[ \frac{2}{21} \beta(\bfk_1,\bfk_{23}) \Big\{\beta(\bfk_2,\bfk_3)+\frac{3}{4}\left(\alpha(\bfk_2,\bfk_3)+\alpha(\bfk_3,\bfk_2)\right)\Big\} \right.  \nonumber \\
& + \frac{1}{42} \alpha(\bfk_1,\bfk_{23}) \Big\{\beta(\bfk_2,\bfk_3)+\frac{5}{2}\left(\alpha(\bfk_2,\bfk_3)+\alpha(\bfk_3,\bfk_2)\right)\Big\}  \nonumber \\
&\left.+ \frac{1}{21} \alpha(\bfk_{23},\bfk_1) \Big\{\beta(\bfk_2,\bfk_3)+\frac{3}{4}\left(\alpha(\bfk_2,\bfk_3)+\alpha(\bfk_3,\bfk_2)\right)\Big\} \right. \nonumber   \\ 
&+ (\mbox{cyclic perm.}) \Big].
\label{eds3rdorderb} 
\end{align}
The symmetrised EdS kernels $F_n$ and $G_n$ have the following properties 
\begin{enumerate}
\item 
As  $\bfk \to 0$ but the integrated vectors $\bfk_i$ do not, $F_i \propto k^2$, which is expected by conservation of momentum. 
\item 
As some $\bfk_i \to 0$ we have $F_n \to \bfk_i\cdot \bfk_j/k_i^2$ which causes a divergence. This low energy divergence, aptly called an \emph{infra-red (IR) divergence}, can be treated by using resummation techniques \cite{Anselmi:2012cn,Matsubara:2007wj,Matarrese:2007wc,Crocce:2005xy}. In practice we can avoid this by having a suitable IR cut-off in the integral. 
\end{enumerate}
In the early universe, when matter dominates, these solutions become valid. It can be shown that  for more general cosmologies, for example when $\Omega_m<1$ and $\Omega_\Lambda >0$, the separability of time and spatial solutions is still a very good approximation with under a few percent deviation in the most extreme cases \cite{Fosalba:1997tn,Bernardeau:1993qu,Bose:2016qun}. Using this we can extend the EdS perturbative expansion (Eq.\ref{edsexp}) to general cosmologies 
\begin{align} 
\delta^{GR}_{NL}(\bfk,a) &= \sum^{\infty}_{n=1}F_1^n(a)\delta_n(\bfk), \\  
\theta^{GR}_{NL}(\bfk,a) &= -f \sum^{\infty}_{n=1}F_1^n(a)\theta_n(\bfk),
\label{grexp}
\end{align}
where we have included the superscript $GR$ to indicate we still work within the GR description of gravity and $F_1$ is the linear growth, found by solving Eq.\ref{linearevol}. Still, as we exit matter domination and/or our description of gravity changes we can no longer fully assume separability of the solutions and in general $\bfk$ and $a$ become entangled \cite{Bernardeau:1993qu,Bouchet:1992uh,Bouchet:1994xp,Catelan:1994kt,Bose:2016qun}. In Appendix A we describe our procedure to obtain solutions (Eq.\ref{nth1} and Eq.\ref{nth2}) of the general evolution equations Eq.\ref{cont2} and Eq.\ref{euler2} with general $\mu(x;a)$, $\gamma_2(\bfk_1, \bfk_2; a)$ and $\gamma_3(\bfk_1, \bfk_2, \bfk_3;a)$ that doesn't assume separability of time and spatial components. This is based off an algorithm proposed in \cite{Taruya:2016jdt}. An optimised version of this algorithm is at the heart of {\tt MG-Copter} and hence this method will be the basis for most of the results obtained in this thesis. For the rest of this section we will then assume we can obtain such general solutions. 


\section{Correlations in the Density and Velocity Fields} 
\begin{figure}[H]
  \captionsetup[subfigure]{labelformat=empty}
  \centering
  \subfloat[]{\includegraphics[width=7.5cm, height=4.5cm]{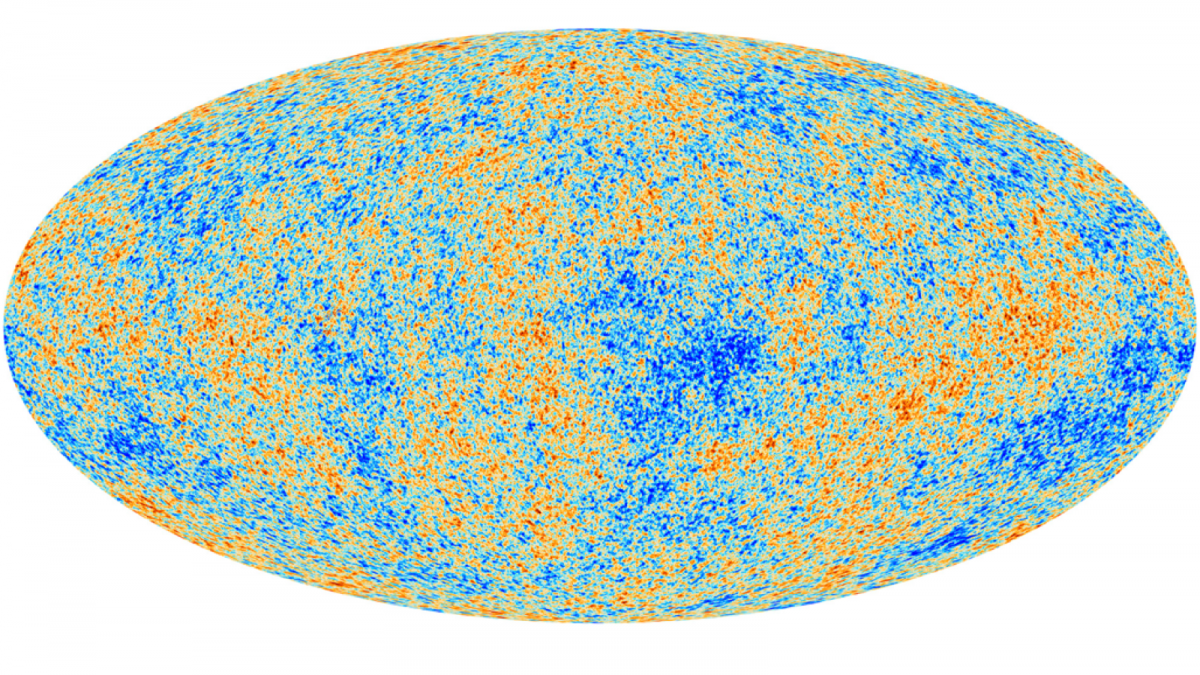}} \quad
  \subfloat[]{\includegraphics[width=7.5cm, height=4.5cm]{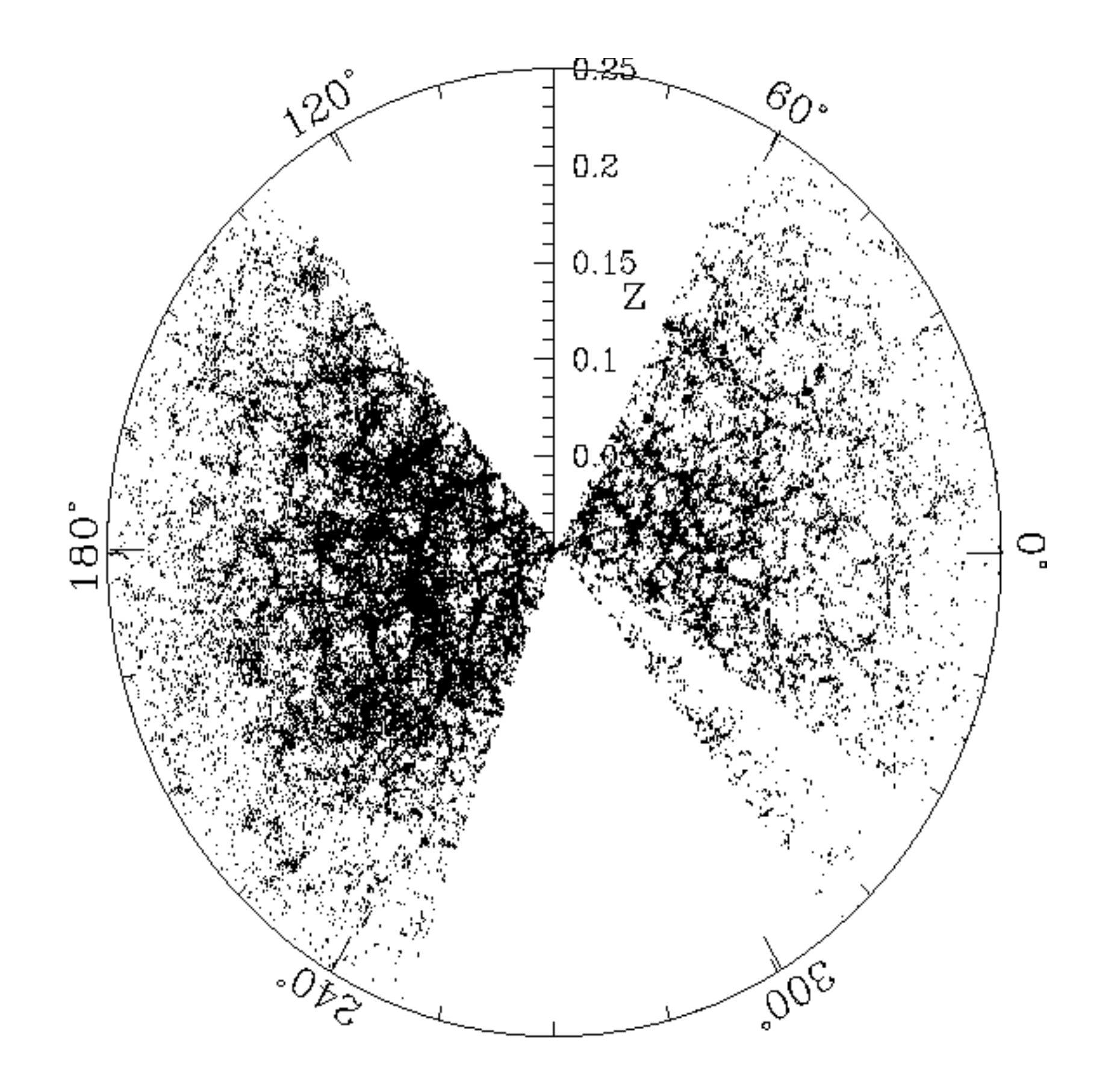}} 
  \caption{Early time (left) and late time (right) measurements of the cosmological density field in the form of the CMB and LSS. Both are indirect measures, temperature fluctuations on the microwave sky and one being redshifted galaxies respectively.}
\label{SDSSvCMB}
\end{figure}

\begin{figure}[H]
  \captionsetup[subfigure]{labelformat=empty}
  \centering
  \subfloat[]{\includegraphics[width=10.5cm, height=1.5cm]{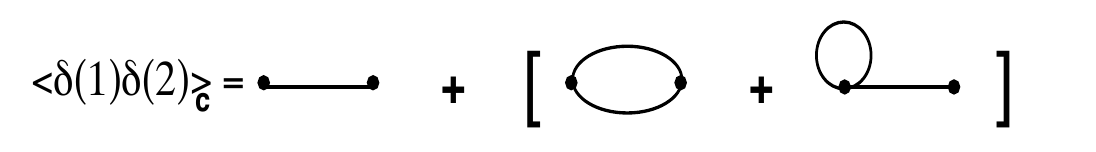}} 
  \caption{The two point correlation of the density contrast at 1-loop level expressed diagrammatically \cite{Bernardeau:2001qr}.}
\label{loopdiagram}
\end{figure}

\begin{figure}[H]
  \captionsetup[subfigure]{labelformat=empty}
  \centering
  \subfloat[]{\includegraphics[width=7.5cm, height=6.6cm]{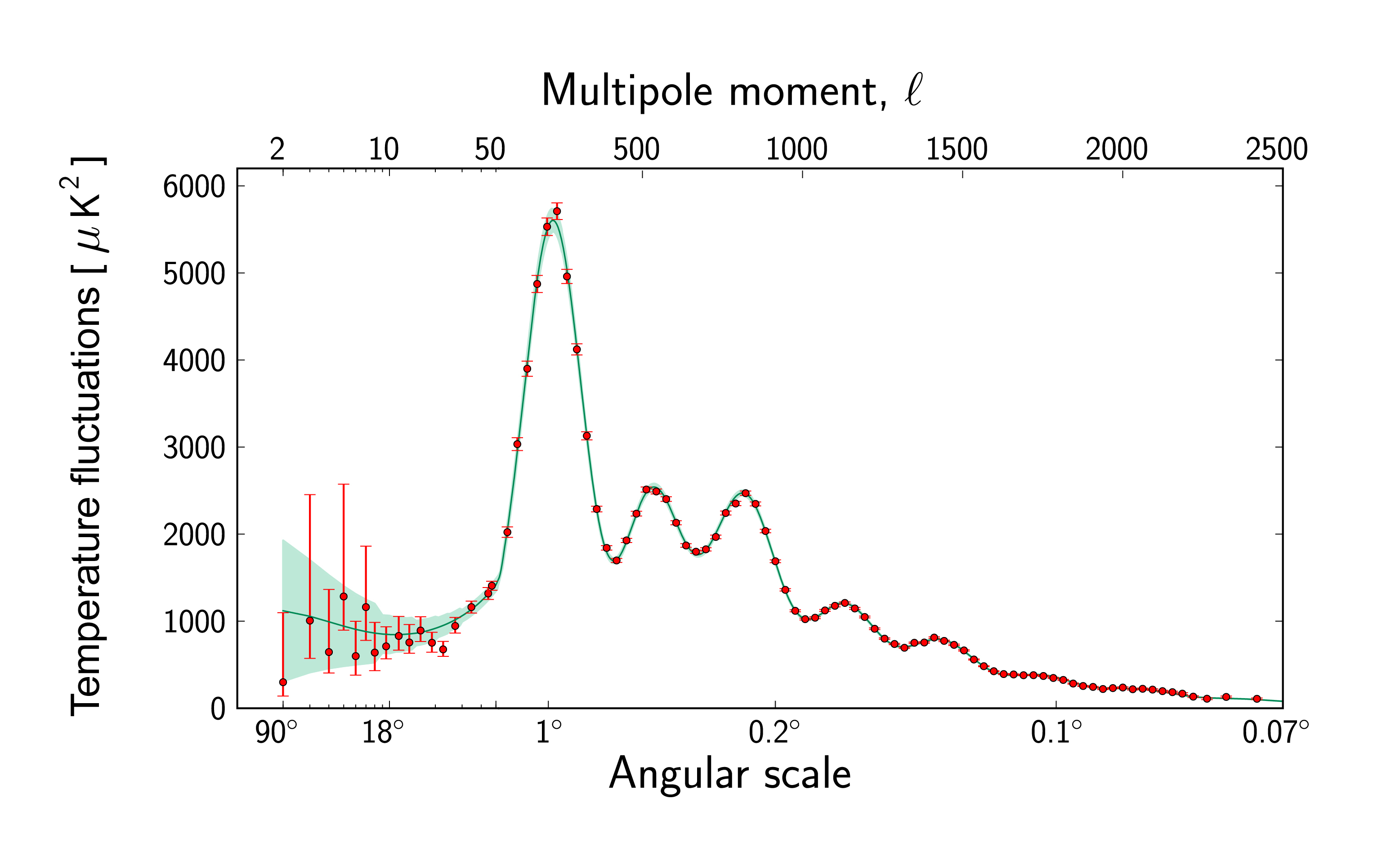}} \quad
  \subfloat[]{\includegraphics[width=7.5cm, height=6.6cm]{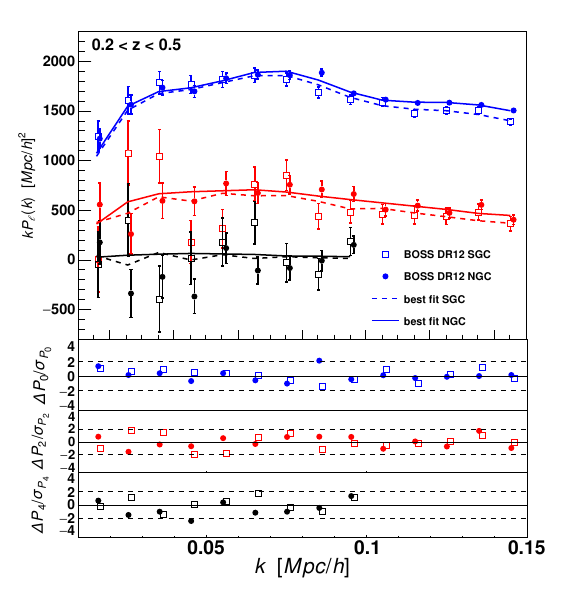}} 
  \caption{Correlations of the early ($z\sim1100$) and late ($0.2<z<0.5$) time perturbations shown in Fig.\ref{SDSSvCMB} in Fourier space. On the left we have the CMB power spectrum as measured by the Planck collaboration \cite{Planck:2015xua} as a function of angular scale on the sky with the best fit linear LCDM power spectrum (see Table.\ref{planckparams}). On the right we have the first 3 redshift space power spectrum multipoles (see Eq.\ref{multipolesl} and Chapter 4 for more details) as measured by the BOSS collaboration within SDSS-III \cite{Beutler:2016arn}. This is given as a function of wave number, with the best fit growth and cosmology using a 2-loop power spectrum modelling.}
\label{SDSSvsCMBps}
\end{figure}

The left hand side of Fig.\ref{SDSSvCMB} shows the full sky temperature map of the CMB at $z\sim1100$ as measured by the PLANCK satellite \cite{Ade:2013sjv} while the right hand side shows a slice on the sky marking 147,986 observed galaxies by SDSS mapped out along the radial axis by redshift up to $z=0.25$ \cite{Blanton:2002wv}. The CMB map shows tiny fluctuations ($\sim mK$) above (red) and below (blue) the average temperature of $2.726$K whereas each point in the SDSS map is a galaxy with on average 100 billion stars. 
\noindent Within the current paradigm, these two images show how the primordial density perturbations evolved with the expansion of the universe and collapsed to form the so called cosmic web. This web is made of large super structures, long filaments joining at nodes, each made of clusters of thousands of galaxies. These galaxies are believed to be fairly reliable tracers of the underlying dark matter density contrast field. In general, a bias is introduced by using galaxies as tracers of the dark matter field and this is the subject of much past and ongoing research \cite{Ishikawa:2013aea,McDonald:2009dh,Berlind:2001xk,Seljak:2000gq,Peacock:2000qk,Mo:1995cs,Landy:1993yu}. We will work with the CDM density field throughout this work. The issue of tracer bias for non-LCDM models is an ongoing work and we touch on this in Chapter 8.
\newline
\newline
These observations encapsulate the peak achievements of experimental cosmology which strongly support the FLRW description with initial \emph{Gaussian} perturbations collapsing under gravitational attraction. Since these are all observations of systems at a single point in their evolution and systems whose initial conditions are not directly accessible, one cannot apply classical deterministic tests to the data. One must instead use the data statistically.  From each of these maps one can extract powerful statistical quantities that make use of the observed distributions. Specifically, cosmologists traditionally concern themselves with the \emph{multipoint correlators}. The 2-point correlator, called the \emph{correlation function}, is defined as 
\begin{equation}
\xi(r) = \langle \delta(\bfx) \delta(\bfx + \bfr) \rangle,
\label{corfunc}
\end{equation}
where $\langle \rangle$ denotes an average taken over an ensemble of possible states that the density contrast field, $\delta(\bfx)$, can assume. The correlation function gives us a measure of how related two points in the density contrast field, separated by distance $r$, are. Roughly large positive values for $\xi(r)$ would indicate there are many over-dense points separated by $r$ in our measurement. We will also define here the \emph{power spectrum} as the FT of the correlation function
\begin{equation}
P(k) \equiv \int d^3r \xi(r) e^{i\bfk\cdot\bfr},
\end{equation}
where $k = |\bfk|$. One can easily show by applying the FT to Eq.\ref{corfunc} that
\begin{equation}
(2\pi)^3 P(k) \delta_D(\bfk + \bfk') =  \langle \delta(\bfk) \delta(\bfk') \rangle.
\label{powerspec}
\end{equation}
Similarly, the three point correlator is given by 
\begin{equation}
\xi_3(\bfx_1,\bfx_2,\bfx_3) = \langle \delta(\bfx_1) \delta(\bfx_2 ) \delta(\bfx_3) \rangle,
\label{corfunc3}
\end{equation}
with its FT given by the so called \emph{bispectrum} 
\begin{equation}
(2\pi)^3  B(\bfk_1,\bfk_2) \delta_D(\bfk_1 + \bfk_2 + \bfk_3) = \langle \delta(\bfk_1) \delta(\bfk_2) \delta(\bfk_3) \rangle.
\end{equation}
We will focus on the correlation function and power spectrum throughout this work. To get a measure of these quantities from the observed galaxy distribution or CMB temperature maps, one must essentially count pairs of galaxies, or measure temperature variations, at different scales on the sky. There are various ways to estimate the correlation function or power spectrum from this data which offer a variety of refinements and we direct the reader to the following for technical details \cite{Kerscher:1999hc,Landy:1993yu,Hamilton:1993fp,Aghanim:2015xee}.
\newline
\newline
Eq.\ref{corfunc} and Eq.\ref{powerspec} are generally well defined quantities for \emph{homogeneous} random fields.  What we mean by homogeneous is that the multipoint probability distribution function $p(\delta(\bfx_1),\delta(\bfx_2), ... )$ is translation invariant in $\bfx_i$. We say random because we model the early perturbations as a random realisation from some set of possibilities. This idea is birthed by the popular \emph{inflation}-description of the early universe. In this picture, the stochasticity comes from initial quantum fluctuations of some scalar field, usually called the inflaton, that are expanded to large scales in a period of rapid spatial expansion. The fluctuations then leave a mark of their energy in the gravitational potential. This then translates into the initial seeds of structure we can see imprinted on the CMB. The rapid expansion during inflation serves many purposes, one being the smoothing out of inhomogeneities. This would explain why the CMB is so uniform in temperature. A full discussion of the inflation paradigm is beyond  the scope of this work but we refer the interested reader to the comprehensive reviews in \cite{Liddle:2000cg,Tsujikawa:2003jp}.
\newline
\newline
One important consequence of starting at quantum fluctuations in the inflationary scenario is that at the end of inflation we typically end up with a set of Gaussian distributed fluctuations \footnote{We note that non-Gaussianities can be predicted in models of inflation  \cite{Chen:2010xka} but primordial non-Gaussianity is well constrained by the latest CMB measurements \cite{Ade:2015ava}.}. It is here that we state \emph{Wick's} or \emph{Isserli's theorem} for Gaussian fields 
\begin{align}
\langle \delta(\bfk_1) ... \delta(\bfk_{2p+1})\rangle &= 0,\\
\langle \delta(\bfk_1) ... \delta(\bfk_{2p})\rangle &= \sum_{pair\mbox{ }permutations} \quad  \prod_{pairs (i,j)} \langle\delta(\bfk_i)\delta(\bfk_j)\rangle, 
\end{align}
which tells us that shortly after inflation the over density field's statistical properties are completely described by the initial power spectrum $P_0(k)$, and all higher order correlations, such as the bispectrum are zero in the absence of non-Gaussianities. This picture changes as the universe expands because gravitational instability pulls the perturbations away from their linear description. Richer statistics are needed to fully describe the fields once gravitational non-Gaussianity has been induced. This said, the 2-point correlations will generally have the smallest statistical errors in survey measurements because of the large number of point pairs at a given scale, over the number of particular 3 point configuration in a bispectrum measurement. We are only concerned with the evolution of these correlators in this work. Modelling higher order statistics for non-LCDM models is the subject of ongoing work and we touch on this in Chapter 8. 
\newline
\newline
One can describe the evolution of the power spectrum at linear level by using the solutions to Eq.\ref{contlin1} and Eq.\ref{eulerlin1} in Eq.\ref{powerspec}
\begin{align}
P^{11}_{\delta \delta}(k,a) &= \frac{1}{(2\pi)^3}\int d^3k' \langle \delta_1(\bfk,a) \delta_1(\bfk',a)\rangle \nonumber \\
& = \int d^3k' F_1(k;a) F_1(k';a) P_0(k') \delta_D(\bfk+\bfk') \nonumber \\ 
& = F_1(k;a)^2 P_0(k), 
\label{treeps}
\end{align}
where $P_0(k)$ is the linear power spectrum, the superscript $(11)$ indicates two 1st order fields are being correlated and the subscript $(\delta \delta)$ indicates we correlate density contrast fields. If $F_1(k;a)$ is not scale dependent, as is for GR, the above relation shows that at the linear level the power spectrum's shape does not change as the universe expands but is only scaled by the linear growth factor. This is called the \emph{tree-level} contribution to the power spectrum which derives its name from Feynman diagrams. These describe the probability of scattering or equivalently the scattering amplitudes of an interaction in quantum field theory. Analogously, when describing structure formation in cosmology we want to find the amplitude of clustering in the cosmic density field. As we've said, the correlation between different points in this field can be described by the power spectrum and can be represented diagrammatically in a Feynman fashion. A tree-level diagram is the simplest configuration one can have and consists of 2 vertices and one edge, shown as the left most diagram on the right hand side of Fig.\ref{loopdiagram}. The vertices represent the two states described by $\delta(\bfk)$ and $\delta(\bfk')$ which `interact' to give the power spectrum.  
On top of the tree level calculation we can have so called \emph{loop corrections}. These are represented diagrammatically by closed loops and are higher order contributions to the amplitude. For a 2-point diagram there are only 2 possible ways of drawing single closed loops. These are shown on the right hand side of Fig.\ref{loopdiagram} and constitute the \emph{1-loop} correction to the tree-level prediction. 
\noindent The 1st diagram in square brackets shows a correction to the overall amplitude that involves both vertices while the 2nd is the tree-level diagram with a loop on top of it. As we will see this correction appears as a $\bfk$ dependent multiplicative factor to the tree-level power spectrum (Eq.\ref{treeps}). Further, we will see that these corrections are quadratic in the linear power spectrum and are the only such corrections to this order. To calculate the 1-loop terms we simply take our perturbative expansion (Eq.\ref{genexp0} and Eq.\ref{genexp}) up to higher order. Using the general solutions given in Eq.\ref{nth1} and Eq.\ref{nth2} and using Wick's theorem, we have the following non-zero terms at 2nd order in $P_0$ 
\begin{align}
(2\pi)^3 \delta_D(\bfk+\bfk') P^{22}_{\delta \delta} (k,a) &=  \langle (\delta_2(\bfk,a) \delta_2(\bfk',a) \rangle, \\ 
 (2\pi)^3 \delta_D(\bfk+\bfk') P^{13}_{\delta \delta} (k,a) &= [\langle (\delta_1(\bfk,a) \delta_3(\bfk',a) \rangle +  \langle (\delta_3(\bfk,a) \delta_1(\bfk',a) \rangle],
\end{align}
which then lets us define the so called \emph{1-loop power spectrum} 
\begin{equation}
P^{1-{\rm loop}}_{ij}(k,a) = P^{11}_{ij}(k,a) + P^{22}_{ij}(k,a) + P^{13}_{ij}(k,a),
\label{loopps}
\end{equation}
where $i,j \in \{ \delta, \theta\}$ and denote which fields we are correlating. $P_{\delta \delta}$ will be called the \emph{matter power spectrum}, $P_{\theta \theta}$ the \emph{velocity power spectrum} and $P_{\delta \theta}$ the \emph{cross power spectrum} for the entirety of this work. Using Wick's theorem and the definition of the linear power spectrum, we can write the loop corrections in terms of the perturbative kernels $F_n$ and $G_n$ as 
\begin{align}
P_{\delta \delta}^{22}(k,a)
&=\frac{2}{(2\pi)^3} \int d^3 \bfk' F_2 (\bfk - \bfk', \bfk';a)^2 
P_0(|\bfk - \bfk'|) P_0(k'),  \label{p22dd} \\
P_{\delta \theta}^{22}(k,a) 
&= \frac{2}{(2\pi)^3}  \int d^3 \bfk' F_2 (\bfk - \bfk', \bfk';a) G_2 (\bfk - \bfk', \bfk';a)
P_0(|\bfk - \bfk'|) P_0(k'), \\
P_{\theta \theta}^{22}(k,a) 
& = \frac{2}{(2\pi)^3}  \int d^3 \bfk' G_2 (\bfk - \bfk', \bfk';a)^2
P_0(|\bfk - \bfk'|) P_0(k'),
\end{align}
and 
\begin{align}
P_{\delta \delta}^{13}(k,a) 
&= \frac{6}{(2\pi)^3}  \int d^3 \bfk' F_1(k;a)F_3(\bfk, \bfk', -\bfk';a) 
P_0(k') P_0(k),   \label{p13dd}  \\
P_{\delta \theta}^{13}(k,a) 
&= \frac{3}{(2\pi)^3}   \int d^3 \bfk'   F_1(k;a)G_3 (\bfk, \bfk', -\bfk';a)
P_0(k') P_0(k) \\
& + \frac{3}{(2\pi)^3}   \int d^3 \bfk' G_1(k;a)F_3 (\bfk, \bfk', -\bfk';a) 
P_0(k')  P_0(k), \\
P_{\theta \theta}^{13}(k,a) 
& = \frac{6}{(2\pi)^3}  \int d^3 \bfk' G_1(k;a) G_3 (\bfk, \bfk', -\bfk';a) 
 P_0(k') P_0(|\bfk - \bfk'|).
\end{align}
Eq.\ref{loopps} gives a \emph{quasi non-linear} prediction for the power spectrum, quasi because we have only included a single loop correction. At infinite order, we would get an exact prediction for the power spectrum \footnote{As long as we're above scales where $\delta(\bfx)<1$!}. In principle one can include higher loop corrections and in the literature 2-loop corrections are used when comparing to data \cite{Beutler:2016arn,Beutler:2013yhm,Blake:2011rj}. We will not consider 2 or higher order loops for reasons outlined in the next section. As a summary of this section, we present the Fourier space statistics of the images in Fig.\ref{SDSSvCMB} in Fig.\ref{SDSSvsCMBps}. This shows the power spectra measurements at early and late times compared to modelling within the standard paradigm of gravitational growth of perturbations described in this chapter. Note that the late time measurement (see right hand side of Fig.\ref{SDSSvsCMBps}) includes an additional complication discussed in Chapter 4. We also note that the errors on the measurements are much larger at large angular scales or low $k$. This is commonly known as {\it cosmic variance} and comes from the fact that we only have one universe to observe. We are limited in the number of galaxies or fluctuations we can observe and at large scales these statistics are very poor which boosts the variance at these scales. 


\section{Successes, Failures and Improvements} 
\begin{figure}[h]
\centering
\centerline{\includegraphics[width=15.5cm,height=9cm]{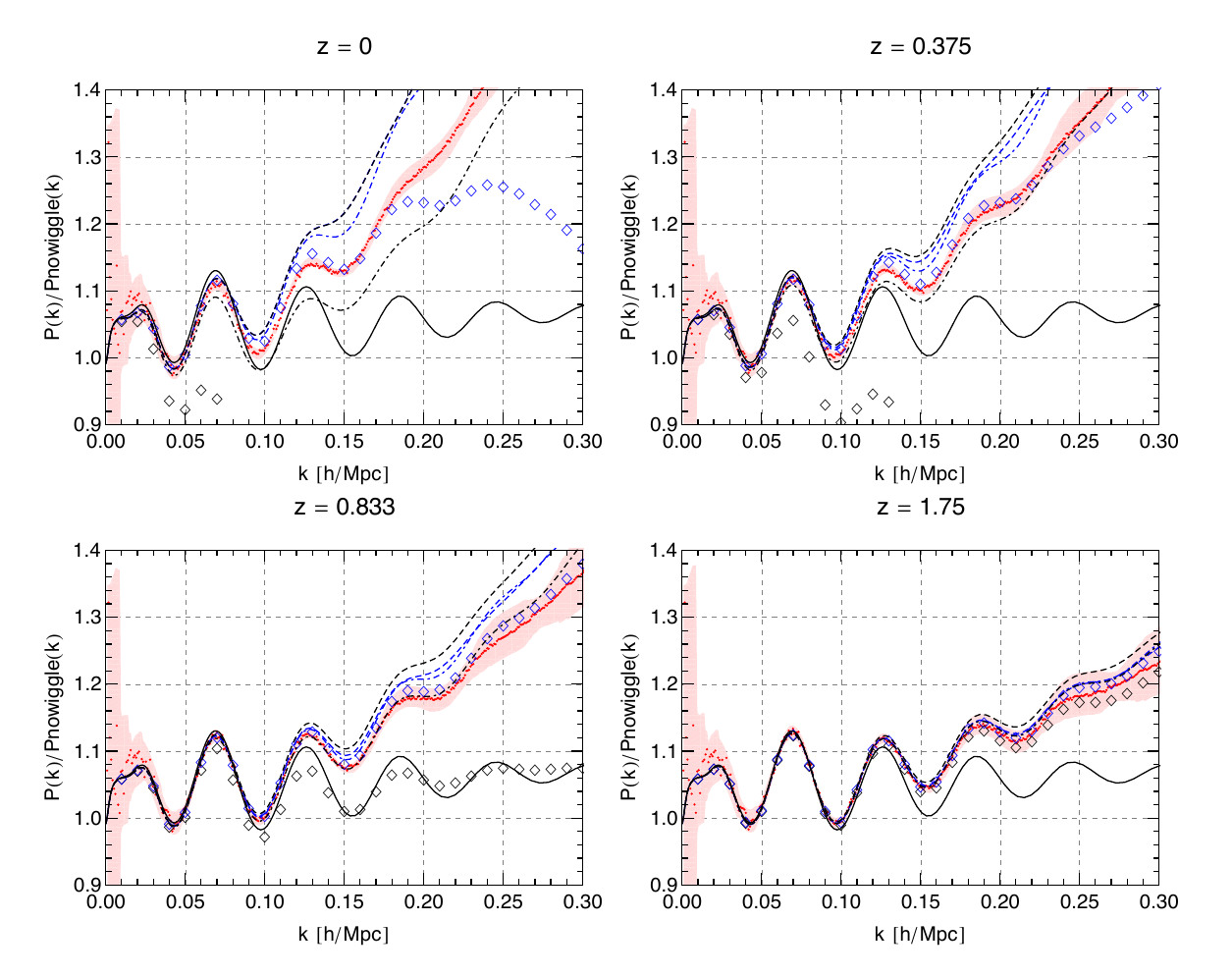}}
\caption{Linear (black; solid), 1-loop (black; dashed), 2-loop (black; dot-dashed) and 3-loop (black; diamonds) SPT power spectrum against N-Body simulation measurement \cite{Kim:2011ab} (red circles) for LCDM at 4 different redshifts. The blue lines and diamonds show the Pad\`e resummation scheme \cite{Blas:2013aba}. Result taken from \cite{Blas:2013aba}}
\label{sptvsobs}
\end{figure}

\begin{figure}[H]
  \captionsetup[subfigure]{labelformat=empty}
  \centering
  \subfloat[]{\includegraphics[width=10.5cm, height=7cm]{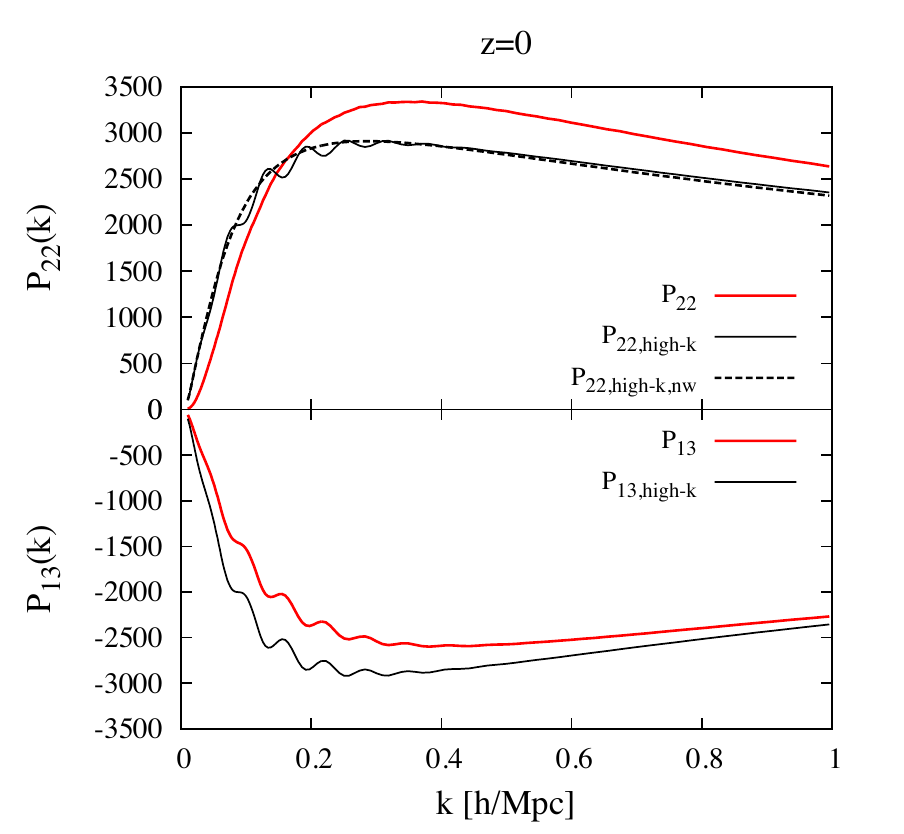}} 
  \caption{The computation of Eq.\ref{p22dd} and Eq.\ref{p13dd} (red) and the computation where we remove large modes from the loop integrations (black). The dashed line shows the $22$ component without its acoustic features \cite{Sugiyama:2013gza}.}
\label{highk}
\end{figure}

At very large scales the LSS of the universe is well described by linear perturbation theory while at scales below around $10$Mpc today the dynamics are highly non-linear and baryonic physics, multi-streaming, velocity dispersions and other effects play large roles. These can be modelled well by so called {\it N-body simulations} which numerically solve the Poisson equation and accordingly displace particles in a computer generated box (see Appendix D).  At intermediate scales, around the BAO ($\mathcal{O}(100)$Mpc), one can hope to accurately model the clustering and velocity of matter using higher order SPT. The power of this approach is its quick computability making it very well suited for data comparisons and theory parameter inference. 
\newline
\newline
Expressions for the power spectrum or correlation function can be easily derived to $n$th order in $P_0$ within SPT and will involve mode-coupling integrals of dimension $3(n-1)$. At 1-loop order we have a 3-dimensional integral which can be solved very quickly \footnote{In the EdS approximation this is reduced to a single integral over the k-vector magnitude because of the separable, analytic solutions.} whereas at 2-loops (3rd order in $P_0$) one has to deal with a 6-dimensional integral. Monte Carlo integration packages such as \cite{Hahn:2004fe} make this problem tractable for low enough dimension integrals, and would provide an ideal means of probing smaller physical scales. The question then becomes, is the loop expansion well behaved so that we may trust higher order loop corrections? Unfortunately, this has been shown not to be true at late times \cite{Blas:2013aba,Carlson:2009it}. Fig.\ref{sptvsobs} shows the comparison of the SPT prediction at varying loop orders up to 3, shown in black, against the full non-linear treatment of the N-body simulation, shown as red dots. We note that at $z=0$ the 3-loop contribution is significantly larger than the 1-loop contribution. Further, at high $k$ and low $z$, where we have most non-linearity, the loop corrections exhibit divergent behaviour and become larger than the tree level contribution. In summary, at low redshifts the loop corrections are divergent and the loop expansion is non-converging with the 3-loop expression doing worse than the tree level. Note that in Fig.\ref{sptvsobs}, each spectrum is divided by a {\it no-wiggle} spectrum \cite{Eisenstein:1997ik} which only captures the broadband shape of the power spectrum. In this way one can isolate the information rich BAO imprint. 
\newline
\newline
 At $z\geq 1.75$ we observe an improvement by including the 3-loop terms, whereas these terms worsen the fit at lower redshifts. This implies that at a given redshift the loop expansion offers improvement only to a certain order, and for $z\leq1$ the 1-loop and 2-loop expressions offer comparable improvement, with the 1-loop result being the optimal fit at $z=0$. 
 \newline
 \newline 
Beyond these small scale ultra-violet problems, there is an issue of IR divergences within the SPT loop integrals. Fig.\ref{highk} shows the equivalence principle at work within the SPT loop integrals themselves. The solid black lines represent the computation of Eq.\ref{p22dd} and Eq.\ref{p13dd} when we take the limit $\bfk > \bfk' \rightarrow 0$. We see they are exactly equal and opposite in sign, cancelling each other out in the loop contribution. What this means is there is no higher order contribution to the power spectrum coming from scales larger than the one's being considered. This is predicted by the equivalence principle as we do not expect signatures of long wavelength density perturbations to show up at smaller scales, just as local experiments cannot probe the existence of a uniform gravitational field. This is the so called {\it IR safe property} of SPT and applies for higher order corrections too \cite{Sugiyama:2013gza}. This is true if the linear spectrum is smooth. If there are oscillatory features in $P_0(k)$ at a scale $k_{\rm osc}$ then the modes $\bfk' \sim \bfk_{\rm osc}$ can affect modes $\bfk>\bfk'$, which effects the IR safe property. To restore this property,  methods that consider small $\bfk'$ contributions from all orders in perturbation theory become necessary. In effect this damps the oscillatory features in the power spectrum.  These are generally called resummation methods. 
  \newline
  \newline
In the next subsection we describe a perturbative prescription that is effective in damping the wiggle part of the power spectrum which is very useful when we wish to perform a FT to obtain the correlation function. Not doing this causes spurious features in the correlation function (see Chapter 4). The method to be discussed  has the added effect of suppressing power at small scales, countering the loop divergences of SPT in the UV, making its performance slightly better than SPT at larger $\bfk$.


\subsection{Regularised Perturbation Theory}
Here we discuss the regularised perturbation theory (RegPT) approach. Being a resummation approach, it effectively sums up IR contributions at all orders in perturbation theory. What we mean is that we will be taking the IR limit, $\bfk_i \rightarrow 0$, of all integrated modes $\bfk_i$, in the loop correction integrands. In this way we will obtain an expression for the quasi non-linear power spectrum which can be transformed to the correlation function safely (see Chapter 4 for more details). 
\newline
\newline
To begin, we go over the {\it multi-point propagator} formalism. This treatment is based on expanding these propagators (to be defined), which contain the entire non-perturbative nature of the field's evolution \cite{Bernardeau:2008fa}. The propagators can be analytically described in terms of the perturbative kernels and allow for the construction of a quasi non-linear power spectrum and correlation function which show excellent agreement with N-body data in real and redshift space as well as for models other than GR \cite{Taruya:2012ut,Taruya:2013my,Taruya:2014faa}. In this framework the fully non-linear power spectrum is given by 
\begin{equation}
P_{NL}(k,a) = \sum^{\infty}_{r=1} P_{\Gamma}^{(r)} (k,a),
\label{regptps1}
\end{equation}
where $P_{\Gamma}^{(r)}$ is the $r$th-order contribution to the power spectrum 
\begin{align}
P_{\Gamma}^{(r)}(k,a) = & r! \int \frac{d^3\bfk_1}{(2\pi)^3} \dots \int \frac{d^3\bfk_r}{(2\pi)^3 } (2\pi)^3 \delta_D(\bfk-\bfk_{1\dots r}) \nonumber \\ & \times \left[\Gamma^{(r)}(\bfk_1,\dots, \bfk_r,a)\right]^2 P_0(k_1) \dots P_0(k_r),
\end{align}
with $\Gamma^{(r)}$ being defined as follows 
\begin{equation}
\Gamma^{(r)}(\bfk_1, \dots, \bfk_r,a) \equiv \Gamma_{tree}^{(r)}(\bfk_1, \dots, \bfk_r,a) + \sum^{\infty}_{n=1}\Gamma_{{n-loop}}^{(r)}(\bfk_1, \dots, \bfk_r,a). 
\label{gamma1}
\end{equation}
The tree level and $n$th-loop propagators are defined in terms of the perturbative kernels $F_n$ (and $G_n$ for a $\theta$ correlation) as 
\begin{align}
\Gamma_{{tree}}^{(r)} (\bfk_1, \dots, \bfk_r) &\equiv F_r(\bfk_1,\dots,\bfk_r;a), \label{gamma2} \\
\Gamma^{(r)}_{{n-loop}}(\bfk_1,\dots, \bfk_r,a) &\equiv \frac{(r+2n)!}{r!2^n n!} \int \frac{d^3\bfp_1}{(2\pi)^3} \dots \int \frac{d^3\bfp_n}{(2\pi)^3} P_0(p_1)\dots P_0(p_n) \nonumber \\ & \times  F_{r+2n}(\bfk_1, \dots, \bfk_r, \bfp_1,-\bfp_1, \dots , \bfp_n, -\bfp_n;a). \label{gamma3} 
\end{align} 
The first order coefficient of the $\Gamma$ expansion is defined as 
\begin{equation} 
\langle \delta(\bfk,a) \delta_1(\bfk',1) \rangle \equiv (2\pi)^3 \delta_D(\bfk + \bfk') \Gamma^{(1)}(k,a) P_0(k).
\end{equation}
This formalism reduces to solving the $F_n$ and $G_n$ kernels, just as in SPT, but here the $r=1$ correction involves kernel computation to infinite order by Eq.\ref{gamma1} and Eq.\ref{gamma3}. Eq.\ref{gamma3} integrates over all scales for the $n$ vector-components of $F_{r+2n}$, but if at $n$th-loop order we only want to keep small scale information for the first $n$ components we can impose $p_m\rightarrow 0$ for $m>n$. It can be shown that in this limit we have the following recursion relation \cite{Sugiyama:2013gza,Sugiyama:2013pwa}
\begin{align}
F_{r+2n}&(\bfk_1, \dots, \bfk_r, \bfp_1, -\bfp_1, \dots, \bfp_n, -\bfp_n ;a)|_{p_{m+1}, \dots, p_n \rightarrow 0} \nonumber \\ \rightarrow &\frac{(r+2m)!}{(r+2n)!}(-1)^{n-m} \left(\frac{\bfk_{1\dots r} \cdot \bfp_{m+1}}{p_{m+1}^2}\right)^2 \dots \left(\frac{\bfk_{1\dots r} \cdot \bfp_n}{p_{n}^2}\right)^2 \nonumber \\  &\times  F_{r+2m}(\bfk_1, \dots, \bfk_r, \bfp_1, -\bfp_1, \dots, \bfp_m, -\bfp_m ;a), 
\end{align}
which when substituted in Eq.\ref{gamma3} we get 
\begin{equation}
\Gamma^{(r)}_{{n-loop}}(\bfk_1,\dots, \bfk_r,a)|_{p_{m+1} \dots p_n \rightarrow 0} \rightarrow \frac{m!}{n!} \left(-\frac{k^2\sigma_d^2}{2}\right)^{n-m}\Gamma^{(r)}_{{m-loop}}(\bfk_1,\dots, \bfk_r,a).
\label{gammarec}
\end{equation}
$\sigma_d^2$ is the dispersion of the linear displacement field
\begin{equation}
\sigma_d^2(k) \equiv \int^{k/2}_0 \frac{dq}{6\pi^2} F_1(q;a)^2P_0(q).
\end{equation}
For $n=1$ (1-loop) we have 
\begin{align}
& \Gamma^{(r)}_{{n-loop}}(\bfk_1,\dots, \bfk_r,a)  \nonumber \\ & =  n \Gamma^{(r)}_{{n-loop}}(\bfk_1,\dots, \bfk_r,a)|_{p_{2} \dots p_n \rightarrow 0} - (n-1)\Gamma^{(r)}_{{n-loop}}(\bfk_1,\dots, \bfk_r,a)|_{p_{1} \dots p_n \rightarrow 0} \nonumber \\ &  \rightarrow \frac{n}{n!} \left(-\frac{k^2\sigma_d^2}{2}\right)^{n-1}\Gamma^{(r)}_{{1-loop}}(\bfk_1,\dots, \bfk_r,a) - \frac{n-1}{n!}\left(-\frac{k^2\sigma_d^2}{2}\right)^{n}\Gamma^{(r)}_{{tree}}(\bfk_1,\dots, \bfk_r,a), 
\label{gammaloop}
\end{align}
where in the second line we multiply by $n$ because we can choose any of the $n$ integrated vectors to go to zero and we must subtract the 2nd term because the first term integrates over the region where all vectors go to zero $n$ times. Finally, in this limit $\Gamma^{(r)}$ (Eq.\ref{gamma1}) can be written as  
\begin{align}
\Gamma^{(r)}(\bfk_1, \dots, \bfk_r,a) = & \exp{\left(-\frac{k^2\sigma_d^2}{2}\right)} \Bigg[ \Gamma_{tree}^{(r)}(\bfk_1, \dots, \bfk_r,a)  \nonumber \\ & \left. + \left(\Gamma_{{1-loop}}^{(r)}(\bfk_1, \dots, \bfk_r,a) + \frac{k^2\sigma_d^2}{2} \Gamma^{(r)}_{tree}(\bfk_1, \dots, \bfk_r,a)\right) \right]. 
\label{gammafinal}
\end{align} 
Now we come back to the power spectrum (Eq.\ref{regptps1}). In terms of the multipoint propagators $\Gamma^{(n)}$ the RegPT 1-loop power spectrum is given by 
\begin{align}
P^{\rm 1-loop,RegPT}_{bc}(k;a) =& \Gamma^{(1)}_b(k;a)\Gamma_c^{(1)}(k;a)P_0(k) \nonumber \\
 & + 2 \int \frac{d^3 \bfq }{(2\pi)^3} \Gamma_b^{(2)}(\bfq,\bfk-\bfq;a)\Gamma_c^{(2)}(\bfq,\bfk-\bfq;a) \nonumber \\ & \quad \times P_0(q)P_0(|\bfk-\bfq|) 
 \label{regptloop}
\end{align}
where $b,c \in \{\delta, \theta\}$ chooses which fields we are correlating. The propagators are given in terms of the perturbative kernels using Eq.\ref{gamma2}, Eq.\ref{gamma3} and Eq.\ref{gammafinal}   
\begin{align}
\Gamma^{(1)}_b(k;a) = & \left[ J_b^{(1)}(k;a)\{ 1+ \frac{k^2\sigma_d^2}{2}\}  \right. \nonumber \\
& \left. + 3\int  \frac{d^3\bfq}{(2\pi)^3} J_b^{(3)}(\bfk,\bfq,-\bfq;a)P_0(q)\right] e^{-k^2\sigma_d^2/2} \\ 
\Gamma^{(2)}_b(\bfq,\bfk-\bfq;a)  = & J_b^{(2)}(\bfq,\bfk-\bfq;a)e^{-k^2\sigma_d^2/2} 
\end{align}
where $J_b^{(n)} = (F_n,G_n)$. Eq.\ref{regptloop} provides us with an expression for the quasi non-linear power spectrum that performs equally well to SPT at relevant scales but damps the power spectrum oscillations through the inclusion of IR contributions at all orders. This will be useful when we wish to perform Fourier transforms of the power spectrum in Chapter 4. This concludes our discussion of perturbation theory. We will discuss alternatives to GR in the next section. 
 



\chapter{Modelling Gravity}
\begin{figure}[H]
  \captionsetup[subfigure]{labelformat=empty}
  \centering
  \subfloat[]{\includegraphics[width=15.5cm, height=7.6cm]{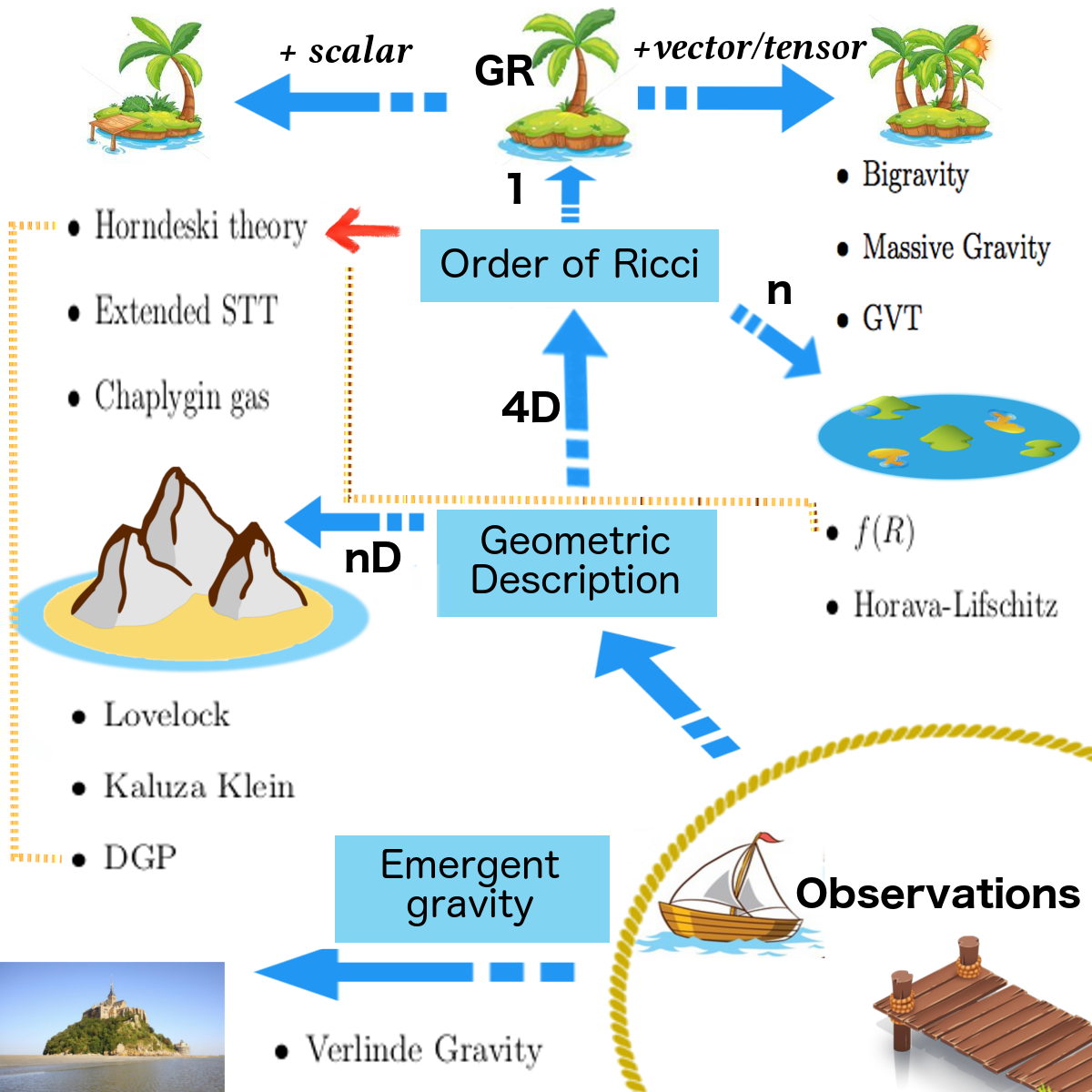}} 
  \caption{A partial archipelago of gravitational theories. From top left moving anti-clockwise: Horndeski theory \cite{Horndeski:1974wa}, Extended scalar tensor theories (eSTT) \cite{Crisostomi:2016czh}, Chaplygin gas \cite{Kamenshchik:2001cp}, Lovelock theory \cite{Lovelock:1971yv}, Kaluza Klein theory \cite{Kaluza:1921tu}, Dvali, Gabadadze and Porrati (DGP) gravity \cite{Dvali:2000hr}, Eric Verlinde's emergent gravity \cite{Verlinde:2010hp}, Horava-Lifschitz theory \cite{Horava:2009uw}, $f(R)$ gravity \cite{Sotiriou:2008rp}, gauge vector-tensor theory (GVT) \cite{Exirifard:2011bz}, massive gravity \cite{Fierz:1939ix,deRham:2010kj}, bigravity \cite{Rosen:1940zza} and GR. }
\label{mgsum}
\end{figure}

\begin{chapquote}{J.D}
``Birth is a fading lighthouse, the glory of its beam and the sturdiness of the shore it loves fading further and further as time blows our sail wayward, forward, to the shores beyond the night.''
\end{chapquote}
In Chapter 1 we discussed the energy ingredients for the standard model of cosmology which are necessary to fit a plethora of cosmological and galactic data. A number of supporting observations and selected references are summarised in Table.\ref{evidences} at the end of this chapter \footnote{For a full set of observational evidence for dark matter see \cite{Roos:2010wb} and for the accelerated expansion see \cite{Copeland:2006wr}.}. If we were to have cited the full set of analyses the table would be many times larger. This goes to show the great support the LCDM picture has garnered within the age of sophisticated astronomical observations. The model also has simple origins - it is a combination of our best model of gravity (GR) extrapolated to the cosmological observations of expansion and isotropy, with the additional philosophical ingredient of the Copernican principle. But with such overwhelming evidence for dominating exotic energy components, this picture is nevertheless put into doubt. 
\newline
\newline
It was John Wheeler who said, 
\newline
\newline
``{\it Matter tells space-time how to curve, and curved space-time tells matter how to move.}" 
\newline
\newline
In terms of mathematics, the left hand side of the Einstein field equations (Eq.\ref{EFES}) gives us the curvature of spacetime while the right hand side gives us the energy content. To modify the standard picture one needs to address one (or both) sides of this coin: either tell curvature to respond differently to an energy content or change the energy content itself (for example the inclusion of $\Lambda$). GR has been shown to be the unique 4 dimensional theory with a single tensor degree of freedom that respects invariance under Lorentz transformations \cite{Weinberg:1965rz}. What this means is that to go beyond LCDM one must either introduce additional degrees of freedom, extend to higher dimensions or break Lorentz invariance. We will strictly assume Lorentz invariance is not violated in this work although many modifications to gravity have forgone this symmetry (for example \cite{Alfaro:2013uga,Horava:2009uw,Jacobson:2000xp}). Further, in this chapter we discuss avenues to get around problems associated with the cosmological constant and do not concern ourselves with dark matter. Much ongoing research is devoted to solving the dark matter issue, both in terms of modifications to the gravitational force (for example \cite{Milgrom:2014usa,Verlinde:2010hp}) and in looking for candidate particles for dark matter (see for example \cite{Liu:2017drf,Kahlhoefer:2017dnp}). 
\newline
\newline
Before we jump into details, Fig.\ref{mgsum} gives a (non-exhaustive) schematic along with some examples of the work that has been done in moving away from pure GR (plus $\Lambda$). At the bottom left we have the non-geometric route to describing gravitational phenomena such as Eric Verlinde's emergent gravity \cite{Verlinde:2010hp}. On the geometric route, one can extend to higher dimensions or include general terms in the curvature invariants (left and right midway up). From the standard GR description, one can also introduce additional degrees of freedom in the form of tensors, vectors and the simplest case, scalars. One should note that combinations of the routes are also common such as Tensor-Vector-Scalar theory \cite{Bekenstein:2004ne} which includes additional tensor, vector and scalar degrees of freedom. Further note that upon any of these routes, one can also drop the assumption of homogeneity (and isotropy to some extent) and assume other metrics beyond FLRW. We direct the interested reader to this review \cite{Hellaby:2009vz}. 
\newline
\newline
As GR is able to describe local phenomena exquisitely, it is reasonable not to stray too far, and to first consider the simplest extension. Our destination will be arguably the closest theories to GR, found at the top left of the chart indicated by the red arrow. Specifically, a very general class of models with a single additional scalar degree of freedom called {\it Horndeski theory} \cite{Horndeski:1974wa} including a generalised potential term. The inclusion of a generalised potential term allows for the inclusion of other theories such as $f(R)$ gravity, joined by the dotted line. We will also be able to relate the 5-dimension brane world theory by Dvali, Gabadadze and Porrati (DGP) \cite{Dvali:2000hr}, indicated in the diagram, to this class for perturbations on small scales. These particular theories will be discussed towards the end of this chapter. We begin with a general introduction to theories involving a single extra scalar degree of freedom - scalar-tensor theories (STT).
\begin{table}[b]
\centering
\caption[Dark Sector Evidence]{Dark Sector Evidence}
    \begin{tabular}{ | p{7cm} | p{8cm} |}
 \hline
   \bf{Observation} & \bf{Summary} \\ \hline
   CMB \cite{Planck:2015xua} & No measured curvature component implying $\Omega_\Lambda>0$ and BAO wiggles imply CDM picture\\ \hline
   Supernovae  \cite{Perlmutter:1998np,Riess:1998cb} &  Direct measurement of $H(z)$ supports LCDM \\ \hline
   BAO \cite{Blake:2011wn} & Measurement of integrated $H(z)$ using BAO peaks consistent with LCDM \\ \hline
   Galaxy Clustering \cite{Reid:2012sw} & Growth of structure consistent with LCDM \\ \hline
  Integrated Sachs-Wolfe effect \cite{Nadathur:2016hky} & Measurement consistent with LCDM  \\ \hline
   Galaxy rotation curves \cite{Rubin:1978kmz,Bosma:1981zz} & Flattening of galaxy velocity curves supports CDM \\ \hline
   Galaxy velocity dispersions \cite{Faber:1976sn} & Virial theorem implies larger mass than observed for galaxy clusters   \\ \hline
   Gravitational Lensing \cite{Collett:2014ola} & Lensing from unobserved matter \\ \hline
 \end{tabular}
  \label{evidences}
 \end{table}

\section{Modified Gravity}
As said, the inclusion of a constant in the GR field equations can give an accelerating FLRW universe. The most straight forward interpretation of this constant energy density is one of a vacuum energy, a long standing prediction of quantum field theory. The identification of these two measurements has led to one of the largest problems in modern physics. The standard model of particle physics predicts a contribution to the vacuum energy density from the known particles that is 120 orders of magnitude larger than the LCDM best fit value. This can be removed by tuning the bare Lagrangian value of $\Lambda$ so that it cancels the predicted one, a cancellation that needs to be accurate to $\sim$120 decimal places if we trust GR up to the Planck scale \footnote{If we trust GR down to the realm of quantum chromodynamics this fine tuning {\it only} needs to be accurate to $\sim$ 40 decimal places!} \cite{Weinberg:1988cp}, which by any standards would be astoundingly fine tuned! We direct the interested reader to \cite{Carroll:2000fy} for a detailed review. This is the old cosmological constant problem, mentioned in Chapter 1. The modifications to gravity we consider are broad enough to tackle this issue in more than one way. For example self-tuning models of $\Lambda$ have been proposed \cite{Kaloper:2013zca,Charmousis:2011bf,Charmousis:2011ea} within the considered class. Other modifications we consider remain agnostic as to how such a large vacuum energy prediction remains unobserved and rather aim at removing the vacuum energy interpretation of an accelerated expansion. In this way, no fine tuning is needed between prediction and observation as well as doing away with the coincidence problem discussed in Chapter 1. Beyond theories possessing these desirable properties, general extensions to GR should be tested on the basis of agnosticism of the large scale nature of gravity.
\newline
\newline
The first classical modification to Einstein's theory, described by the Einstein-Hilbert action (Eq.\ref{ehaction}), came just under 6 years after the publication of GR. Theodor Kaluza and Oskar Klein \cite{Kaluza:1921tu} attempted to unify electromagnetism and gravity by considering a compactified 5th dimension in 4-dimensional spacetime. Their theory led to a set of equations which included the original Einstein field equations as well as Maxwell's equations. It also included a 5th metric component $g_{44}$ which behaved like a scalar field in 4 dimensions. This component, sometimes called the `compacton', could be seen as just that, a scalar field which measured the scale of the fifth dimension. 
\newline
\newline
Drawing on this idea, P. Jordan considered 5-dimensional flat spacetime in which a 4-dimensional curved manifold was embedded. This led to 4-dimensional gravitational equations and a scalar field, $\varphi_J (x)$, which based on a suggestion by Dirac \cite{dirac1921}, could be described as a time-dependent gravitational constant. This theory saw the introduction of the so called \emph{non-minimal coupling term}, $R\varphi_J (x)$, replacing $R$ in Eq.\ref{ehaction}. Non-minimal coupling has found motivation in quantum corrections to scalar field theory \cite{birrell1980} and is necessary for the renormalizability of the scalar field theory in curved space \cite{Freedman:1974ze,Freedman:1974gs}. Apart from theoretical motivation, a non-trivial manner of interaction is compelling since it properly distinguishes the theory from GR giving it ability to address fundamental shortcomings of the classical theory.  
\newline
\newline
In 1961, Carl H. Brans and Robert H. Dicke \cite{Brans:1961sx} further modified the Jordan Lagrangian. We quote the Jordan-Brans-Dicke (JBD) Lagrangian below
\begin{equation}
L=  \frac{1}{2\kappa}\left[ \phi \tilde{R} -\frac{\omega_c}{\phi} \tilde{g}^{\mu \nu} \partial_\mu \phi \partial_\nu \phi \right]  + L_M (\Psi_M), 
\label{BDL2}
\end{equation}
with $\phi$ being the JBD scalar field that is a function of spacetime coordinates and $\omega_c$ being a coupling constant. In this form, the Lagrangian is in what is known as the Jordan Frame and $\tilde{g}_{\mu \nu}$ is called the Jordan Frame metric. We touch more on this in the next subsection. Eq.\ref{BDL2} would serve as the archetype for STT.  One can vary the action with respect to the additional degree of freedom, $\phi$, and find the minimum to obtain the \emph{Klein-Gordon equation} (KGE)
\begin{equation}
 \tilde{\Box}  \phi  = \frac{1}{3+2\omega_c} T, 
\label{kg1}
\end{equation}
where $\tilde{\Box}$ is the spacetime D'Alembertian operator with respect to $\tilde{g}_{\mu \nu}$. Eq.\ref{kg1} describes the dynamics of the scalar field and its interaction with the matter fields within the spacetime. We also note that the right hand side of Eq.\ref{kg1} is given in terms of the matter energy-momentum tensor. This ensures that the force mediated by $\phi$ respects the weak equivalence principle (WEP), a fact that can be seen by considering the case of zero momentum. In this case the source of the scalar field is given by the integrated $T_{00}$ component, which is the total energy of the system, whatever the distribution may be. Eq.\ref{kg1} also suggests that the scalar field mediates a long-range force between massive objects just as the classical gravitational force does in the weak field limit of GR. For coupling constants of the order of unity, the strength of this force is of the same order of magnitude as the Newtonian gravitational force. This is obviously problematic because all solar system tests (see \cite{Bertotti:2003rm,Williams:1995nq,Anderson:1995df,Dickey:1994zz,Talmadge:1988qz} for example or \cite{Bean:2010zq,Reynaud:2008yd} for general reviews) show no signs of deviations from the standard GR force. For example, the Cassini mission has constrained $ \omega_c > 40,000$ using Shapiro time delays of radio emissions, at the $2\sigma$ level \cite{EspositoFarese:2004cc}, and an early work \cite{Salmona:1967zz} in the high gravitational field regime (where the effects of the field could be expected to be most evident since the field is coupled to curvature) showed negligible differences from GR for values of $\omega_c$ as small as $1.5$. This left the desire for a STT which would pass all the existing solar-system tests but would differ noticeably in the high field or large scale limit, for example around neutron stars or at cosmic scales. This is the focus of the next section.
\newline
\newline
One important question we may ask is, how do we obtain acceleration from STT? In JBD theory, we note that by comparing Eq.\ref{ehaction} and Eq.\ref{BDL2} we can define a varying gravitational constant 
\begin{equation}
G \rightarrow \frac{1}{\phi}.
\end{equation}
If the scalar field is time varying we can hope to get a weakening of gravity at late times which can cause accelerated expansion. But let us look at the, flat FLRW metric dynamical equations and the KGE. If the matter content is governed by a pressureless perfect fluid, these are given by \footnote{These are given using the Jordan Frame.} 
\begin{align}
H^2 &= \frac{\rho}{3 \phi} + \frac{\omega_c}{2}\frac{\ddot{\phi}^2}{\phi^2} - 3H \frac{\dot{\phi}}{\phi},  \\ 
2\dot{H} + 3H^2  &= - \frac{\omega_c}{2} \frac{\dot{\phi}^2}{\phi^2} - 2H \frac{\dot{\phi}}{\phi} - \frac{\ddot{\phi}}{\phi}, \\
\ddot{\phi} + 3H\dot{\phi} &= \frac{\rho}{3+2\omega_c},
\end{align} 
where we assumed the scalar field is homogeneous and varies with time. First we note that setting $\phi = 1/\kappa = \mbox{constant}$ we get the usual Friedmann equations (Eq.\ref{friedmann1} and Eq.\ref{friedmann2}). If we require acceleration we look for a solution for $a$ that gives $\ddot{a}>0$. This has been shown to be possible with the power law solution 
\begin{equation}
a = a_0 t^{2(\omega_c+1)/(3\omega_c +4)},
\end{equation}
which satisfies the dynamical equations and for values of $ -2 \leq \omega_c \leq -3/2$  \cite{Bertolami:1999dp} we get accelerating solutions. Of course there are a number of problems here. One problem is that a negative value for $\omega_c$ gives the wrong sign for the kinetic term introducing unphysical, negative energies for the field making it a so called {\it ghost}. Another issue is that the choice of $\omega_c$ gives the wrong evolution during the radiation dominated epoch and consequently runs into problems with big-bang nucleosynthesis. Further, far more obviously, we just said solar system constraints put $\omega_c>40,000$! 
\footnote{Beyond these problems, this proposal predicts an age of the universe much older than $H_0^{-1}$ which is in clear tension with observations.} Of course, JBD constitutes a very basic extension of GR and since then a lot has been done to get around these problems. In fact, in 1974 Gregory Horndeski \cite{Horndeski:1974wa} wrote down the most general extension to GR that includes a single extra scalar degree of freedom, later rediscovered by Cedric Deffayet et al \cite{Deffayet:2011gz} \footnote{Their equivalence was shown in \cite{Kobayashi:2011nu}.}. This theory includes four general functions of the scalar field and its kinetic term in canonical form 
\begin{equation}
X = -\frac{1}{2}g^{\mu \nu} \partial_\mu \phi \partial_\nu \phi,
\label{cankin}
\end{equation}
as well as higher order powers of scalar field derivative terms. Here one must be careful. By introducing these higher order derivatives into the Lagrangian one expects to get higher than 2nd order equations of motion for $\phi$, and by Ostragradsky's theorem \cite{Ostrogradsky:1850fid},  this will introduce instabilities. By instability we mean, the Hamiltonian of the field will contain extra degrees of freedom which will be unbounded from below. This can lead to negative energy states which are unphysical. Theories with this instability are usually called ghostly because of this (see \cite{Woodard:2006nt} for a discussion). One must note that Ostragradsky's theorem applies to theories where the highest order derivative can be written as a function of canonical variables, i.e. non-degenerate theories. By considering degenerate theories one can obtain healthy theories with higher than 2nd order equations of motion \cite{Langlois:2015skt,Langlois:2015cwa,Crisostomi:2016czh}. We do not consider these classes here. 
\newline
\newline
There are ways to avoid the so called Ostragradsky ghost when including higher order derivatives in the Lagrangian. One way is by imposing certain field symmetries, such as the Galilean symmetry, $\phi \rightarrow \phi + b_\mu x^\mu +c$, and then choosing the subset of these theories which give second order equations of motion. This is used to effect in the {\it Galileon} theories of modified gravity \cite{Nicolis:2008in}, which include cubic, quartic, quintic and higher order powers of scalar field derivatives in their Lagrangians. 
\newline
\newline
Thankfully Horndeski ensured that his theory was `healthy' and no Ostragradsky ghost modes were present in the class of theories it described.
The Horndeski action is given by
\begin{equation}
S = \int d^4x \sqrt{-g}(L_{GG} + L_M),
\end{equation}
where $L_M$ is the usual, minimally coupled, matter Lagrangian and 
\begin{align}
L_{GG} = & G_2(\phi,X) - G_3(\phi,X)\Box\phi +G_4(\phi,X)R  \nonumber \\
&+ G_{4,X}(\phi,X)[(\Box \phi)^2-(\nabla_\mu \nabla_\nu \phi)^2]   \nonumber \\
&+ G_5(\phi,X)G_{\mu \nu}\nabla^\mu\nabla^\nu\phi  \nonumber \\
&- \frac{1}{6}G_{5,X}(\phi,X) [(\Box\phi)^3-3\Box\phi(\nabla_\mu\nabla_\nu\phi)^2+2(\nabla_\mu\nabla_\nu\phi)^3].
\label{Horndeski1}
\end{align}
$G_i(\phi,X)$ with $i \in \{ 2,3,4,5\} $, are free functions of a scalar field $\phi$ and its canonical kinetic term X (Eq.\ref{cankin}), and $G_{i,X}(\phi,X) = \partial G_{i}/\partial X$. Below are the functional forms for some of the previously mentioned models.
\begin{center}
{\bf LCDM}
\begin{equation}
G_2 = -\Lambda,  \quad \quad G_4=1/(2\kappa), \quad \quad G_3 = G_5 =0.  
\end{equation}
{\bf STT} 
\begin{equation}
G_2=G_2(\phi,X), \quad \quad G_4=F(\phi), \quad \quad G_3=G_5 =0. 
\end{equation}
{\bf Galileons}
\begin{equation}
G_2=-c_2X, \quad \quad G_3=c_3X/\Lambda_0^3, \quad \quad G_4=1/2\kappa+c_4X^2/\Lambda_0^6, \quad \quad  G_5=3c_5X^2/\Lambda_0^9, 
\label{gallileon}
\end{equation}
\end{center}
where $F(\phi)$ is a general function of $\phi$ \footnote{See the next section for details.}. The $c_i$ in Galileon models are coupling constants and $\Lambda_0$ represents the energy scale of the theory. Recently, the LIGO/VIRGO consortium observed gravitational waves from a neutron-neutron star merger \cite{TheLIGOScientific:2017qsa} and its electromagnetic counterpart  \cite{Monitor:2017mdv,GBM:2017lvd}. These observations have constrained the speed of gravitational waves $c_T$ to be almost identical to that of light $c_T^2-1 \leq 6\times10^{-15}$. In Galileon theories the speed of gravitational waves is given by \cite{Brax:2015dma}
\begin{equation}
\Big|\frac{4c_4 x^2}{1-3c_4 x^2}\Big| = |c_T^2-1|  \leq 6\times10^{-15},
\end{equation}
where $x^2 = \dot{\phi}^2/(H^2\kappa)$. We see that the combination $c_4x^2$ needs to close to zero, ruling out very large values of $c_4$. Supermassive black hole data constrain values at the lower end of parameter space \cite{Sakstein:2017bws}. Together, the complementary tests result in the quartic Galileon being ruled out in terms of explaining the accelerated expansion \cite{Sakstein:2017xjx}.
\newline
\newline
With such generality at hand, theories with desirable or interesting properties should be relatively easy to find. What is now more pressing is how do these theories evade the stringent solar system constraints that GR has passed? 


\section{Screening Mechanisms} 
After Brans and Dicke proposed their theory, works by Bergmann \cite{Bergmann:1968ve}, Nordtvedt \cite{Nordtvedt:1970uv} and Wagoner \cite{Wagoner:1970vr} adjusted the JBD model to incorporate a wider spectrum of scalar fields and couplings. They included a coupling function ($\omega_c \Rightarrow \omega_c (\phi)$) as opposed to a coupling constant as well as allowing self interaction of the scalar field through a potential term $V(\phi)$ in the Lagrangian. This could take the form of a mass term for $\phi$. This gave a possible mechanism to pass all the local tests, while providing a deviation from GR in the relatively untested, strong field and cosmological regimes. Such mechanisms are now known as {\it screening mechanisms}. The upgraded Lagrangian is written as 
\begin{equation}
L_{\rm JF}=  \frac{1}{2\kappa}\left[ \phi_{\rm JF} \tilde{R} -\frac{\omega_c(\phi_{\rm JF})}{\phi_{\rm JF}} \tilde{g}^{\mu \nu} \partial_\mu \phi_{\rm JF} \partial_\nu \phi_{\rm JF} - V(\phi_{\rm JF}) \right] + L_M (\Psi_M). 
\label{BDL3}
\end{equation}
When the Lagrangian does not include a scalar field coupling to the matter fields it is referred to as being in the \emph{Jordan frame} (now indicated by the subscript `JF') and $\tilde{g}_{\mu \nu}$ is called the Jordan frame metric. Note that $\tilde{R}$ is computed using this metric. Before talking about screening within these models, let us first consider a suitable {\it conformal transformation} to the metric ($\tilde{g}_{\mu \nu} = A^2(\phi) {g}_{\mu \nu}$)  and a field redefinition, both as follows 
\begin{equation}
A= \frac{1}{\sqrt{\phi_{JF}}} ,
\end{equation}
\begin{equation}
\phi_{JF} = e^{ \phi/ \sqrt{2\omega_c +3}}.
\end{equation}
The conformal transformation factor $A$ is a function of spacetime coordinates and such transformations preserve the angles between spacetime vectors \footnote{One can think of such a transformation as a local rescaling of the spacetime interval.}. Upon applying these transformations we arrive at the so called {\it Einstein frame} Lagrangian 
\begin{equation}
L_{BDE}=  \left[ \frac{R}{2\kappa}- \frac{1}{2 }{g}^{\mu \nu} \partial_\mu \phi \partial_\nu \phi - V(\phi)\right] +L_M (\tilde{g}_{\mu \nu}, \Psi_M).
\label{BDLE}
\end{equation} 
Note that this form looks just like the Einstein-Hilbert Lagrangian with additional components in the energy-momentum sector. One important difference is that now the metric is coupled to the matter fields in $L_M$. In this frame the field equations are the same as those in GR with the following energy-momentum tensor 
\begin{equation}
T_{\mu \nu} = T^{\rm GR}_{\mu \nu} + T^{\phi}_{\mu \nu},
\end{equation}
where 
\begin{equation}
 T^{\phi}_{\mu \nu}=\frac{1}{\kappa}\left(\partial_\mu \phi \partial_\nu \phi - \frac{1}{2} g_{\mu \nu} \partial^\lambda \phi \partial_\lambda \phi\right).
\end{equation}
The KGE is given by 
\begin{equation}
 \Box \phi = -\kappa T \frac{d\ln A}{d \phi} + V'(\phi),
\label{kg2}
\end{equation}
with the prime denoting a derivative with respect to $\phi$. These forms of the equations will be of use shortly when we look at two main types of screening mechanisms, starting with a type that the potential used above can provide. We also note that the next two subsections closely follow the discussion found in \cite{Sakstein:2015oqa}. 


\subsection{Self Interaction Screening}
We will here discuss the type of screening that a potential term can provide. To start, let's consider the Poisson equation in the Newtonian limit upon the perturbed Minkowski spacetime (Eq.\ref{perturbedmink}), which is the relevant limit for any physics within the Solar System 
\begin{equation}
\nabla^2 \Phi_N = 4\pi G_N \rho.
\label{poissonN}
\end{equation}
This very special equation gives us a relation between the classical gravitational force and the density distribution. We can characterise the solar system limit by $\Phi_N \sim 10^{-6}$ which is the Newtonian gravitational potential of the Sun \cite{Sakstein:2015oqa}. This makes Eq.\ref{poissonN} an essential benchmark in constraining modified gravity. If we now look at the Einstein frame Poisson equation for JBD theory, we find an additional source term on the right hand side
\begin{equation}
\nabla^2 \Phi_N = 4\pi G_N \rho - \frac{ \nabla^2 \phi}{2}.
\label{poissonBD}
\end{equation}
This additional source leads to an additional force, the so called {\it 5th force}, which is not observed in any of the solar system tests. In the self interaction type of screening, one attempts to remove the source in high density regions by the inclusion of a suitable self interaction potential $V(\phi)$ in the Lagrangian. The KGE for a pressureless perfect fluid matter content is given by\footnote{ We also neglect the time derivative of $\phi$ as we are not considering cosmological timescales here.}
\begin{equation}
\nabla^2 \phi = 8 \pi G \alpha \rho- V'(\phi) \equiv V_{eff}'(\phi),
\label{KGBD}
\end{equation}
where we have defined an effective potential and $\alpha$ is defined as 
\begin{equation}
\alpha \equiv \frac{d \ln{A}}{d\phi}.
\end{equation}
When this potential finds a minimum ($V_{eff}'(\phi)=0$) the source for the scalar field is killed off and we recover Newtonian gravity. 
\newline
\newline
Consider now a spherically symmetric object with radius $R$. Say further that within some radius we will call the {\it screening radius}, $r_S$,  the effective potential is minimised so that we recover Newtonian gravity. For $r \ge r_S$ we can neglect the $V'(\phi)$ term in Eq.\ref{KGBD} since this is generically of the form $V'(\phi) \sim m^2_0\phi$ where $m_0^2 = V''(\phi) \sim H_0^{2}$ is the background field's mass, and is negligible compared to the field's Laplacian.  We can then integrate Eq.\ref{KGBD} to get  
\begin{equation}
F_5 = \nabla \phi = \alpha \frac{d\phi}{dr} = 2 \alpha^2 \frac{GM(R)}{r^2} \left[ 1- \frac{M(r_S)}{M(r)} \right],
\label{F5}
\end{equation}
where 
\begin{equation}
M(r)= 4\pi \int_0^r r^2 \rho(r) dr.
\end{equation}
It is apparent now from Eq.\ref{F5} that when the screening radius $r_S \sim R$ we have $F_5 \sim 0$ and the object is {\it self screened}. This all relies on us being able to find some potential $V(\phi)$ and conformal factor $A(\phi)$ such that $V_{eff}(\phi)$ finds a minimum at some $r_S$ which isn't too much smaller than $R$ so that the scalar field is only sourced by a very thin shell of the mass and $F_5$ remains small by Eq.\ref{F5}. Such forms have been found in $f(R)$ theories for example. 
\newline
\newline
$f(R)$ gravity (see \cite{Sotiriou:2008rp,Capozziello:2007ec} for reviews) is one of the most straightforward extensions to Einstein's theory, replacing the scalar curvature $R$ in the Einstein Hilbert action (Eq.\ref{ehaction}) with a general function $f(R)$
\begin{equation}
S = \int d^4x \sqrt{-g} \left[ \frac{R +f(R)}{2\kappa} + L_m \right].
\end{equation}
These models can utilise self interaction screening with the following forms for the self interaction potential and conformal factor
\begin{equation}
V(\phi)= \frac{m_0^{4+n}}{\phi^n}, \indent A(\phi)=e^{B\phi},
\end{equation}
$B$ being a constant and $m_0$ being the mass of the field. These are called {\it Chameleon type} \cite{Khoury:2003rn,Navarro:2006mw,Faulkner:2006ub} models. Such models have been well constrained using astrophysical tests (see \cite{Burrage:2017qrf} for a comprehensive review). Next we will look at another type of screening mechanism which has received a lot of attention in recent years, the Vainshtein mechanism. 
 
 
\subsection{Vainshtein Screening}
This type of screening finds its origins back in the early 70s, when Arkady Vainshtein proposed a screening coming from non-linearities which could satisfy Solar system constraints for Fierz-Pauli theory \cite{Vainshtein:1972sx}. 
\newline
\newline
Without a lengthy detour into history and the massive gravity motivations, we can start by considering Eq.\ref{poissonBD}. Essentially, what the previously discussed mechanism did was to use a scalar field self interaction term to cancel the scalar field's Laplacian on the right hand side of Eq.\ref{poissonBD}, leaving us with Newtonian gravity. The next obvious way to get rid of the fifth force term in some limit is to tweak the form of the Laplacian term itself. We can do this by introducing higher order derivatives of the scalar field in the Lagrangian Eq.\ref{BDL3}. We remind the reader that within Horndeski's theory we are protected against Ostragradsky ghost instabilities when doing this. 
\newline
\newline
Consider Eq.\ref{KGBD} without the self interaction term $V'(\phi)$, and a spherically symmetric object of mass M and radius $R$.  Integrating, this time from $0$ to $R$, we get the fifth force $F_5 = 2 \alpha^2 F_N$, where $F_N = GM/R^2$. 
Consider now replacing the Laplacian in Eq.\ref{KGBD} with some other operator, for instance in {\it cubic Galileon theory} (see Eq.\ref{gallileon} with $c_i=0$ for $i>3$ and $c_2 = -1/\kappa$, $c_3 =2 \Lambda_0/\kappa$)
\begin{equation}
L=  \frac{1}{\kappa}\left[\frac{R}{2} -\frac{1}{2} g^{\mu \nu} \partial_\mu \phi \partial_\nu \phi - \frac{1}{\Lambda_0^2} g^{\mu \nu} \partial_\mu \phi \partial_\nu \phi \Box \phi \right]+ L_M (A^2(\phi)g_{\mu \nu},\Psi_M),
\end{equation}
with $A(\phi) = e^{\alpha \phi}$. Taking the variation of this action with respect to $\phi$ gives the following KGE
\begin{equation}
\Box \phi + \frac{2}{\Lambda_0^2} \left[(\Box \phi)^2 - \nabla_\mu \nabla_\nu \phi \nabla^\mu \nabla^\nu \phi \right] = 8\pi\alpha G\rho.
\end{equation}
Again, considering a static, spherically symmetric mass distribution we get
\begin{equation}
\frac{1}{r^2} \frac{d}{dr} \left[ r^2 \frac{d\phi}{dr} + \frac{r}{\Lambda_0^2} \left(\frac{d\phi}{dr}\right)^2 \right] = 8\pi\alpha G\rho.
\label{cubicG1}
\end{equation} 
Integrating this equation 
\begin{equation}
\frac{d\phi}{dr} + \frac{1}{\Lambda_0^2 r} \left( \frac{d\phi}{dr} \right)^2 = \frac{F_5}{\alpha} + \frac{1}{\Lambda_0^2 r}\left(\frac{F_5}{\alpha}\right)^2 =  2\alpha F_N,
\label{cubicG2}
\end{equation}
where we substituted the scalar field derivative using the first part of Eq.\ref{F5}. Dividing Eq.\ref{cubicG2} by $F_N$ we get the following expression  
\begin{equation}
\frac{F_5}{F_N} + \left(\frac{r_V}{r}\right)^3 \left(\frac{F_5}{F_N}\right)^2 = 2\alpha^2,
\end{equation}
where 
\begin{equation}
r_V=\left(\frac{GM}{\alpha \Lambda_0^2}\right)^{\frac{1}{3}}.
\end{equation}
Screening is now apparent by taking the small and large limits of $r$  
\begin{align}
\frac{F_5}{F_N} = & 2\alpha^2 \left( \frac{r}{r_V}\right)^{3/2} ,     &r \ll r_V, \\ 
\frac{F_5}{F_N} =&  2 \alpha^2,   &r \gg r_V.
\end{align} 
$r_V$ is called the Vainshtein radius and it governs the regime of screening. An approximate calculation can show that for the Sun and for $\alpha  \sim 1$,  $r_V \sim 1pc$, which gives complete screening of the Solar System making the mechanism extremely efficient. 
\newline
\newline
To conclude we look at the forms of $\mu$, $\gamma_2$ and $\gamma_3$ (Eq.\ref{eq:poisson1} and Eq.\ref{eq:Perturb3}) in Horndeski theory and we take a better look at two examples of each type of screening mechanism. This will allow us to make predictions for the power spectrum and correlation function within these models. These forms have not been derived in the literature and so the next subsection are part of the original results presented in this thesis. They can be found in \cite{Bose:2016qun}. 

\section{Generalised Cosmological Perturbations} 
We have seen how modifications to GR can be crafted cleverly to avoid over 100 years of stringent local experimental verifications. Further, such models can provide a large distance modification of gravity allowing for accelerating solutions. This takes us back to the cosmological setting. Let us start by considering the perturbed flat FLRW metric
\begin{equation}
ds^2 = - (1 + 2 \Phi) dt^2 +  a(t)^2 (1 - 2 \Psi) \delta_{ij} dx^i dx^j,
\label{pertFLRW}
\end{equation}
and apply it to Horndeski's theory. Specifically, we will study the effect scalar field perturbations have on the matter and velocity perturbations
\begin{equation}
\phi  = \bar{\phi}(t) + \delta \phi(t,\bfx),
\end{equation}
where the overbar represents a background value. Let us also make the simplifying assumption that the perturbations are slowly varying with time when compared to their spatial variation allowing us to neglect their time derivatives. This assumption is called the {\it quasi-static appoximation} and is well justified far within the Hubble horizon where the cosmological evolution is slow. This can be seen by considering the KGE for Eq.\ref{BDL3} with $V=0$ in an FLRW background with a matter content described by a perfect fluid 
\begin{equation} 
\ddot{\phi} + 3H\dot{\phi} - c_s^2 \nabla^2 \phi = \frac{\kappa}{2} \bar{\rho}_m \delta.
\end{equation} 
$c_s^2$ is the speed of the perturbations, in many theories being around unity. In Fourier space we have $\ddot{\phi} + 3H \dot{\phi} - c_s^2 k^2 \phi$ on the left hand side. Then we see for small scales (corresponding to large $k$) and late times (corresponding to small $H$) we see the third term dominates. This approximation's validity slowly comes into question as astronomical surveys begin to explore larger volumes of the universe and also for particular theories where $c_s^2$ is significantly less than unity (see \cite{Sawicki:2015zya} for limits of its validity in modified gravity theories). For models, epochs and scales considered in this work, we find it is a good approximation \cite{Bose:2014zba,Winther:2015pta}. 
 \newline
 \newline
The equations of motion for the metric perturbations are found by solving the field equations for the Horndeski Lagrangian (Eq.\ref{Horndeski1}) using the metric given in Eq.\ref{pertFLRW}. In Fourier space these are given by \cite{Takushima:2015iha}
\begin{align}
-k^2&({\cal F}_{\cal T} \Psi(\bfk;a) -{\cal G}_{\cal T} \Phi(\bfk;a) - A_1 Q(\bfk;a)) \nonumber \\ 
= & \frac{B_1}{2a^2H^2}\Gamma[\bfk,Q,Q;a] + \frac{ B_3}{a^2 H^2} \Gamma[\bfk,Q,\Phi;a], \label{eomhorn1} \\ \nonumber \\ 
-k^2&({\cal G}_{\cal T} \Psi(\bfk;a) + A_2 Q(\bfk;a)) -\frac{a^2}{2}\bar{\rho}_m \delta(\bfk;a) \nonumber \\
 =& -\frac{B_2}{2a^2H^2}\Gamma[\bfk,Q,Q;a] - \frac{ B_3}{a^2 H^2} \Gamma[\bfk,Q,\Psi;a] -\frac{C_1}{3a^4H^4}\Xi_1[\bfk,Q,Q,Q;a], \label{eomhorn2}  
 \end{align}
along with the KGE
 \begin{align}
-k^2 &(A_0 Q(\bfk;a)-A_1 \Psi(\bfk;a) - A_2 \Phi(\bfk;a)) = \nonumber \\
  & -\frac{B_0}{a^2H^2}\Gamma[\bfk,Q,Q;a] + \frac{ B_1}{a^2 H^2} \Gamma[\bfk,Q,\Psi;a] +  \frac{ B_2}{a^2 H^2} \Gamma[\bfk,Q,\Phi;a] \nonumber \\ &+ \frac{ B_3}{a^2 H^2} \Gamma[\bfk,\Psi,\Phi;a]  
 +\frac{C_0}{3a^4H^4}\Xi_1[\bfk,Q,Q,Q;a] +\frac{C_1}{3a^4H^4}\Xi_1[\bfk,Q,Q,\Phi;a] \nonumber \\
& + M_1Q(\bfk;a) + \frac{M_2}{a^2H^2} Q(\bfk;a)^2 + \frac{M_3}{a^4H^4}Q(\bfk;a)^3,
\label{eomhorn3}
\end{align}
where we have parametrized the scalar field perturbation as $Q=\delta \phi/ (a\bar{\phi}')$, again the prime denoting scale factor derivative. We have defined
\begin{align}
\Gamma[\bfk,Z_1,Z_2;a] =& \frac{1}{(2\pi)^3}\int d^3\bfk_1 d^3 \bfk_2 \delta_D(\bfk-\bfk_{12})\lambda(\bfk_1,\bfk_2)  Z_1(\bfk_1;a) Z_2(\bfk_2;a), \\ 
\Xi_1[\bfk,Z_1,Z_2,Z_3;a] = & \frac{-1}{(2\pi)^6}\int d^3\bfk_1 d^3 \bfk_2 d^3\bfk_3\delta_D(\bfk-\bfk_{123}) \nonumber \\
& \times \left[k_1^2 k_2^2 k_3^2 - 3k_1^2(\bfk_2\cdot\bfk_3)^2 +2(\bfk_1\cdot \bfk_2)(\bfk_2 \cdot \bfk_3) (\bfk_3 \cdot \bfk_1) \right] \nonumber \\ & \times Z_1(\bfk_1;a) Z_2(\bfk_2;a)Z_3(\bfk_3;a), \\
\Xi_2[\bfk,Z_1,Z_2,Z_3;a] = & \frac{-1}{(2\pi)^6}\int d^3\bfk_1 d^3 \bfk_2 d^3\bfk_3\delta_D(\bfk-\bfk_{123}) \left[ k_1^2 k_2^2 k_3^2 - k_3^2(\bfk_1\cdot\bfk_2)^2 \right. \nonumber \\
&  \left. -2k_1^2(\bfk_2 \cdot \bfk_3)^2 +2(\bfk_1\cdot \bfk_2)(\bfk_2 \cdot \bfk_3) (\bfk_3 \cdot \bfk_1) \right] \nonumber \\ & \times Z_1(\bfk_1;a) Z_2(\bfk_2;a)Z_3(\bfk_3;a), 
\end{align}
where $ \lambda(\bfk_1, \bfk_2) =k_1^2 k_2^2 - (\bfk_1 \cdot \bfk_2)^2 $. Scalar field potential terms $M_1(a),M_2(a)$ and $M_3(a)$ have been introduced into the KGE so that we can accommodate chameleon models such as $f(R)$ gravity. The functions $A_0$, $A_1$, $A_2$, $B_0$, $B_1$, $B_2$, $B_3$, $C_0$ and $C_1$ depend only on the scale factor and are given in Appendix B along with definitions for ${\cal F}_{\cal T}$ and ${\cal G}_{\cal T}$. These are reproduced from the Appendices of \cite{Takushima:2015iha}. One can also look at \cite{Takushima:2015iha,Takushima:2013foa,Kimura:2011dc,DeFelice:2011hq} for the pure Horndeski case with $M_1 = M_2 = M_3 =0$. We have also defined the $Z_i$ as 
\begin{align}
Z_0(k) &= \frac{2 a^2}{k^2} \frac{C_1 {\cal T}}{{\cal Z}}, \quad  
Z_1(k) =  \frac{2a^2 }{3k^2} \frac{3 C_0 {\cal T}
+ C_1 {\cal R} }{ {\cal Z}}, \quad
Z_2(k) = \frac{a^2 }{k^2} \frac{2 {\cal T}}{{\cal Z}}, \nonumber \\  
Z_3(k) & = \frac{a^2 }{k^2} \frac{2 B_3 {\cal T}}{{\cal Z}}, \quad Z_4(k) = \frac{2 a^2 }{k^2} 
\frac{B_1 {\cal T}+ B_3 {\cal R}}{{\cal Z}}, \quad 
Z_5(k) = \frac{2 a^2 }{k^2} 
\frac{B_3 {\cal T}}{{\cal Z}}, \nonumber \\  
Z_6(k) & = \frac{2 a^2 }{k^2} 
\frac{B_2 {\cal T}+ B_3 {\cal S} }{{\cal Z}}, \quad
Z_7(k)  = \frac{2 a^2 }{k^2} 
\frac{B_2 {\cal T}+B_3 {\cal S}}{{\cal Z}}, \quad \nonumber \\ 
Z_8(k) & = \frac{2 a^2 }{k^2} 
\frac{B_1 {\cal T} + B_3 {\cal R}}{{\cal Z}},  \nonumber \\  
Z_9(k) & = \frac{2 a^2 }{k^2} 
\frac{-2 B_0 {\cal T} +B_1 {\cal S}+B_2 {\cal R}}{{\cal Z}},\quad
Z_{10}(k)  = \frac{4 a^2 }{k^2} 
\frac{{\cal T}}{{\cal Z}}.
\end{align} 
\newline
\newline
In \cite{Takushima:2015iha}, the dynamical equations above, together with the continuity and Euler equations are solved perturbatively up to third order using the separability ansatz discussed in Chapter 2. {\tt MG-Copter} calculates the SPT kernels by solving the evolution equations numerically as discussed in Appendix A. For this purpose, we follow a perturbative approach where as usual we assume $\delta \ll 1$ and then expand our field perturbations, $\Psi,\Phi$ and $Q$, in terms of increasing orders of $\delta$ in order to find forms for $\mu(k;a)$, $\gamma_2(\bfk, \bfk_1,\bfk_2;a)$ and$\gamma_3(\bfk, \bfk_1,\bfk_2,\bfk_3;a)$ initially introduced in Eq.\ref{eq:Perturb3} and Eq.\ref{eq:poisson1} in Chapter 2, justifying the form proposed there. 
\newline
\newline
At first order, the solutions to the field equations and KGE are given by
\begin{align}
\Phi_1 & = -\frac{{\cal R}}{{\cal Z}} \frac{a^2}{k^2} \bar{\rho}_m \delta 
\equiv  \frac{\bar{\rho}_m \delta}{\Upsilon_{\Phi}(k)}, \\
\Psi_1 &=-\frac{{\cal S}}{{\cal Z}} \frac{a^2}{k^2} \bar{\rho}_m \delta 
 \equiv  \frac{\bar{\rho}_m \delta}{\Upsilon_{\Psi}(k)},  \\
Q_1 &= -\frac{{\cal T}}{{\cal Z}}
 \frac{a^2}{k^2} \bar{\rho}_m \delta 
\equiv  \frac{\bar{\rho}_m \delta}{\Upsilon_{Q}(k)},
\end{align}
where we have defined the quantities $\Upsilon_{I}(k)$, $I\in \{ \Phi, \Psi$ , $Q$\}. The reliant quantities are 
\begin{align}
{\cal R}= \tilde{A}_0 {\cal F}_{\cal T} & - A_1^2, \quad 
 {\cal S}= \tilde{A}_0 {\cal G}_{\cal T} + A_1 A_2, \quad 
{\cal T}= A_1 {\cal G}_{\cal T} +  A_2 {\cal F}_{\cal T}, \quad  \nonumber \\ 
&{\cal Z}= 2 (A_2^2 {\cal F}_{\cal T} + 2 A_1 A_2 {\cal G}_{\cal T} + \tilde{A}_0 {\cal G}_{\cal T}^2),
\end{align}
and
\begin{equation}
\tilde{A}_0 = A_0 + \frac{M_1}{k^2}. 
\end{equation}
At second order, the solutions are 
\begin{align}
\Phi_2 &= 
\int\frac{d^3\bfk_1d^3\bfk_2}{(2\pi)^3}\,
\delta_{\rm D}(\bfk-\bfk_{12}) \Gamma_{2 \Phi} (\bfk, \bfk_1, \bfk_2, k_{12})
\delta(\bfk_1)\,\delta(\bfk_2), \\
\Psi_2 &=
\int\frac{d^3\bfk_1d^3\bfk_2}{(2\pi)^3}\,
\delta_{\rm D}(\bfk-\bfk_{12}) \Gamma_{2 \Psi} (\bfk, \bfk_1, \bfk_2, k_{12})
\delta(\bfk_1)\,\delta(\bfk_2), \\
Q_2 &=
\int\frac{d^3\bfk_1d^3\bfk_2}{(2\pi)^3}\,
\delta_{\rm D}(\bfk-\bfk_{12}) \Gamma_{2 Q} (\bfk, \bfk_1, \bfk_2, k_{12})
\delta(\bfk_1)\,\delta(\bfk_2), 
\end{align}
where
\begin{align}
\Gamma_{2 I}(\bfk, \bfk_1, \bfk_2, k_{12}) 
=
\frac{\rho_m^2}{a^4H^2} \Big[ & \left(
\frac{W_{1 I}(k_{12})}{\Upsilon_{\Psi}(k_1) \Upsilon_{Q}(k_2)} 
+ \frac{W_{2 I}(k_{12})}{\Upsilon_{\Phi}(k_1) \Upsilon_{Q}(k_2)}  
 \right. \nonumber \\ & \left.+ \frac{W_{3 I}(k_{12})}{\Upsilon_{\Psi}(k_1) \Upsilon_{\Phi}(k_2)}  
+\frac{W_{4 I}(k_{12})}{\Upsilon_{Q}(k_1) \Upsilon_{Q}(k_2)} 
\right) 
\lambda(\bfk_1, \bfk_2) \nonumber \\
&+ \frac{W_{5 I}(k_{12})}{\Upsilon_{Q}(k_1) \Upsilon_{Q}(k_2)} M_2
\Big],
\end{align}
and 
\begin{align}
W_{1 \Phi}(k) &= \frac{2 a^2}{k^2} 
\frac{B_1 {\cal T} + B_3 {\cal R}}{{\cal Z}}, \quad
W_{2 \Phi}(k) = \frac{2 a^2}{k^2}
\frac{B_2 {\cal T} + B_3 {\cal S}}{{\cal Z}}, \quad \nonumber \\
W_{3 \Phi}(k) & = \frac{2 a^2 }{k^2}
\frac{B_3 {\cal T}}{{\cal Z}},
\nonumber  \\
W_{4 \Phi}(k) &=\frac{a^2}{k^2}
\frac{-2 B_0 {\cal T}+ B_1 {\cal S} + B_2 {\cal R} }{{\cal Z}},\quad
W_{5 \Phi}(k) =\frac{2 a^2}{k^2}
\frac{{\cal T}}{{\cal Z}},   \\ \nonumber \\
W_{1 \Psi}(k) &= \frac{2 a^2}{k^2} 
\frac{ A_2 B_1 {\cal G}_{\cal T} +  B_3 {\cal S}}{{\cal Z}}, \quad
W_{2 \Psi}(k) = \frac{2 a^2}{k^2}
\frac{A_2 B_2 {\cal G}_{\cal T} - A_2^2 B_3}{{\cal Z}}, 
\nonumber  \\
W_{3 \Psi}(k) &= \frac{2 a^2}{k^2}
\frac{A_2 B_3 {\cal G}_{\cal T}}{{\cal Z}},
\nonumber \\
W_{4 \Psi}(k) &=\frac{a^2}{ k^2}
\frac{ - 2 A_2 B_0 {\cal G}_{\cal T} - A_2^2 B_1 + B_2 {\cal S} }{{\cal Z}},\quad
W_{5 \Psi}(k) =\frac{2 a^2}{k^2}
\frac{ A_2 {\cal G}_{\cal T}}{{\cal Z}}, \\ \nonumber \\
W_{1 Q}(k) &= \frac{2 a^2}{k^2} 
\frac{ -B_1 {\cal G}_{\cal T} + B_3 {\cal T}}{{\cal Z}}, \quad
W_{2 Q}(k) = \frac{2 a^2}{k^2}
\frac{-B_2 {\cal G}_{\cal T}^2 + A_2 B_3 {\cal G}_{\cal T}}{{\cal Z}}, 
\nonumber  \\
W_{3 Q}(k) & = -\frac{2 a^2}{k^2}
\frac{B_3 {\cal G}_{\cal T}^2}{{\cal Z}},
\nonumber \\
W_{4 Q}(k) &=\frac{a^2 }{ k^2}
\frac{ 2 B_0 {\cal G}_{\cal T}^2 +  A_2 B_1 {\cal G}_{\cal T} +B_2 {\cal T}}{{\cal Z}},\quad
W_{5 Q}(k) =-\frac{2 a^2 }{k^2}
\frac{{\cal G}_{\cal T}^2}{{\cal Z}}.
\end{align} 
Lastly, at third order we only give the relevant potential perturbation for the Poisson equation
\begin{equation}
\Phi_3=\int\frac{d^3\bfk_1d^3\bfk_2d^3\bfk_3}{(2\pi)^6}
\delta_{\rm D}(\bfk-\bfk_{123})
\Gamma_{3 \Phi}(\bfk, \bfk_1, \bfk_2, \bfk_3)
\delta(\bfk_1)\,\delta(\bfk_2)\,\delta(\bfk_3),
\end{equation}
where 
\begin{align}
&\Gamma_{3 \Phi}(\bfk, \bfk_1, \bfk_2, \bfk_3)
= \nonumber \\
& \frac{ \rho_m^3}{3a^6H^4 } 
\left[    \frac{Z_0(k)}{\Pi_Q(k_1) \Upsilon_{Q}(k_2) \Upsilon_{\Phi}(k_3)} \xi_h(\bfk_1, \bfk_2, \bfk_3) + 
\frac{3Z_1(k)}{\Pi_Q(k_1) \Pi_Q(k_2) \Pi_Q(k_3)}
\xi_h (\bfk_1, \bfk_2, \bfk_3)
 \right.
 \nonumber  \\
&\left.
+\frac{3Z_2(k)}{\Pi_Q(k_1) \Pi_Q(k_2) \Pi_Q(k_3)} M_3  + (\mbox{cyclic perm.}) \right] \nonumber \\
&  + \frac{\rho_m}{3a^4H^2} \left[ Z_3(k) \frac{\lambda(\bfk_1, \bfk_{23}, k) \Gamma_{2 \Psi} (\bfk, \bfk_2, \bfk_3, k_{23})}{\Upsilon_{\Phi}(k_1)}  + Z_4(k) \frac{\lambda(\bfk_1, \bfk_{23}, k) \Gamma_{2 \Psi} (\bfk, \bfk_2, \bfk_3, k_{23})}{\Upsilon_{Q}(k_1)}  \right.  \nonumber \\  
&\left. +
Z_5(k) \frac{\lambda(\bfk_1, \bfk_{23}, k) \Gamma_{2 \Phi} (\bfk, \bfk_2, \bfk_3, k_{23})}{\Upsilon_{\Psi}(k_1)} 
+ 
Z_6(k) \frac{\lambda(\bfk_1, \bfk_{23}, k) \Gamma_{2 \Phi} (\bfk, \bfk_2, \bfk_3, k_{23})}{\Upsilon_{Q}(k_1)} \right. \nonumber \\
& \left.+ 
Z_7(k) \frac{\lambda(\bfk_1, \bfk_{23}, k) \Gamma_{2 Q} (\bfk, \bfk_2, \bfk_3, k_{23})}{\Upsilon_{\Phi}(k_1)} 
+
Z_8(k) \frac{\lambda(\bfk_1, \bfk_{23}, k) \Gamma_{2 Q} (\bfk, \bfk_2, \bfk_3, k_{23})}{\Upsilon_{\Psi}(k_1)} \right. \nonumber  \\
&\left. +
Z_9(k) \frac{\lambda(\bfk_1, \bfk_{23}, k) \Gamma_{2 Q} (\bfk, \bfk_2, \bfk_3, k_{23})}{\Upsilon_{Q}(k_1)} 
+
Z_{10}(k) \frac{M_2 \Gamma_{2 Q} (\bfk, \bfk_2, \bfk_3, k_{23})}{\Upsilon_{Q}(k_1)} \right. \nonumber \\ & + (\mbox{cyclic perm.}) \Big],
\end{align}
where
\begin{align}
\xi_h(\bfk_1, \bfk_2, \bfk_3) = &   
k_1^2 k_2^2 k_3^2- k_1^2 (\bfk_2 \cdot \bfk_3)^2 + k_2^2 (\bfk_3 \cdot \bfk_1)^2 + k_3^2 (\bfk_1 \cdot \bfk_2)^2 \nonumber \\ &-2 (\bfk_1 \cdot \bfk_2) (\bfk_2 \cdot \bfk_3)(\bfk_3 \cdot \bfk_1).
\end{align}
Finally, the Euler equation modification parameters given in Chapter 2, $\mu(k;a)$,  $\gamma_2(\bfk, \bfk_1,\bfk_2;a)$ and $\gamma_3(\bfk, \bfk_1,\bfk_2,\bfk_3;a)$ can then be related to $\Upsilon_\Phi(k;a)$,  $\Gamma_{2\Phi}(\bfk, \bfk_1,\bfk_2;a)$ and $\Gamma_{3\Phi}(\bfk, \bfk_1,\bfk_2,\bfk_3;a)$ through order by order comparisons. We find the relations are 
\begin{align}
\mu(k;a) & = -\frac{k^2}{a^2} \frac{2}{\kappa \Upsilon_\Phi(k;a)}, \label{hornmu} \\
\gamma_2(\bfk, \bfk_1,\bfk_2;a) &= -\frac{k^2}{a^2H^2} \Gamma_{2\Phi} (\bfk, \bfk_1,\bfk_2;a), \label{horng2} \\ 
\gamma_3(\bfk, \bfk_1,\bfk_2,\bfk_3;a) &=  -\frac{k^2}{a^2H^2} \Gamma_{3\Phi} (\bfk, \bfk_1,\bfk_2,\bfk_3;a). \label{horng3}
\end{align}
The functions on the right hand sides are expressed in terms of $A_0$,$A_1$,$A_2$,$B_0$,$B_1$,  $B_2$,$B_3$,$C_0$ and $C_1$  which are in turn relatable to the Horndeski Lagrangian functions. Within this framework we can solve for the density contrast and velocity divergence up to 3rd order in any theory within the Horndeski class with generalised potential terms $M_1,M_2$ and $M_3$, allowing us to construct the power spectrum and correlation function for any of these theories. This is the first step in having a quasi non-linear prediction that can be compared with real galaxy clustering data and realising cosmological constraints on viable competing theories. Next we look at two theories which exhibit screening of the two different types discussed in the previous section and which fall under the theoretical umbrella discussed here. We provide specific forms for $\mu$,$\gamma_2$ and $\gamma_3$ in these models which will be used in later chapters to compare to computer generated cosmological simulations. 
 

\subsubsection{Chameleon Example: Hu-Sawicki $f(R)$ Gravity}
\begin{figure}[H]
  \captionsetup[subfigure]{labelformat=empty}
  \centering
  \subfloat[]{\includegraphics[width=15cm, height=6.6cm]{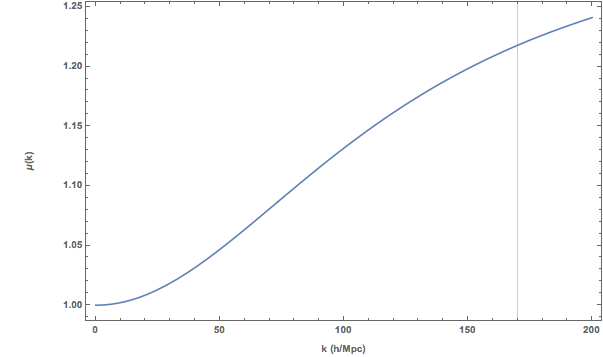}} \quad
  \caption{A plot of the linear enhancement of the Newtonian potential in the Hu-Sawicki model with $p_1=0.0001 \mbox{Mpc}^2/h^2$ against wave mode $k$ at $z=0$. The dotted line indicates the scale of the Milky Way galaxy. }
\label{muf4}
\end{figure}

Here we provide a more detailed look at the $f(R)$ Chameleon screened model by Hu and Sawicki \cite{Hu:2007nk} which provides a rare and well studied \cite{Song:2015oza,Hammami:2015iwa,Dossett:2014oia,Lombriser:2013wta,Hellwing2013,Zhao:2013dza,Okada:2012mn,Li:2011pj,Lombriser:2010mp,Schmidt:2009am,Brax:2008hh,Song:2007da} example of self-screening. Firstly, the $f(R)$ Lagrangian is given by 
\begin{equation}
L_{fR} = \frac{1}{2\kappa}\left[R + f(R)\right] + L_M(\Psi_M),
\end{equation}
where $f(R)$ is a general function of the Ricci scalar $R$. The Hu-Sawicki form for $f(R)$ is given by 
\begin{equation}
f(R)  = -m^2 \frac{c_1 (R/m^2)^n}{c_2(R/m^2)^n+1},
\label{husawicki}
\end{equation}
where $m$ is a mass scale for the theory taken to be 
\begin{equation} 
m^2 \equiv \frac{\kappa \rho_{m,0}}{3} = \Omega_{m,0} H_0^2, 
\label{massscale}
\end{equation}
with $\rho_{m,0}$ being the average density today. $c_1$ and $c_2$ are dimensionless parameters to be determined soon. We will take $n=1$ for the remainder of this work. $f(R)$ models are known to be equivalent to a STT with a non-trivial potential, which becomes evident when we look at the trace of the modified field equations
\begin{equation}
\Box f_R = \frac{1}{3} \left[R - f_R R + 2f(R) -\kappa \rho_m\right] \equiv \frac{\partial V_{eff}}{\partial f_R},
\label{kgeforfr}
\end{equation}
where $f_R = df(R)/dR$ and $\rho_m$ is the matter density of the universe (again we have assumed a pressureless perfect fluid, matter dominated universe). Eq.\ref{kgeforfr} can be seen as the KGE for $f_R$ with a potential term, $V_{eff}$. For a background expansion close to LCDM, we require screening. The condition for this is an extremum of the potential (see previous section)
\begin{equation}
0 = R-f_R R + 2f - \kappa \rho_m.
\end{equation} 
and we also look for environments where $R\gg1$ ($f_R \ll 1$ or $f\ll R$ by Eq.\ref{husawicki}) which results in 
\begin{equation}
R = \kappa \rho_m -2f \approx -\kappa \rho_m-m^2 c_1/c_2,
\label{riccifr}
\end{equation}
where we have used Eq.\ref{husawicki} in the small $m^2/R$ limit in the approximate equality. In these limits, the KGE in the cosmological setting is 
\begin{equation} 
\Box f_R = \frac{1}{3} \left[R -\kappa \rho_m\right].
\end{equation}
We consider perturbations of $f_R$ as 
\begin{equation}
\phi = \delta f_R \equiv f_R - \bar{f}_R,
\end{equation}
where the bar represents the quantity is calculated on the background. Again assuming quasi-static perturbations, this results in the following dynamical equation for $\phi$
\begin{equation}
\frac{3}{a^2} \nabla^2 \phi = -\kappa \bar{\rho}_m \delta + \delta R, \quad \delta R \equiv R(f_R) - R(\bar{f}_R).
\end{equation}
By Taylor expanding $\delta R$ about the background up to 3rd oder in $\phi$ in Fourier space and using the general STT form of the Poisson equation
\begin{equation}
-\frac{k^2}{a^2} \Phi = \frac{\kappa \bar{\rho}_m \delta }{2} - \frac{k^2}{2a^2} \phi, 
\end{equation}
 we can then compare to Eq.\ref{eq:poisson1} to get 
\begin{align}
\mu(k;a) = &1 + \left(\frac{k}{a}\right)^2\frac{1}{3\Pi(k;a)},\\
\gamma_2(k,\bfk_1,\bfk_2;a) =& -\left(\frac{k}{a}\right)^2\frac{1}{12\Pi(k;a)}\left(\frac{\kappa \rho_m}{3H}\right)^2 \frac{M_2(\bfk_1,\bfk_2;a)}{\Pi(k_1;a)\Pi(k_2;a)}, 
\label{frmugam2}
\end{align}
and
\begin{align}
&\gamma_3(k,\bfk_1,\bfk_2,\bfk_3;a)=  -\left(\frac{k}{a}\right)^2\frac{1}{32H^3\Pi(k;a)}\left(\frac{\kappa \rho_m}{3}\right)^3 \nonumber \\    & \times  \Big[\frac{M_2(\bfk_2,\bfk_3;a)M_2(\bfk_1,\bfk_{23};a) - M_3(\bfk_1,\bfk_2,\bfk_3;a) \Pi(k_{23};a)}{\Pi(k_{23};a)\Pi(k_1;a)\Pi(k_2;a)\Pi(k_3;a)} \Big],
\label{frgam3}
\end{align}
where
\begin{align}
\Pi(k;a) =& \left(\frac{k}{a}\right)^2 + \frac{\bar{R}_f(a)}{3},\\
M_2(\bfk_1,\bfk_2;a) = & \bar{R}_{ff}(a), \\
M_3(\bfk_1,\bfk_2,\bfk_3;a) = & \bar{R}_{fff}(a),
\end{align}
where $\bar{R}_{f} \equiv d\bar{R}(f_R)/df_R$. To determine these values let's first look at the value of the Ricci scalar for a flat FLRW metric
\begin{equation}
\bar{R} = 12 H^2 +6 \dot{H} = 3H_0^2\left(\frac{\Omega_{m,0}}{a^3}+ 4(1-\Omega_{m,0})\right),
\label{backr}
\end{equation}
where the 2nd equality is valid for LCDM. Comparing the 2nd equality to Eq.\ref{riccifr} and using Eq.\ref{massscale}, we note that to achieve an expansion history close to that of LCDM we should choose 
\begin{equation}
\frac{c_1}{c_2} \approx 6\frac{\Omega_{\Lambda,0}}{\Omega_{m,0}}.
\end{equation}
This gives a value for $\bar{R}_0$ ($a=1$) of 
\begin{equation}
\bar{R}_0 =  H_0^2 \Omega_{m,0}\left(\frac{12}{\Omega_{m,0}} -9\right).
\label{R0}
\end{equation} 
Second, taking the derivative of Eq.\ref{husawicki} with the necessary condition $R\gg m^2$ gives a value for the scalar field $f_R$ 
\begin{equation}
f_R = -\frac{c_1}{c_2^2} \left(\frac{m^2}{R}\right)^2,
\label{fofr}
\end{equation}
which we combine with Eq.\ref{R0} to get a useful parametrisation for the parameter combination $c_1/c_2^2$ 
\begin{equation}
\frac{c_1}{c_2^2} = -\bar{f}_{R0} \left(\frac{R_0}{m^2}\right)^2 = -f_{R0} \left(\frac{12}{\Omega_{m,0}} -9\right)^{2}.
\end{equation}
Commonly, astronomical surveys aim to constrain the present day value of the background scalar field $\bar{f}_{R0}$, and in later chapters we choose values of this parameter when testing our framework for $f(R)$. The scalar field is then given by 
\begin{align}
{f}_R = &  \bar{f}_{R0} \left(\frac{R_0}{m^2}\right)^2 \left(\frac{m^2}{{R}}\right)^2, \nonumber \\ 
 =&  \bar{f}_{R0} \left(\frac{R_0}{R}\right)^2,  \nonumber \\ 
 \bar{f}_R = & \frac{\bar{f}_{R0}}{9} \left(\frac{12}{\Omega_{m,0}} -9\right)^2 \left[\frac{1}{a^3}+ 4\left(\frac{1}{\Omega_{m,0}}-1\right)\right]^{-2} \label{backfr},
 \end{align}
where we have used Eq.\ref{backr} and Eq.\ref{R0} in the last equality to give an expression for the background. Inverting the 2nd line above and taking consecutive derivatives with respect to $f_R$ gives 
\begin{align}
R = & +\bar{R}_0\sqrt{\frac{\bar{f}_{R0}}{f_R}},  \\
R_f = &  -\frac{\bar{R}_0}{2} \sqrt{\frac{\bar{f}_{R0}}{f_{R}^3}},  \\
R_{ff} = & +\frac{3\bar{R}_0}{4} \sqrt{\frac{\bar{f}_{R0}}{f_{R}^5}}, \\
R_{fff} = & -\frac{15\bar{R}_0}{8} \sqrt{\frac{\bar{f}_{R0}}{f_{R}^7}}.
\end{align} 
Now all we need to do is plug in Eq.\ref{backfr} to get the background expressions and use these in Eq.\ref{frmugam2}
 and Eq.\ref{frgam3}. Finally we obtain the following linear and non-linear interaction terms  
 \begin{align}
\mu(k;a) = &1 + \left(\frac{k}{a}\right)^2\frac{1}{3\Pi(k;a)}, \\ 
\gamma_2(k,\bfk_1,\bfk_2;a)  = &- \frac{9}{48}\left(\frac{k}{aH}\right)^2\left(\frac{\Omega_{m,0}}{a^3}\right)^2 \nonumber \\ & \times \frac{[\Omega_{m,0} -4a^3(\Omega_{m,0}-1)]^5}{a^{15}p_1^2 (3\Omega_{m,0}-4)^4}\frac{1}{\Pi(k;a)\Pi(k_1;a)\Pi(k_2;a)}, \label{frg2} \\ 
\gamma_3(k,\bfk_1,\bfk_2,\bfk_3;a)   = &  \left(\frac{k}{aH}\right)^2 \left(\frac{\Omega_{m,0}}{a^3}\right)^3 \nonumber \\  \times & \frac{1}{36\Pi(k;a)\Pi(k_1;a)\Pi(k_2;a)\Pi(k_3;a)\Pi(k_{23};a)}  \nonumber \\ 
 \times  & \left[-\frac{45}{8} \frac{\Pi(k_{23};a)}{a^{21}p_1^3}\left( \frac{[\Omega_{m,0} - 4a^3(\Omega_{m,0}-1)]^7}{(3\Omega_{m,0}-4)^6} \right) \right. \nonumber  \\ & \left. +\left( \frac{9}{4a^{15} p_1^2} \frac{[\Omega_{m,0}-4a^3(\Omega_{m,0}-1)]^5}{(3\Omega_{m,0}-4)^4} \right)^2\right],\label{frg3}
\end{align}
where
\begin{equation}
\Pi(k;a) = \left(\frac{k}{a}\right)^2+\frac{[\Omega_{m,0} - 4a^3(\Omega_{m,0}-1)]^3}{2p_1a^9(3\Omega_{m,0}-4)^2},
\end{equation}
and $p_1 = |\bar{f}_{R0}|/H_0^2$. This is the chosen parametrisation of this theory within {\tt MG-Copter} and we will use values of this in later chapters. We note that the linear Poisson function $\mu(k;a)$ is scale dependent which prevents us from making a separability assumption as in the EdS case. In this case the perturbations must be solved numerically (see Appendix A). The higher order functions $\gamma_2$ and $\gamma_3$ are also functions of scale $k$ as well as integrated vectors $\bfk_1,\bfk_2$ and $\bfk_3$. These functions are responsible for screening via the Chameleon mechanism which can be seen by considering only $\mu(k;a)$. At large scales $k\ll1$, $\mu \rightarrow 1$ and we recover GR, but at small scales the $k$-dependent term becomes important and we are left with a small scale modification of gravity. In fact, one can easily take the $k\gg1$ limit of this function for which one gets $\mu \rightarrow 4/3$, enhancing the Newtonian gravitational force by $1/3$. This is in obvious violation of solar system tests. To get around this, one requires non-linear interaction terms, quantified at leading order by $\gamma_2$ and $\gamma_3$, to suppress this enhancement. Fig.\ref{muf4} shows the value of $\mu$ for $z=0,\Omega_{m,0}=0.313$ and $p_1 =0.0001 \mbox{Mpc}^2/h^2$. The dotted line marks $k=170 h$/Mpc. This corresponds to a length scale of $r\approx 55$kpc \footnote{$k\approx 2\pi/r$.}, an estimate for the size of the Milky Way. It is here that we expect screening to completely kill the enhancement of the Newtonian potential. We will finish this section with a Vainshtein screened example. 

\subsubsection{Vainshtein Example: DGP Gravity}

Another interesting modified gravity theory besides the Galileon models which employs Vainshtein screening was proposed by Dvali, Gabadadze and Porrati (DGP) in \cite{Dvali:2000hr}. The model assumes we live on a 4-dimensional manifold embedded in a 5D spacetime called the bulk.  At the time, this theory gained a lot of attention for not requiring a cosmological constant to explain cosmic acceleration. It does this by having gravity `dilute' at large distances through the 5th dimension. The DGP action can be written as follows
\begin{equation}
S_{DGP} = \frac{1}{32 \pi r_c} \int d^4x \sqrt{-g_5} R_5 + \int d^4 x \sqrt{-g}( \frac{R}{2 \kappa} + L_M),
\label{DGP}
\end{equation}
where $R_5$ and $g_5$ are the Ricci Scalar and metric in 5D, while $L_M$ is the matter Lagrangian confined to the 4D manifold. $r_c$ is the model's free parameter which represents the scale at which we cross from the 4D gravity to the 5D gravity regime. Applying this model to a FLRW cosmology we obtain the Friedman equation  
\begin{equation}
\epsilon \frac{H}{r_c} = H^2 - \frac{\kappa}{3} \rho_m,
\label{DGPfried}
\end{equation}
where $\epsilon = \pm 1$ . This gives us two distinct solutions.  The $+$ solution is called the {\it self-accelerating branch}. In this branch we have $H=1/r_c$ at late times when $\rho_m$ becomes negligible (see \cite{Deffayet:2000uy} for a derivation) which gives an accelerating solution ($a \propto e^{t/r_c}$) . The $-$ solution requires a cosmological constant to achieve this and so does not solve the problem we are concerned with. It is interesting nevertheless because of its screening properties.
\newline
\newline
In the 4D gravity limit, one recovers a JBD-like theory with the constant $\omega_c$ being replaced by a parameter given below. 
\begin{equation}
\omega(\tau)=\frac{3}{2} ( \beta(\tau) -1), \quad \quad \quad \beta(\tau)=1-2\epsilon H r_c \left(1+\frac{\dot{H}}{3H^2} \right).
\label{DGPpar1}
\end{equation}
Unfortunately, it is well known that the self accelerating branch has a ghost mode. Despite this, the DGP model provides a great toy model,  which offers both analytic solutions and is Vainshtein screened. Below we simply present the Poisson functions for this model \cite{Koyama:2009me} 
\begin{equation}
\mu(k;a)  = 1 + \frac{1}{3\beta(a)}, 
\label{mudgp}
\end{equation}
\begin{equation}
\gamma_2(k,\bfk_1,\bfk_2;a) = -\frac{H_0^2}{24 H^2 \beta(a)^3 \Omega_{rc}} \left(\frac{\Omega_{m,0}}{a^3}\right)^2 (1-\mu_{1,2}^2), 
\label{g2dgp}
\end{equation}
\begin{equation}
\gamma_3(k,\bfk_1,\bfk_2,\bfk_3;a) = \frac{H_0^2}{144 H^2 \beta(a)^5 \Omega_{rc}^2} \left(\frac{\Omega_{m,0}}{a^3}\right)^3 (1-\mu_{2,3}^2) (1-\mu_{1,23}^2),
\label{g3dgp}
\end{equation}
where we have rephrased $\beta(\tau)$ as 
\begin{equation}
\beta(a)= 1+\frac{H}{H_0 \sqrt{\Omega_{rc}}}\left(1+\frac{aH'}{3H}\right),
\end{equation}
and $\mu_{i,j}$ is the cosine of the angle between $\bfk_i$ and $\bfk_j$. We use the parameter $\Omega_{rc} = 1/(4r_c^2H_0^2)$ as the model degree of freedom in {\tt MG-Copter} and values of this will be used in later chapters. Unlike Hu-Sawicki theory the linear order function is not scale dependent allowing us to solve the perturbations under the separability assumption. This has been shown to be a good approximation \cite{Koyama:2009me} and we provide a comparison of this with the full numerical solutions in Chapter 5. Again we see the need for the higher order functions to suppress small scale modifications to Newtonian predictions \footnote{For example with $\Omega_{rc} = 1$ we get $\mu = 1.196$. As $\Omega_{rc}\rightarrow \infty$, $\mu \rightarrow 4/3$ as before and we get a 1/3 enhancement of Newton's constant.}.
\newline
\newline
This concludes our discussion of modified gravity. Before we can apply the predictions for growth to measurements from the real universe and begin constraining prospective theories in a consistent way, there is another vital piece of the pipeline to add. It is essentially a systematic in the way we observe the constituents of the universe, and holds a lot of vital information about the growth of structure.


\chapter{Redshift Space Distortions}
\begin{figure}[h]
\centering
\centerline{\includegraphics[width=15.5cm,height=8.3cm]{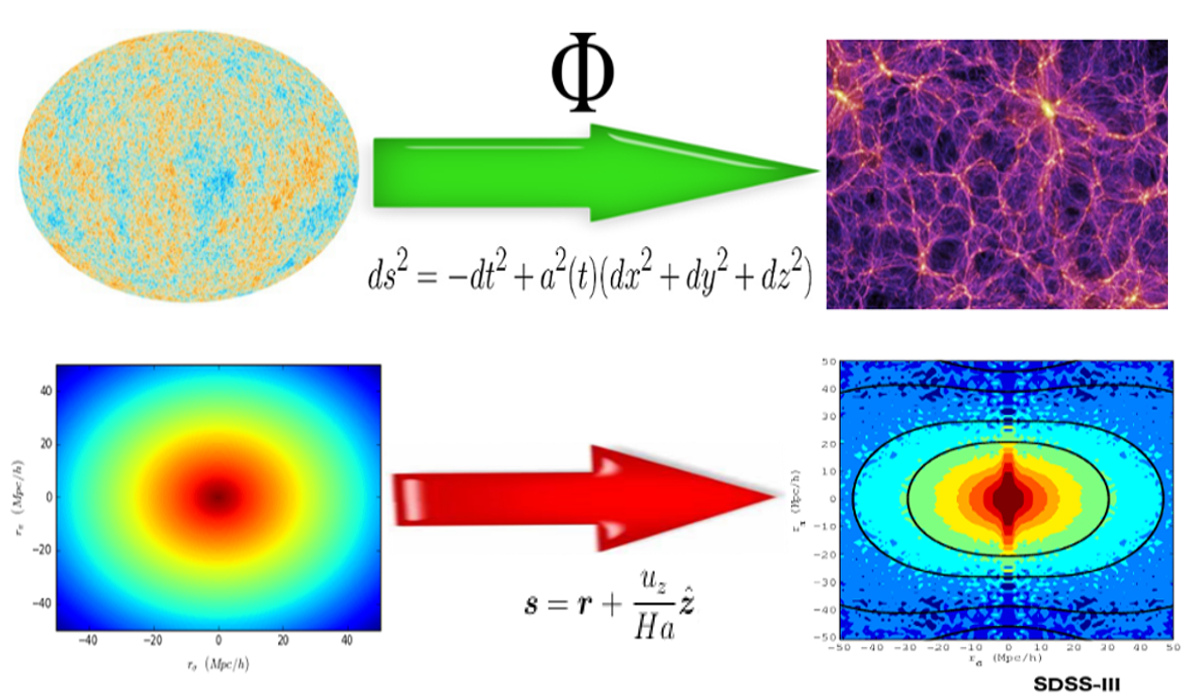}}
\caption{The theoretical journey to data comparisons and theoretical constraints. {\bf Top left:} The temperature fluctuation map of the CMB \cite{Planck:2015xua}. {\bf Top right:} A snapshot of the Millennium N-body simulation \cite{Springel:2005nw}. {\bf Bottom left:} Correlation function for spherical collapse. {\bf Bottom right:} The redshift space correlation function as measured by SDSS \cite{Reid:2012sw}.}
\label{real2red}
\end{figure}

\begin{chapquote}{J.D}
``The cliffs at his back outran the moonlight, merging with night at unreachable heights, just an illusion of some higher end."
\end{chapquote}
In light of the previous chapter, along with the power spectrum construction discussed in Chapter 2, we are now in a position to construct LSS statistics for a large variety of gravity models using forms for $\mu$,$\gamma_2$ and $\gamma_3$. This is all well and good, but there is one large issue that musn't be overlooked - we do not observe the true positions of galaxies that trace the LSS.
\newline
\newline
In principle, by Hubble's law $z = Hd$, spectroscopic data is able to give us depth information over a 2D galaxy survey on the sky. This would be very straightforward if cosmic expansion was the only factor to consider. Unfortunately, galaxies have their own peculiar velocities. If a galaxy has a peculiar velocity $\boldsymbol{u} = d \boldsymbol{x}/dt$   and true position $\boldsymbol{r}$ (relative to the observer), then its true position and position in redshift space $\boldsymbol{s}$ are related by 
\begin{equation}
\boldsymbol{s}=\boldsymbol{r} + \frac{u_z}{Ha} \hat{\boldsymbol{z}},
\label{redtoreal}
\end{equation}
where $u_z= \boldsymbol{u} \cdot \hat{\boldsymbol{z}}$ and we have chosen the $z$-coordinate axis to be along our line of sight (LOS). Eq.\ref{redtoreal} describes a non-linear mapping between real and redshift space and shows us that in redshift space the galaxy distribution will be distorted by the peculiar velocities. Fig.\ref{real2red} serves as a summary, showing a schematic of the two mappings we've described so far and the equations that we've assumed (Eq.\ref{FLRWmetric} and Eq.\ref{redtoreal}). The top mapping shows how we first map initial perturbations to their final state through growth of structure governed by a spatially expanding metric and a gravitational interaction described effectively by the potential $\Phi$.  The bottom mapping shows how spherical collapse under gravity in reality is actually observed with anisotropies when converting redshift measurements to depth on the 2d sky in the FLRW paradigm. These two mappings constitute a large part of the theoretical journey to predictions that can be compared with astronomical survey data! In this chapter we will look at some models for the redshift space power spectrum and correlation function. For further reading, we direct to this excellent introduction to the redshift space anisotropy \cite{Hamilton:1997zq}. 
\section{Fourier Space}
It was pointed out by Kaiser \cite{Kaiser:1987qv} that the real space power spectrum $P(\boldsymbol{k})$ (Eq.\ref{treeps}) is scaled by a factor of $[F_1 - G_1\mu^2]^2$ in redshift space at first order, so that we get the {\it Kaiser formula} \cite{Kaiser:1987qv}\footnote{Note we have omitted time dependence for brevity.} 
\begin{equation}
P_{K}^S(k,\mu) = [F_1(k) - G_1(k)\mu^2]^2P_0(k),
\label{linkais}
\end{equation}
where the superscript $S$ is used to indicate a redshift space quantity and $\mu = \boldsymbol{k} \cdot \boldsymbol{\hat{z}}/k$ is the cosine of the angle between the LOS and $\bfk$\footnote{The use of $\mu$ here should not be confused with the function $\mu(k;a)$ (see Eq.\ref{poissonmglin}) which will always include its arguments.}. The Kaiser model accounts well for the {\it Kaiser effect} which is a flattening effect of the distribution occurring at large scales when the collapse velocity is small. This is shown in the outer contours of the bottom right image of Fig.\ref{real2red}. 
\newline
\newline
At smaller, non-linear scales we have larger peculiar velocities and we get what is called the {\it Fingers of God} (FoG) which is an elongation of the distribution along the LOS. This is shown in the bottom right, inner red region of Fig.\ref{real2red}. To treat this, phenomenological models have proposed the prefactor, $D_{FoG}$, introduced to Eq.\ref{linkais}. It usually takes a Lorentzian or Gaussian form \cite{Scoccimarro:2004tg,Percival:2008sh,Cole:1994wf,Peacock:1993xg,Park:1994fa,Ballinger:1996cd,Magira:1999bn}. In addition, the real space power spectra in the expanded expression of Eq.\ref{linkais} are taken up to 1-loop order to better account for non-linearities. This was first proposed by Scoccimaro \cite{Scoccimarro:2004tg} 
\begin{equation}
P^S(k,\mu) = D_{FoG}(\sigma_vk\mu)[P_{\delta \delta}(k) - 2\mu^2  P_{\delta \delta}(k)  + \mu^4 P_{\delta \delta}(k)], 
\label{nonlinkais}
\end{equation}
where $\sigma_v$ is the linear velocity dispersion of the galaxy distribution 
\begin{equation}
\sigma_v^2(k) \equiv \frac{1}{3}\langle \bfu(\bfx) \cdot \bfu(\bfx) \rangle =  \int^{\infty}_0 \frac{dq}{6\pi^2} G_1(q;a)^2P_0(q).
\label{sigmaveqn}
\end{equation} 
An improved model of RSD was developed by Taruya, Nishimichi and Saito (TNS) \cite{Taruya:2010mx} in 2010 which accounts for both the FoG and Kaiser effects as well as non-linearities. To begin, consider conservation of mass between redshift and real space 
\begin{equation}
[1+\delta^{S}(\boldsymbol{s})] d^3 \boldsymbol{s} = [1+\delta(\boldsymbol{r})] d^3 \boldsymbol{r},
\end{equation}
which gives 
\begin{equation}
\delta^S(\boldsymbol{s}) = \left| \frac{\partial \boldsymbol{s}}{\partial \boldsymbol{r}} \right|^{-1} [1+\delta(\boldsymbol{r})]-1,
\end{equation}
where $\left| \partial \boldsymbol{s} / \partial \boldsymbol{r} \right| = 1+ \nabla_z u_z/ (Ha)$ is the Jacobian moving from $\bfr$ coordinates to $\boldsymbol{s}$ coordinates. In Fourier space we have
\begin{equation}
\delta^S(\boldsymbol{k})= \int d^3 \boldsymbol{r} \left[ \delta(\boldsymbol{r}) - \frac{ \nabla_z u_z}{Ha} \right] e^{-i(\frac{k\mu u_z }{Ha} + \boldsymbol{k} \cdot \boldsymbol{r})},
\end{equation}
where we have used Eq.\ref{redtoreal} in the exponential. From this we can construct the power spectrum in redshift space using Eq.\ref{powerspec}. Expanding this out gives
\begin{equation}
P^S(\boldsymbol{k}) = \int d^3 \boldsymbol{x} e^{-i \boldsymbol{k} \cdot \boldsymbol{x}} \langle e^{j_1 A_1} A_2 A_3 \rangle,
\label{PSred1}
\end{equation}
with 
\begin{align}
j_1= &-i k \mu, \\ \nonumber 
A_1 = &u_z(\boldsymbol{r}) - u_z(\boldsymbol{r}'),  \\ \nonumber 
A_2 = &\delta(\boldsymbol{r})  -\nabla_z u_z(\boldsymbol{r}),  \\ \nonumber 
A_3 = &\delta(\boldsymbol{r}') - \nabla_z u_z(\boldsymbol{r}'), \nonumber 
\end{align}
where $\boldsymbol{x}=\boldsymbol{r}-\boldsymbol{r}'$ is the separation in real space. We also make the scaling $u_z(\boldsymbol{r}) \rightarrow  u_z(\boldsymbol{r})/(a H)$ as the LOS component of the velocity field, consistent with the definition in Eq.\ref{thetadef}. 
\newline
\newline
The next step is to write this expression in terms of {\it cumulants}, which will be indicated by $\langle  \rangle_c$. The $n$th order cumulant can be thought of as a measure of the interaction between $n$ of the variables. For example with Gaussian random variables, by Wick's theorem, all cumulants of order greater than 2 are zero. The cumulants of a random variable $X$ are found via the cumulant generating function 
\begin{equation}
K(t) \equiv \log(\mbox{E}[e^{tX}]) = \sum^{\infty}_{n=1} \kappa_n \frac{t^n}{n!}, 
\end{equation}
where E is the expectation value, $t$ is a real number and $\kappa_n$ is the $n$th order cumulant which is used to define the power series expansion of $K(t)$. By taking the $n$th order derivative of this expansion with respect to $t$ and setting $t=0$ we obtain $\kappa_n$ in terms of the moments of the distribution, found by expanding $E[e^{tX}]$     
\begin{equation}
E[e^{tX}]  = 1+ t m_1 +  \frac{t^2}{2} m_2 + \cdots, 
 \end{equation}
where $m_1=\mu$ is the mean and $m_2 = \sigma^2$ is the variance. We can extend this to higher dimensions (see \cite{Matsubara:2007wj} and \cite{Scoccimarro:2004tg} for a derivation) 
\begin{equation}
\langle e^{\boldsymbol{j}\cdot\boldsymbol{A}} \rangle = \exp \{ \langle e^{\boldsymbol{j}\cdot\boldsymbol{A}} \rangle_c \}, 
\end{equation}
where $\boldsymbol{A}=\{ A_1,A_2,A_3\}$ is a stochastic vector field and $\boldsymbol{j}=\{j_1,j_2,j_3\}$ is some constant vector. Taking the derivative twice with respect to $j_2$ and $j_3$ and then setting them to $0$ we get the following expression for the kernel of Eq.\ref{PSred1}
\begin{equation}
\langle e^{j_1 A_1} A_2 A_3 \rangle = \exp \{ \langle e^{j_1A_1} \rangle_c \} [  \langle e^{j_1A_1} A_2 A_3 \rangle_c +  \langle e^{j_1A_1} A_2 \rangle_c  \langle e^{j_1A_1} A_3\rangle_c ].
\label{tns1}
\end{equation}
The exponential factor is known to dampen the redshift space power spectrum at non-linear scales due to virialised, highly uncorrelated motion. The character of this factor is known to be partly non-perturbative and it has been shown to impart  minimal acoustic features to the power spectrum \cite{Taruya:2010mx}. Motivated by this, it can be replaced with the phenomenological functional form of $\mbox{D}_{\mbox{FoG}}(k\mu \sigma_v)$ with the velocity dispersion of the galaxy distribution, $\sigma_v$, now treated as a free parameter. The other terms are treated perturbatively. Taking $j_1$ as our small parameter we can expand the expression in brackets up to quadratic order to get 
\begin{align}
 \langle e^{j_1A_1} A_2 A_3 \rangle_c +  \langle e^{j_1A_1} A_2 \rangle_c  \langle e^{j_1A_1} A_3\rangle_c & \simeq \langle A_2 A_3 \rangle + j_1 \langle A_1 A_2 A_3 \rangle_c \nonumber \\ &+ j_1^2 \{ \frac{1}{2} \langle A_1^2 A_2 A_3 \rangle + \langle A_1 A_2 \rangle_c \langle A_1A_3 \rangle_c \}. 
 \end{align}
Note here that the first term on the right hand side at linear order in Eq.\ref{PSred1} gives Eq.\ref{linkais}, where we also need expand the exponential prefactor in Eq.\ref{tns1} and keep only linear order terms. We find the term $\langle A_1^2 A_2 A_3 \rangle $ is  proportional to $\mathcal{O}(P_0(k)^3)$. We now keep terms up to quadratic in the linear power spectrum (consistent with the 1-loop correction) to get the following expression
 \begin{align}
 P^S_{TNS}(k,\mu) = \mbox{D}_{FoG} (k\mu \sigma_v) \{ &P_{\delta \delta} (k) - 2 \mu^2 P_{\delta \theta}(k) + \mu^4 P_{\theta \theta} (k) \nonumber \\ &+ A(k,\mu) + B(k,\mu) + C(k,\mu) \}, 
 \label{redshiftps}
 \end{align}
 \noindent where $P_{\delta \delta}, P_{\delta \theta}$ and $ P_{\theta \theta}$ are all at 1-loop order. The correction terms, $A$ and $B$ are given by 
 \begin{align}
 A(k,\mu) &=  -(k \mu) \int d^3 \boldsymbol{k'} \frac{k_z '}{k'^2} \{B_\sigma(\boldsymbol{k'},\boldsymbol{k}-\boldsymbol{k'},-\boldsymbol{k})-B_\sigma(\boldsymbol{k'},\boldsymbol{k}, -\boldsymbol{k}-\boldsymbol{k'}) \}, 
 \label{Aterm}
 \\
 B(k,\mu)& = (k \mu)^2 \int d^3\boldsymbol{k'} F(\boldsymbol{k'}) F(\boldsymbol{k}-\boldsymbol{k'}),
 \label{Bterm}
 \end{align}
 where $F(k)$ is a function of the linear cross and velocity divergence power spectra
 \begin{equation}
 F(\boldsymbol{k}) = \frac{k_z}{k^2}\left[P_{\delta \theta} (k) - \frac{k_z^2}{k^2}P_{\theta \theta} (k) \right]. 
 \end{equation}
 The cross bispectrum $B_\sigma$ is given by
 \begin{equation}
 \delta_D(\boldsymbol{k}_1+ \boldsymbol{k}_2+ \boldsymbol{k}_3)B_\sigma( \boldsymbol{k}_1,\boldsymbol{k}_2,\boldsymbol{k}_3) = \langle \theta(\boldsymbol{k}_1)\{ \delta(\boldsymbol{k}_2) - \frac{k_{2z}^2}{k_2^2} \theta(\boldsymbol{k}_2)\}\{ \delta(\boldsymbol{k}_3) - \frac{k_{3z}^2}{k_3^2} \theta(\boldsymbol{k}_3)\}\rangle.
 \label{lcdmbi}
 \end{equation}
\newline
We can write $B_\sigma$ up to 2nd order in the linear power spectrum by expanding the perturbations up to 2nd order. In terms of the first and second order kernels (see Eq.\ref{nth1}, Eq.\ref{nth2} and Appendix A)
\begin{align} 
& B_\sigma(\boldsymbol{k}_1,\boldsymbol{k}_2,\boldsymbol{k}_3) = \nonumber \\ & 2\left[ \left(F_1(k_2) -\frac{k_{2z}^2}{k_2^2} G_1(k_2) \right)\left(F_1(k_3) -\frac{k_{3z}^2}{k_3^2} G_1(k_3)\right) G_2(\bfk_2,\bfk_3) P_0(k_2) P_0(k_3) \right. \nonumber \\ +
 &G_1(k_1)\left(F_1(k_3) -\frac{k_{3z}^2}{k_3^2} G_1(k_3)\right) \left(F_2(\bfk_1, \bfk_3) -\frac{k_{2z}^2}{k_2^2} G_2(\bfk_1,\bfk_3)\right)P_0(k_1) P_0(k_3) \nonumber \\ +
 & \left. G_1(k_1)\left(F_1(k_2) -\frac{k_{2z}^2}{k_2^2} G_1(k_2)\right) \left(F_2(\bfk_1, \bfk_2) -\frac{k_{3z}^2}{k_3^2} G_2(\bfk_1,\bfk_2)\right)P_0(k_1) P_0(k_2)  \right]. 
 \label{bispectrum}
 \end{align}
It is useful to note here that using the symmetry $B_\sigma(\bfk_1,\bfk_2,\bfk_3) = B_\sigma(\bfk_1,\bfk_3,\bfk_2) =B_\sigma(-\bfk_1,-\bfk_2,-\bfk_3)$, we can rewrite eq.(\ref{Aterm}) as 
 \begin{align}
 A(k,\mu) =  -(k \mu) \int d^3 \boldsymbol{k'} \left[  \frac{k_z '}{k'^2} B_\sigma(\boldsymbol{k'},\boldsymbol{k}-\boldsymbol{k'},-\boldsymbol{k}) +  \frac{k\mu-k_z'}{|\bfk-\bfk'^2|} B_\sigma(\boldsymbol{k}-\boldsymbol{k'}, \boldsymbol{k'},-\boldsymbol{k}) \right].
 \label{Aterm2}
 \end{align}
$F(\boldsymbol{k})$ is already 2nd order in the linear power spectrum. In terms of the perturbation kernels we can write it as 
 \begin{equation}
 F(\boldsymbol{k}) = \frac{k_z}{k^2}G_1(k) \left[ F_1(k) P_0 (k) - \frac{k_z^2}{k^2} G_1(k)P_0 (k) \right]. \label{tnsfterm}
 \end{equation}
 Finally, the $C$ term is given at 1-loop order by 
 \begin{equation}
  C(k,\mu) = (k \mu)^2 \int d^3\boldsymbol{p} d^3 \boldsymbol{q} \delta_D(\bfk-\bfp-\bfq)\frac{\mu_p^2}{p^2}G_1(p)^2[F_1(q)- G_1(q)\mu_q^2]^2 P_0(p)P_0(q),
  \label{cterm}
\end{equation}
where $\mu_p = \bfp \cdot \boldsymbol{\hat{z}}/p$ and similarly for $\mu_q$. This term has been shown to have very little acoustic structure and so its impact on the power spectrum is one of overall amplitude, only coming into play at quasi non-linear scales as it is a 2nd order contribution. Based on this, one can essentially absorb it into the phenomenological damping function $D_{FoG}$, as this term has been promoted with an extra degree of freedom, $\sigma_v$. By this, we ignore this term in the forthcoming chapters but have included here it for completeness. The general-kernel forms for the correction terms are part of the original results presented in this thesis and can be found in \cite{Bose:2016qun}. In the case of an EdS universe, we refer the reader to \cite{Taruya:2010mx} for the explicit forms of the correction terms.  
\newline
\newline
 The main feature of this model is the inclusion of the $A$ and $B$ correction terms which account for higher-order interactions between the density and velocity fields. This gives the model good predictive power at weakly nonlinear scales, as shown by $N$-body comparisons \cite{Taruya:2010mx,Nishimichi:2011jm,Taruya:2013my}. In GR these terms have been shown to enhance the power spectrum amplitude at the BAO scale and have a non-negligible effect on the shape of the power spectrum \cite{Taruya:2010mx}. The left hand side of Fig.\ref{frecovab} shows the results of an Monte Carlo Markov Chain (MCMC) analysis (see Appendix C for details) conducted in \cite{Taruya:2010mx} that shows the importance of the $A$ and $B$ correction terms in achieving unbiased measures of growth. The analysis was conducted assuming an ideal survey's errors. The survey was taken to observe a volume of $20 \mbox{Gpc}^3/h^3$ and a galaxy density of $\bar{n} = 5\times 10^{-4} \mbox{Mpc}^3/h^3$. Not including these terms prevents the theoretical template in achieving accurate estimates of the N-body simulation's fiducial value of logarithmic growth $f$. The right hand side shows a similar analysis but this time for different perturbative modelling of $A$ and $B$ for an N-body simulation conducted within the context of an $f(R)$ theory \cite{Taruya:2013quf}. Here they use an ideal survey volume of  $10 \mbox{Gpc}^3/h^3$ to model the errors on the simulation data. Again, this shows that incorrect modelling of these terms leads to a failed capture of the simulation's fiducial logarithmic growth factor $f(a)$. 
 \begin{figure}[H]
 \centering
  \subfloat[]{\includegraphics[width=7.5cm, height=8.cm]{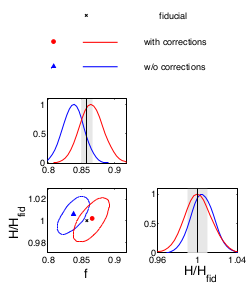}} \quad
  \subfloat[]{\includegraphics[width=7.5cm, height=8.cm]{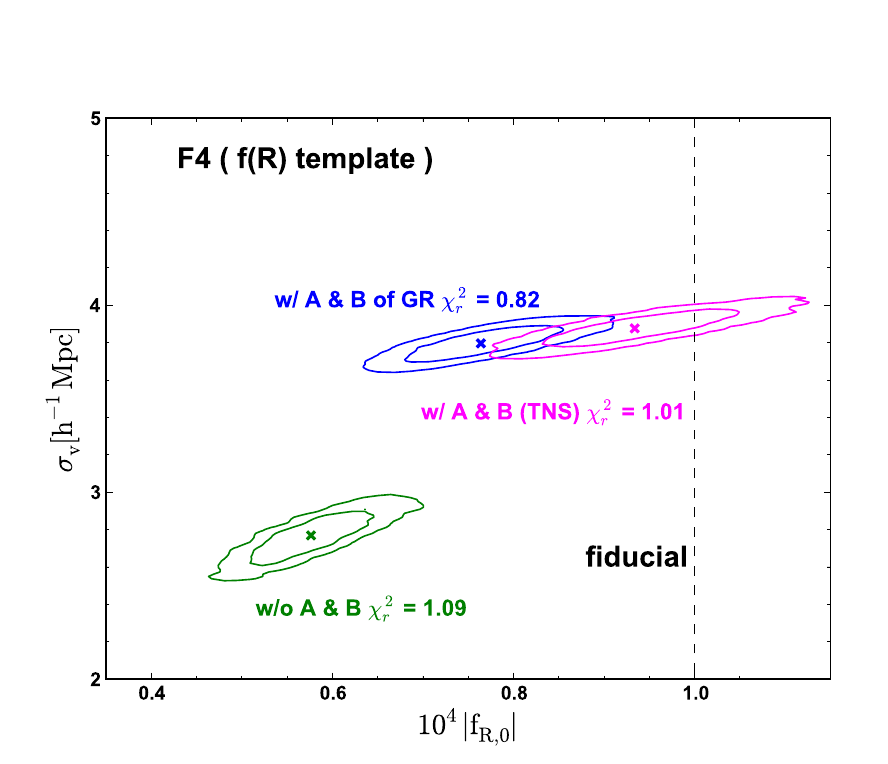}}
   \caption{Results from \cite{Taruya:2010mx} (left) and \cite{Taruya:2013quf} (right) which highlight the importance of including the $A$ and $B$ correction terms as well as their perturbative modelling within the context of modified gravity theories.}  
\label{frecovab}
\end{figure}
 \noindent In modified gravity theories the $B$ term is generally expected to be enhanced because of its linear growth dependance while the $A$ term involves the 2nd order perturbations so it is not obvious how it will change. It has been shown that these terms require correct modelling within a given gravitational model's context to avoid biased constraints as shown in \cite{Taruya:2013quf}, their main result shown on the right of Fig.\ref{frecovab}.
 \newline
\newline
Eq.(\ref{redshiftps}) gives a non-linear prediction for the redshift space power spectrum. In GR this expression has been computed up to 2-loop order in various perturbation theory (PT) schemes \cite{Taruya:2013my,Taruya:2009ir,Okamura:2011nu,Crocce:2007dt,Crocce:2012fa,Taruya:2012ut} which generally have a larger range of validity over the SPT treatment. Despite this, as well as the convergence problems of the SPT treatment, the scheme still gives us a good working range of scales in the quasi non-linear regime making it well suited for the goal of probing gravity. We note that so far no dynamics for the perturbations have been taken into account and so the forms of all expressions in this section are gravity model  independent. Our attention is now turned to a different route in describing matter clustering in redshift space. We next discuss non-linear expressions for the correlation function.


\section{Configuration Space} 
\begin{figure}[h]
\centering
\centerline{\includegraphics[width=15.5cm,height=8cm]{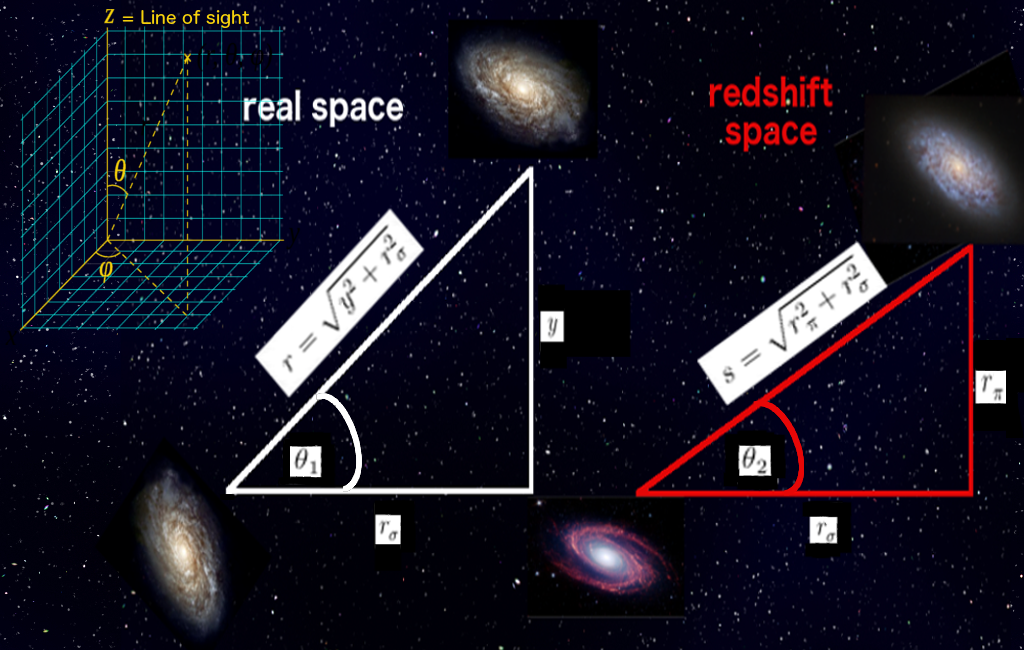}}
\caption{A schematic of the real (left) to redshift space (right) mapping of galaxies within configuration space with the variables used in the text. The LOS is taken to be along the $z$-axis.}
\label{real2redconfig}
\end{figure}

Let us now consider things in configuration space. Fig.\ref{real2redconfig} shows a schematic of the shift of galaxy positions along the LOS due to peculiar velocities with variables we will use in the text. As a starting point, we can proceed by a simple FT of the power spectrum discussed in the previous section. 
\subsection{FT of TNS Spectrum}
To move to configuration space, one can simply take the FT of Eq.\ref{linkais} or Eq.\ref{redshiftps} although there are some problems associated with the non-linear case as discussed in the final section of Chapter 2. We can remedy this by using the RegPT 1-loop power spectrum instead of SPT. One would also need to treat the $A$ and $B$ TNS correction terms in the RegPT approach. The $A$ and $B$ correction terms are evaluated at tree-level since they are treated as next to leading order and are given in terms of the propagators as 
\begin{align}
P_{bc,tree}(k;a) = & \Gamma_b^{(1)}(k;a)\Gamma_c^{(1)}(k;a)P_0(k), \\
B_{bcd,tree}(\bfk_1,\bfk_2,\bfk_3) = &2 \Gamma_b^{(2)}(\bfk_2,\bfk_3;a)\Gamma^{(1)}_c(k_2;a)\Gamma_d^{(1)}(k_3;a) P_0(k_2)P_0(k_3) \nonumber \\& + ({\rm cyc. perm}),
\end{align}
where now the propagators are evaluated at tree level too 
\begin{equation}
\Gamma_{b,tree}^{(n)}(\bfk_1 ,\dots ,\bfk_n;a) = J_a^{(n)}(\bfk_1, \dots ,\bfk_n;a)e^{-k_{1...n}^2\sigma_d^2/2}.
\end{equation}
\subsection{Streaming Models}
\begin{figure}[H]
  \captionsetup[subfigure]{labelformat=empty}
  \centering
  \subfloat[]{\includegraphics[width=15.5cm, height=6.6cm]{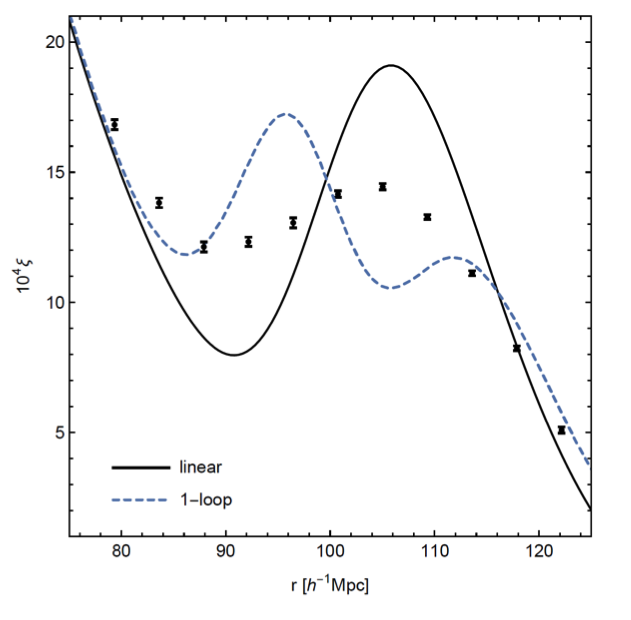}} \quad
  \caption{Real space correlation function calculated using the 1-loop spectrum (blue) and linear spectrum (black solid) against simulation measurements (points) \cite{Baldauf:2015xfa}.}
\label{regptmerit}
\end{figure}

Alternatively, the redshift space correlation function can be constructed within a {\it streaming model} approach. This approach expresses the redshift space correlation function as a convolution of the real space correlation function and the probability distribution of particle velocities along the LOS, taken around some mean pair-wise infall velocity
\begin{equation}
\xi^S_{\rm SM} (r_\sigma, r_\pi) = \int dy F_u\left[\frac{u_z}{Ha} - \frac{y}{r} u_{12}(r)\right] \xi(r).
\label{xistream}
\end{equation}
Here $y$ and $r_\pi$ are the separations of galaxies parallel to the LOS  in real space and redshift space respectively, $r_\sigma$ is the separation perpendicular to the LOS, $u_z$ is the LOS component of the particle's peculiar velocities \footnote{ $u_z/Ha=r_\pi-y$ by Eq.\ref{redtoreal}} and $\xi(r)$ is the real space correlation function. $F_u$ is the particle velocity probability distribution about the LOS component of the mean particle pair-wise infall velocity 
\begin{equation}
\boldsymbol{u}_{12} = u_{12} \hat{\bfr} = \overline{\bfu(\bfx) - \bfu(\bfx')} = \langle [\bfu(\bfx) - \bfu(\bfx')][1+\delta(\bfx)][1+\delta(\bfx']\rangle,
\label{v12def}
\end{equation}
where the bar indicates an average over particle positions. This average is weighted by the number of particles at each position $\bfx$ which is proportional to $[1+\delta(\bfx)]$, explaining these factors in the 2nd equality. 
\newline
\newline
We can derive Eq.\ref{xistream} from the probabilistic definition of correlation. $\xi^S(r_\sigma,r_\pi)$ is a measure of the probability above the uncorrelated Poisson probability of finding a particle pair at some separation $s = \sqrt{r_\sigma^2+r_\pi^2}$. Mathematically, the probability $dP$ of finding a particle within volume $d^3\boldsymbol{s}_1$ and a particle within volume $d^3\boldsymbol{s}_2$ separated by $s$ is given by 
\begin{equation}
dP = \bar{n}^2 [1+\xi^S(r_\sigma,r_\pi)]d^3\boldsymbol{s}_1 d^3\boldsymbol{s}_2,
\label{dp1}
\end{equation}
where $\bar{n}$ is the particle number density and the first term in square brackets represents the base Poisson probability. The streaming model approach proposes the expression of this probability in the following form 
\begin{equation}
dP = \bar{n}^2 [1+\xi(r)] F_u(V) \delta_D\left(\frac{u_z}{Ha} - \frac{y}{r} u_{12}(r) - V\right) dV dy d^3\boldsymbol{s}_1 d^3\boldsymbol{s}_2,
\label{dp2}
\end{equation}
where $V$ is the velocity between the pairs. What Eq.\ref{dp2} says is that we have an excess probability of finding the pair at separation $r=\sqrt{y^2 + r_\sigma^2}$ given by $\xi(r)$ which can be converted to redshift space separation $s$ by incorporating the probability of the shift parallel to the LOS ($r_\pi$). This is given by the velocity probability distribution $F_u$ along with the Dirac delta. By equating Eq.\ref{dp1} and Eq.\ref{dp2} and integrating over the pair-wise velocity $V$ we get Eq.\ref{xistream}.   
\newline
\newline
Working in the linear regime, Karl Fisher \cite{Fisher:1994ks} developed the so called linear streaming model (LSM) which uses the mean infall velocity between pairs ($u_{12}$) and the velocity dispersion along the LOS ($\sigma_{12}^2 $) as ingredients connecting theory to the RSD phenomenon. Specifically, it is the scale dependence of $u_{12}$ and $\sigma_{12}^2 $ which drives the distribution of galaxies away from isotropy. By assuming a Gaussian form for $F_u$, the linear streaming model (LSM) is given by 
\begin{equation}
1+\xi^s_{\rm LSM} (r_\sigma, r_\pi) = \int G(r,y) e^{-[r_\pi - y]^2/2\sigma_{\rm 12,lin}^2(r,\mu)} \frac{dy}{\sqrt{2\pi\sigma_{\rm 12,lin}^2(r,\mu)}},
\label{lsmmodel}
\end{equation}
where $\mu = \hat{\bfr} \cdot \hat{z} = y/r$ and 
\begin{equation}
G(r,y) = \left[1 + \xi^r_L(r) + \frac{y}{r} \frac{(r_\pi -y)u_{\rm 12,lin}(r)}{\sigma_{\rm 12,lin}^2(r,\mu)} - \frac{1}{4}\frac{y^2}{r^2}\frac{u_{\rm 12,lin}^2(r)}{\sigma_{\rm 12,lin}^2(r,\mu)}\left(1-\frac{(r_\pi-y)^2}{\sigma_{\rm 12,lin}^2(r,\mu)}\right) \right].
\end{equation} 
$\xi^r_L$ is the linear real space galaxy correlation function determined by Fourier transforming the linear power spectrum, $u_{\rm 12,lin} = \langle \delta(\bfx) \bfu(\bfx')\rangle $  is the linear mean infall velocity of a particle pair with real space separation $r = y^2 + r_\sigma^2$ and $\sigma_{\rm 12,lin}^2(r,\mu) = \langle (\bfu_{\rm LOS}(\bfx) -\bfu_{\rm LOS}(\bfx') )^2 \rangle$ is the linear velocity dispersion. The linear predictions for these are given below in terms of the generalised 1st order perturbative kernels $(F_1,G_1)$
\begin{equation}
\bfu_{\rm 12,lin}(r) = u_{\rm 12,lin}\hat{\bfr} = \frac{1}{\pi^2} \int dk k  j_1(kr) G_1(k;a) F_1(k;a) P_0(k),
\label{linv12}
\end{equation}
 where $j_1(k)$ being the 1st order spherical Bessel function. We also have 
  \begin{equation} 
 \sigma_{\rm 12,lin}^2(r, \mu^2) = 2\left[\sigma_v^2 - \frac{1}{2\pi^2}\int dk G_1(k;a) \mathcal{J}(kr,\mu^2) P_0(k)  \right],
 \label{lins12}
 \end{equation}
where $\sigma_v^2$ is given by Eq.\ref{sigmaveqn} and we define
  \begin{equation} 
 \mathcal{J}(kr,\mu^2) = \mu^2\left(j_0(kr)-\frac{2j_1(kr)}{kr}\right) + (1-\mu^2)\frac{j_1(kr)}{kr}.
 \label{mybessel}
 \end{equation}
Moving away from the linear regime, we consider the non-linear redshift space correlation function developed in \cite{Reid:2011ar}, known as the Gaussian streaming model (GSM) from its core assumption that the matter's pairwise velocity probability distribution is again of a Gaussian form
\begin{equation}
1+\xi^s_{\rm GSM}(r_\sigma, r_\pi) = \int [1+\xi(r)] e^{-[r_\pi - y - \mu u_{12}(r)]^2/2\sigma_{12}^2(r,\mu)} \frac{dy}{\sqrt{2\pi\sigma_{12}^2(r,\mu)}}. 
\label{redshiftcor}
\end{equation} 
Here $\xi$ is the non-linear real space correlation function, $u_{12}(r)$ is the non-linear mean infall velocity of a particle pair and $\sigma_{12}^2(r,\mu)$ is the non-linear, non-isotropic velocity dispersion.  
\newline
\newline
In \cite{Reid:2011ar} the authors use the Lagrangian perturbation theory (LPT) model of \cite{Matsubara:2008wx} for the real space correlation function . Here we use a RegPT prescription where our real space correlation function is produced by Fourier transforming the RegPT 1-loop matter power spectrum (Eq.\ref{regptloop})
\begin{equation}
\xi^r(r) = \int \frac{d^3k}{(2\pi)^3} e^{i\bfk\cdot \bfx} P^{\rm 1-loop,RegPT}_{\delta \delta} (k).
\label{pregab}
\end{equation}
\noindent  Fig.\ref{regptmerit} shows the result of performing the FT in Eq.\ref{loopps} using the 1-loop spectrum (blue), linear spectrum (black solid) and a simulation measurement (points)\cite{Baldauf:2015xfa}. The 1-loop transform clearly shows an incorrect BAO feature which the resummation of IR modes in the RegPT approach remedies very well (see Fig.\ref{xir1}). As discussed in Chapter 2, $P^{{\rm1-loop, RegPT}}_{ab}(k)$ can be readily constructed for general models of gravity. 
\noindent Finally, the mean infall velocity and velocity dispersion are given by correlations between the density field and the velocity field. Using a perturbative treatment of the fields we can derive expressions for these ingredients for general models of gravity in the linear and quasi non-linear regime. Here we give expressions for $u_{12}$ and $\sigma_{12}^2$ appearing in Eq.\ref{redshiftcor} in terms of the generalised kernels $(F_n,G_n)$ (see Eq.\ref{nth1} and Eq.\ref{nth2}). These forms have not been derived in the literature and so the next two subsections are part of the original results presented in this thesis and can be found in \cite{Bose:2017dtl}. In the case of GR, using the EdS approximation for the kernels, one can follow Appendices A1 and A2 of \cite{Reid:2011ar}. 


\subsection{Mean Infall Velocity $u_{12}(r)$}
The mean infall velocity arises from correlating the density field with the velocity. In terms of these correlations we can write (Eq.27 of \cite{Reid:2011ar}) 
\begin{align}
[1+\xi_L^r(r)] u_{12}(r)\hat{r} = & 2\langle \delta_1(\bfx)\bfu_1(\bfx+\bfr)\rangle + 2\sum_{i>0} \langle \delta_i(\bfx)\bfu_{4-i}(\bfx+\bfr)\rangle \nonumber \\ & +2\sum_{i,j>0} \langle \delta_i(\bfx) \delta_j(\bfx+\bfr)\bfu_{4-i-j}(\bfx+\bfr) \rangle.
\end{align}
 $\xi_L$ is the linear matter correlation function and $\bfu$ is the peculiar velocity field perturbation. The summation is performed up to consistent order for the $1-loop$ calculation.The correlations in the above expression up to 2nd order in the linear power spectrum are given below. 
\begin{align}
 2 &\left( \langle \delta_1(\bfx)\bfu_1(\bfx+\bfr)\rangle + \sum_{i>0} \langle \delta_i(\bfx)\bfu_{4-i}(\bfx+\bfr)\rangle \right) \nonumber \\ & =   \frac{\hat{r}}{\pi^2} \int dk k P_{\delta \theta}^{{\rm 1- loop}}(k,a) j_1(kr), 
 \end{align}
where $P_{\delta \theta}^{{\rm 1- loop}}$ is given by Eq.\ref{regptloop}. The last term has three contributions at loop order, $i,j=(1,1),(1,2),(2,1)$. The contribution from the first two of these (A5 of \cite{Reid:2011ar}) is given as
\begin{align} 
2 \left( \langle \delta_1(\bfx) \delta_1(\bfx+\bfr)\bfu_{2}(\bfx+\bfr) \rangle + \langle \delta_1(\bfx) \delta_2(\bfx+\bfr) \bfu_{1}(\bfx+\bfr) \right. & \left. \rangle \right) = \nonumber \\
   \frac{1}{2\pi^4} \int_0^\infty dk dy \int_{-1}^1 dx k^4 y j_1(kr)  P_0(k) P_0(ky)F_1(k;a) & \nonumber \\
  \times  \bigg(\frac{F_1(ky;a) G_2(ky,k,-x;a) y(1-yx)}{1+y^2-2yx} + G_1(ky;a) & F_2(ky,k,-x;a)x  \bigg),  \label{gsmt1}
\end{align}
and the (2,1) contribution (A6 of \cite{Reid:2011ar}) is given as 
\begin{align}
 2  \langle \delta_2(\bfx) \delta_1(\bfx+\bfr)\bfu_{1}(\bfx+\bfr) \rangle = & \nonumber \\ 
 \frac{1}{2\pi^4} \int_0^\infty dk dy  \int_{-1}^1 dx k^4 y x j_1(kr) & P_0(ky)P_0(k\sqrt{1+y^2-2yx}) \nonumber \\ 
  \times  F_1(k\sqrt{1+y^2-2yx};a) & G_1(ky;a)F_2(ky,k\sqrt{1+y^2-2yx}, u ;a),  \label{gsmt2}
\end{align}
where we have written the kernels in terms of the integrated vector's magnitudes and angle between them:  $|\bfk| = k$, $|\bfp| =  ky$ and $|\bfk - \bfp| = k\sqrt{1+y^2-2yx}$ with $x = \hat{\bfk}\cdot\hat{\bfp}$ and $u = \hat{(\bfk-\bfp)}\cdot \hat{\bfp}$. This notation is used for the velocity dispersion expressions below.


\subsection{Velocity Dispersion $\sigma_{12}^2(r,\mu^2)$}
The velocity dispersion depends on both the separation of the pair $r$ and the angle the separation vector $\bfr$ makes with the LOS $\phi_{lr}$ expressed through the argument $\mu^2 = {\rm cos}^2 (\phi_{lr})$. One can combine the perpendicular and parallel components of $\sigma_{12}^2$ to get the expression  (Eq.29 to Eq.32 of \cite{Reid:2011ar}) 
\begin{align}
&[1+\xi_L^r(r)] \sigma_{12}^2(r,\mu^2) =\\ \nonumber & 2 \left( \langle (u^{\ell}(\bfx))^2\rangle - \langle u^{\ell}(\bfx)  u^{\ell}(\bfx+\bfr) \rangle \right)  + 2 \langle \delta(\bfx)( u^{\ell}(\bfx))^2\rangle  \nonumber \\
& + 2 \left[\langle \delta(\bfx) (u^{\ell}(\bfx+\bfr))^2 - 2 \langle \delta(\bfx)  u^{\ell}(\bfx) u^{\ell}(\bfx+\bfr) \rangle \right] \nonumber \\
& + 2 \left[ \langle \delta(\bfx) \delta(\bfx + \bfr)  (u^{\ell}(\bfx))^2 \rangle - \langle \delta(\bfx) \delta(\bfx + \bfr)  u^{\ell}(\bfx)  u^{\ell}(\bfx + \bfr) \rangle \right], 
\label{sig12m}
\end{align} 
where $\ell$ denotes the component of $\bfu$ along the LOS. We give these component by component below. 
\begin{align} 
 2 \langle (u^{\ell}(\bfx))^2\rangle = 6 \sigma_v^2 = &  \frac{1}{3\pi^2} \int dk \frac{j_1(kr)}{kr} G_1(k;a)^2 P_0(k), \\ 
 - 2 \langle u^{\ell}(\bfx)  u^{\ell}(\bfx+\bfr) \rangle = & -\frac{1}{\pi^2} \int dk P_{\theta \theta}^{\rm 1-loop} (k;a),\mathcal{J}(kr,\mu^2),  
 \end{align}
 where again $P_{\theta \theta}^{\rm 1-loop}$ is evaluated using Eq.\ref{regptloop} and $\mathcal{J}(kr,\mu^2)$ is given in Eq.\ref{mybessel}. The third term contributes a constant to $\sigma_{12}^2(r,\mu^2)$. This is treated as a free parameter ($\sigma^2_{\rm iso}$) in our later analysis in Chapter 6 (see \cite{Satpathy:2016tct} for example) but we give the PT prediction for this isotropic contribution below 
 \begin{align}  
 2\langle \delta(\bfx)( u^{\ell}(\bfx))^2\rangle = \frac{1}{6\pi^4} \int dk dy \int_{-1}^1 & dx k^3 y^2 P_0(k) P_0(ky) G_1(k;a) \nonumber \\
  \times \left(\frac{2G_2(ky,k,x;a) F_1(ky;a)(1+yx)}{\sqrt{1+y^2+2yx}} \right. &\left.-\frac{xG_1(ky;a)F_2(ky,k,x;a)}{y}\right).
 \label{sigmaiso}
 \end{align} 
We can expand the 2nd line of Eq.\ref{sig12m} as 
\begin{align}
 2 \left[\langle \delta(\bfx) (u^{\ell}(\bfx+\bfr))^2\rangle \right. &\left.- 2 \langle \delta(\bfx)  u^{\ell}(\bfx) u^{\ell}(\bfx+\bfr) \rangle \right] =   \nonumber  \\ & 4 \langle \delta_1(\bfx)  u^{\ell}_1(\bfx+\bfr) u^{\ell}_2(\bfx+\bfr) \rangle + 2 \langle \delta_2(\bfx) (u^{\ell}_1(\bfx+\bfr))^2\rangle \nonumber \\ 
 & -4  \langle \delta_1(\bfx)  u^{\ell}_1(\bfx) u^{\ell}_2(\bfx+\bfr) \rangle -4  \langle \delta_1(\bfx)  u^{\ell}_2(\bfx) u^{\ell}_1(\bfx+\bfr) \rangle \nonumber \\ 
 &  - 4\langle \delta_2(\bfx)  u^{\ell}_1(\bfx) u^{\ell}_1(\bfx+\bfr) \rangle. 
 \end{align}
The integrals of these terms are given below
\begin{align}
4 \langle \delta_1(\bfx)  u^{\ell}_1(\bfx+\bfr) u^{\ell}_2(\bfx+\bfr) \rangle =& \nonumber \\  \frac{1}{2\pi^4} \int dk dy \int_{-1}^1  dx k^3 y P_0(k)P_0(ky) & \frac{F_1(k;a) G_1(ky;a) G_2(ky,k,x;a)}{\sqrt{1+y^2+2yx}} \nonumber \\  \times \left( j_0(kr)(y-2x-3x^2y) \right. & \left. - \mathcal{J}(kr,\mu^2) y(1-x^2) \right), \label{gsmt3}
\end{align}
\begin{align} 
 & 2 \langle \delta_2(\bfx) (u^{\ell}_1(\bfx+\bfr))^2\rangle = \nonumber \\
  & \frac{1}{16\pi^6} \int dk dy k^3 y G_1(ky;a)G_1(k;a) P_0(k) P_0(ky) \int_{-1}^1 dx_1 dx_2  \cos{(kyrx_1 + krx_2)}
\nonumber \\ & \times \int_0^{2\pi} d\phi_1 d\phi_2  F_2(k,y,\bar{x};a)  \left[ \mu^2(\bar{x}-2x_1x_2) - \bar{x}+x_1 x_2 \right],
\label{a13term}
 \end{align}
 where $\bar{x} = x_1x_2 + \sqrt{(1-x_1^2)(1-x_2^2)}\sin{\phi_1}\sin{\phi_2}$. The 4 dimensional angular integration in this expression is performed using the Monte Carlo integration algorithm Cuba \cite{Hahn:2004fe}. Next we have  
\begin{align}
& -4 \langle \delta_1(\bfx)  u^{\ell}_1(\bfx) u^{\ell}_2(\bfx+\bfr) \rangle = \nonumber \\ & -\frac{1}{\pi^4} \int dk dy \int_{-1}^1  dx k^3 y x P_0(ky) P_0(k\sqrt{1+y^2-2yx}) \mathcal{J}(kr,\mu^2) \nonumber \\ & \times G_1(ky;a) F_1(k\sqrt{1+y^2-2yx};a) G_2(ky,k\sqrt{1+y^2-2yx},u;a), 
\end{align}
and 
\begin{align}
& -4 \langle \delta_1(\bfx)  u^{\ell}_2(\bfx) u^{\ell}_1(\bfx+\bfr) \rangle 
 - 4\langle \delta_2(\bfx)  u^{\ell}_1(\bfx) u^{\ell}_1(\bfx+\bfr) \rangle =  \nonumber \\ &  -\frac{1}{\pi^4} \int dk dy \int_{-1}^1 dx k^3 y P_0(k) P_0(ky) \mathcal{J}(kr,\mu^2) \nonumber \\ & \times G_1(k;a)\left(G_2(ky,k,x;a)F_1(ky;a)\frac{y(1+yx)}{1+y^2+2yx} - x G_1(ky;a)F_2(ky,k,x;a)\right). \label{gsmt4}
\end{align}
Finally, the last term in Eq.\ref{sig12m} evaluates to 
\begin{align}
2\left[ \langle \delta(\bfx) \delta(\bfx + \bfr)  (u^{\ell}(\bfx))^2 \rangle \right.  & \left. - \langle \delta(\bfx) \delta(\bfx + \bfr)  u^{\ell}(\bfx)  u^{\ell}(\bfx + \bfr) \rangle \right] = \nonumber \\ & \xi(r) \sigma_{12,{\rm lin}}^2(r,\mu^2) + \frac{1}{2}u_{12,{\rm lin}}^2(r) \mu^2,
\label{sig12last}
\end{align} 
where $\sigma_{12,{\rm lin}}^2(r,\mu^2) $ and $u_{12,{\rm lin}}^2(r)$ are the linear predictions for the velocity dispersion (Eq.\ref{lins12}) and mean infall velocity (Eq.\ref{linv12}).  At leading order the first term in Eq.\ref{sig12last} cancels with the 2nd term on the LHS of Eq.\ref{sig12m} and so we omit it in our calculations and simply deduct the linear mean infall velocity term.
\newline
\newline
Perturbation theory predicts a constant contribution to the velocity dispersion, $\sigma^2_{\rm iso}$ (Eq.\ref{sigmaiso}) given in units of (Mpc/$h$)${}^2$. As mentioned, this is treated as a free parameter allowing us to describe deviations to the predicted scale dependance on small scales where non-linear FoG effects are strong and unable to be treated perturbatively.  
\newline
\newline 
This concludes our discussion on redshift space observables. Together Eq.\ref{redshiftps} and Eq.\ref{redshiftcor} provide theoretical estimates of clustering statistics for dark matter in a spatially expanding universe in the quasi non-linear regime. They are both modelled within the framework of perturbation theory but along two different routes, one from conservation of mass moving from real to redshift space and the other employs a probabilistic approach with an ansatz for the velocity distribution function. In the following 2 chapters we compare these predictions to fully non-linear measurements coming from suites of N-body simulations within three models of gravity: GR, Hu-Sawicki $f(R)$ and nDGP. This serves to quantify the applicable scales of the modelling as well as a validation of {\tt MG-Copter}. The next chapter is concerned with these comparisons in Fourier space.  
\newpage


\chapter{Beyond the Standard Model: The Redshift Space Power Spectrum}
\begin{chapquote}{J.D}
``Fog flooded the pass and the shrouded sun cast new light upon the mountainside."
\end{chapquote}


In this chapter we provide validation of {\tt MG-Copter} on two fronts. First, we compare the performance of the numerical algorithm discussed in Appendix A with the analytic forms for the real and redshift space power spectra in GR and nDGP gravity. Second, we compare results of the code for GR and $f(R)$ with simulation data. The results in this chapter rely heavily on the algorithm described in Appendix A and we suggest the reader looks at this before going through this and the following chapter. Importantly, we refer to sampling size $n_1$ introduced in Appendix A which quantifies the accuracy of the numerical algorithm. Further, we concern ourselves with dark matter distributions and no tracer bias is considered. All results in this section can be found in \cite{Bose:2016qun}. 

\section{Comparing to Analytic Forms}
\begin{figure}[H]
  \captionsetup[subfigure]{labelformat=empty}
  \centering
  \subfloat[]{\includegraphics[width=7.5cm, height=7cm]{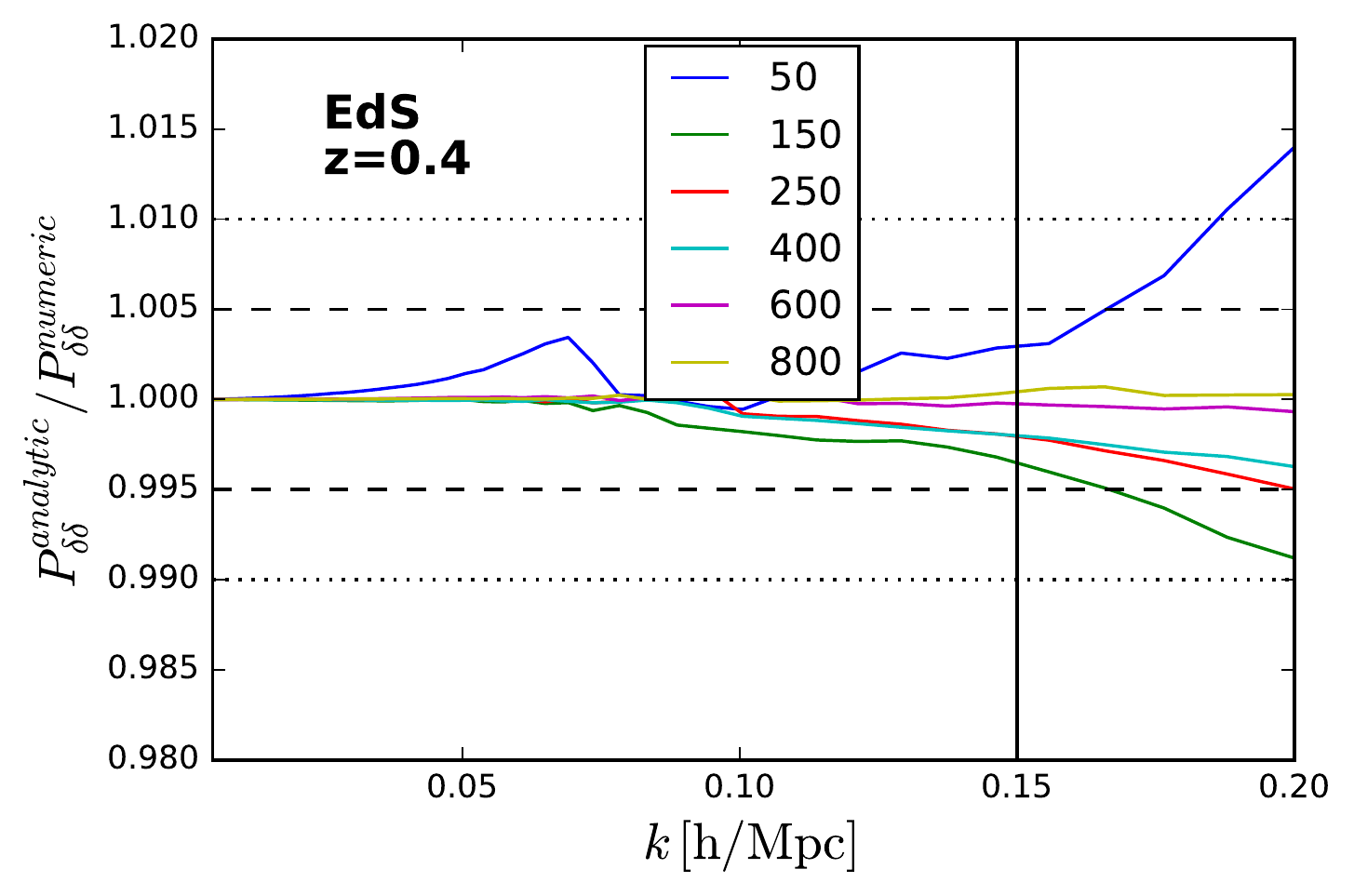}} \quad
  \subfloat[]{\includegraphics[width=7.5cm, height=7cm]{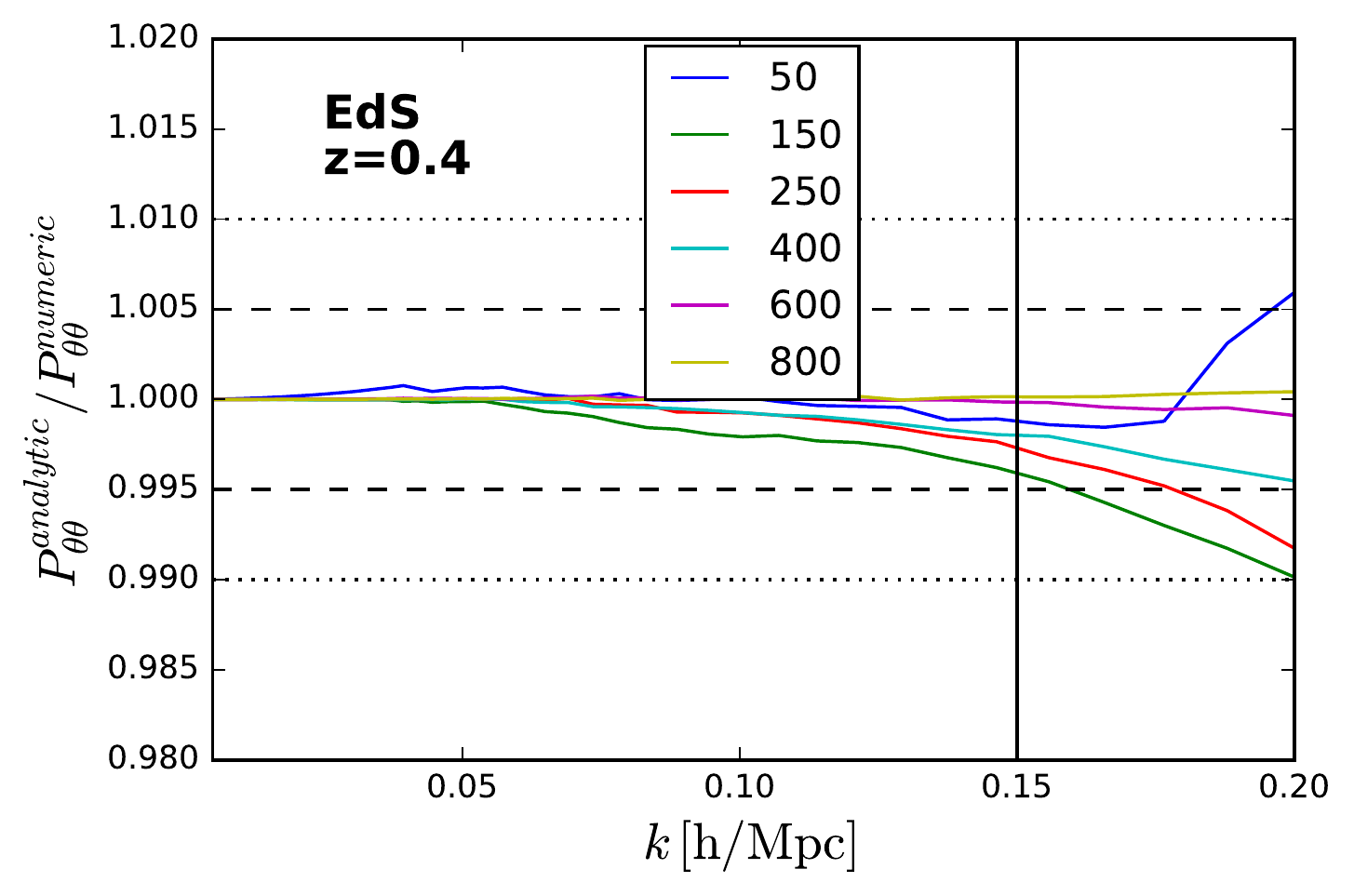}} 
  \caption{Test for convergence of the numerical to analytical 1-loop matter  (left) and velocity (right) power spectrum in the EdS cosmology for $n_1=50,150,250,400$ and $800$ at $z=0.4$.}
\label{convergence1}
\end{figure}

\begin{figure}[H]
  \captionsetup[subfigure]{labelformat=empty}
  \centering
  \subfloat[]{\includegraphics[width=7.5cm, height=7cm]{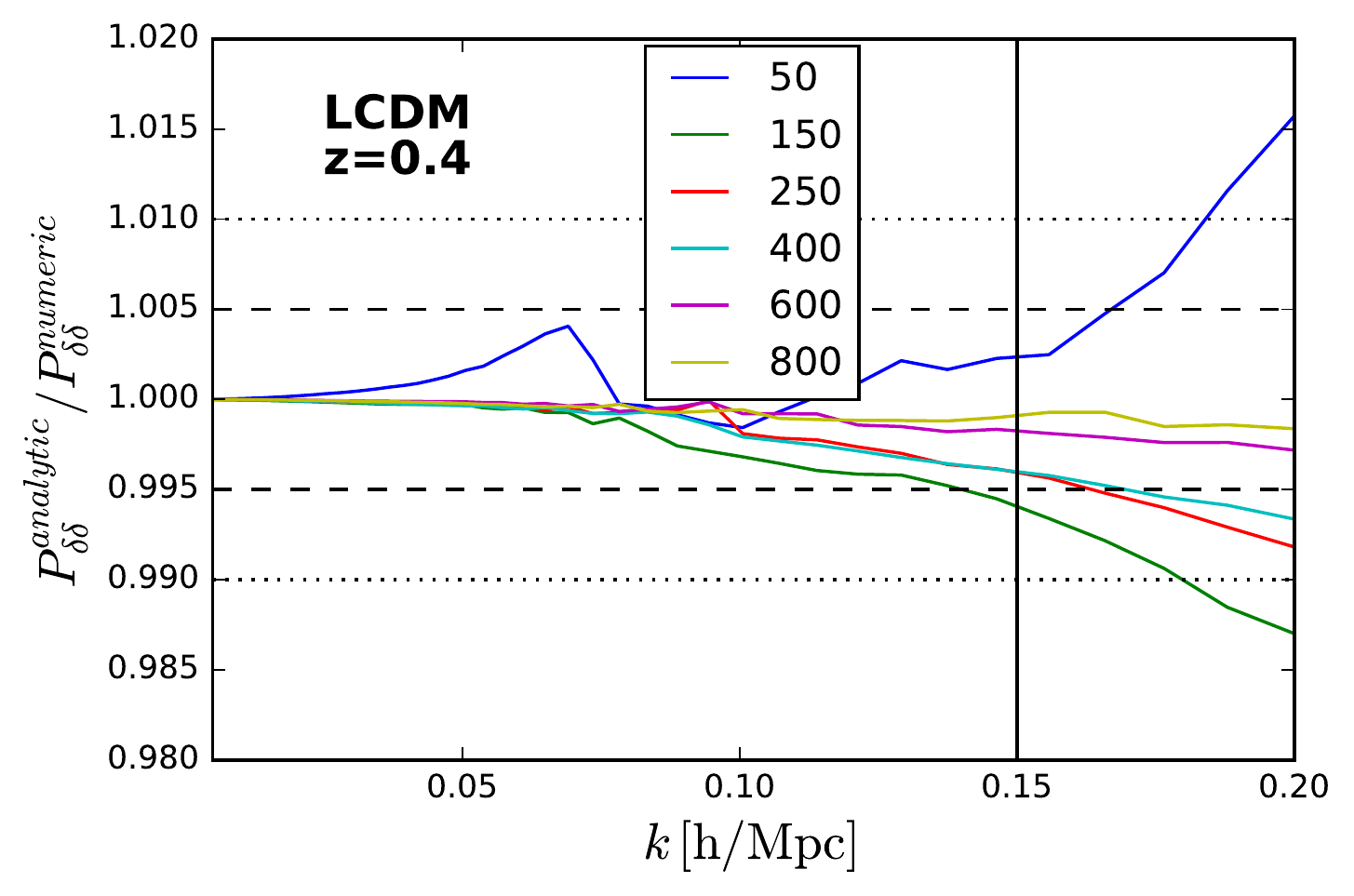}} \quad
  \subfloat[]{\includegraphics[width=7.5cm, height=7cm]{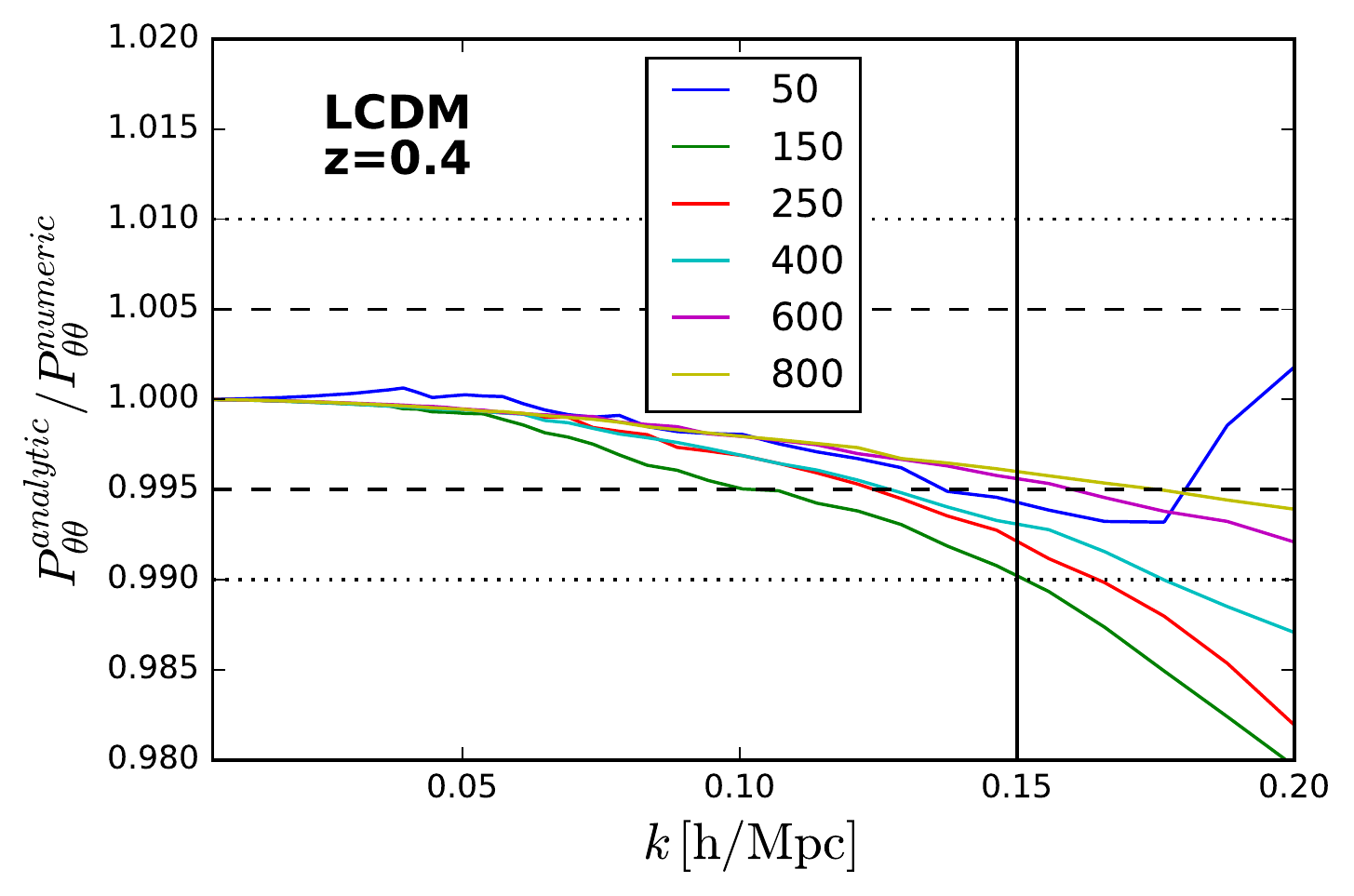}} 
  \caption{Test for convergence of the numerical to analytical 1-loop matter  (left) and velocity (right) power spectrum in the LCDM cosmology for $n_1=50,150,250,400$ and $800$ at $z=0.4$.}
\label{convergence2}
\end{figure}

\begin{figure}[H]
  \captionsetup[subfigure]{labelformat=empty}
  \centering
  \subfloat[]{\includegraphics[width=7.5cm, height=7cm]{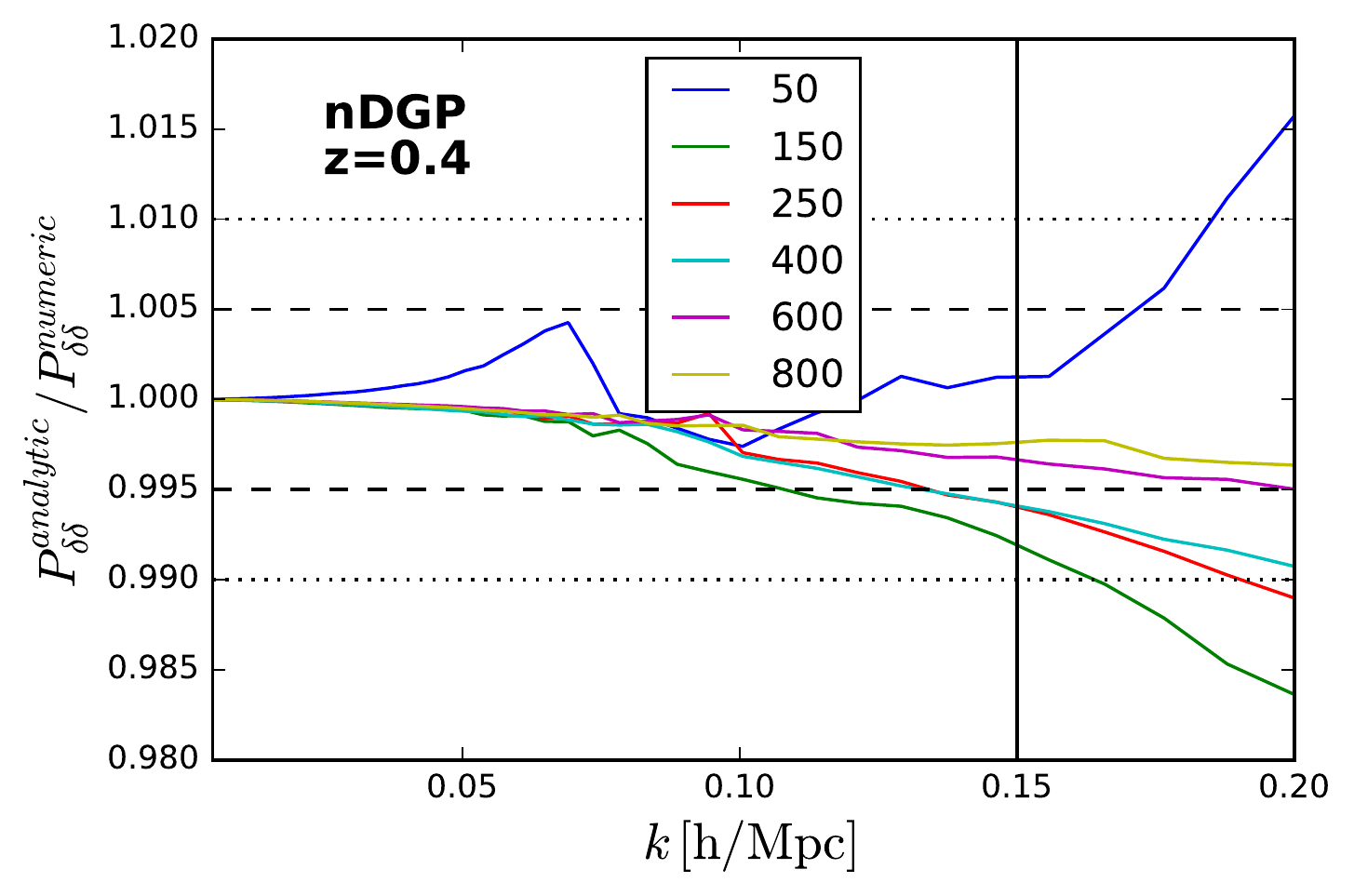}} \quad
  \subfloat[]{\includegraphics[width=7.5cm, height=7cm]{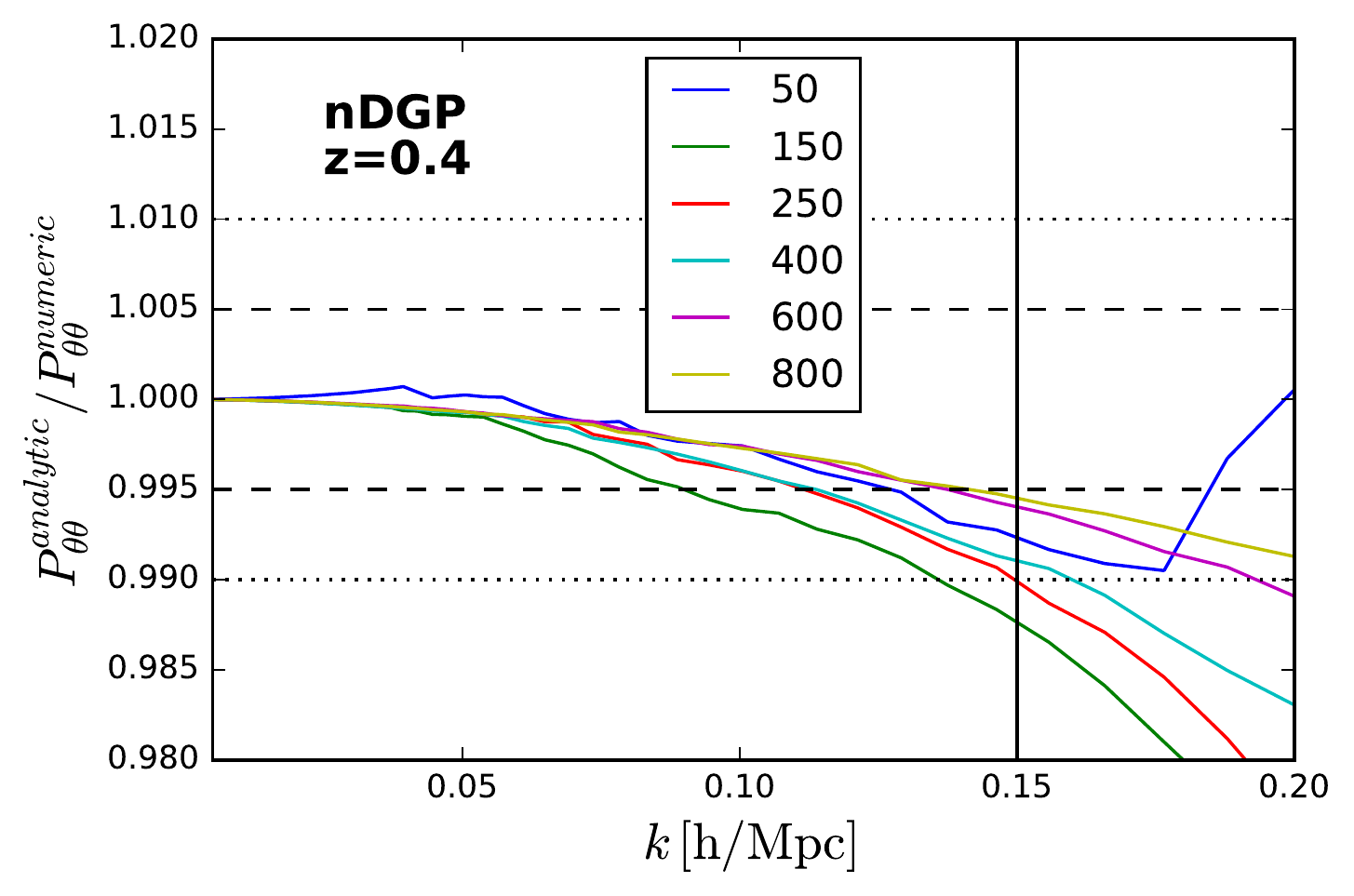}} 
  \caption{Test for convergence of the numerical to analytical 1-loop matter  (left) and velocity (right) power spectrum   in nDGP gravity for $n_1=50,150,250,400$ and $800$ at $z=0.4$.}
\label{convergence3}
\end{figure}

\begin{figure}[H]
  \captionsetup[subfigure]{labelformat=empty}
  \centering
  \subfloat[]{\includegraphics[width=7.5cm, height=7cm]{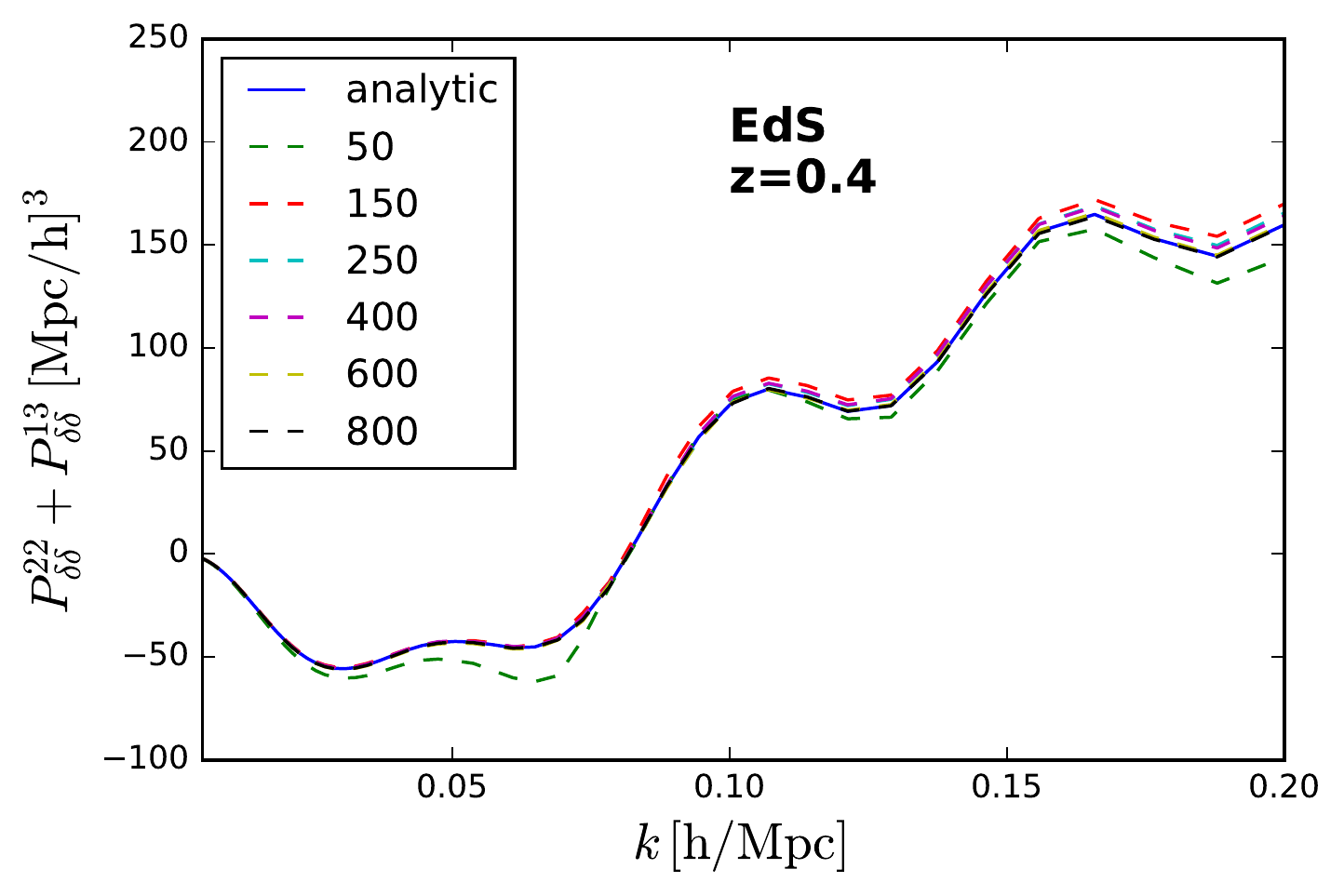}} \quad
  \subfloat[]{\includegraphics[width=7.5cm, height=7cm]{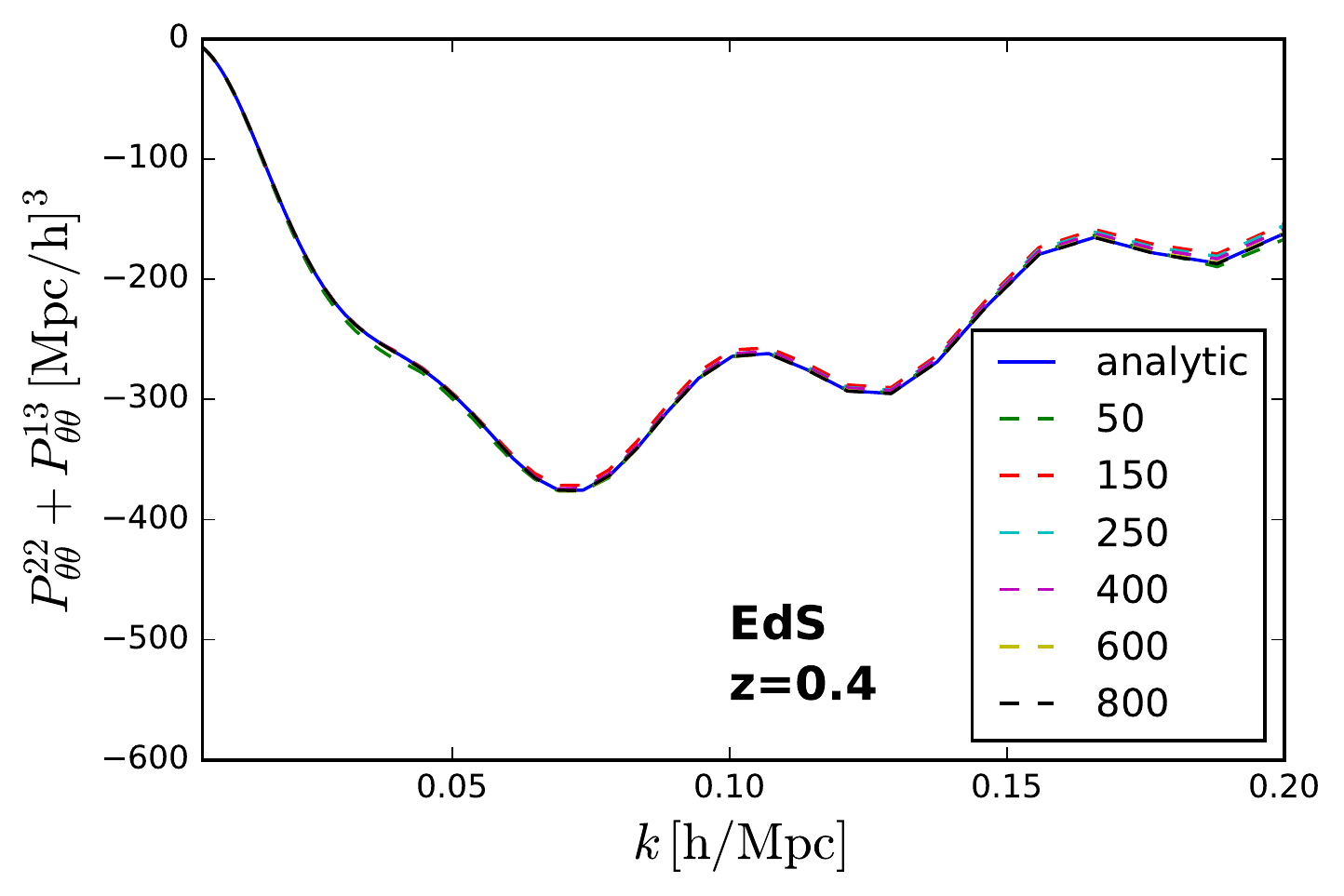}} 
  \caption{Test for convergence of the numerical to analytical matter  (left) and velocity (right) 1-loop contributions in the EdS cosmology for $n_1=50,150,250,400,600$ and $800$ at $z=0.4$.}
\label{convergence4}
\end{figure}

\begin{figure}[H]
  \captionsetup[subfigure]{labelformat=empty}
  \centering
  \subfloat[]{\includegraphics[width=7.5cm, height=7cm]{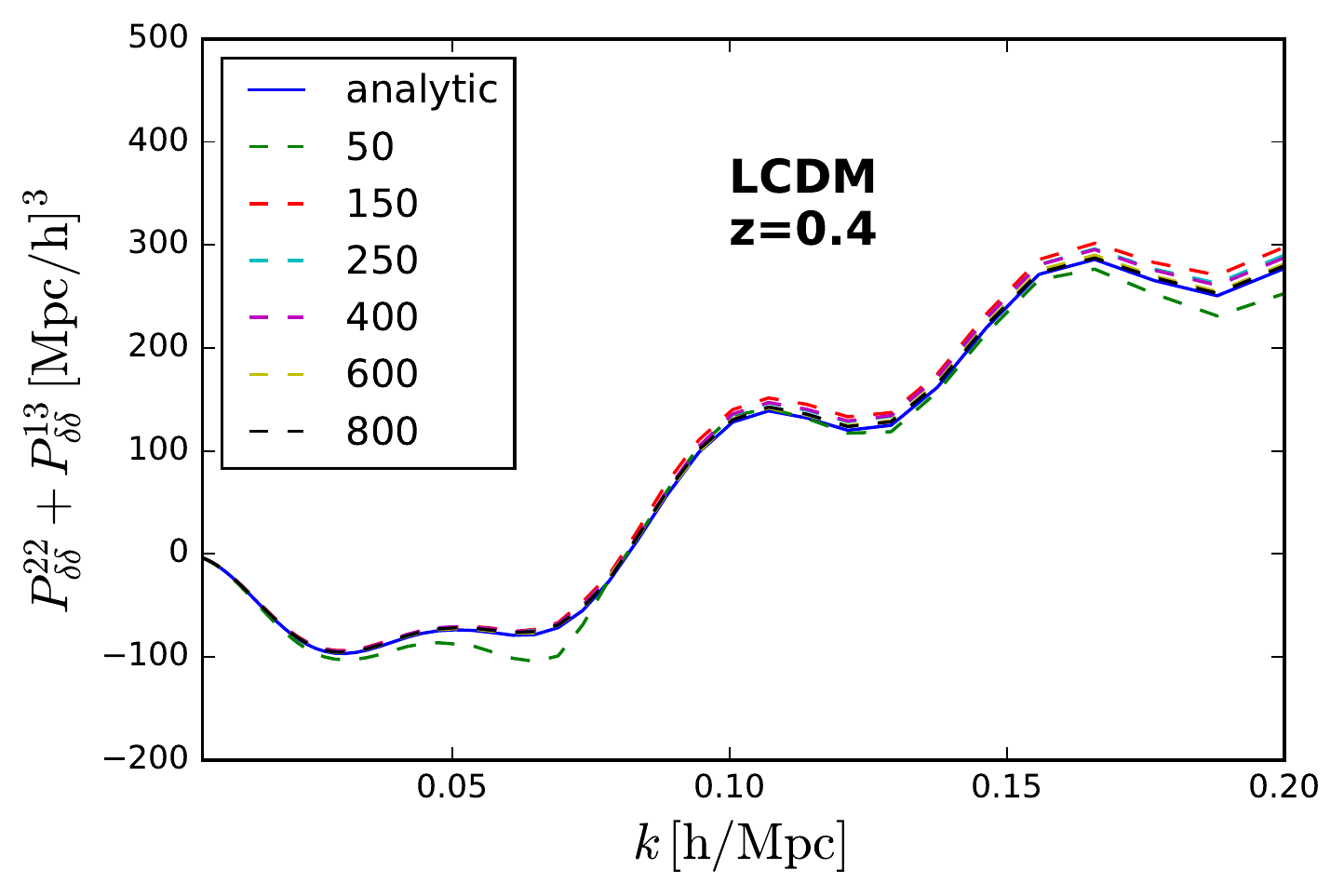}} \quad
  \subfloat[]{\includegraphics[width=7.5cm, height=7cm]{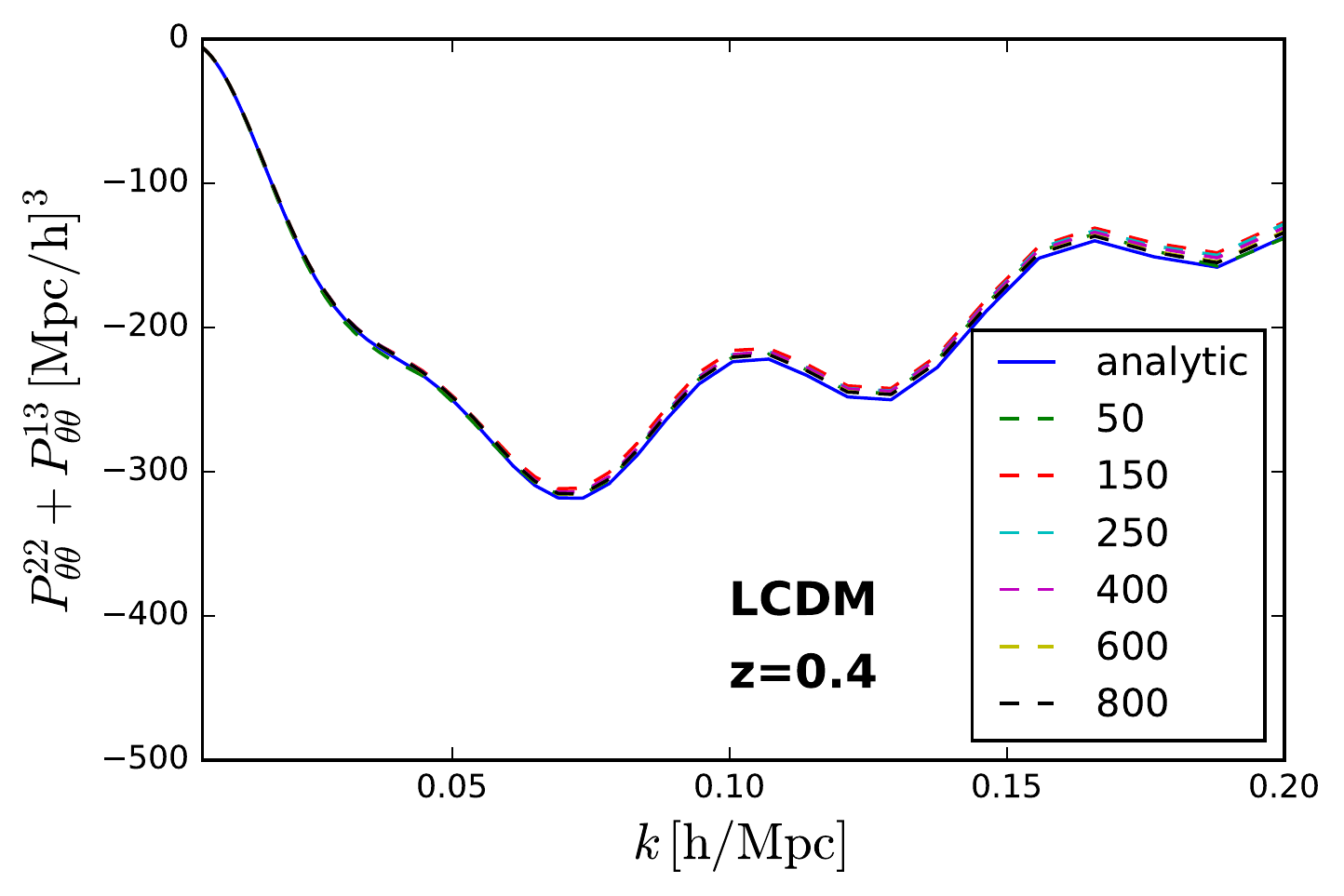}} 
  \caption{Test for convergence of the numerical to analytical matter (left) and velocity  (right) 1-loop contributions the LCDM cosmology for $n_1=50,150,250,400,600$ and $800$ at $z=0.4$.}
\label{convergence5}
\end{figure}

\begin{figure}[H]
  \captionsetup[subfigure]{labelformat=empty}
  \centering
  \subfloat[]{\includegraphics[width=7.5cm, height=7cm]{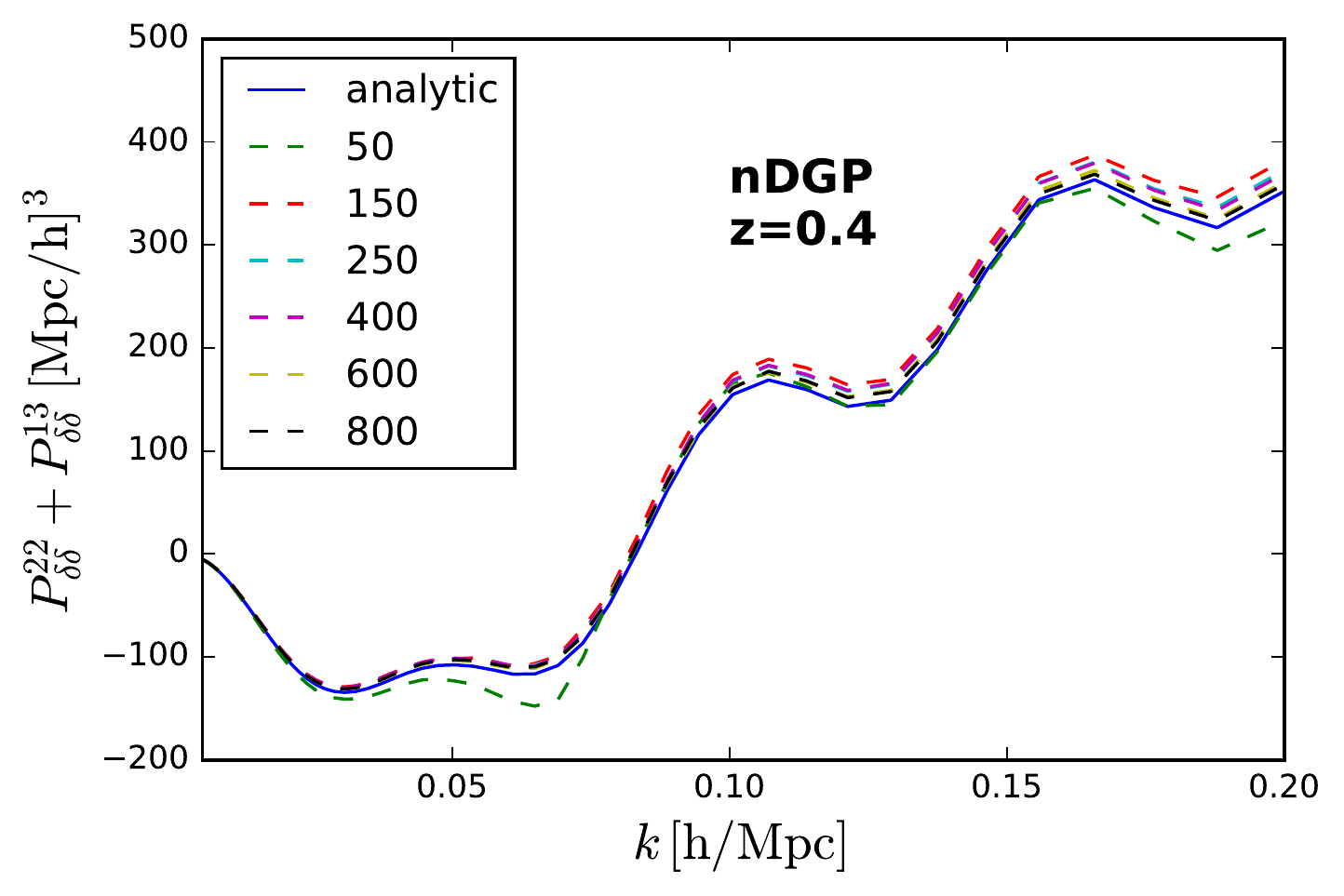}} \quad
  \subfloat[]{\includegraphics[width=7.5cm, height=7cm]{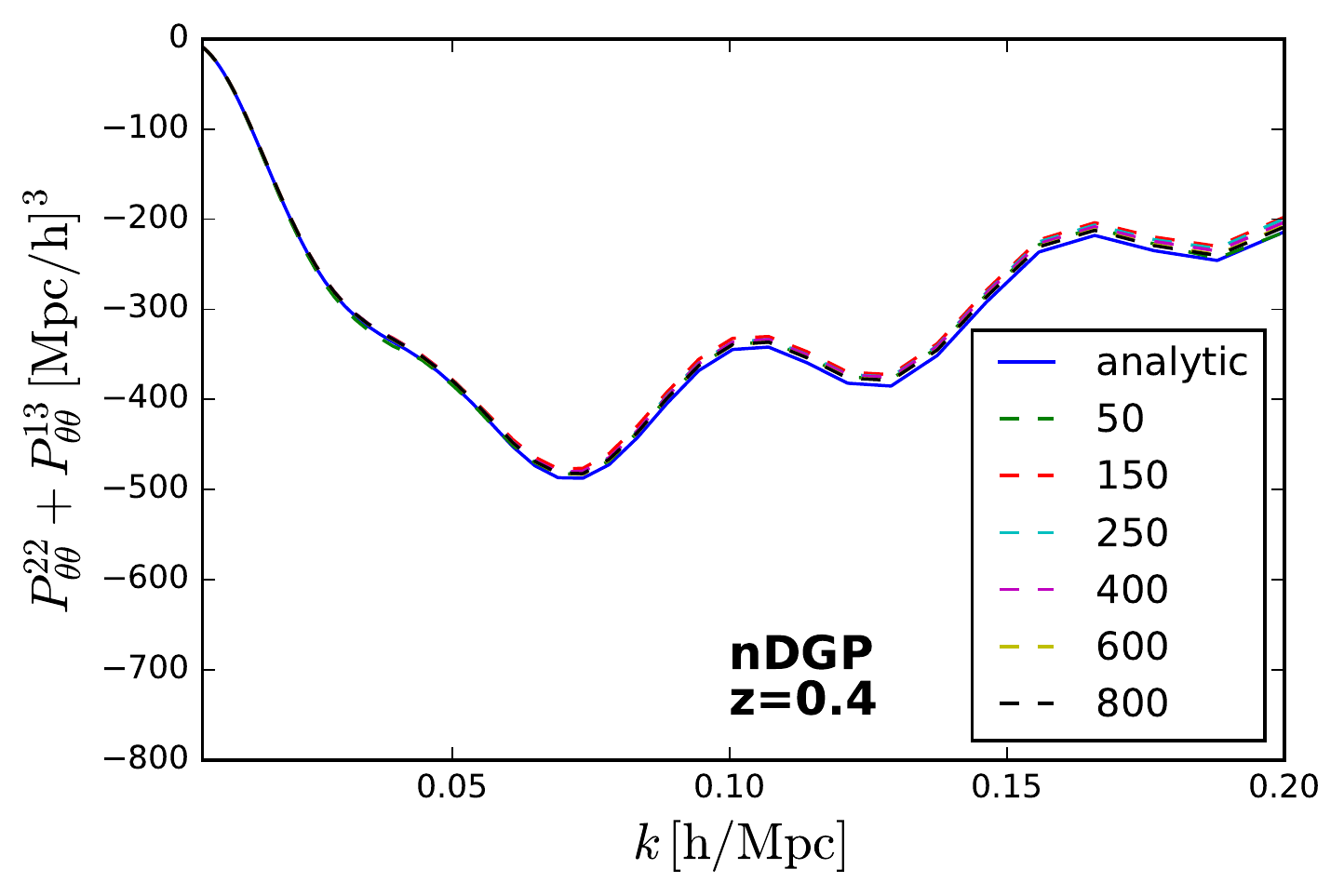}} 
  \caption{Test for convergence of the numerical to analytical matter (left) and velocity (right)1-loop contributions in nDGP gravity for $n_1=50,150,250,400,600$ and $800$ at $z=0.4$.}
\label{convergence6}
\end{figure}

\begin{figure}[H]
  \captionsetup[subfigure]{labelformat=empty}
  \centering
  \subfloat[]{\includegraphics[width=10.5cm, height=7cm]{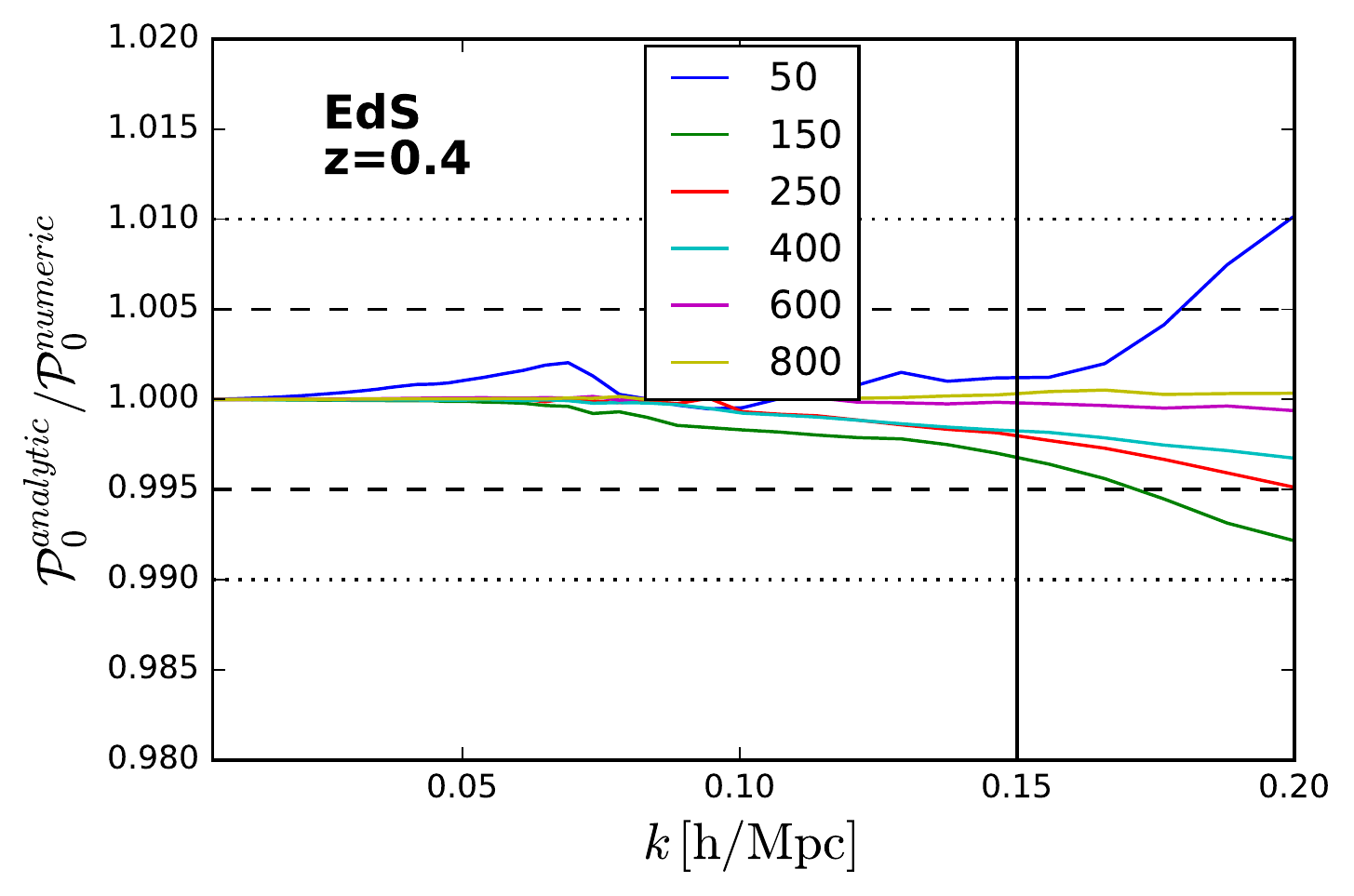}} 
  \caption{Test for convergence of the numerical to analytical TNS redshift space monopole of the power spectrum in the EdS cosmology for $n_1=50,150,250,400$ and $800$ at $z=0.4$.}
\label{convergence7}
\end{figure}

\begin{figure}[H]
  \captionsetup[subfigure]{labelformat=empty}
  \centering
  \subfloat[]{\includegraphics[width=7.5cm, height=7cm]{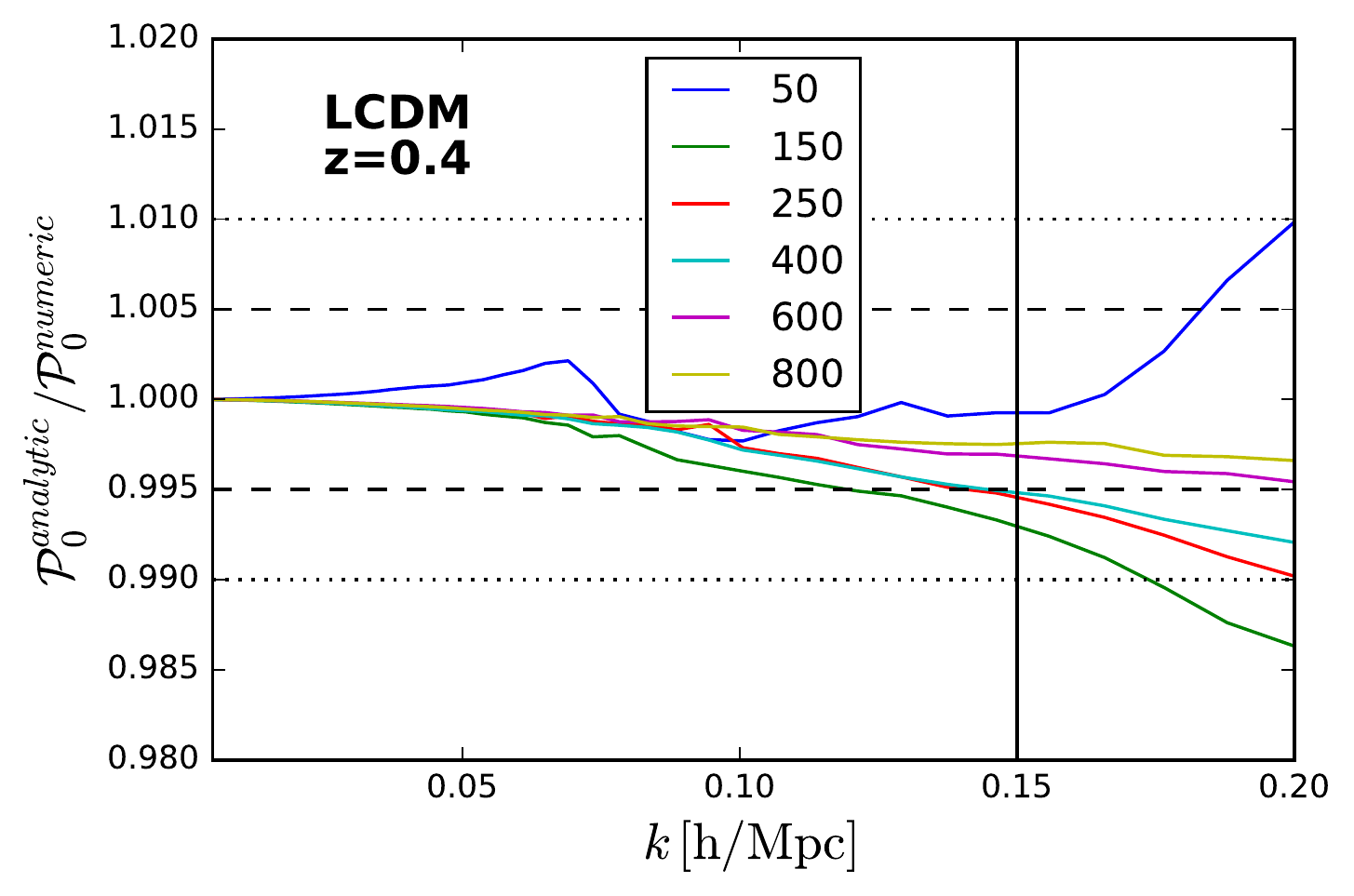}} \quad
  \subfloat[]{\includegraphics[width=7.5cm, height=7cm]{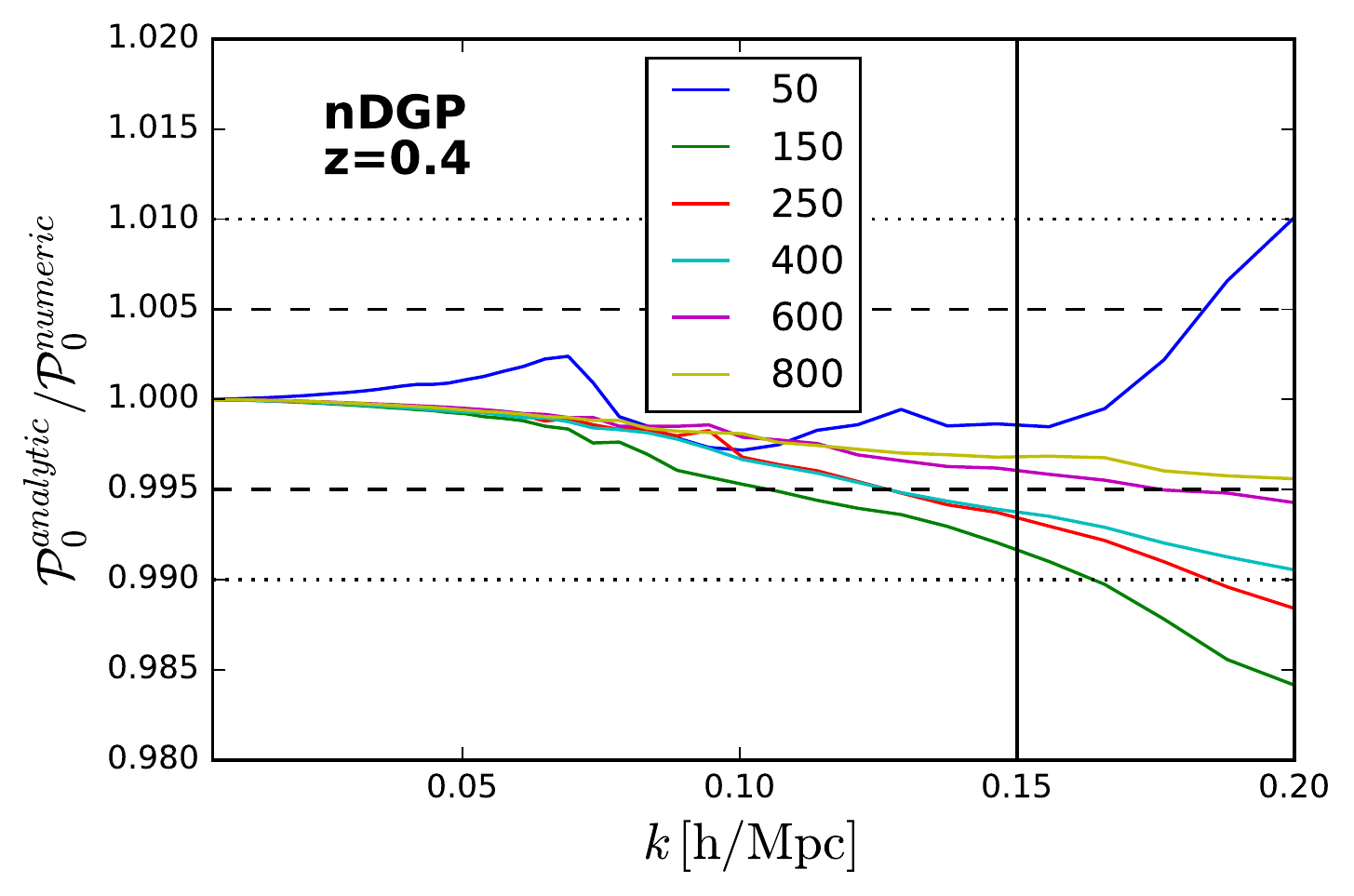}} 
  \caption{Test for convergence of the numerical to analytical TNS redshift space monopole of the power spectrum in the LCDM cosmology (left) and in nDGP gravity (right) for $n_1=50,150,250,400$ and $800$ at $z=0.4$.}
\label{convergence8}
\end{figure}

Here we discuss the performance of the code  in reproducing analytic results. In this section all results are based on a primordial linear power spectrum generated by the Boltzman solver CLASS \cite{Blas:2011rf} with a flat LCDM cosmology with the following parameters $h=0.697$, $n_s=0.971$, $\Omega_b= 0.046$, $\Omega_m = 0.281$ and  $\sigma_8=0.82$ \footnote{$\sigma_8$ gives the amplitude of the density contrast at a scale of $8$Mpc/$h$, giving the power spectrum normalisation. $n_s$ gives the tilt of the primordial power spectrum.}. We take $z=0.4$ corresponding to $a=0.71$. We compute the 1-loop correction terms $P^{22}(k)$ and $P^{13}(k)$ as well as the monopole $P_0(k)$ for 30 $k$ modes taken logarithmically from $k=0.005$h/Mpc to $k=0.2$ $h$/Mpc, with the power spectrum multipoles being given by 
\begin{equation}
P_l^{S}(k) = \frac{2l+1}{2}\int^1_{-1}d\mu P_{TNS}^{S}(k,\mu)\mathcal{P}_l(\mu),
\label{multipolesl}
\end{equation}
where $\mathcal{P}_l(\mu)$ denote the Legendre polynomials and $P_{TNS}^{S}(k)$ is given by Eq.\ref{redshiftps}. This is done for 3 different cases; the LCDM cosmology describing our linear power spectrum, an EdS cosmology and for the normal branch of the DGP model (nDGP) using the LCDM case background cosmology. For the nDGP case we take $\Omega_{rc} = 1/(4 H_0^2 r_c^2)=0.438$.  
\newline
\newline
We expect the numerical algorithm to exactly reproduce the analytic expressions for the EDS cosmology where no scale dependence is involved in the density and velocity field's evolution (see Chapter 2). We also expect a very close match for both the LCDM and nDGP cosmologies where scale and time separation of the perturbations is known to be a very good approximation. We refer the reader to \cite{Koyama:2009me} for the analytic forms of the nDGP perturbations up to 3rd order. Further, to determine the realm of validity for the SPT calculations \footnote{Which will be important for determining sampling size $n_1$.} we solve the following equation 
\begin{equation}
\frac{k_{\rm max}^2}{6\pi^2} \int^{k_{\rm max}}_0 dq P^{11}_{\delta \delta} (q;a) = 0.18
\label{validityrange}
\end{equation}
This range is the 1 \% accuracy regime, found empirically by comparing perturbation theory with N-body simulations for GR \cite{Nishimichi:2008ry}. It is useful in providing a rough realm of validity in the general model case. For our convergence tests we plot up to a $k_{max} = 0.2$ which was calculated using the above relation in a LCDM cosmology for $z = 0.4$  which is the upper redshift on DESI's bright galaxy survey \cite{Levi:2013gra}, the lowest redshift survey it will undertake. It will also look at 18 million emission line galaxies in the redshift range $0.6 \leq z \leq 1.6$. Perturbation theory is known to do better at higher redshift (see Fig.\ref{sptvsobs}), giving a larger realm of validity for this and other upcoming surveys. We also note here that the validity range ($k_{max}$) is expected to be smaller for the nDGP case because of the fifth force's enhancement of velocities and clustering at larger scales. This statement is generically true for most modified models of gravity. 
\newline
\newline
Fig.\ref{convergence1}, Fig.\ref{convergence2} and Fig.\ref{convergence3} illustrate the convergence of the numerical auto 1-loop  power spectra to the analytic results for the EdS, LCDM and nDGP cases respectively as we increase the amount of sampling \footnote{See Appendix A for information on sampling.}. The horizontal lines show the $0.5\%$ and $1\%$ deviations while the vertical line at $k=0.15h$/Mpc denotes a rough realm of validity for the DGP case, determined using Eq.\ref{validityrange}. The numerical results were computed using a single initialisation of the kernels since the Euler and continuity equations in these models are $k$ independent. We also plot the convergence for the 1-loop part of the power spectra ($P^{22}(k)+P^{13}(k)$) in Fig.\ref{convergence4}, Fig.\ref{convergence5} and Fig.\ref{convergence6}. Fig.\ref{convergence7} and Fig.\ref{convergence8} show the same convergence for the TNS monopole.
\newline
\newline
Based on the EdS case, we find that $n_1=150$ is sufficient sampling to achieve below percent accuracy within the validity range and can be safely used as a standard for the general scale independent case. With this sampling size the numerical algorithm computes 30 1-loop power spectra values  in 5 seconds. If further accuracy is required the sampling can easily be increased at a time cost. Parallelisation over $k$ using MPI and {\tt OpenMP} makes it possible to get around this time cost very easily. In the next section we look at $f(R)$ gravity as a $k$-dependent test of the code. 


\section{Scale Dependent Perturbations in $f(R)$}
\begin{figure}[H]
  \captionsetup[subfigure]{labelformat=empty}
  \centering
  \subfloat[]{\includegraphics[width=7.5cm, height=7cm]{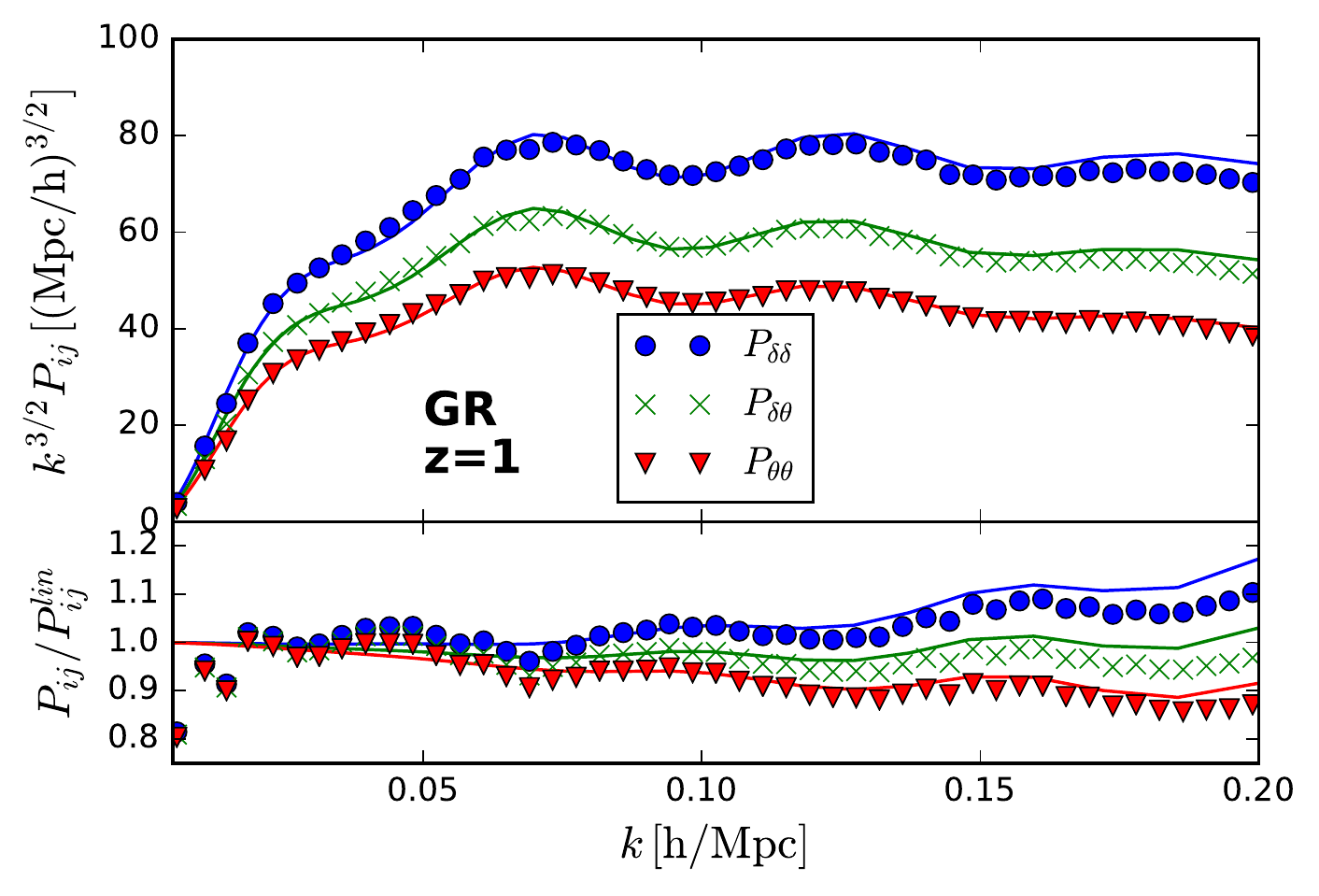}} \quad
  \subfloat[]{\includegraphics[width=7.5cm, height=7cm]{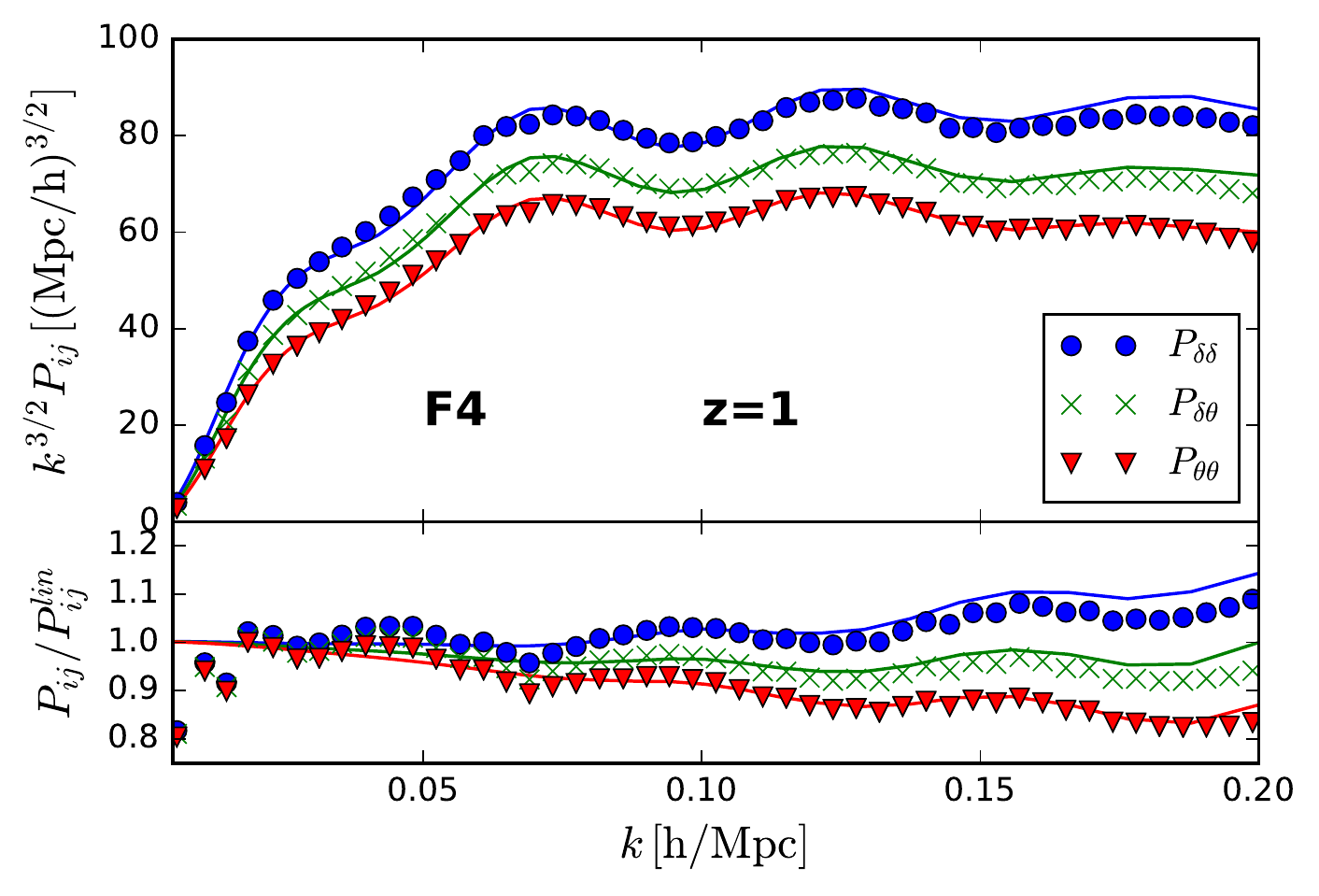}} 
  \caption{Comparing {\tt MG-Copter} predictions (lines) with the N-body predictions (points) of the auto and cross power spectra of density and velocity fields in real space at $z=1$ for GR (left) and $f(R)$ (right).  The top panels show the power spectra multiplied by $k^{3/2}$ and the bottom panels show the deviations from the linear predictions.  }
\label{frps}
\end{figure}

\begin{figure}[H]
  \captionsetup[subfigure]{labelformat=empty}
  \centering
  \subfloat[]{\includegraphics[width=7.5cm, height=7cm]{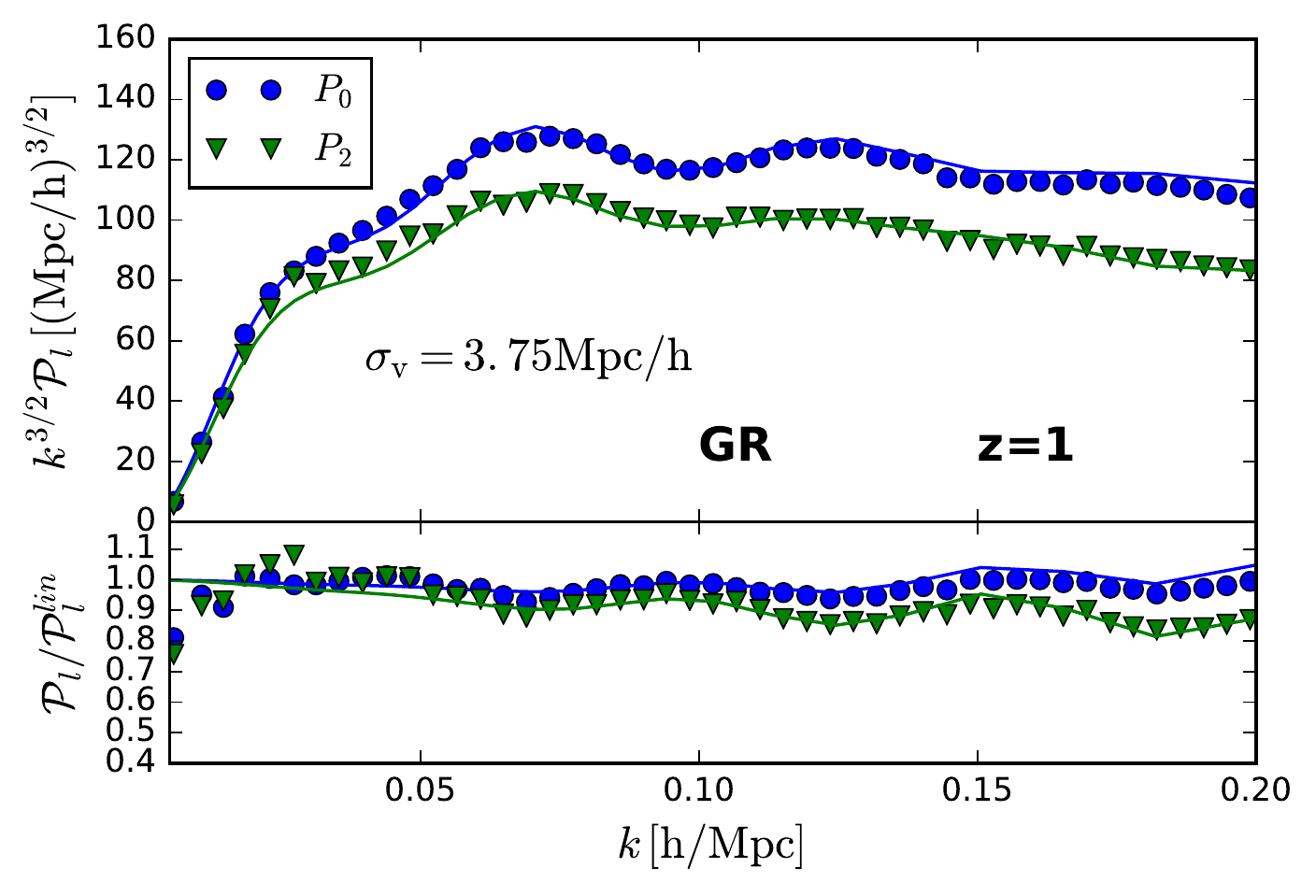}} \quad
  \subfloat[]{\includegraphics[width=7.5cm, height=7cm]{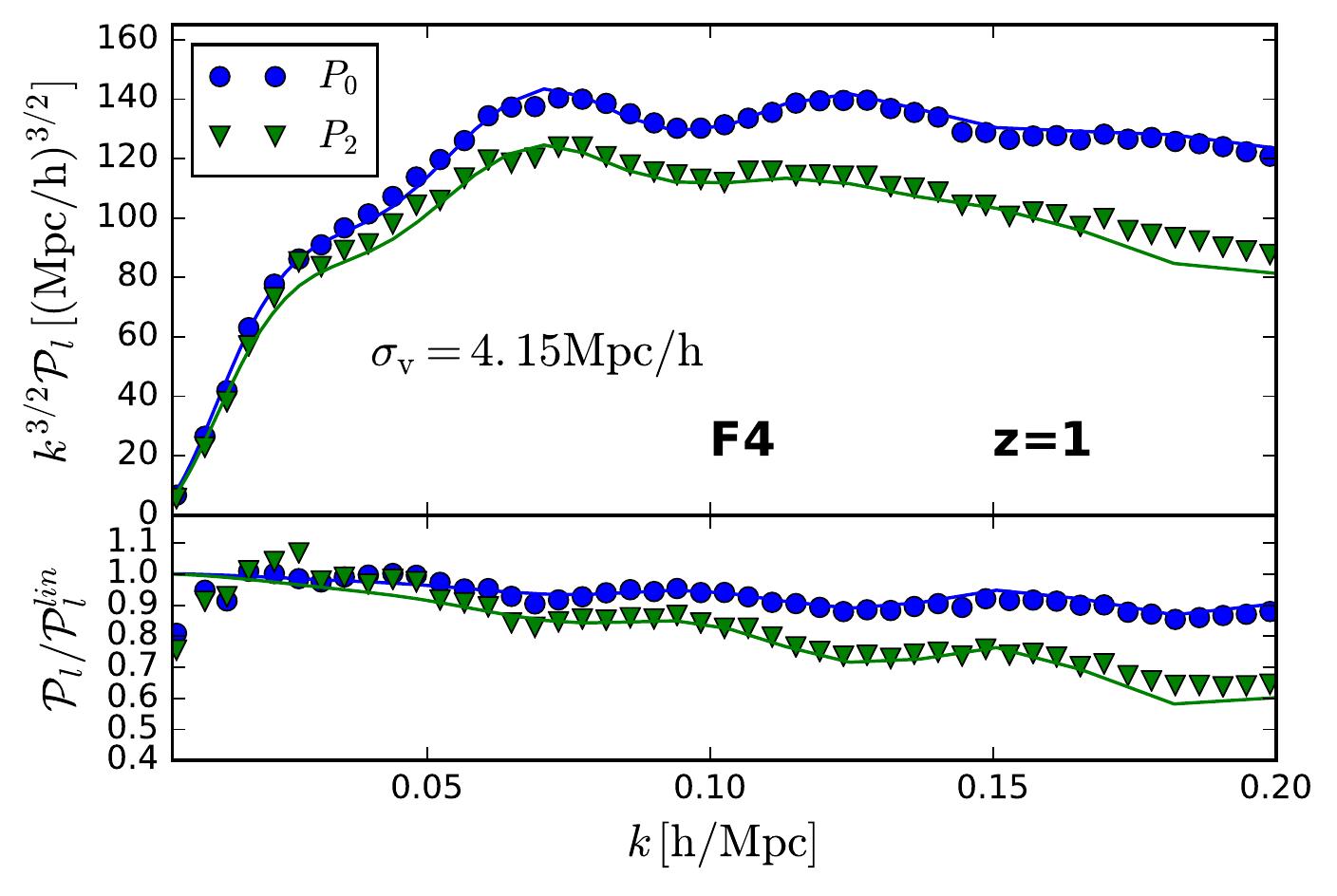}} 
  \caption{Comparing {\tt MG-Copter} predictions (lines) with the N-body predictions (points) of the TNS redshift space monopole (blue) and quadrupole (green) power spectra at $z=1$ for GR (left) and $f(R)$ (right).  The top panels show the multipoles multiplied by $k^{3/2}$ and the bottom panels show the deviations from the linear predictions. }
\label{frmulti}
\end{figure}

Here we attempt to reproduce some results of \cite{Taruya:2013quf}. In that work the authors compute the  SPT predictions for the 1-loop power spectra as well as the TNS redshift space monopole ($\ell=0$) and quadrupole ($\ell=2$) moments  for $f(R)$ and LCDM, and compare them with N-body results. The N-body results they use are from a subset of simulations presented in \cite{Li:2012by,Jennings:2012pt} which are an average of 6 realisations. Each realisation uses a cubic box of size $1.5 $ Gpc/$h$ and $1024^3$ particles, with initial conditions generated at $z=49$ using a linear matter power spectrum generated with the cosmological parameters: $\Omega_m=0.24, \Omega_\Lambda = 0.76, \Omega_b=0.0481, h=0.73, n_s=0.961$ and $\sigma_8=0.801$. We use this linear power spectrum to generate the SPT results. 
\newline
\newline
The authors consider the model of Hu and Sawicki (see Sec.3.3) with $n=1$ and $|f_{R0}|=10^{-4}$, henceforth called $F4$. The results are computed at $z=1$ corresponding to $a=0.5$. 
\newline
\newline
Fig.\ref{frps} compares the real space power spectra ($P_{\delta \delta}, P_{\delta \theta}$ and $P_{\theta \theta}$) from {\tt MG-Copter} using $n_1=250$ with the N-body results for both GR and F4. We find the N-body results match the SPT calculations very well within the considered range of scales. The bottom panels of the plot also compare the non-linearity coming from the 1-loop corrections with the full non-linearity of the simulations. The corrections agree with the N-body results at the percent level up to $k\sim 0.12 h$/Mpc and at the 3\% level at $k\sim 0.15 h$/Mpc.  We note here that the range of validity of SPT in $f(R)$ as dictated by Eq.(\ref{validityrange}) is less than in LCDM. This is because we have stronger growth at linear order because of fifth force effects. Fig.\ref{frps} and Fig.\ref{frmulti} both exhibit the earlier breakdown of SPT in this model. 
\newline
\newline
Moving to redshift space, we look at the monopole and quadrupole in the TNS model. For the FoG function we adopt a Gaussian damping function 
\begin{equation}
D_{FoG}(k\mu \sigma_v) = \exp[-(k\mu \sigma_v)^2], 
\end{equation}
where $\sigma_v$ is the free parameter to be fit to the N-body data. We fit $\sigma_v$ using a $\chi^2$ fit
\begin{equation}
\chi^2 = \sum_{l=0,2}\sum_i\frac{ \left[P_{l,N-body}^{(S)}(k_i)-P_{l,TNS}^{(S)}(k_i)\right]^2}{[\Delta P_{l}^{(S)}(k_i)]^2},
\end{equation}
where we sum up to $k_i=0.15h$/Mpc. $\Delta P_{l}^{(S)}(k_i)$  is given by the variance over the 6 realisations. We find that for F4 $\sigma_v=4.15$Mpc/$h$ while for GR we find a  best fit of $\sigma_v=3.75$Mpc/$h$. The $f(R)$ value is larger because of fifth force enhancements of velocities. Fig.\ref{frmulti} shows that the TNS model can accurately fit the data within the realm of validity highlighted in Fig.\ref{frps}. We note some deviations in the quadrupole at low $k$, but this is still within the cosmic variance errors of the simulation.  
\newline
\newline
This concludes our validation of {\tt MG-Copter}. Appendix A discusses time costs for various sampling selections and optimisations of the code. The next two chapters are dedicated to applications of the code.
 \begin{figure}[H]
  \captionsetup[subfigure]{labelformat=empty}
  \centering
  \subfloat[]{\includegraphics[width=10.5cm, height=7cm]{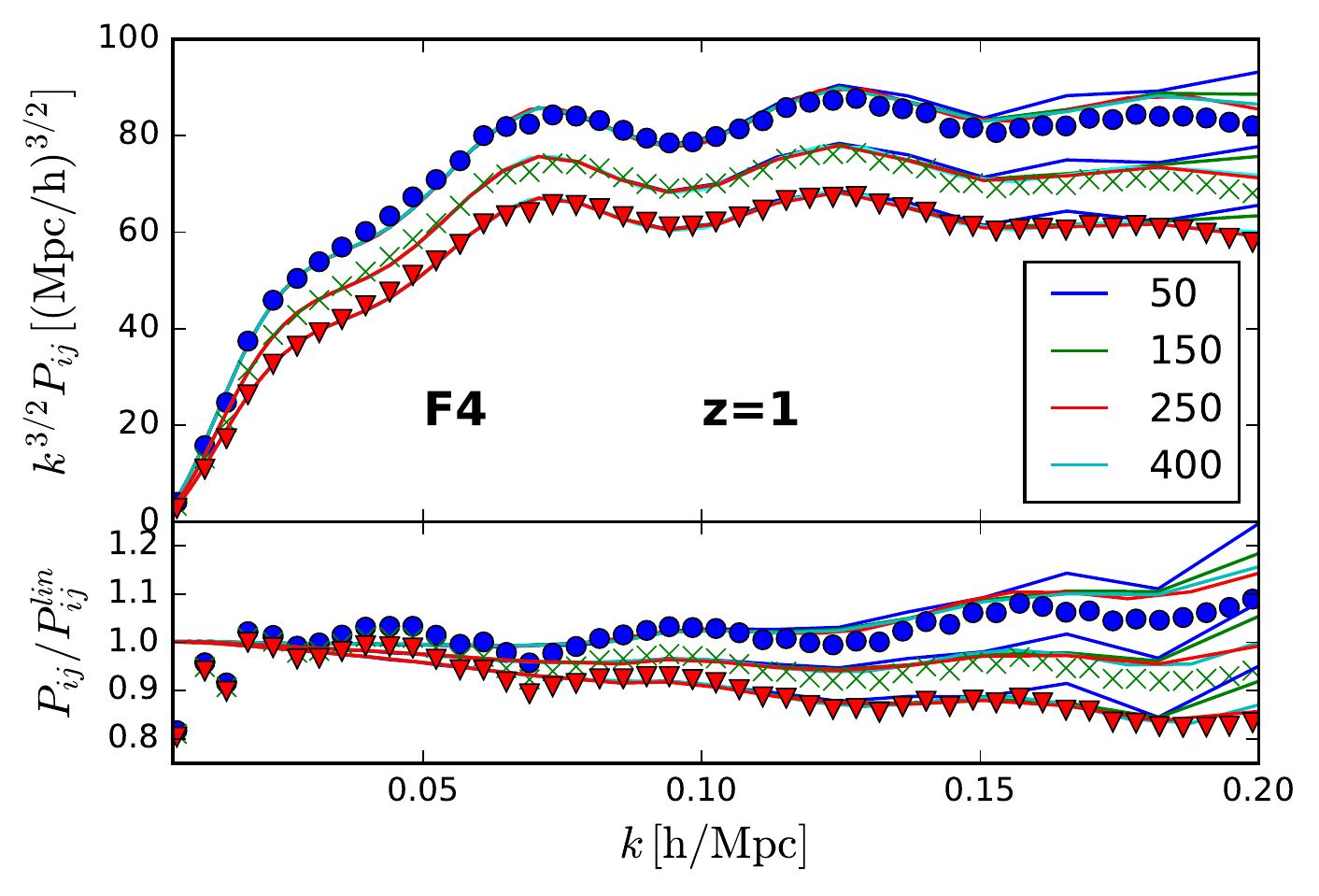}} \quad
  \caption{Replica of Fig.\ref{frps} (right)  for for $n_1=50, 150, 250$ and $400$.}
\label{frconv}
\end{figure}


\chapter{Beyond the Standard Model: The Redshift Space Correlation Function}
\begin{chapquote}{J.D}
``The fog receded and the cold, hard, physical, familiar mountain floor expanded into view."
\end{chapquote}

In this Chapter we will present predictions using Eq.\ref{redshiftcor} for the three models used in the last chapter; nDGP, Hu-Sawicki $f(R)$ and GR. We will compare these results with the FT of Eq.\ref{redshiftps} where the free parameter of the model, $\sigma_v$, is fit to N-body simulations. Again, this is done for dark matter only and no tracer bias is included. The Fourier space comparisons for nDGP can be found in Appendix A while for GR and $f(R)$ we use the best fit $\sigma_v$ found in \cite{Taruya:2014faa}.  All results in this section can be found in \cite{Bose:2017dtl}. 
\newline
\newline
Traditionally the correlation function measured from surveys has been used for theory-data comparisons. Here we focus on the GSM. The GSM has been shown to fit N-body data to percent level on and below the BAO scale in the case of GR and has been widely applied to data \cite{Samushia:2012iq,Reid:2012sw,Wang:2013hwa,Satpathy:2016tct,Alam:2015qta}. Work has also been done in extending and modifying the model to achieve a better range of applicability \cite{Uhlemann:2015hqa,Bianchi:2016qen,Bianchi:2014kba}. When considering the first two multipole moments, the GSM can accurately probe down to scales $ s \geq 30 \mbox{Mpc}/h$ in the GR case when compared to N-body simulations for halos and matter and does very well in comparison to other redshift space correlation function models \cite{White:2014naa}. 
\newline
\newline
Together with the TNS model, the GSM is one of the most widely used theoretical templates applied to observational data. By having these two models at one's disposal it is possible to make independent comparisons to data where observational systematics are different for each statistic. This will help beat down systematic uncertainty in growth measurements. Further, by comparing the GSM predictions with the FT of the TNS power spectra, we can get a handle on their consistency. We do expect the TNS transform to be the more accurate of the two because it makes use of the free parameter $\sigma_v$ to capture small scale dispersion effects whereas the pure GSM model contains no phenomenological ingredients. To merit the GSM approach, the TNS approach requires resummation techniques such as RegPT to perform the Fourier transform. These techniques are currently under scrutiny in light of the effective field theory of large scale structure (EFToLSS)\cite{Baumann:2010tm,Carrasco:2012cv}. These techniques are not needed in the GSM modelling. Of course one needs simulation measurements to see which approach does better, but their comparison has merit as a consistency check in their ability to model the observations as well as deviations from GR. This consistency is explored at and around the BAO scale ($50\mbox{Mpc}/h \leq s \leq 180 \mbox{Mpc}/h$) in this chapter. 
\newline
\newline
Our background cosmology is taken from WMAP9 \cite{Hinshaw:2012aka}: $\Omega_b = 0.046$, $\Omega_m = 0.281$, $h=0.697$, and $n_s=0.971$. The simulations used to measure $\sigma_v$ have the following specifications. The box width is $1024 \mbox{Mpc}/h$ with $1024^3$ dark matter particles used and a starting redshift of $49$. The linear theory power spectrum normalisation was set to be $\sigma_8=0.844$. The nDGP simulation uses $\Omega_{rc}=1/4r_c^2H_0^2=0.438$ while the $f(R)$ simulation uses $|f_{R0}|=10^{-4}$. We consider the redshift of $z=0.5$ where SPT benefits from a good range of validity while still being relevant for upcoming surveys such as Euclid and DESI. We will start with a comparison of linear and non-linear predictions for the real space correlation function.


\section{Real Space}
\begin{figure}[H]
  \captionsetup[subfigure]{labelformat=empty}
  \centering
  \subfloat[]{\includegraphics[width=10.3cm, height=7cm]{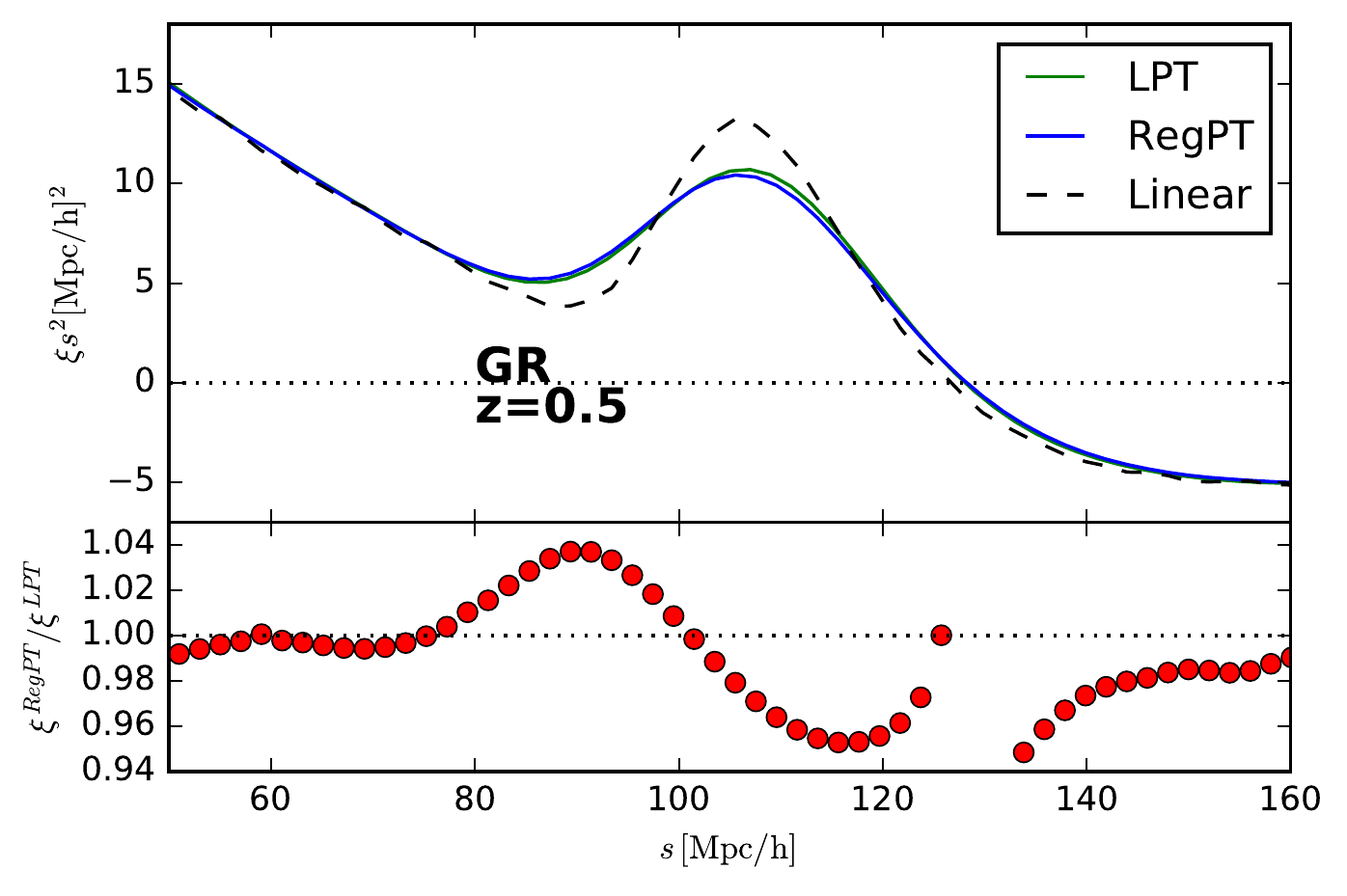}}
  \caption{Comparison of real space predictions for the correlation function using LPT (green), FT of RegPT (blue) and Linear (black,dashed). The bottom panel shows the fractional difference between the LPT and FT of RegPT. The reader should keep in mind that there is a 0-crossing at $r=130$Mpc/$h$ causing large fractional differences.}
\label{xir1}
\end{figure}

\begin{figure}[H]
  \captionsetup[subfigure]{labelformat=empty}
  \centering
  \subfloat[]{\includegraphics[width=10.3cm, height=7cm]{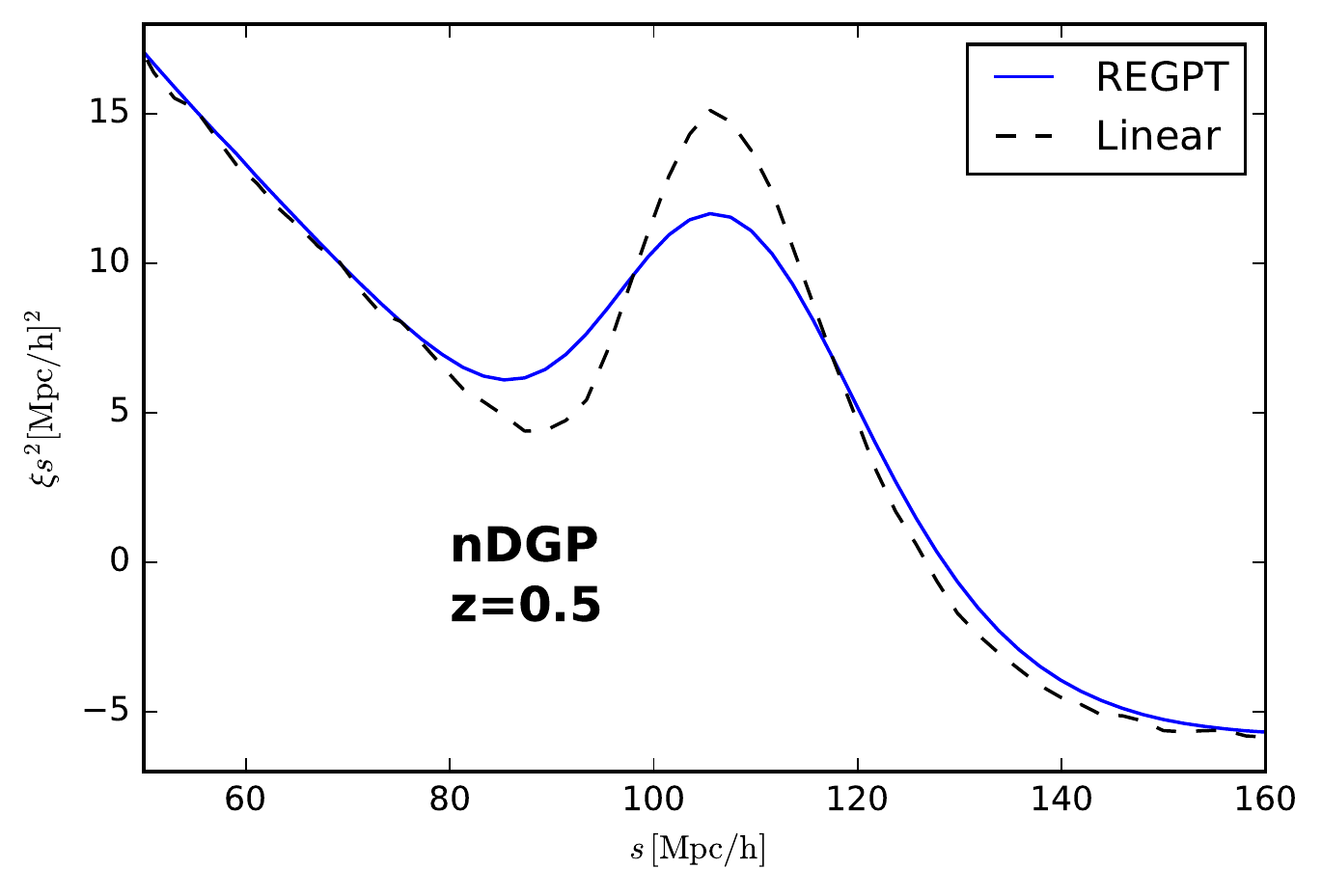}} 
  \caption{Comparison of real space predictions for the correlation function using the FT of the 1-loop power spectrum RegPT (blue) and FT of linear power spectrum (black,dashed) for nDGP. }
\label{xir2}
\end{figure}

\begin{figure}[H]
  \captionsetup[subfigure]{labelformat=empty}
  \centering
    \subfloat[]{\includegraphics[width=10.3cm, height=7cm]{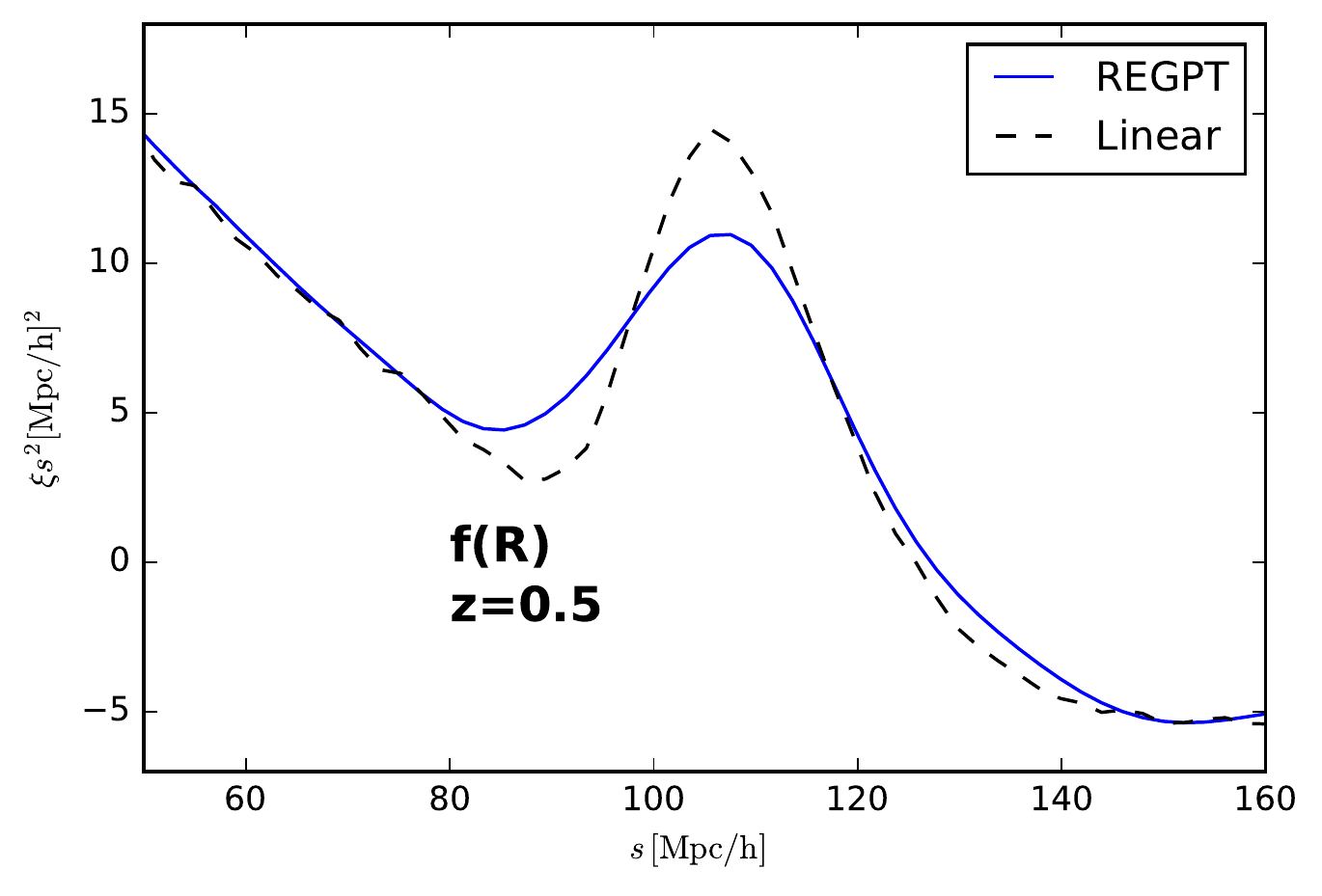}} 
  \caption{Comparison of real space predictions for the correlation function using the FT of the 1-loop power spectrum RegPT (blue) and FT of linear power spectrum (black,dashed) for $f(R)$. }
\label{xir3}
\end{figure}

We compare the FT of the RegPT 1-loop expressions with the LPT model of \cite{Matsubara:2008wx}. This model has been tested against N-body simulations in the GR case and has shown to be percent level accurate at scales of $r>25$ Mpc/$h$ \cite{Reid:2011ar}. It has been employed in spectroscopic survey analyses within the Baryon Oscillation Spectroscopic Survey (BOSS) \cite{Reid:2012sw}. Although we can only do this for GR, this will give us a handle on the accuracy of our FT approach to the multipoles in the next section. The transform of the RegPT power spectrum was compared to N-body results for GR and $f(R)$ in \cite{Taruya:2014faa} showing  good agreement above and around the BAO scale.
\newline
\newline
Fig.\ref{xir1} shows the real space correlation function as predicted by Eq.\ref{pregab} (using matter-matter 1-loop) and the LPT prediction of  \cite{Matsubara:2008wx} for dark matter. The FT of the linear power spectrum is also shown as a dashed line. We see both RegPT and LPT give a smoothing of the BAO bump - a well known non-linear effect - and that they agree on small and large scales at the percent level while around the BAO bump they show up to a $4\%$ difference with the RegPT treatment showing slightly more damping around this scale. For completeness we  also show the RegPT predictions against the linear predictions for the other models of gravity considered (Fig.\ref{xir2} and Fig.\ref{xir3}). We notice that the non-linear RegPT predictions for these models show more damping of the BAO bump when compared to the GR case, an expected effect of enhanced linear structure formation as well as enhanced 2nd and 3rd order non-linearities. 


\section{Redshift Space: TNS vs GSM}
\begin{figure}[H]
  \captionsetup[subfigure]{labelformat=empty}
  \centering
  \subfloat[]{\includegraphics[width=7.5cm, height=7cm]{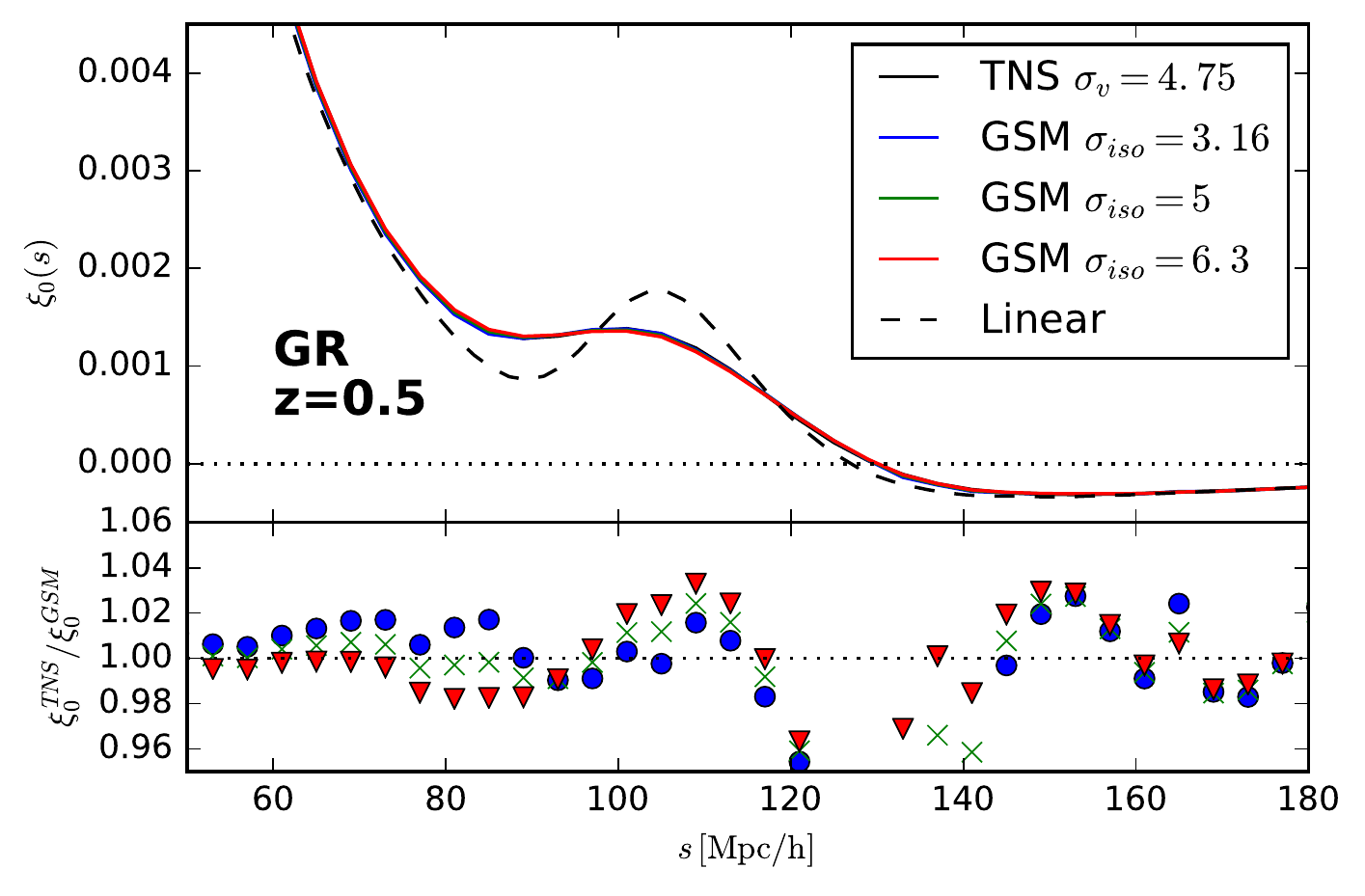}} \quad
    \subfloat[]{\includegraphics[width=7.5cm, height=7cm]{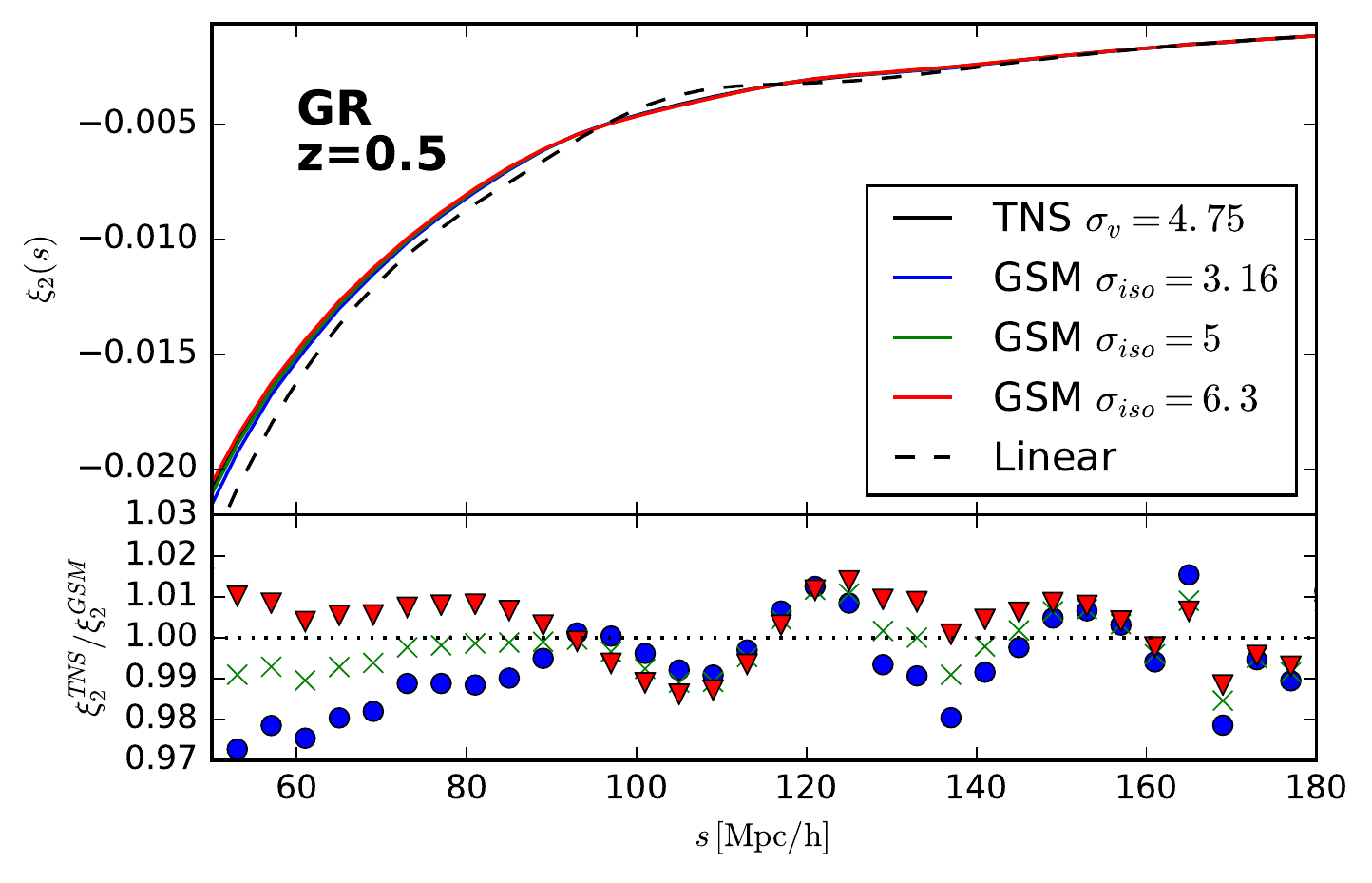}} 
  \caption{Comparison of the redshift space predictions for the correlation function using the FT of the TNS power spectrum (black solid) with $\sigma_v=4.75$ Mpc/$h$ and the GSM for three values of $\sigma_{\rm iso}$ (in units of Mpc/$h$) for GR. The PT prediction for $\sigma_{\rm iso}$ is given by the blue curve. The LSM prediction is shown as a dashed black curve. The left plot shows the monopole while the right plot shows quadrupole. The bottom panels shown the fractional difference between the TNS transform and the GSM. Keep in mind the zero crossing indicated by the dotted line in the top panel of the monopole giving large fractional differences. }
\label{xirs1}
\end{figure}

\begin{figure}[H]
  \captionsetup[subfigure]{labelformat=empty}
  \centering
  \subfloat[]{\includegraphics[width=7.5cm, height=7cm]{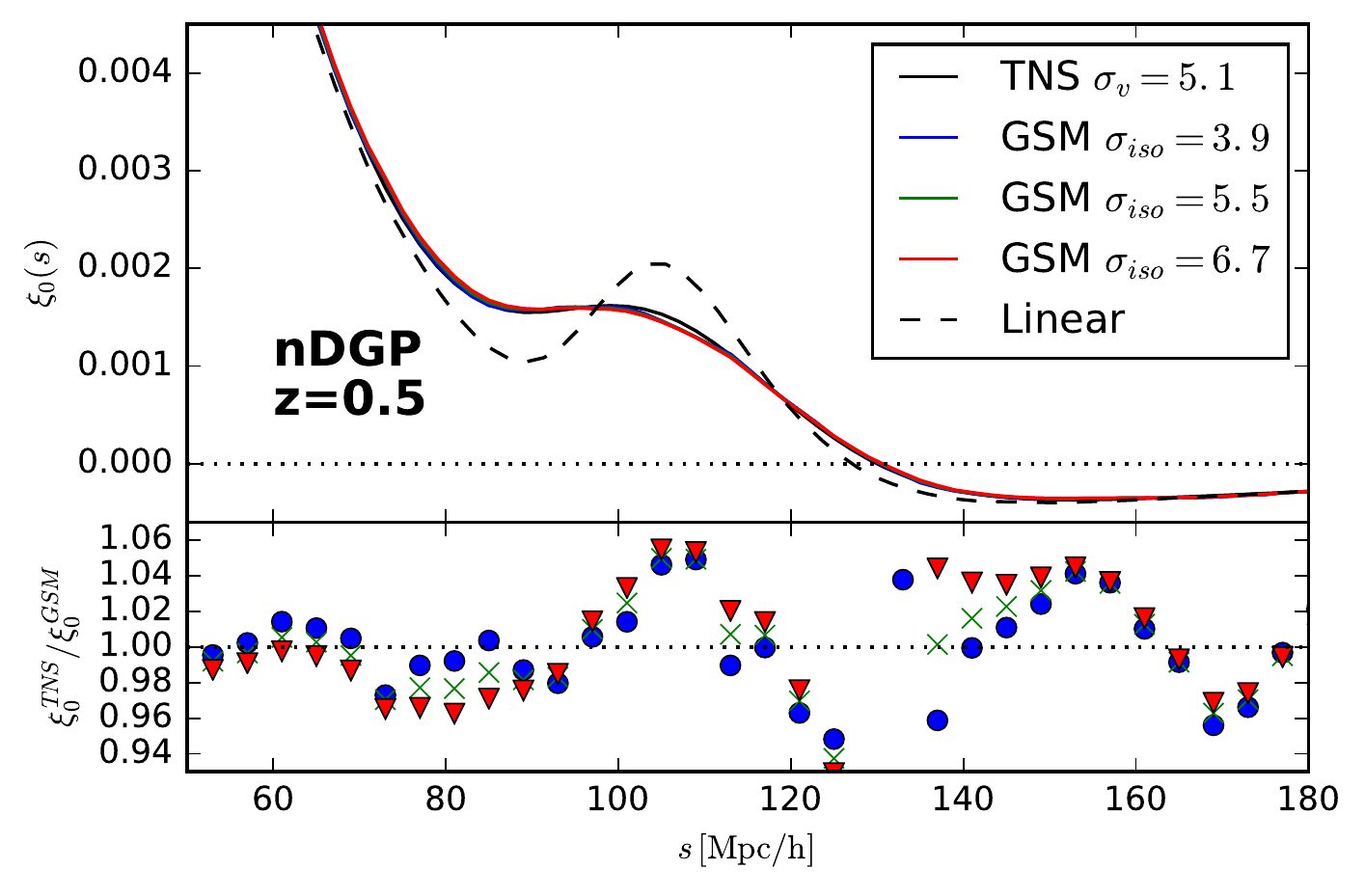}} \quad
    \subfloat[]{\includegraphics[width=7.5cm, height=7cm]{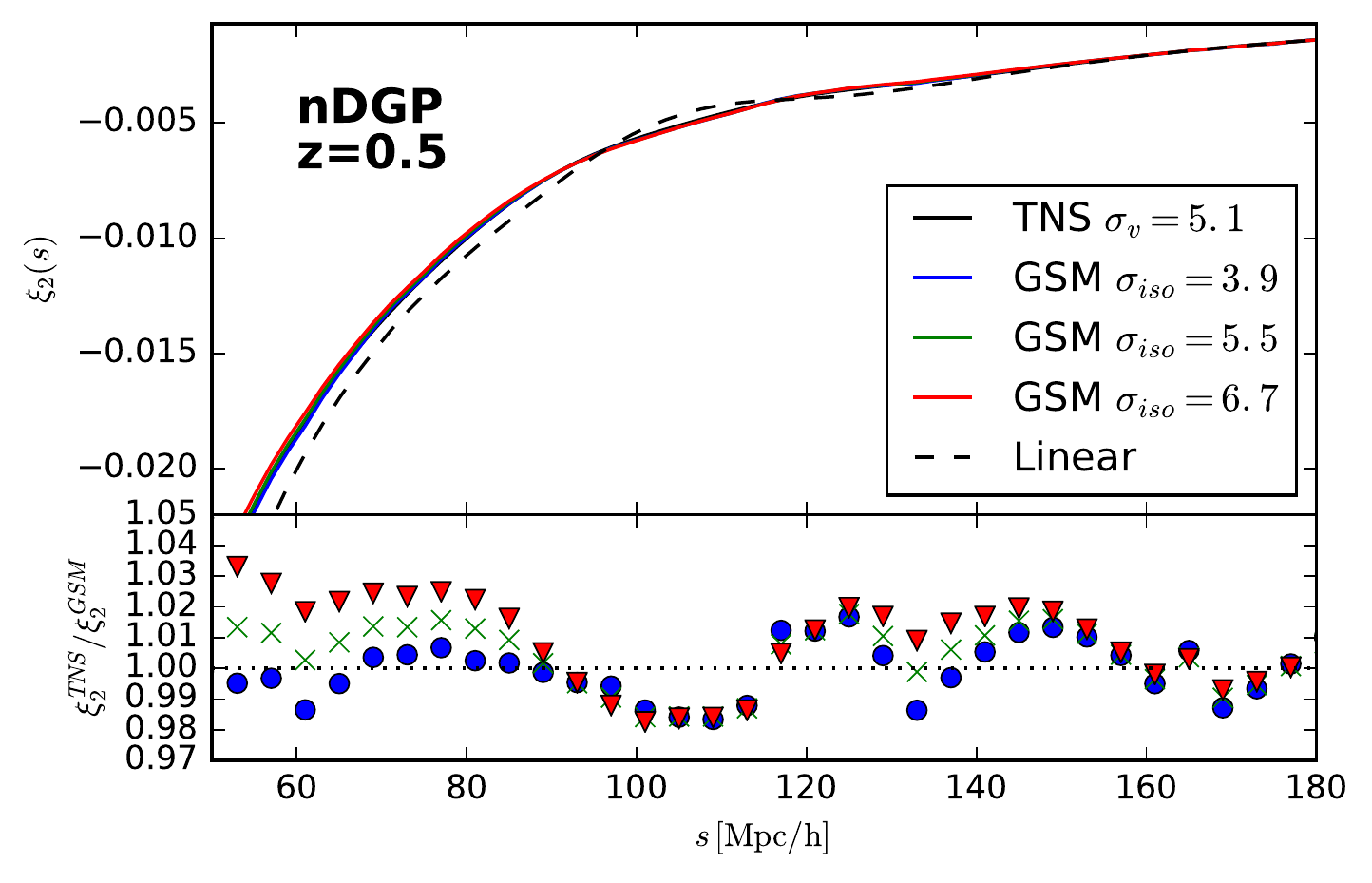}} 
  \caption{Same as Fig.\ref{xirs1} but for the nDGP model of gravity with $\Omega_{rc} = 0.438$. The TNS transform uses $\sigma_v=5.1$ Mpc/$h$.}
\label{xirs2}
\end{figure}

\begin{figure}[H]
  \captionsetup[subfigure]{labelformat=empty}
  \centering
  \subfloat[]{\includegraphics[width=7.5cm, height=7cm]{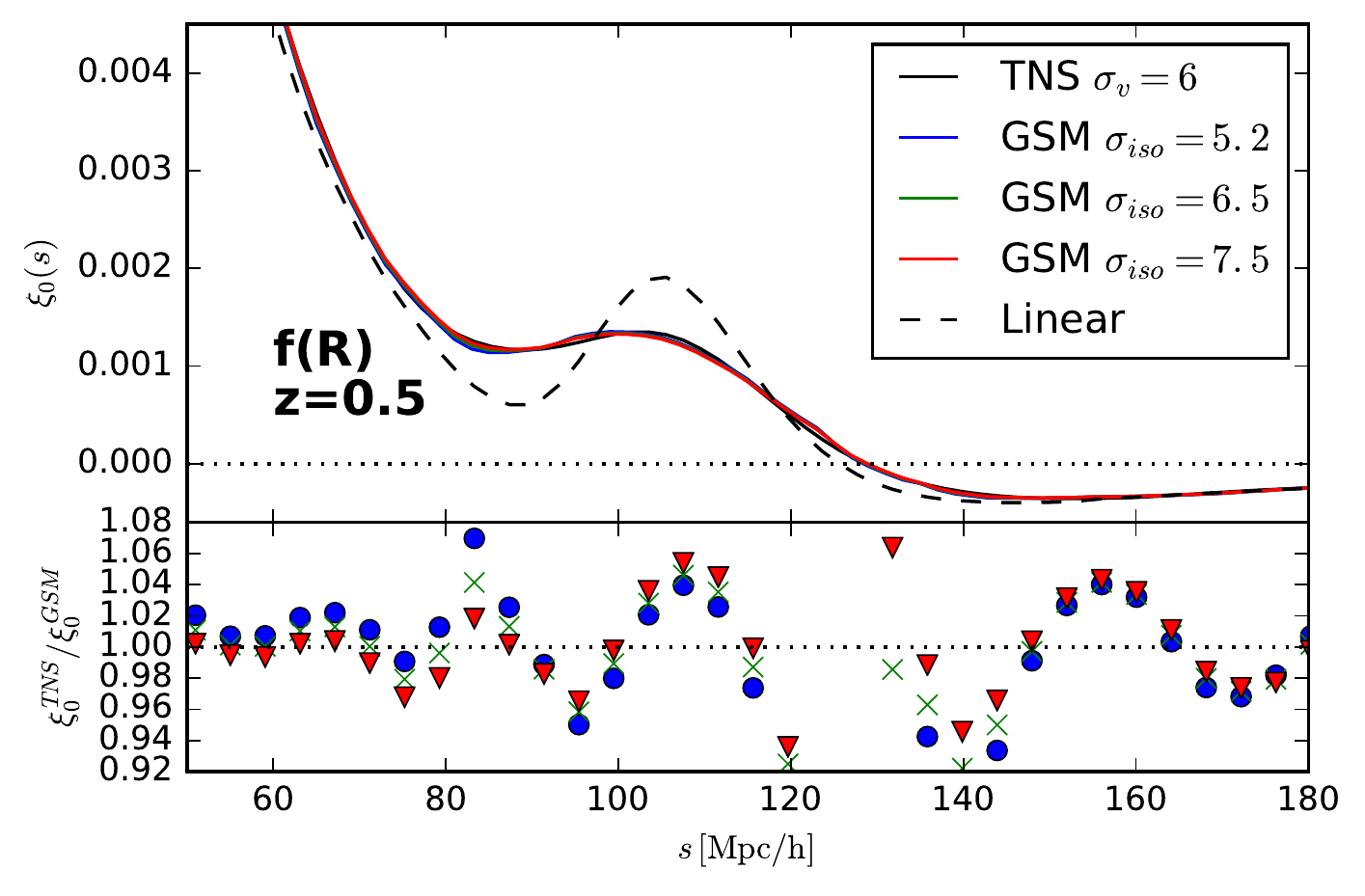}} \quad
    \subfloat[]{\includegraphics[width=7.5cm, height=7cm]{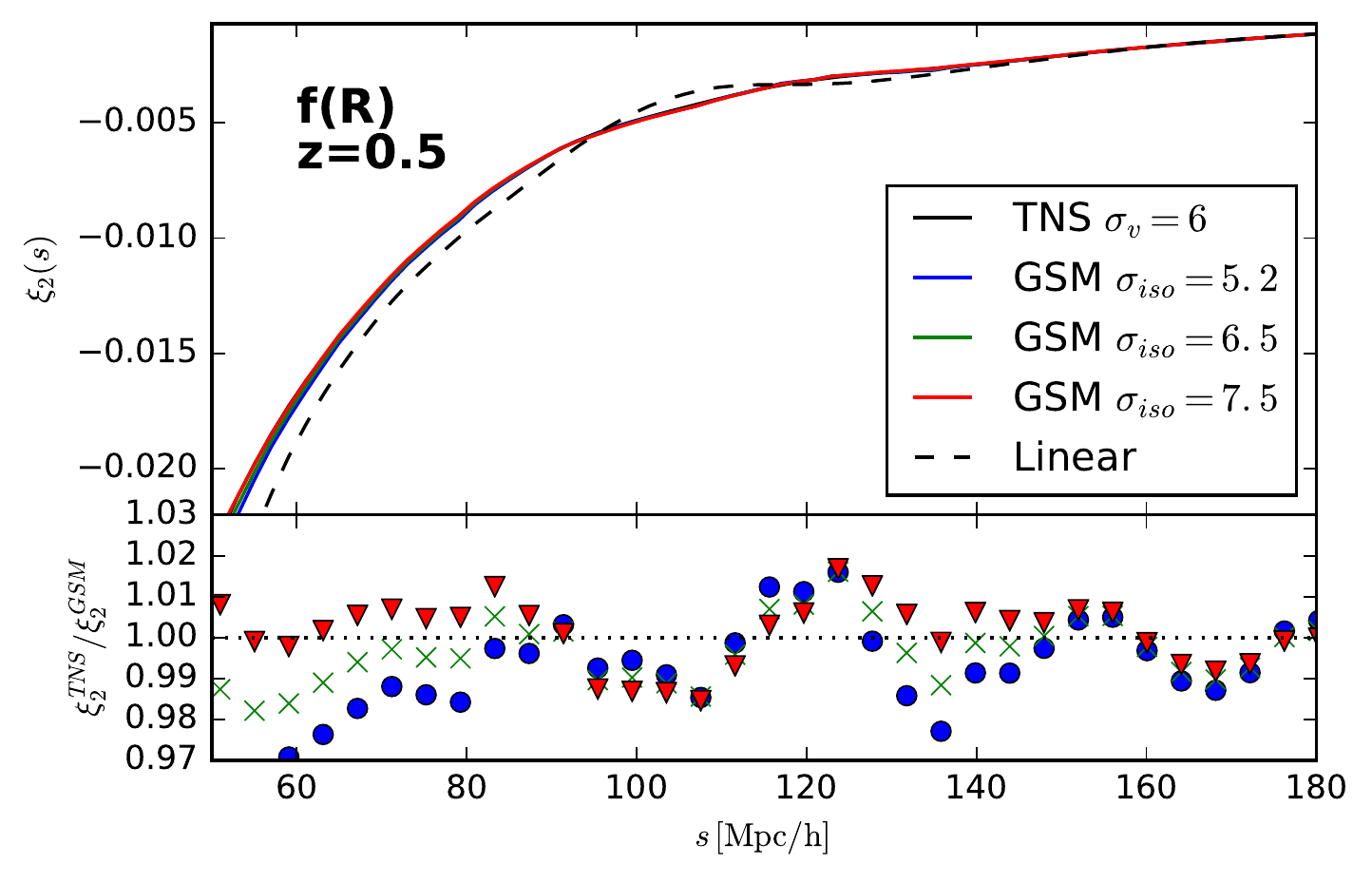}} 
  \caption{ Same as Fig.\ref{xirs1} but for the Hu-Sawicki model of $f(R)$ gravity with $|f_{R0}|=10^{-4}$. The TNS transform uses $\sigma_v=6$ Mpc/$h$.}
\label{xirs3}
\end{figure}

\begin{figure}[H]
  \captionsetup[subfigure]{labelformat=empty}
  \centering
  \subfloat[]{\includegraphics[width=7.5cm, height=7cm]{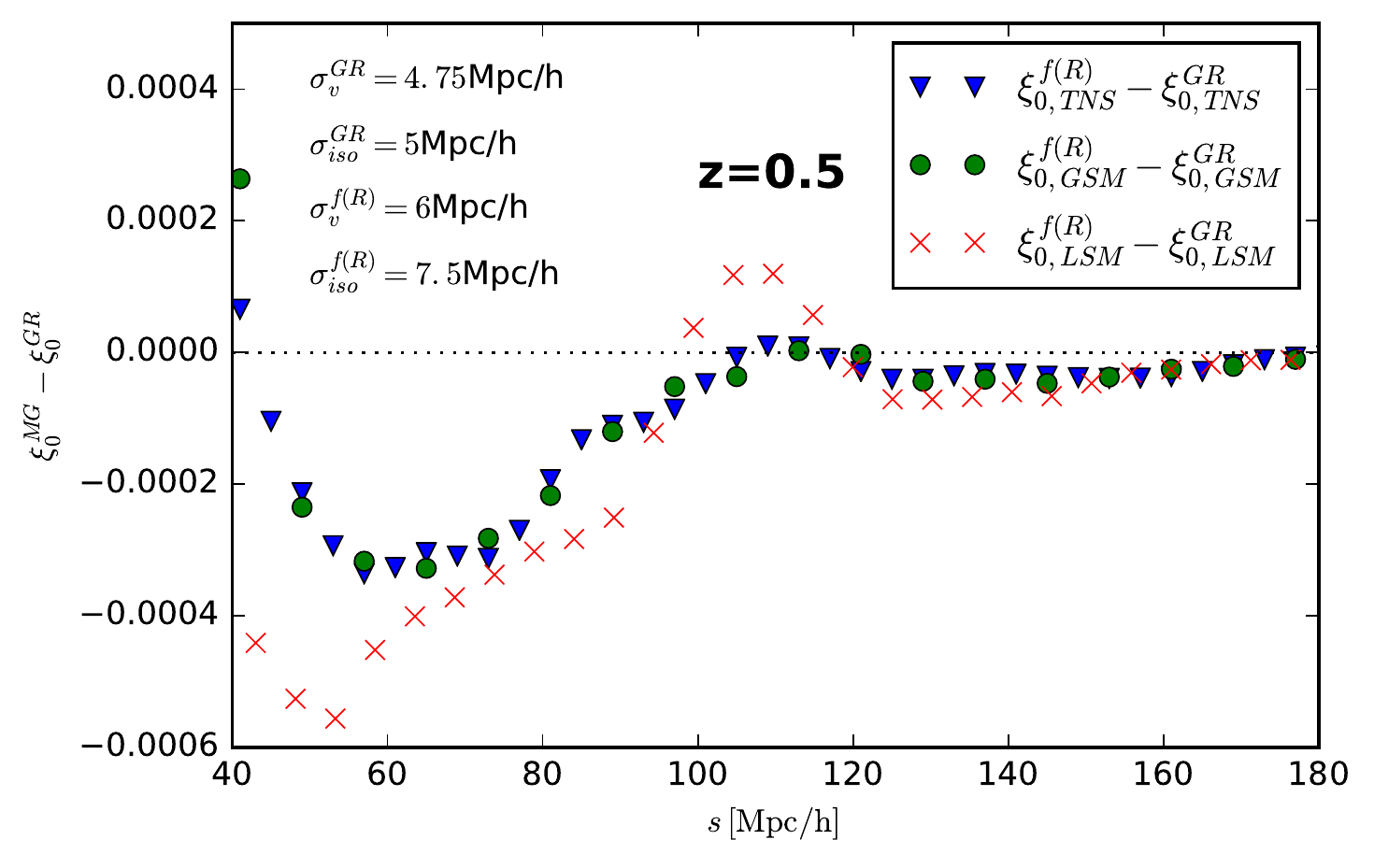}} \quad
    \subfloat[]{\includegraphics[width=7.5cm, height=7cm]{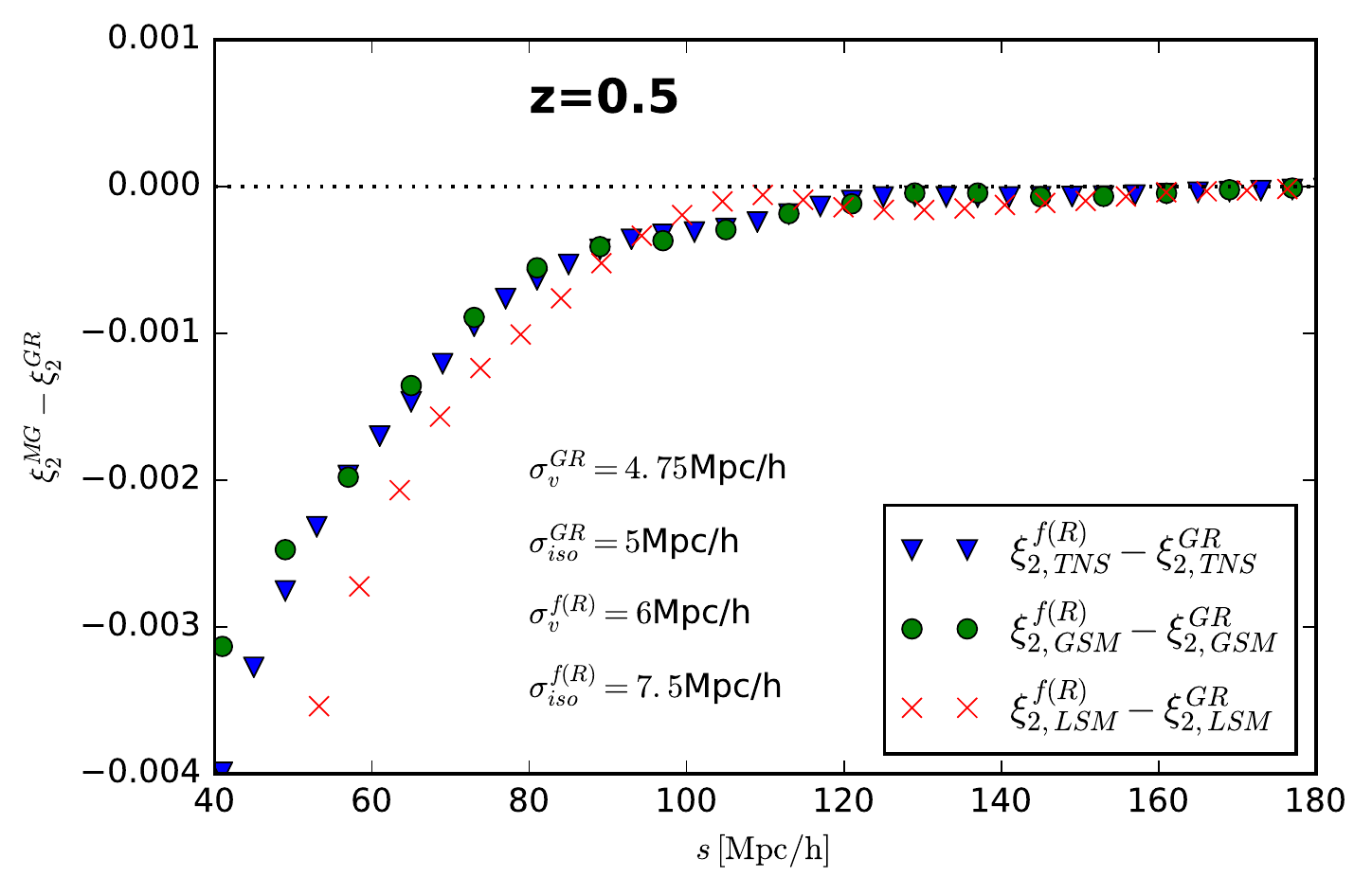}} 
  \caption{The difference in the GSM (green), FT of the TNS best fit (blue) and LSM(red) multipoles between $f(R)$ and GR at $z=0.5$. The monopole difference is shown on the left and the quadrupole difference is shown on the right. }
\label{frvgr}
\end{figure}

\begin{figure}[H]
  \captionsetup[subfigure]{labelformat=empty}
  \centering
  \subfloat[]{\includegraphics[width=7.5cm, height=7cm]{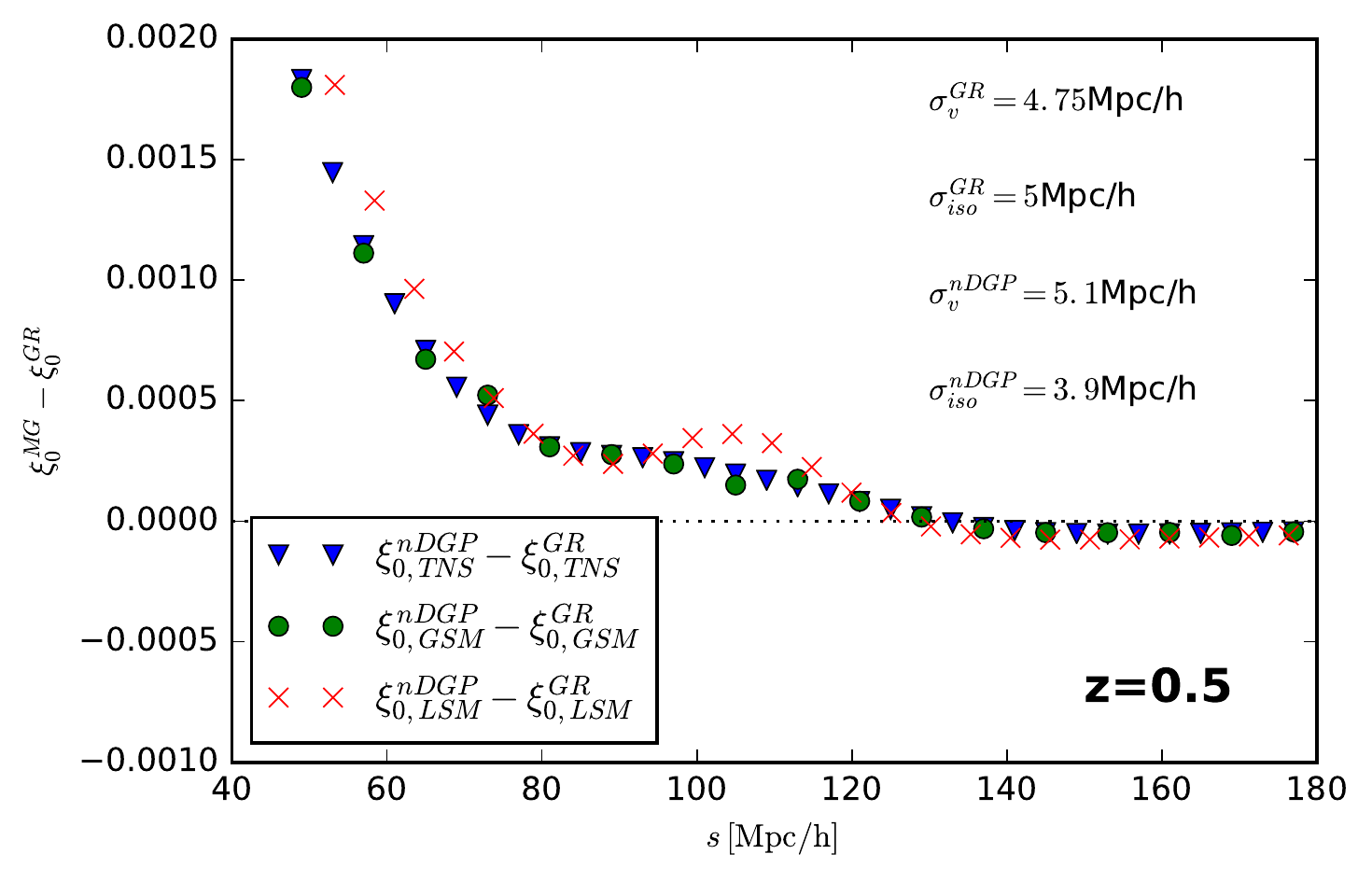}} \quad
    \subfloat[]{\includegraphics[width=7.5cm, height=7cm]{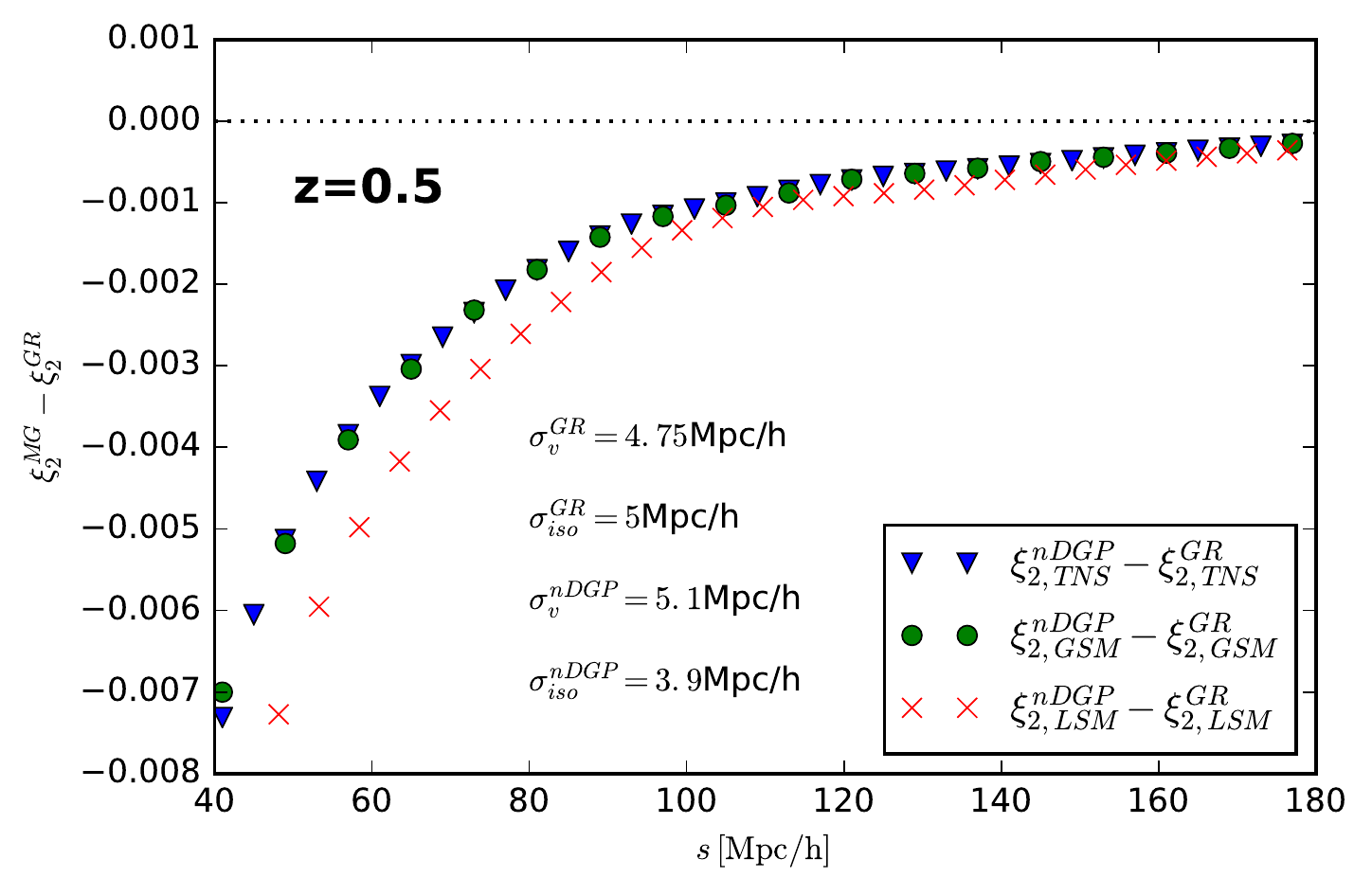}} 
  \caption{The difference in the GSM (green), FT of the TNS best fit (blue) and LSM(red) multipoles between nDGP and GR at $z=0.5$. The monopole difference is shown on the left and the quadrupole difference is shown on the right. }
\label{dgpvgr}
\end{figure}

Moving to redshift space, we will use the FT of the best fit TNS multipoles shown on the left of Fig.\ref{pab1} and Table.II of \cite{Taruya:2014faa}. Because of the TNS's extra degree of freedom ($\sigma_v$), the model has a large advantage compared to the GSM which can be completely determined by SPT. In general the correlation function needs to be measured many times from independent N-body realisations and averaged because of the small imprint of the acoustic features which can be greatly hidden by scatter. MG simulations are more computationally expensive than GR ones and so only a few are available. Thus, a clean configuration space measurement in MG theories is not readily available. This makes the TNS transform a good and practical benchmark to compare the GSM predictions to in the absence of averaged N-body correlation function measurements. The configuration space multipoles are given by \cite{Cole:1993kh,Hamilton:1997zq} 
\begin{equation}
\xi_\ell^{S}(s)= \frac{i^\ell}{2\pi^2} \int dk k^2 P_\ell^{S}(k)j_\ell(ks),
\label{multipolesR}
\end{equation}
where $j_\ell$ is the $\ell^{\rm th}$ order spherical Bessel function and $P_\ell^{S}$ is given by Eq.\ref{multipolesl}. Again we will only consider the first two multipoles, $\ell = 0,2$.  
\newline
\newline
The top panel of Fig.\ref{xirs1} shows the monopole (left) and quadrupole (right) predictions for the redshift space correlation function within GR. We have plotted the TNS transform with $\sigma_v = 4.75$Mpc/$h$ in black against the GSM predictions for three different values of the isotropic contribution to the velocity dispersion parametrised by $\sigma^2_{\rm iso}$. The blue curve is the GSM prediction where $\sigma_{\rm iso}$ takes the PT predicted value given by Eq.\ref{sigmaiso}. The predictions look very reasonable with significant smearing of the BAO bump over the LSM prediction (Eq.\ref{lsmmodel}), mostly seen in the monopole. 
\newline
\newline
The bottom panels of Fig.\ref{xirs1} show the fractional differences between the TNS transform and the GSM predictions. Fractional differences go up to $4\%$ in the monopole around the BAO scale and slightly less for the quadrupole, with slightly more damping of the BAO bump by the GSM predictions. We find that around this scale the PT prediction (Eq.\ref{sigmaiso}) for the isotropic contribution to the velocity dispersion does well for the monopole, whereas for the quadrupole the higher valued green curve ($\sigma_{\rm iso} = 5 $Mpc/$h$) does better, which is consistent with the TNS best fit velocity dispersion.  
\newline
\newline
Similar results are found for the nDGP model of gravity shown in Fig.\ref{xirs2}. The deviations of the GSM predictions from the TNS transform are only slightly larger than in the GR case, going up to $6\%$ in the monopole at the BAO scale. The PT prediction for $\sigma_{\rm iso}$ ($\sigma_{\rm iso} = 3.9 $Mpc/$h$) does the best over both multipoles at smaller scales with the green ($\sigma_{\rm iso} = 5.5 $Mpc/$h$) doing a bit better around the BAO bump. Both these values are consistent with the TNS best-fit value. 
\newline
\newline
The $f(R)$ predictions are shown in Fig.\ref{xirs3}. In this case the monopole's fractional differences are significantly larger with up to $8\%$ more damping in the GSM model. The quadrupole differences remain $\leq 3\%$ around the BAO scale. In this case the PT predicted value for $\sigma_{\rm iso}$ ($5.2 $Mpc/$h$) seems to underestimate the value with  $\sigma_{\rm iso} = 7.5 $Mpc/$h$ being more consistent with the TNS transform. This being said, to really tell which treatment performs better we wait for comparisons with simulation data. As mentioned earlier, many realisations are needed to get a converged measurement of the correlation function. This can be done for GR but for MG theories simulations are expensive computationally. By using COmoving Lagrangian Acceleration (COLA) approaches such as those described in \cite{Winther:2017jof}, this problem becomes tractable.
\newline
\newline
Fig.\ref{frvgr} and Fig.\ref{dgpvgr} show the differences between the modified gravity predictions and the GR ones for both theoretical predictions for the correlation function as well as the linear prediction. We see that in both the FT of TNS and GSM the effect of modifying gravity is very similar indicating both approaches give comparable signals of deviations from GR.  In the monopole, around the acoustic bump, both non-linear approaches also reduce the MG-Signal with a larger difference seen in linear modelling. The LSM also shows larger differences at scales below the BAO in the quadrupole. One other feature is that $f(R)$ gravity shows a suppression compared to GR around $40$Mpc$/h<s<100 $Mpc$/h$ while nDGP shows an enhancement over GR for the monopole. 
\newline
\newline
This concludes the presentation of configuration space comparisons. The next chapter deals with theoretical template comparisons to simulation data. In particular, we investigate the relevance of consistent gravitational modelling when considering upcoming large volume spectroscopic galaxy surveys.
\newpage


\chapter{Validating Theoretical Templates for Stage IV Surveys}

\begin{chapquote}{J.D}
``Where was the peak? Was there a peak? The man dug his foot deeper to sturdy himself. Then sharpened his sight, cutting into the torrents of snow that only wished to swallow him." 
\end{chapquote}

\begin{figure}[H]
 \centering
  \subfloat[]{\includegraphics[width=7.5cm, height=7cm]{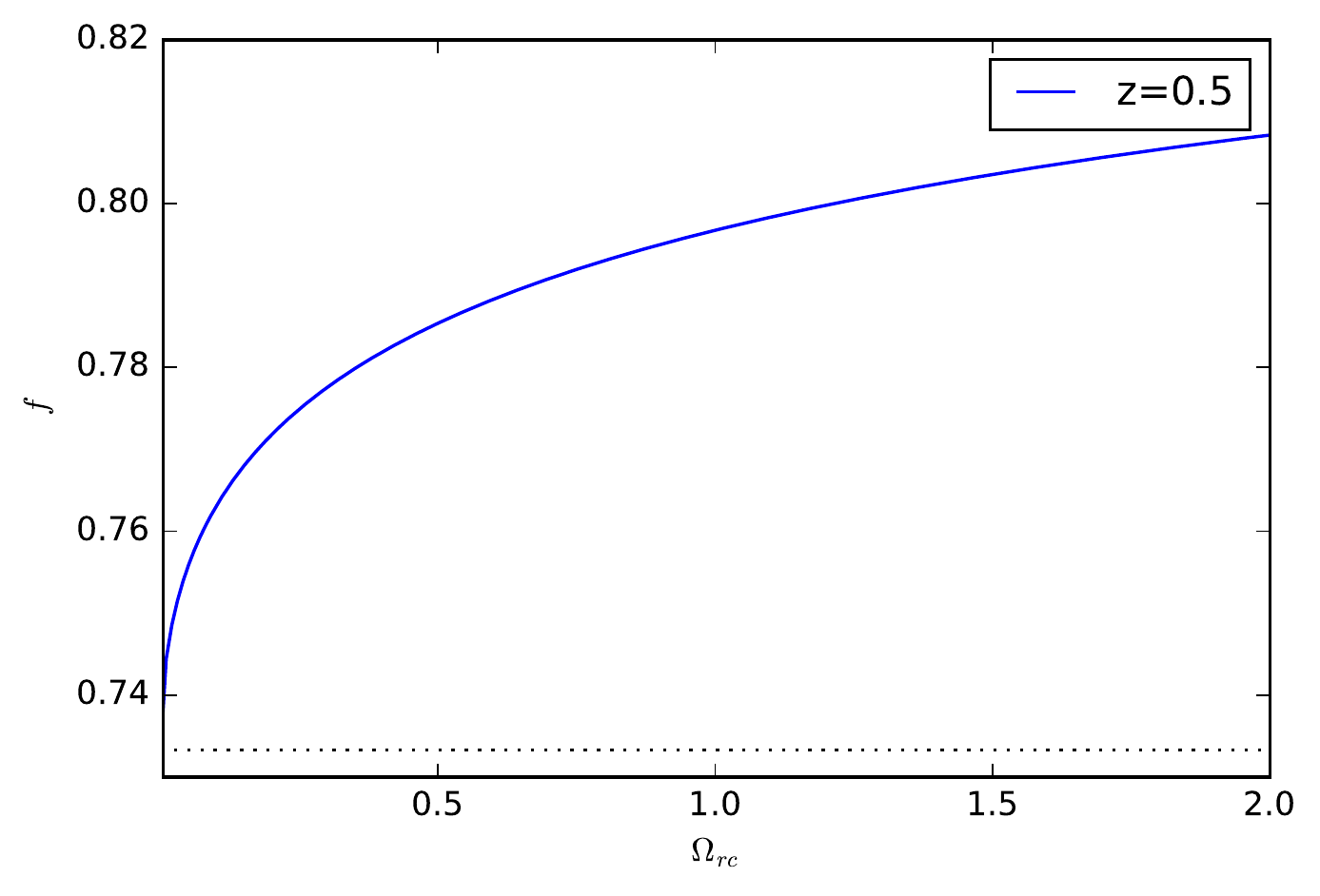}} \quad
  \subfloat[]{\includegraphics[width=7.5cm, height=7cm]{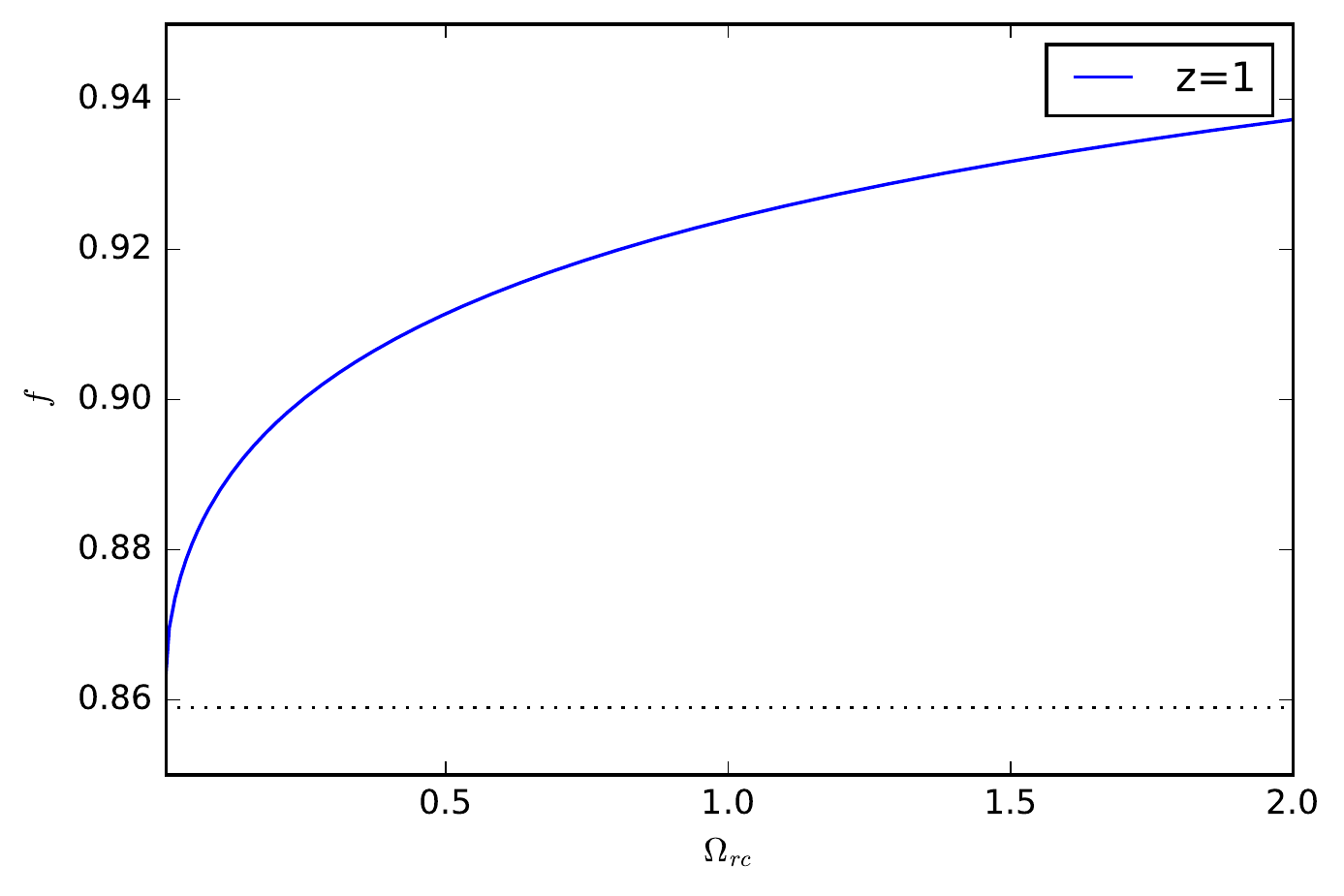}}
   \caption{The logarithmic growth $f$ as a function of $\Omega_{rc}$ at $z=0.5$ (left) and $z=1$ (right). $\Omega_{rc}=0$ corresponds to GR and is marked by the dotted line.}
\label{forc}
\end{figure}

In this chapter we present predictions generated using the framework described in chapters 2 and 4 as well as investigate the issue of model bias, extending the work of \cite{Barreira:2016mg}. Further we provide a means of quickly testing the validity of constraining non-GR models using the standard GR template. Overall, our aim will be to assess the level and scales at which gravitational modelling becomes important and starts to affect the parameter constraints in a significant way, specifically in the context of the next generation of spectroscopic surveys. To do this, we compare the predictions of two theoretical templates with redshift space data from COLA (COmoving Lagrangian Acceleration) simulations in nDGP and GR. Using COLA for these comparisons is validated using a suite of full N-body simulations for the same theories. The two theoretical templates correspond to the standard general relativistic perturbation equations and those same equations modelled within nDGP\footnote{All simulation data is not part of the original work of this thesis.}. 
\newline
\newline
Focus will be on the estimation of the logarithmic growth rate, $f$. Thus, inspired by the approach presented in \cite{Oka:2013cba}, we will fix the amplitude of linear density perturbations, or equivalently $\sigma_8$, to the fiducial value determined by the simulation's cosmology. We will be dealing with data for which all cosmological parameters (\emph{ie}. $\Omega_m,\Omega_{DE},H_0$ etc) are already known \emph{a priori}, and therefore choose to keep all of them fixed during the analysis. Allowing them all to vary would just decrease the statistical significance of the estimates and fixing them does not introduce systematics. In this way we end up with only two free parameters $\{\sigma_v, f \}$ where $\sigma_v$ is the 1-dimensional velocity dispersion in the TNS model which needs to be fit to data. 
\newline
\newline
In the analysis, a LCDM background is assumed for the nDGP modelling so that the difference between nDGP and LCDM appears only in the structure growth. $f$ is derived from the linear versions of Eq.\ref{cont2} and Eq.\ref{euler2} (the right hand side as well as  $\gamma_2$ and $\gamma_3$ being set to 0) and because $\Omega_m=\Omega_m^{fiducial}$, the only free parameter is $\Omega_{rc}$ and so we will opt to parametrise $f$ by $\Omega_{rc}$. Since $\Omega_{rc}>0$, a lower bound for $f$ is also imposed. Otherwise the priors for both $\sigma_v$ and $f$ are flat. For clarity, Fig.\ref{forc} shows the relationship between $f$ and $\Omega_{rc}$ at $z=0.5$ and $z=1$. We note that $f(\Omega_{rc} = 0)$ corresponds to the logarithmic growth in LCDM. 
\noindent To get an estimate of $f$ we consider the multipoles of Eq.\ref{redshiftps} given by Eq.\ref{multipolesl}. Further, a robust estimation of the higher order multipoles requires simulations with large volumes and high mass resolution. Since our simulations are limited in size and resolution we will limit our modelling and analysis to the monopole and quadrupole.
\newline
\newline
To test the ability of the theoretical templates to recover the fiducial $f$ we perform a MCMC analysis using the following likelihood function 
\begin{align}
-2\ln{\mathcal{L}} =  \sum_n \sum_{\ell,\ell'=0,2}& \left(P^{S}_{\ell,{\rm data}}(k_n)-P^{S}_{\ell,{\rm model}}(k_n)\right) \mbox{Cov}^{-1}_{\ell,\ell'}(k_n) \nonumber \\ & \times \left(P^{S}_{\ell',{\rm data}}(k_n)-P^{S}_{\ell',{\rm model}}(k_n)\right),
\label{covarianceeqn}
\end{align}
where $\mbox{Cov}_{\ell,\ell'}$ is the covariance matrix between the different multipoles. Expressions for the covariance components can be found in Appendix G which give a good estimate of cosmic variance. For our analysis we do not consider non-Gaussianity in the covariance (see \cite{Takahashi:2009bq} for justification of this treatment) but include the effect of shot-noise assuming the number density of an ideal future survey, $\bar{n}= 4 \times 10^{-3} h^3 /\mbox{Mpc}^{3}$. Further, we use linear theory to estimate the covariance matrix components. This approximation has been checked to work well for scales $k\leq0.3h$/Mpc in the LCDM simulations used in \cite{Taruya:2010mx}. Given this, we have found it sufficient for our purposes to check that the covariance of both the density and velocity divergence spectra are the same in both our LCDM and nDGP simulations within the scales of interest. We have estimated that the LCDM and nDGP scaled covariance (the so-called decoherence function, see \cite{Chodorowski:2001id}) are in agreement to sub percent levels at $k\leq0.2h$/Mpc.
\newline
\newline
We consider two redshifts, $z=1$ and $z=0.5$. This is to give an idea of the trade off between enhanced non-linearity but decreased realm of validity of SPT at lower $z$. For both these redshifts, we assume in the MCMC analysis that the errors are those characteristic for a survey with a volume of $V_s=10 \mbox{Gpc}^3/h^3$, which is conservatively smaller than the upcoming DESI survey which is further around 3 times smaller than the Euclid spectroscopic survey. Our number density, $\bar{n}= 4 \times 10^{-3} h^3 /\mbox{Mpc}^{3}$ and volume are comparable to that of the BOSS MGS sample \cite{Ross:2014qpa} or DESI's BGS \cite{Zhao:2013dza,Aghamousa:2016zmz}. 
\newline
\newline
Since we only have a $1\mbox{Gpc}^3/h^3$ box realisation of a full N-body run at our disposal, using solely this data set could hamper our analysis through uncertainties connected with cosmic variance that become severe at the box-scale. To overcome this, an additional 20 $\times$ $1\mbox{Gpc}^3/h^3$ box realisations of COLA are used. We do this by employing a rather inexpensive modified gravity COLA approach (hereafter MG-PICOLA) recently  presented in \cite{Winther:2017jof}. MG-PICOLA is based on a parallel COLA implementation (PICOLA) (see \cite{Howlett:2015hfa} for details). As we have mentioned MG-PICOLA is relatively computationally inexpensive, but this advantage comes at the price of significantly limited accuracy in the non-linear regime of structure formation. We have carefully performed many tests to ascertain that MG-COLA is sufficiently accurate for the purpose of the analysis presented in this work. The reader is referred to Appendix F for the details of the MG-PICOLA tests performed. 


\section{Theory vs Simulations}
Here we provide comparison of the full N-body measurements and MG-PICOLA measurements to the SPT predictions. The two N-body simulations (GR and nDGP) were run using the AMR code {\tt ECOSMOG} \cite{Li:2011vk}. The background cosmology is taken from WMAP9 \cite{Hinshaw:2012aka}: $\Omega_m = 0.281$, $h=0.697$, and $n_s=0.971$. The box length is $1024 \mbox{Mpc}/h$ with $1024^3$ dark matter particles used and a starting redshift of $49$. This design sets the resulting mass resolution at $m_p\cong7.8\times10^{10}M_{\odot}h^{-1}$ and the Nyquist fluid approximation limit of $k_{Nyq}\cong\pi h$/Mpc. The most refined AMR grid were at level 16, setting a maximal force resolution at $\epsilon=1024/2^{16}=0.015 \mbox{Mpc}/h$. The initial conditions were generated using {\tt MPGrafic}\footnote{Available at \url{http://www2.iap.fr/users/pichon/mpgrafic.html}} and both nDGP and LCDM simulations begin with the same initial seeds. The linear theory power spectrum normalisations was set to be $\sigma_8=0.844$. The nDGP simulation uses $\Omega_{rc}=1/4r_c^2H_0^2=0.438$. 
\newline
\newline 
To begin, we compare the real space N-body auto and cross power spectra with the SPT 1-loop predictions. This provides a measure of the non-linearity captured by SPT as well as realm of validity of the PT treatment. We then compare the TNS multipoles to MG-PICOLA measurements.
 
\subsection{Real Space Comparisons: N-body}
\begin{figure}[H]
  \captionsetup[subfigure]{labelformat=empty}
  \centering
  \subfloat[]{\includegraphics[width=7.5cm, height=7cm]{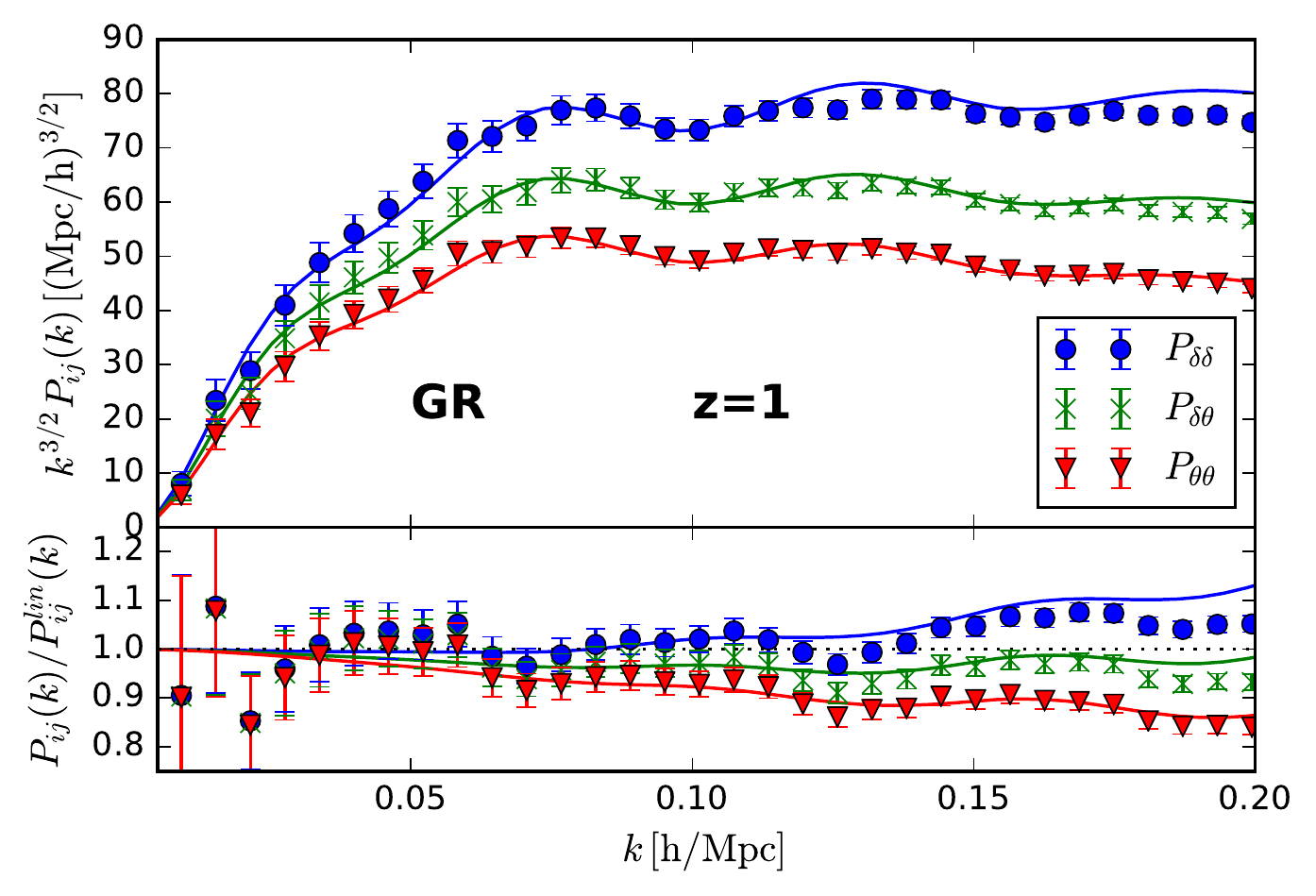}} \quad
  \subfloat[]{\includegraphics[width=7.5cm, height=7cm]{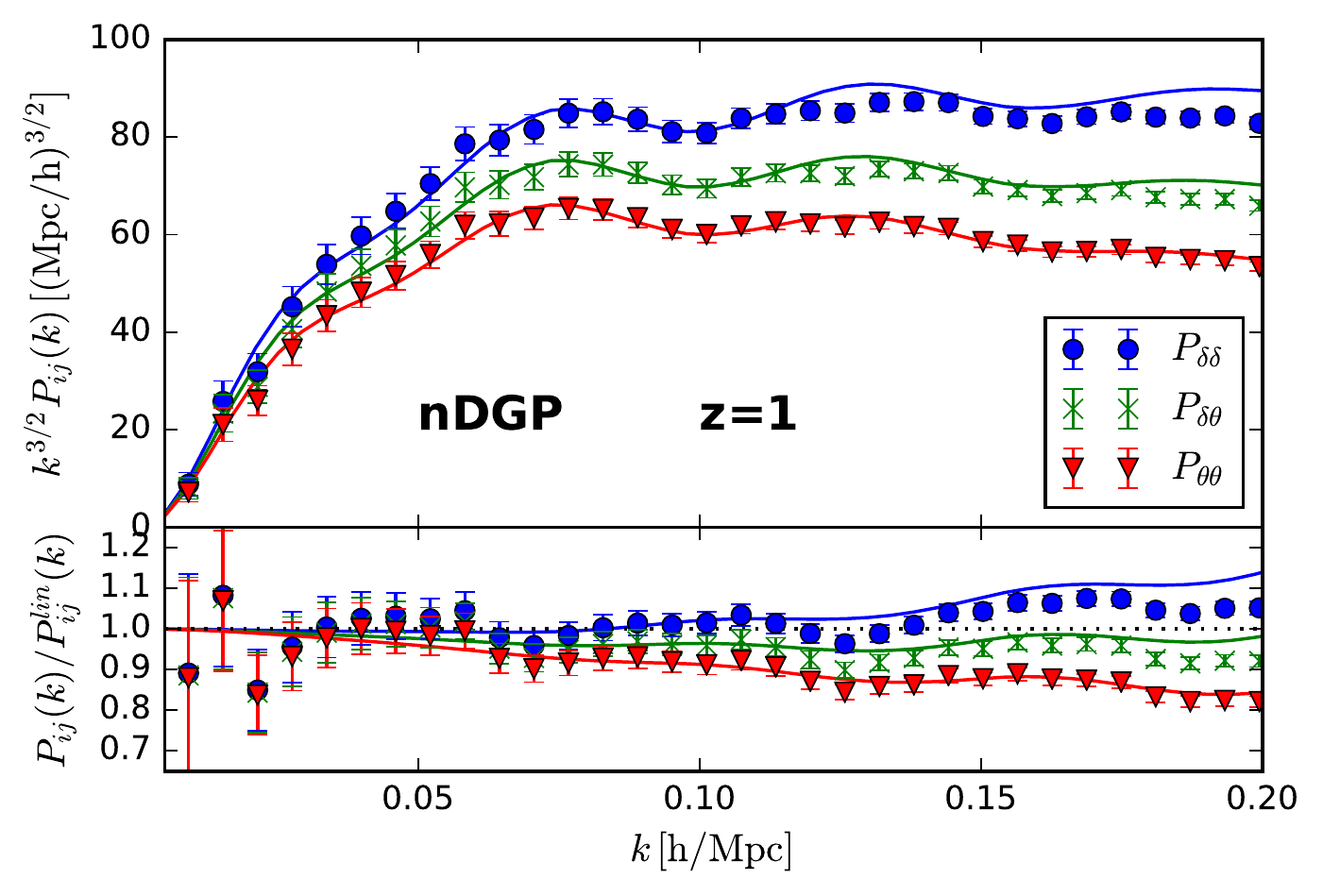}} 
  \caption{SPT predictions (solid) and N-body measurements (points) of the auto and cross power spectra of density and velocity fields in real space at $z=1$ for GR (left) and nDGP (right) fitted with Poisson errors assuming a $1 \mbox{Gpc}^3/h^3$ volume. The top panels show the power spectra scaled by $k^{3/2}$ and the bottom panels show the deviations from the linear predictions. }
\label{nbodypsz1}
\end{figure}

\begin{figure}[H]
  \captionsetup[subfigure]{labelformat=empty}
  \centering
  \subfloat[]{\includegraphics[width=7.5cm, height=7cm]{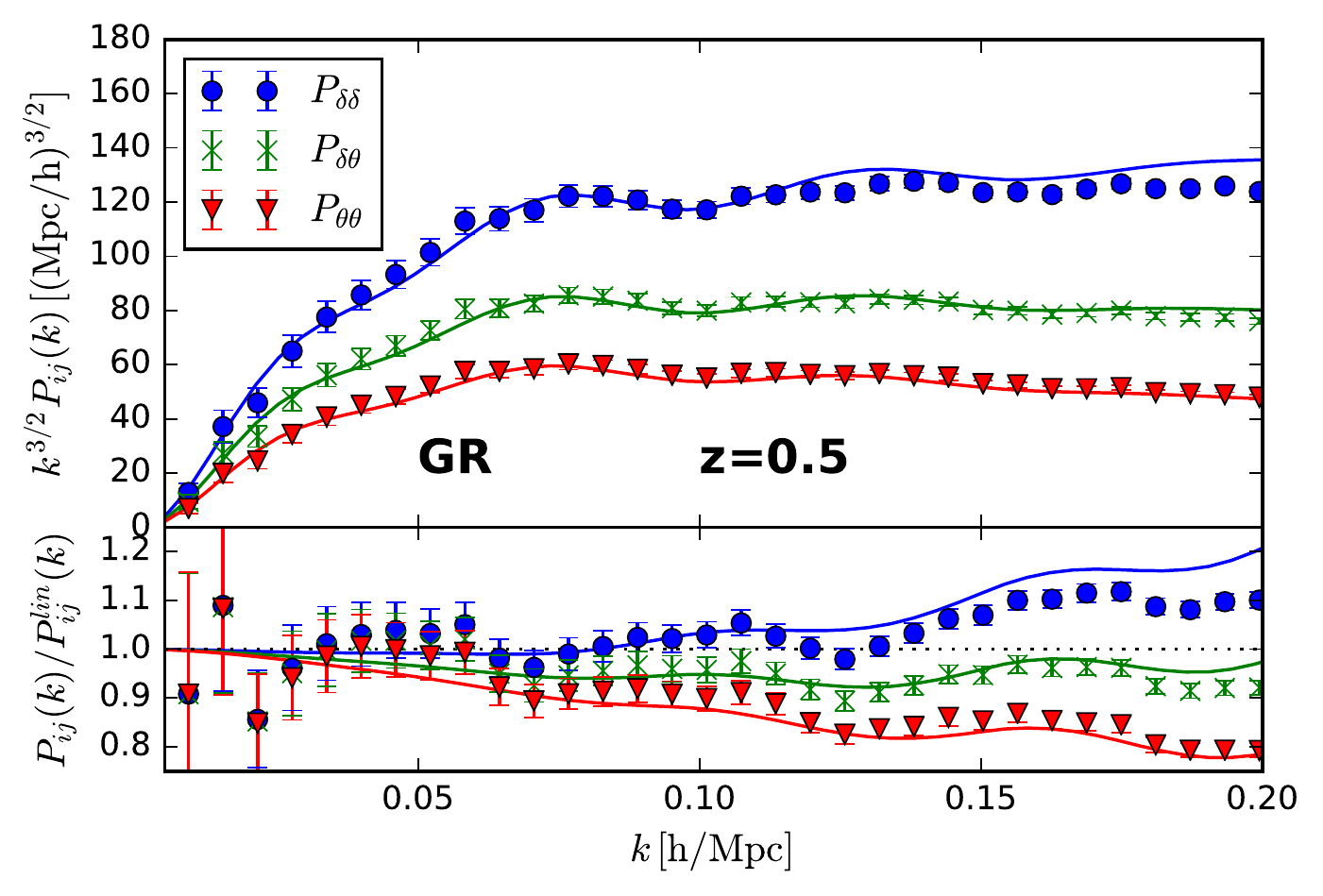}} \quad
  \subfloat[]{\includegraphics[width=7.5cm, height=7cm]{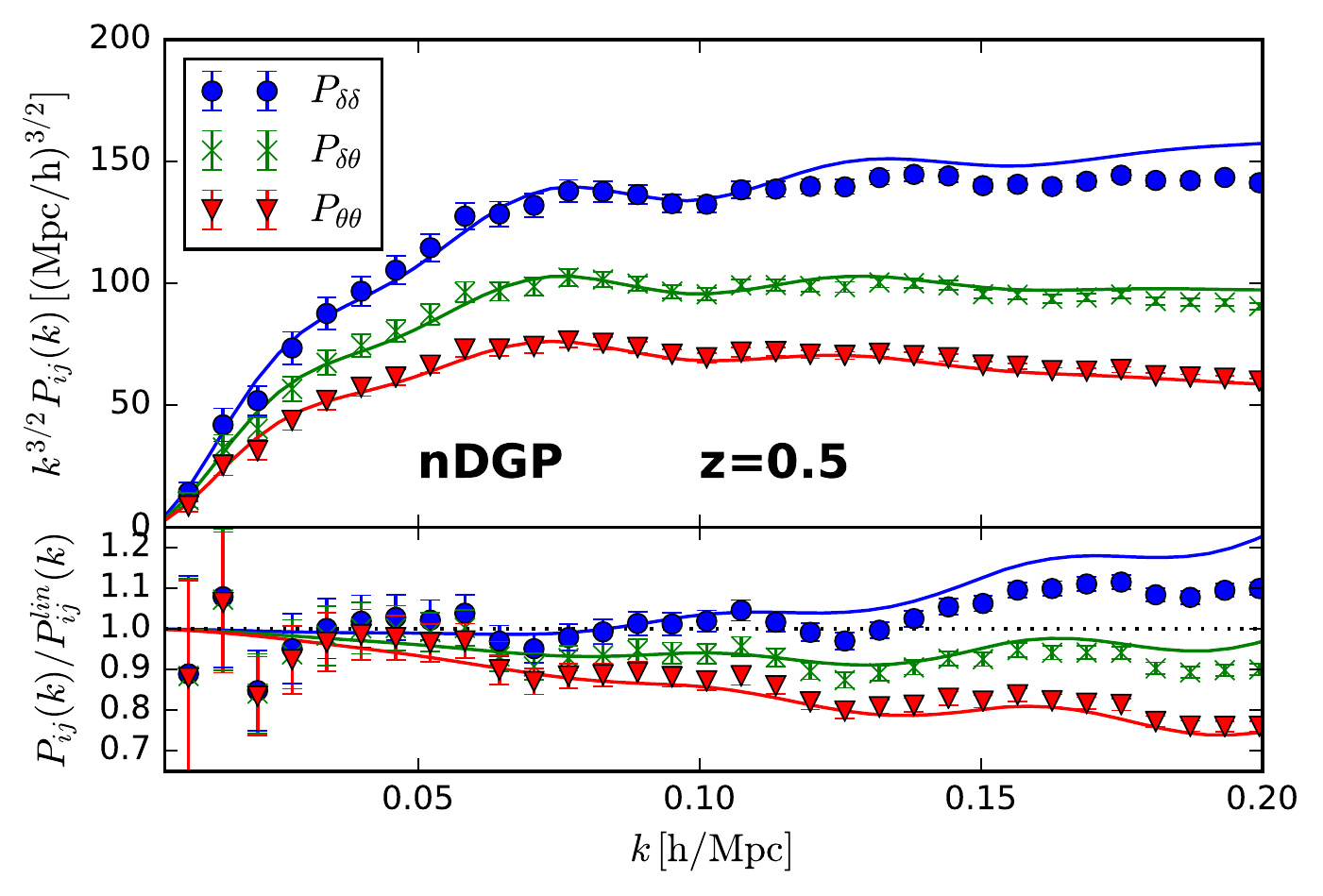}} 
  \caption{SPT predictions (solid) and N-body measurements (points) of the auto and cross power spectra of density and velocity fields in real space at $z=0.5$ for GR (left) and nDGP (right) fitted with Poisson errors assuming a $1 \mbox{Gpc}^3/h^3$ volume.  The top panels show the power spectra scaled by $k^{3/2}$ and the bottom panels show the deviations from the linear predictions.}
\label{nbodypsz05}
\end{figure}

\begin{figure}[H]
  \captionsetup[subfigure]{labelformat=empty}
  \centering
  \subfloat[]{\includegraphics[width=7.5cm, height=7cm]{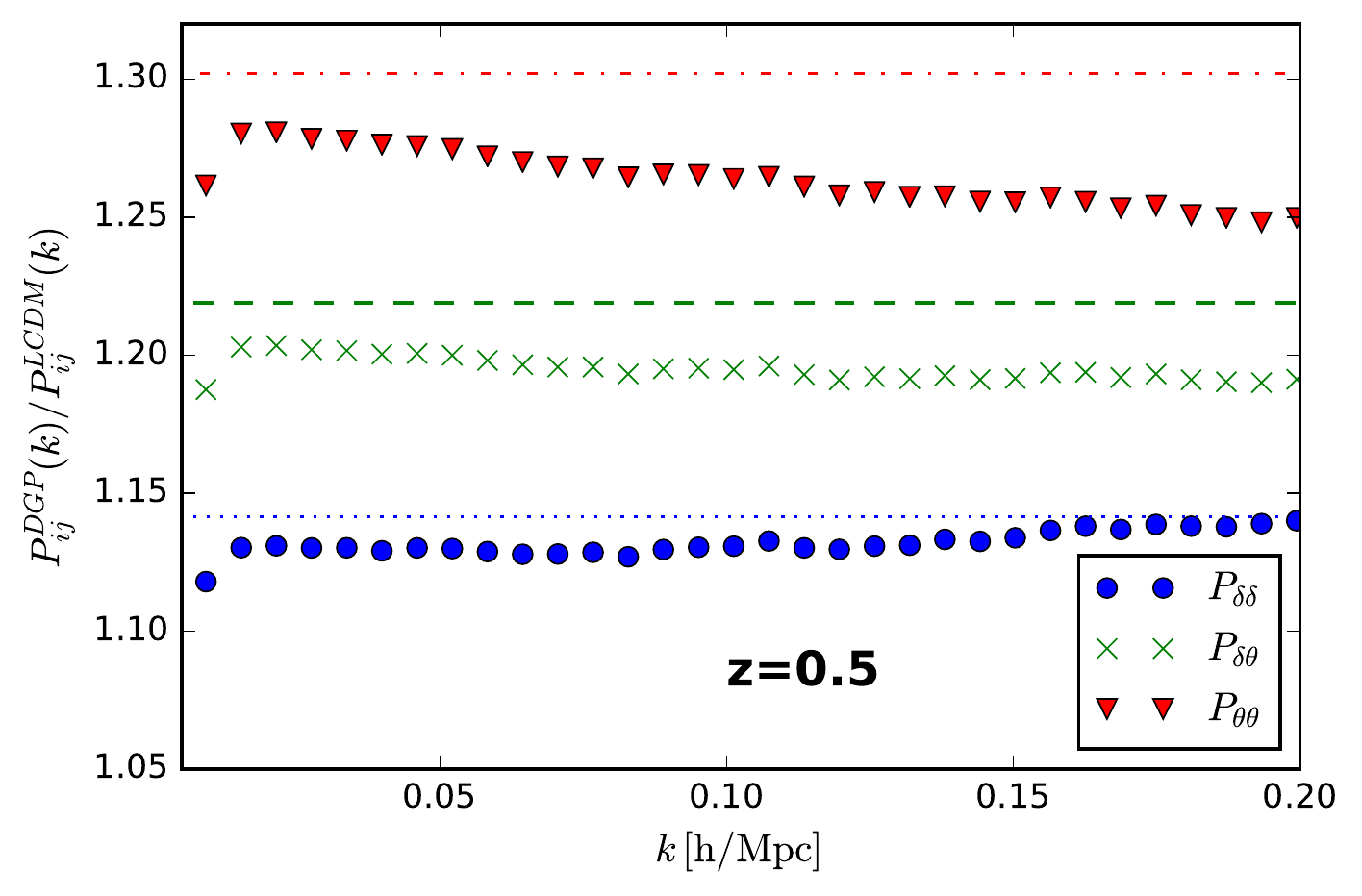}} \quad
  \subfloat[]{\includegraphics[width=7.5cm, height=7cm]{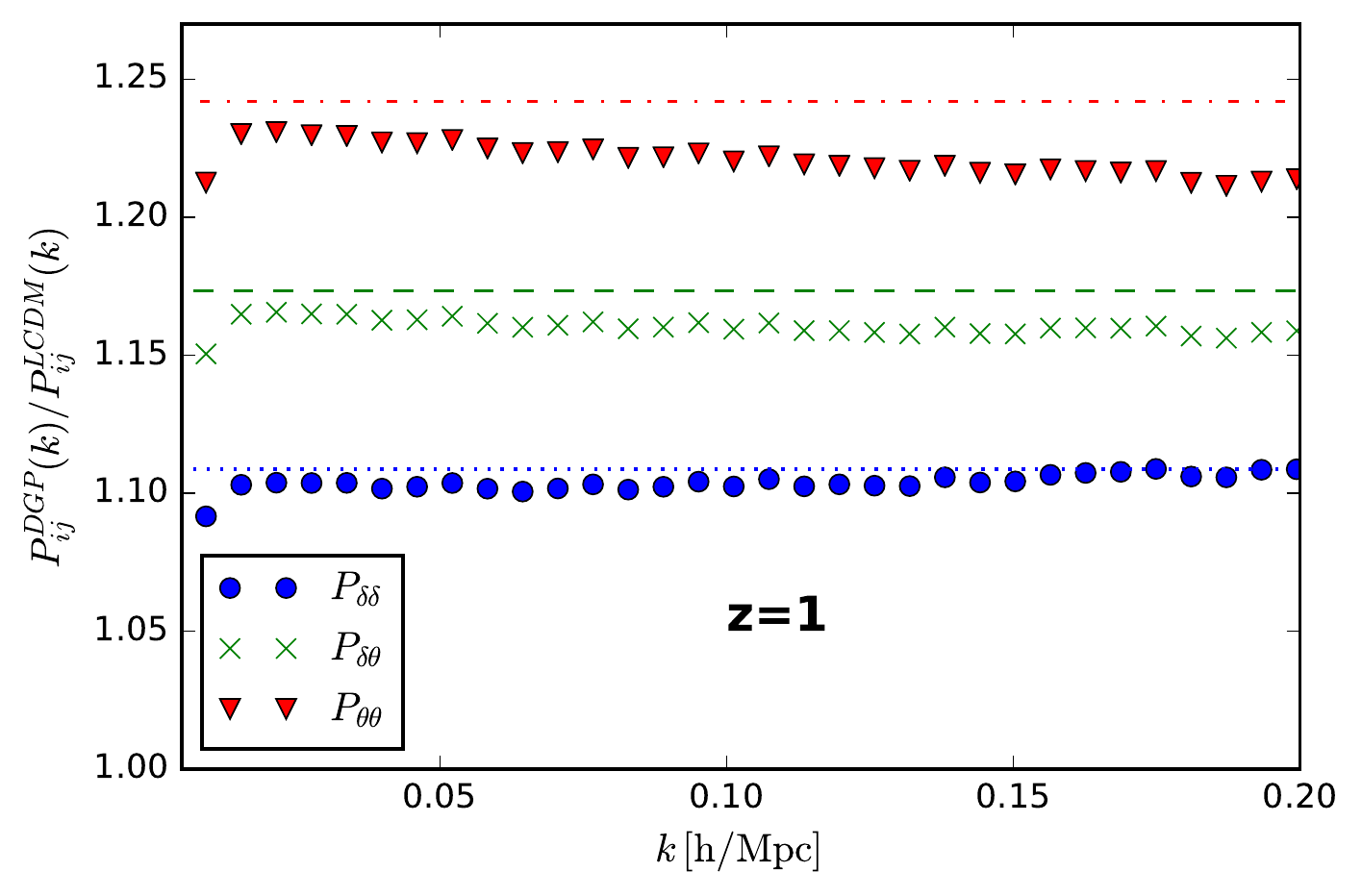}} 
  \caption{The ratio of the nDGP to LCDM N-body real space spectra (points) at $z=0.5$ (left) and $z=1$ (right). The linear ratios are shown as dotted, dashed and dot-dashed lines.}
\label{nbodyrat1}
\end{figure}

\begin{figure}[H]
  \captionsetup[subfigure]{labelformat=empty}
  \centering
  \subfloat[]{\includegraphics[width=7.5cm, height=7cm]{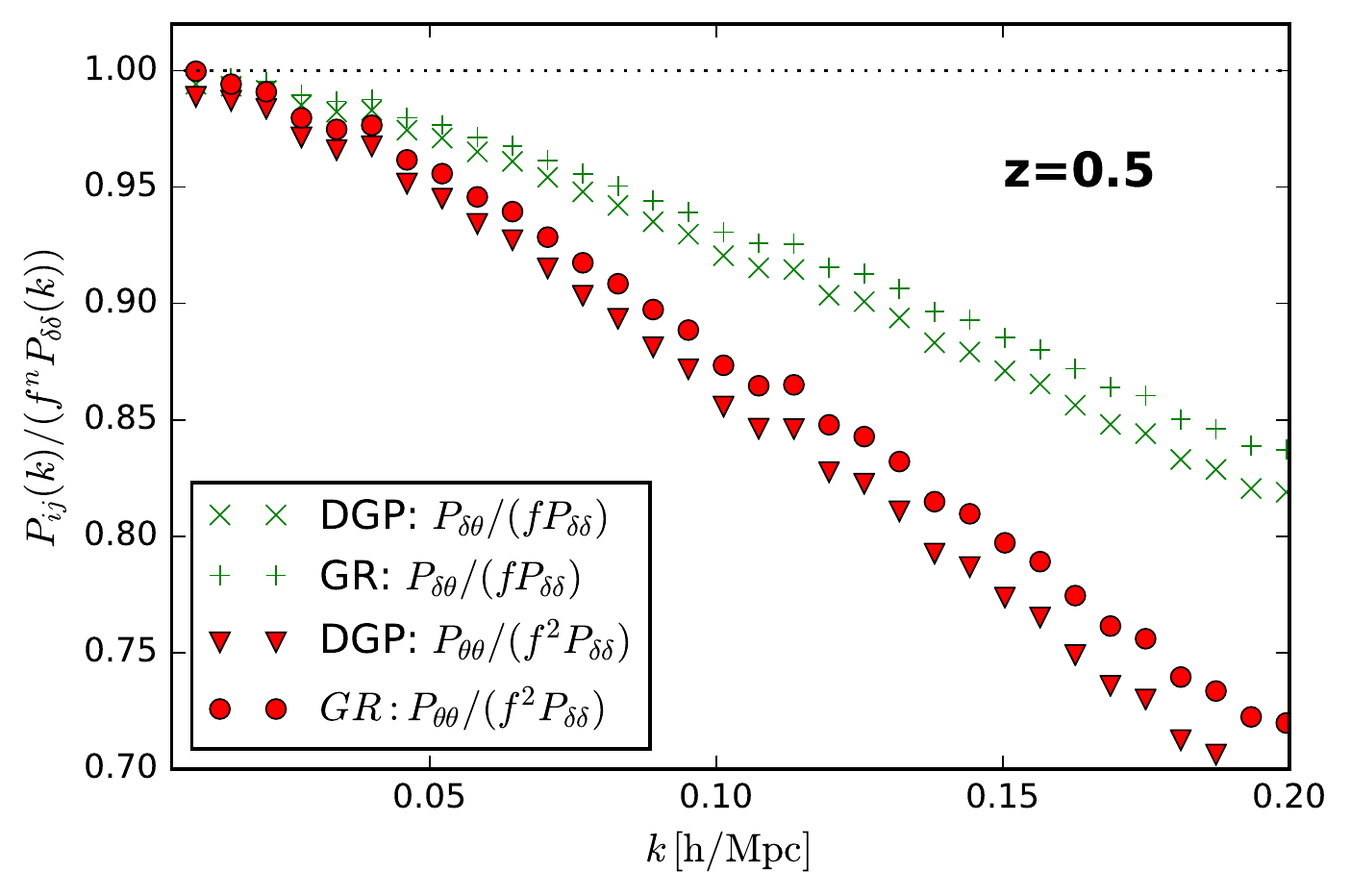}} \quad
  \subfloat[]{\includegraphics[width=7.5cm, height=7cm]{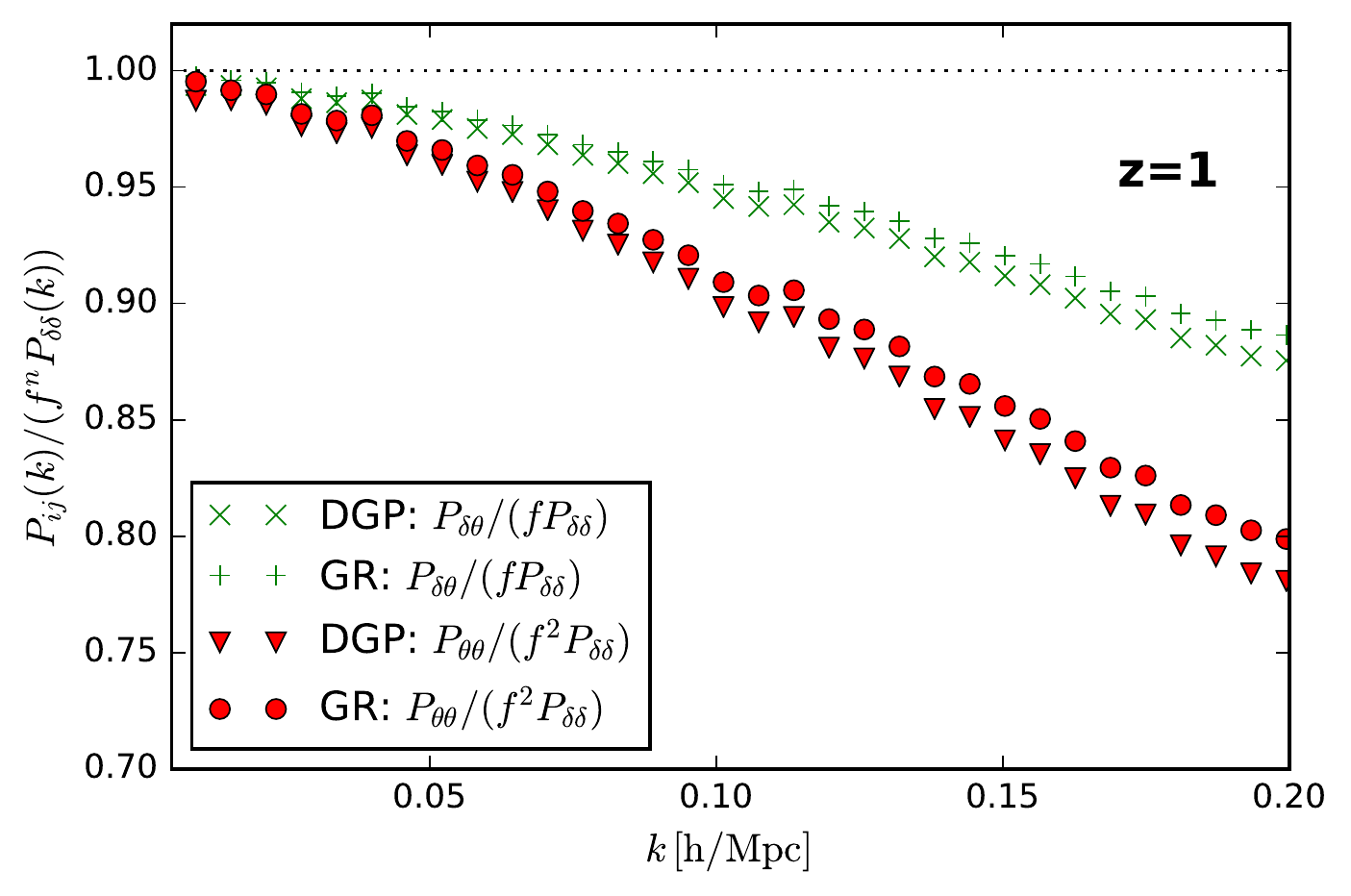}} 
  \caption{The ratio of the N-body real space auto velocity (red) and cross spectra (green) to their linear predictions in LCDM (circles and pluses) and in nDGP (triangles and crosses) at $z=0.5$ (left) and $z=1$ (right).}
\label{nbodyrat2}
\end{figure}

To obtain a spectra measurement from the simulations, we use the {\it Delaunay Tesselation Field Estimator} (DTFE) method implemented in publicly avilable DTFE code by \cite{cv2011}. The DTFE code employs {\it the Delaunay Tesselation Field Estimation}, a method described in details in \cite{sv2000,vs2009}, which assures that the resulting smooth fields have the highest attainable resolution, are volume weighted and have suppressed sampling noise. The fields are then smoothed using top-hat filtering and we proceed to obtain density $P_{\delta\delta}(k)=\langle\delta(\bfk)\delta^{*}(\bfk)\rangle$ and velocity divergence $P_{\theta\theta}(k)=\langle\theta(\bfk)\theta^*(\bfk)\rangle$ power spectra following the method of \cite{Li:2012by,Hellwing2013}. It is well known that energetic processes connected to the highly non-linear physics of galaxy formation affect the cosmic density field up to scales even of tens of Megaparsecs. Thus our simplistic N-body only approach could introduce some additional scatter and bias in the analysis. However, as recently shown by \cite{Hellwing2016} with a use of the state-of-the-art galaxy formation
simulation, the EAGLE suite \cite{Schaye:2014tpa}, the energetic baryonic processes have a negligible impact on both velocity and density fields
on the scales we consider in this study. Hence, we can be assured that our analysis will not be affected by the fact that we
ignore baryons completely in our modelling.
\newline
\newline
Fig.\ref{nbodypsz1} shows SPT does very well in modelling non-linearities at $z=1$. For the nDGP simulation we find agreement at $1(3)\%$ level up to $k=0.175 (0.18), 0.18(0.2)$ and $0.22(0.26)h$/Mpc for $P_{\delta \delta}$, $P_{\delta \theta}$ and $P_{\theta \theta}$ respectively. Given this we consider scales up to $k_{\rm max} = 0.195h$/Mpc at $z=1$. We have fit Poisson errors to this data assuming a volume of $1\mbox{Gpc}^3/h^3$ with a shot noise term of $\bar{n} = 3\times10^{-4}h^3/\mbox{Mpc}^3$ (see Eq.27 of \cite{Zhao:2013dza} for example). 
\newline
\newline
On the other hand, Fig.\ref{nbodypsz05} shows the decline in the accuracy of the SPT approach at later times.  Again for the nDGP simulation, now at $z=0.5$, we find agreement at $1(3)\%$ level up to $k=0.12 (0.18), 0.18(0.2)h$/Mpc for $P_{\delta \delta}$ and $P_{\delta \theta}$ respectively. $P_{\theta \theta}$ is found to be very noisy around the $1$-$3\%$ band within $0.1 \leq k \leq 0.2$. Given this, we consider scales up to a $k_{\rm max} = 0.147h$/Mpc at $z=0.5$. The stated ranges of applicability of our SPT modelling in nDGP are found to be very similar to the GR simulation comparisons. 
\newline
\newline
In Fig.\ref{nbodyrat1} we have plotted the ratio of the real space spectra in nDGP to LCDM simulations for $z=0.5$ and $z=1$ along with the linear predictions as dotted and dashed lines. This figure captures the effects of modified gravity on the real space spectra and growth. We note that the density spectra remain very much the same as we proceed into the quasi non-linear regime. On the other hand we find that the DGP velocity spectra becomes more suppressed as we go to smaller scales. This scale dependence is expected as large fifth force enhancements to the velocity field act to increase the velocity dispersion, which effectively reduces the correlation between particle velocities. The effect is larger at lower redshift where there is more non-linear structure growth. In Fig.\ref{nbodyrat2} we further elucidate this point where we have plotted the ratio of velocity and cross spectra to their linear predictions for both LCDM and nDGP simulations. A first point is that clearly non-linearity becomes very important in the scales considered. Secondly, the nDGP simulation shows an enhanced non-linearity (and hence suppression of velocity correlations) over the LCDM simulation. 
\newline
\newline
These results are consistent with previous results found for other MG simulations such as in Fig.4 and Fig.7 of \cite{Winther:2015wla} and Fig.7 and Fig.8 of \cite{Winther:2017jof}. They found a pattern of a constant boost at the linear scales (reflecting enhanced growth rate) with the difference in spectra beginning to be suppressed closer to non-linear scales. At non-linear scales the 5th force in nDGP theories starts to be screened effectively  by the Vainshtein mechanism, recovering the Newtonian value inside most of the virialised structures (see also \cite{Falck:2015rsa}). Our analysis is limited to only quasi-linear scales where the complicated scale-dependent patterns of the Vainshtein mechanism are still not well developed. 


\subsection{Redshift Space Comparisons : MG-PICOLA}
\begin{figure}[H]
  \captionsetup[subfigure]{labelformat=empty}
  \centering
  \subfloat[]{\includegraphics[width=7.5cm, height=7cm]{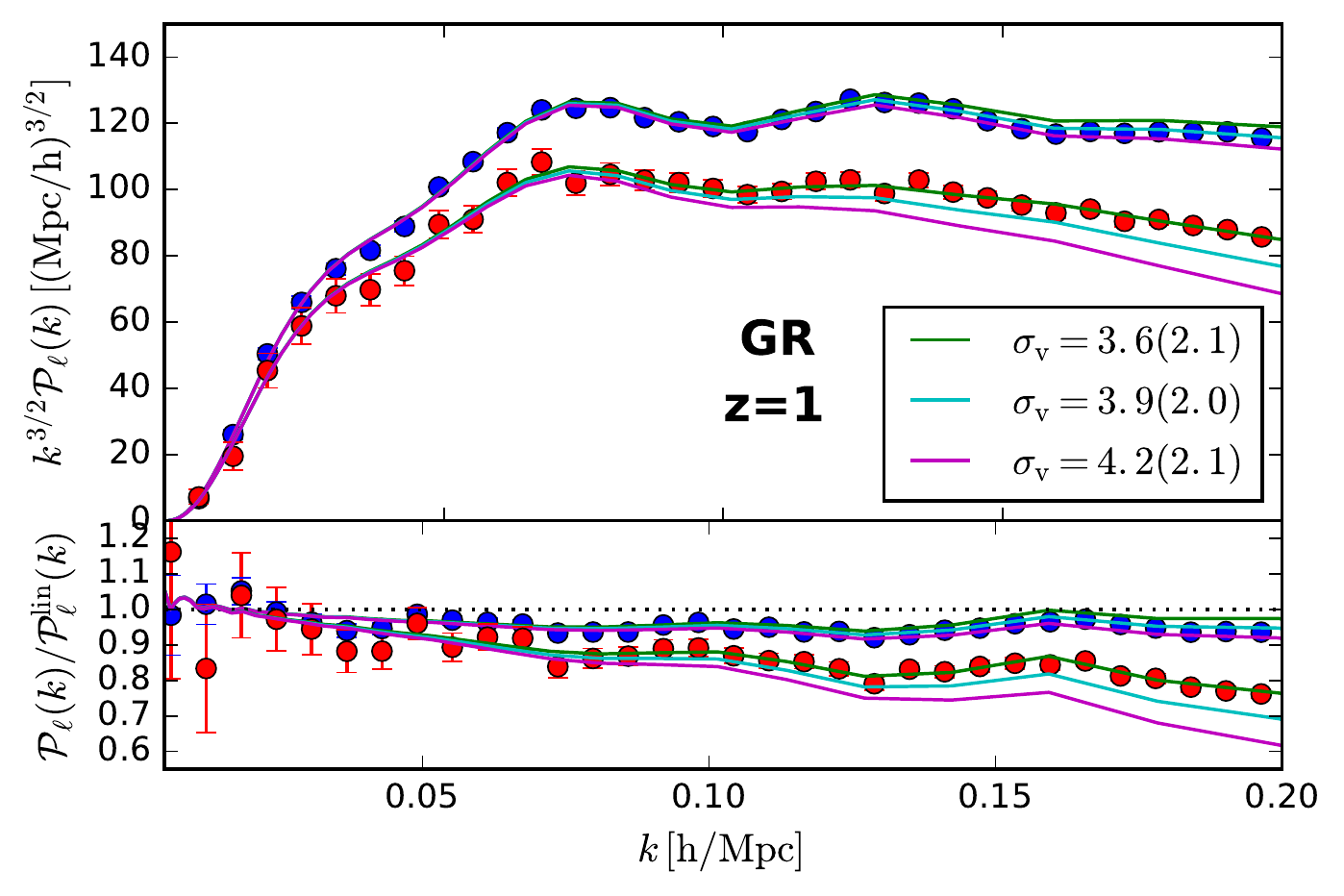}} \quad
  \subfloat[]{\includegraphics[width=7.5cm, height=7cm]{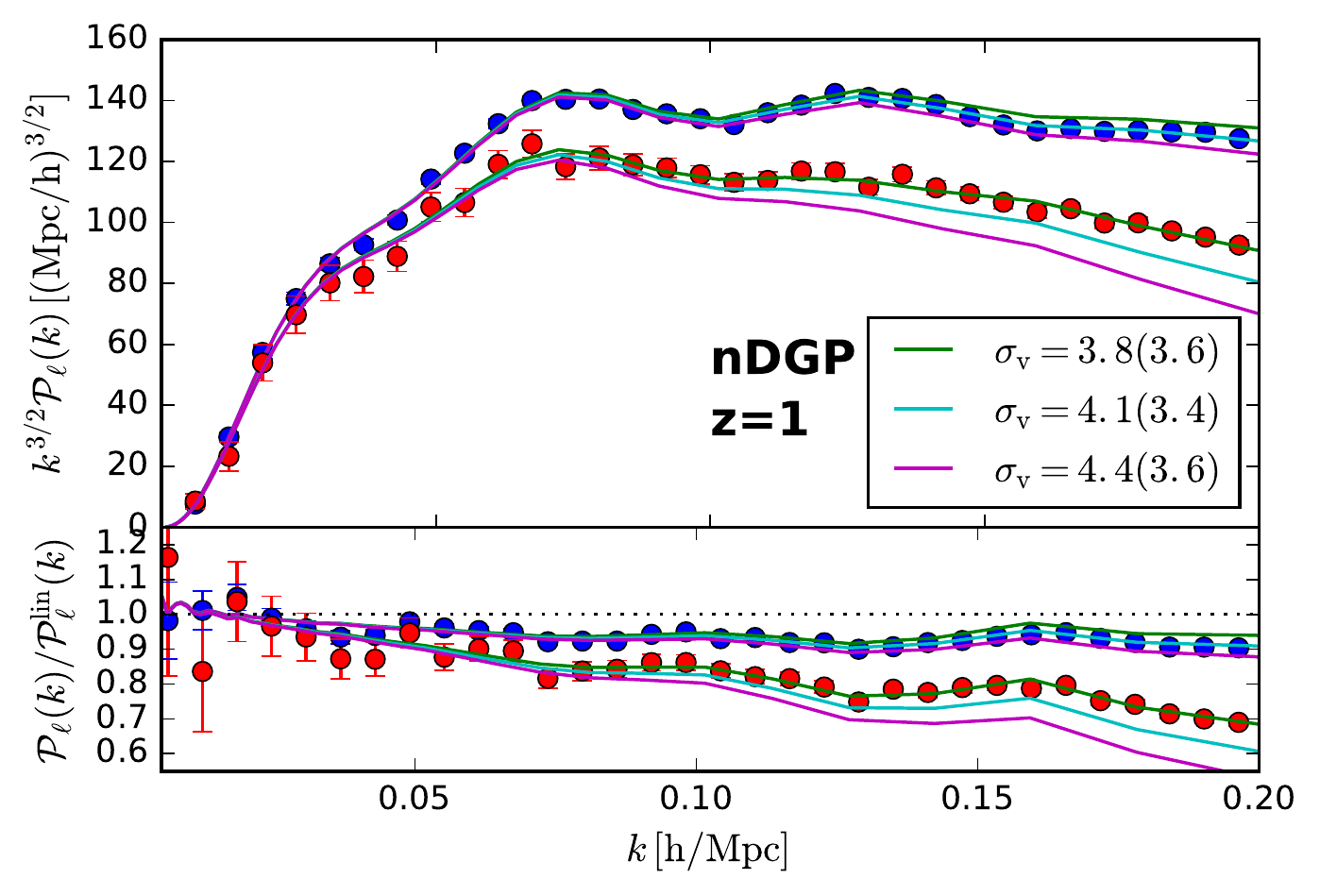}} 
  \caption{TNS predictions (solid) and MG-PICOLA measurements (points) of the monopole (the upper group of points/lines) and quadrupole (the lower group of points/lines)  at $z=1$ for GR (left) and nDGP (right). Three values of the TNS model parameter $\sigma_v$ are shown (in units of Mpc/$h$) with their respective reduced $\chi^2$ in brackets. The top panels show the multipoles multiplied by $k^{3/2}$ and the bottom panels show the deviations from Kaiser's linear prediction. The error bars are those from an ideal survey of $V_s=10 \mbox{Gpc}^3/h^3$. }
\label{nbodyplz1}
\end{figure}

\begin{figure}[H]
  \captionsetup[subfigure]{labelformat=empty}
  \centering
  \subfloat[]{\includegraphics[width=7.5cm, height=7cm]{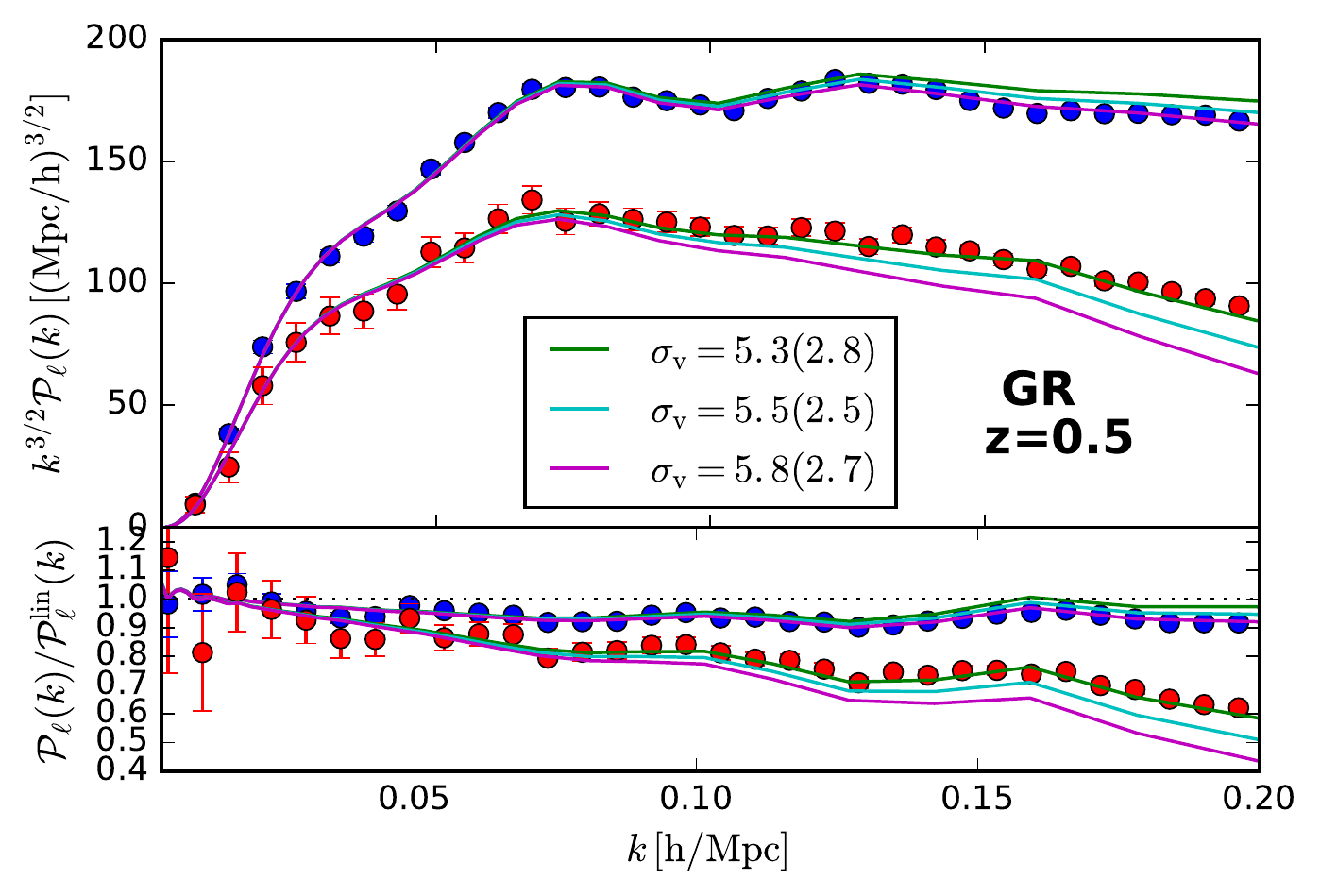}} \quad
  \subfloat[]{\includegraphics[width=7.5cm, height=7cm]{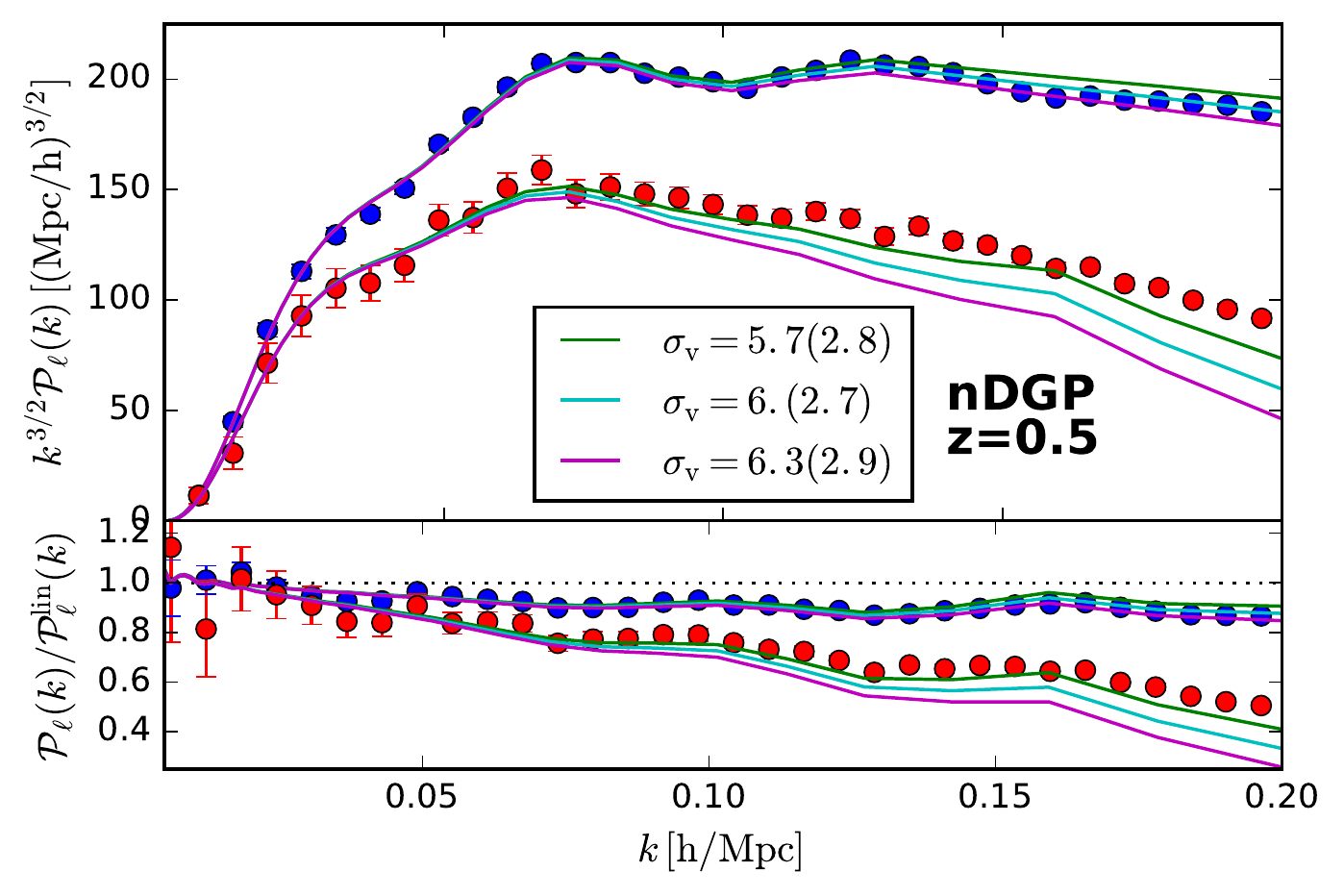}} 
  \caption{TNS predictions (solid) and MG-PICOLA measurements (points) of the monopole  (the upper group of points/lines) and quadrupole (the lower group of points/lines) at $z=0.5$ for GR (left) and nDGP (right). Three values of the TNS model parameter $\sigma_v$ (in units of Mpc/$h$)  are shown with their respective reduced $\chi^2$ in brackets. The top panels show the multipoles multiplied by $k^{3/2}$ and the bottom panels show the deviations from Kaiser's linear prediction. The error bars are those from an ideal survey of $V_s=10 \mbox{Gpc}^3/h^3$.}
\label{nbodyplz05}
\end{figure}

Next we take a look at the predictive power of the TNS multipoles, providing the realm of validity of the RSD modelling. The MG-PICOLA multipoles are measured using the distant-observer approximation \footnote{i.e we assume that the observer is located at a distance much greater than the boxsize ($r\gg 1024 \mbox{Mpc}/h$), and so we treat all the lines-of-sight as parallel to the chosen Cartesian axes of the simulation box. Next, we use an appropriate velocity component ($v_x, v_y$ or $v_z$) to disturb the position of a matter particle.} and averaged over three LOS directions. We further average over 20 MG-PICOLA simulations each of $1\mbox{Gpc}^3/h^3$ volume thus ignoring the mode covariance at and above box-size scales. This should correspond to an ideal survey with a resulting volume of $20\mbox{Gpc}^3/h^3$. 
\newline
\newline
Fig.\ref{nbodyplz1}  shows the monopole and quadrupole predictions at $z=1$ for three different values of $\sigma_v$ where we have fit using 32 $k$-bins up to $k_{\rm max}=0.195h$/Mpc. The reduced $\chi^2_{\rm red} = -2\ln{\mathcal{L}}/{\rm N_{dof}}$ is shown in brackets, where $N_{dof}$ is the total degrees of freedom at that $k_{\rm max}$ which equals twice the number of bins minus the number of parameters. In this case $N_{dof}=62$($=32\times2-2$), since we have 2 parameters. Similarly, Fig.\ref{nbodyplz05} shows the same results at $z=0.5$ where we have fit using 24 $k$-bins ($N_{dof}=46$) up to $k_{\rm max}=0.147h$/Mpc as dictated by the real space power spectra comparisons. 
\newline
\newline
\noindent We find that the fit of PT is comparable between GR and nDGP simulations at both redshifts, although GR does slightly better at small scales at $z=0.5$. The quadrupole also does worse than the monopole over both models at $z=0.5$ which we can attribute to the increased dependency on the velocity auto power spectrum for which the theoretical template does worse in reproducing (see Fig.\ref{nbodypsz05}). 
\newline
\newline
As was done with the power spectra, we identify the scale at which deviations from theory are within the $1(3)\%$ region. We do this using the values of $\sigma_v$ with the lowest $\chi^2$ shown in Fig.\ref{nbodyplz1} and Fig.\ref{nbodyplz05}. This will be used to set a maximum Fourier mode for our statistical analysis. For the nDGP simulation at $z=1$ we find agreement at $1(3)\%$ level up to $k=0.24(0.25)h$/Mpc for $P_{0}$. $P_{2}$ is significantly noisier around the $1\%$ deviation region but matches PT up to $k= 0.16h$/Mpc within $3\%$. Similarly, for $z=0.5$ we find a theory-data agreement of $1(3)\%$ up to $k=0.147(0.2)h$/Mpc for $P_{0}$ while a much worse $k=0.09(0.1)h$/Mpc for $P_{2}$. Since $P_{2}$'s contribution to Eq.\ref{covarianceeqn} for the likelihood is significantly smaller than $P_{0}$ because of their respective errors, we decide to use $k_{\rm max}=0.147h$/Mpc at $z=0.5$ and $k_{\rm max}=0.195h$/Mpc at $z=1$ for the MCMC analysis, despite the poor quadrupole fit. In the case of GR the range of validity is found to be similar.
\newline
\newline
By having assessed and compared the range of validities of SPT for both a LCDM and nDGP cosmologies, we have gained a handle on how enhanced dynamics due to fifth force interactions degrade SPT's performance. Our comparisons indicate that the non-linearity generated by $\Omega_{rc}=0.438$ is mild enough not to significantly effect the range of validity. We caution however, this may not be the case for other models of gravity characterised by a larger degree of deviation from GR dynamics. Next we will test the capabilities of the theoretical templates when matching to the MG-PICOLA multipoles. 
\newpage


\section{Template Performance}
\begin{figure}[H]
  \captionsetup[subfigure]{labelformat=empty}
  \centering
  \subfloat[]{\includegraphics[width=7.5cm, height=7cm]{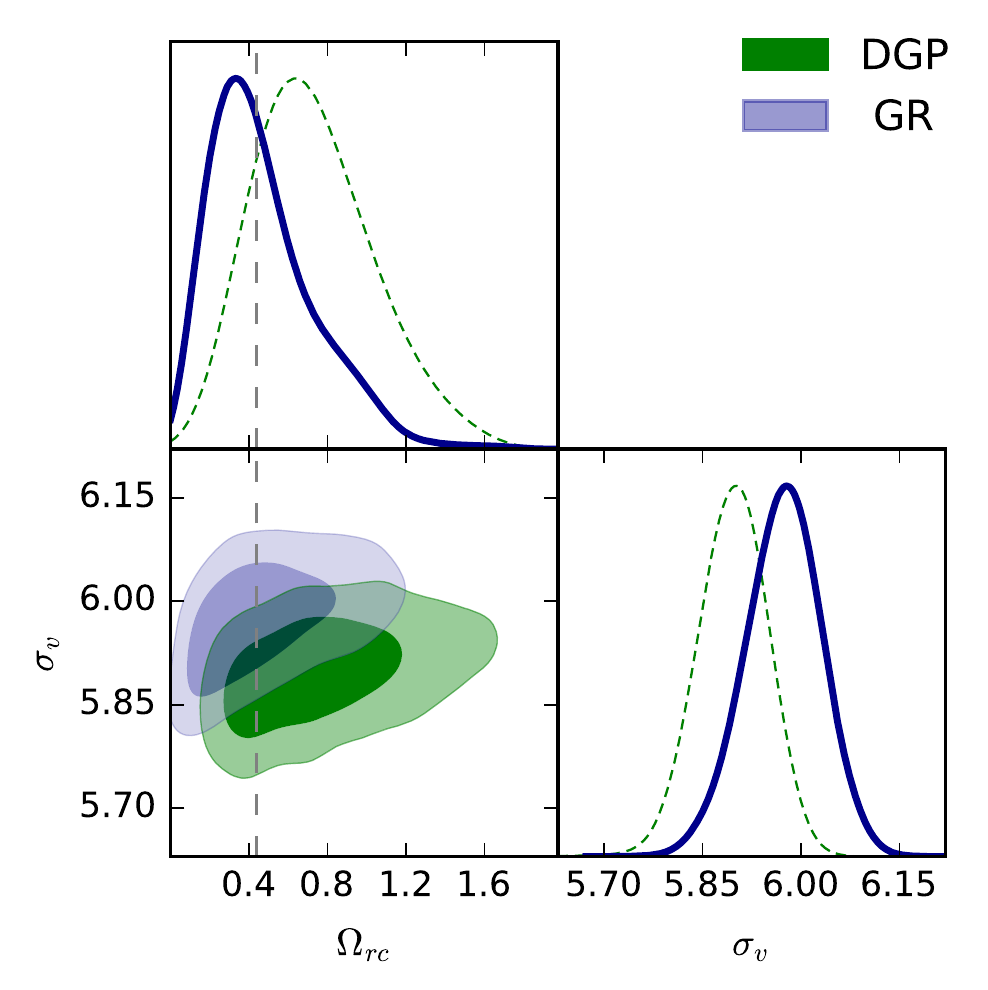}} \quad
  \subfloat[]{\includegraphics[width=7.5cm, height=7cm]{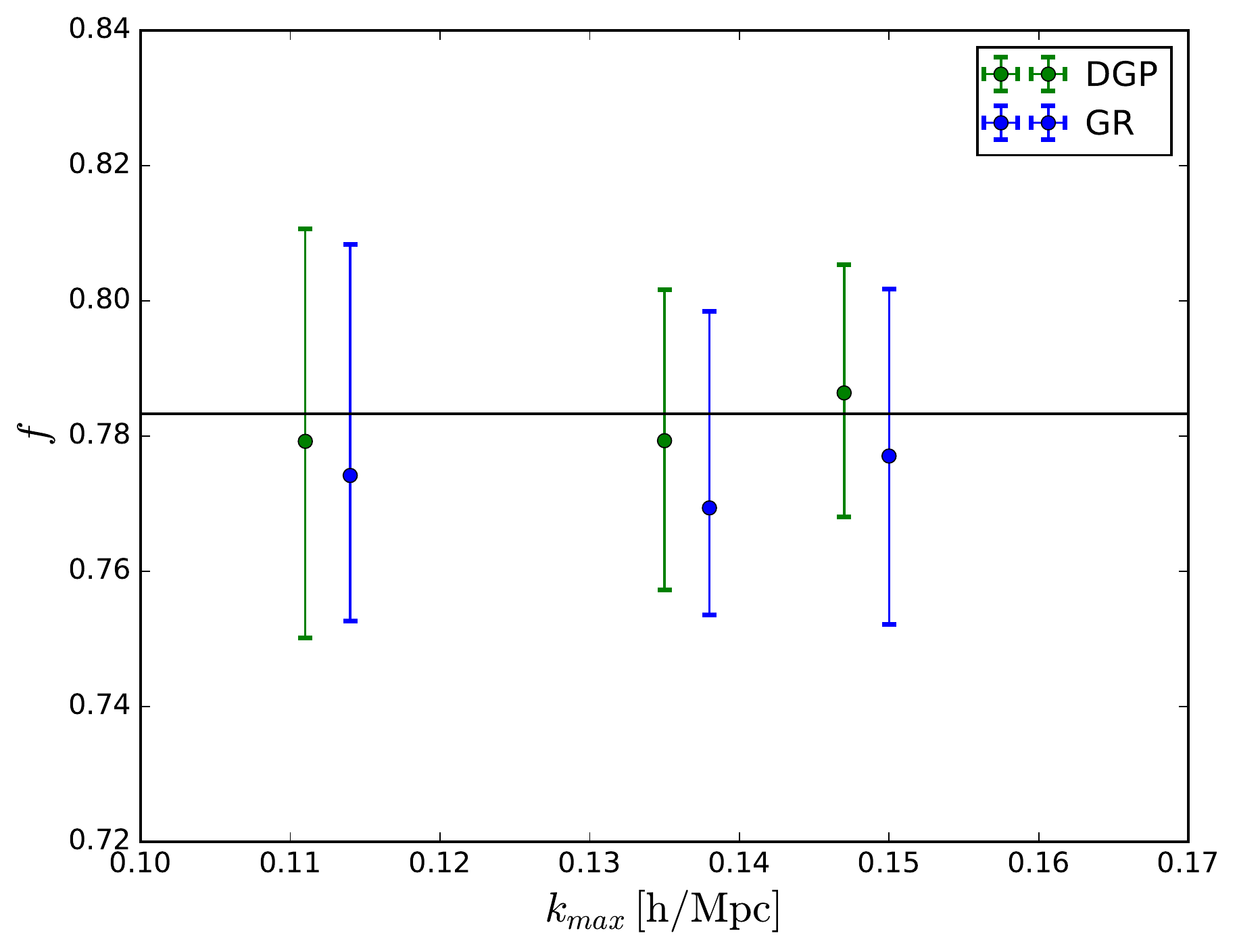}} 
  \caption{{\bf Left:} The $1\sigma$ and $2\sigma$ confidence contours for the DGP template and the GR template at $z=0.5$ fitting up to $k_{\rm max}=0.147h$/Mpc using 24 bins with the simulation's fiducial value for $\Omega_{rc}$ indicated by the dashed line. {\bf Right:} The best fit value for $f$ as a function of $k_{\rm max}$ with the $2\sigma$ errors for the DGP and GR template. The GR template values (blue) have been slightly shifted for better visualisation. A survey of volume of $10\mbox{Gpc}^{3}/h^3$ is assumed. }
\label{nbodyz05}
\end{figure}

\begin{figure}[H]
  \captionsetup[subfigure]{labelformat=empty}
  \centering
  \subfloat[]{\includegraphics[width=7.5cm, height=7cm]{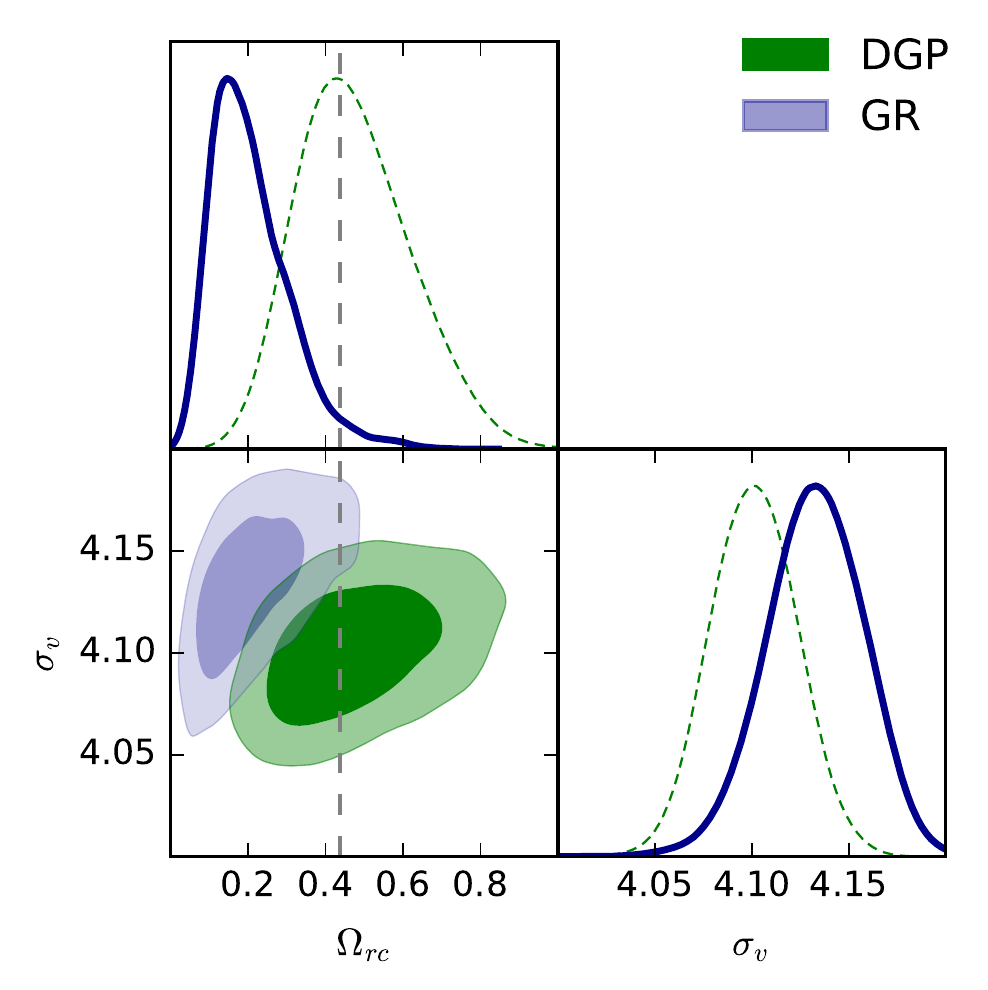}} \quad
  \subfloat[]{\includegraphics[width=7.5cm, height=7cm]{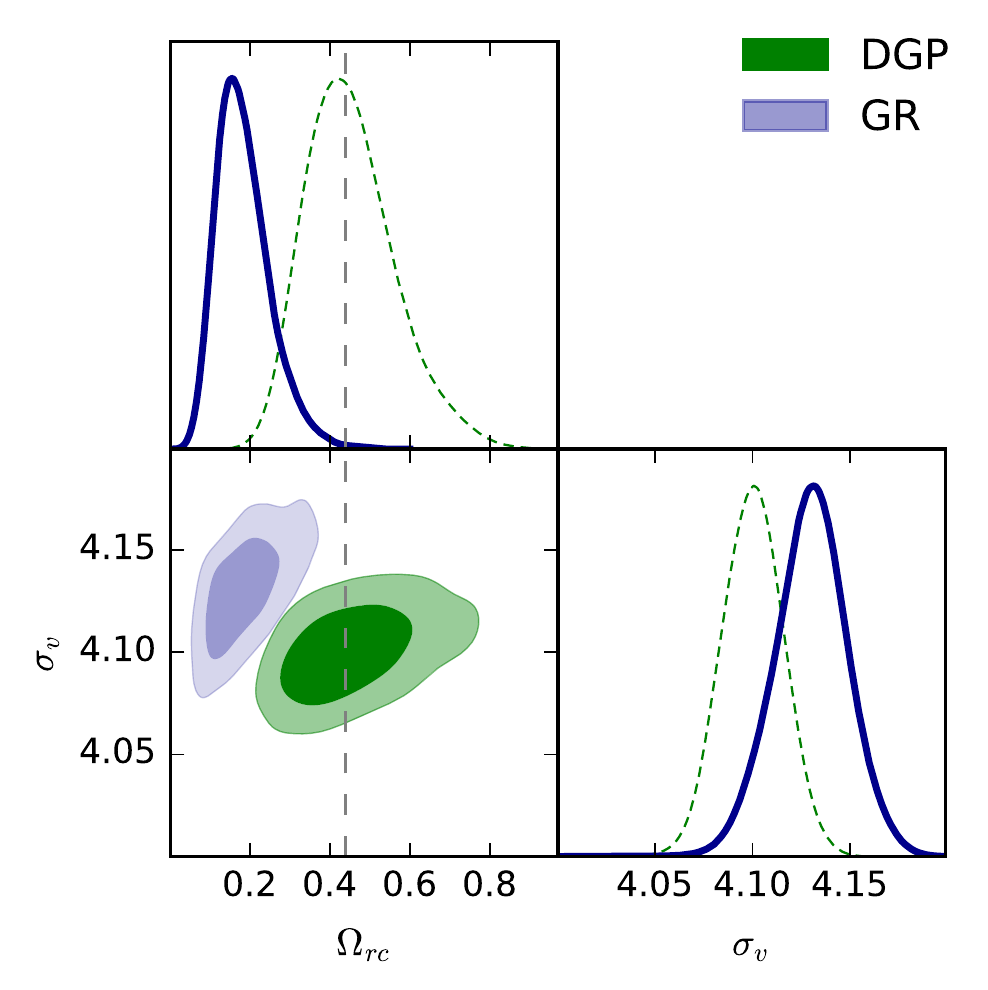}} 
  \caption{ {\bf Left:} The $1\sigma$ and $2\sigma$ confidence contours for the DGP template and the GR template at $z=1$ fitting up to $k_{\rm max}=0.195h$/Mpc using 32 bins with the simulation's fiducial value for $\Omega_{rc}$ indicated by the dashed line. The survey volume is taken to be $10\mbox{Gpc}^{3}/h^3$. {\bf Right:} Same as left plot but with the survey volume is taken to be $20 \mbox{Gpc}^{3}/h^3$ . }
\label{nbodyz1a}
\end{figure}

\begin{figure}[H]
  \captionsetup[subfigure]{labelformat=empty}
  \centering
 \includegraphics[width=12.3cm, height=8.39cm]{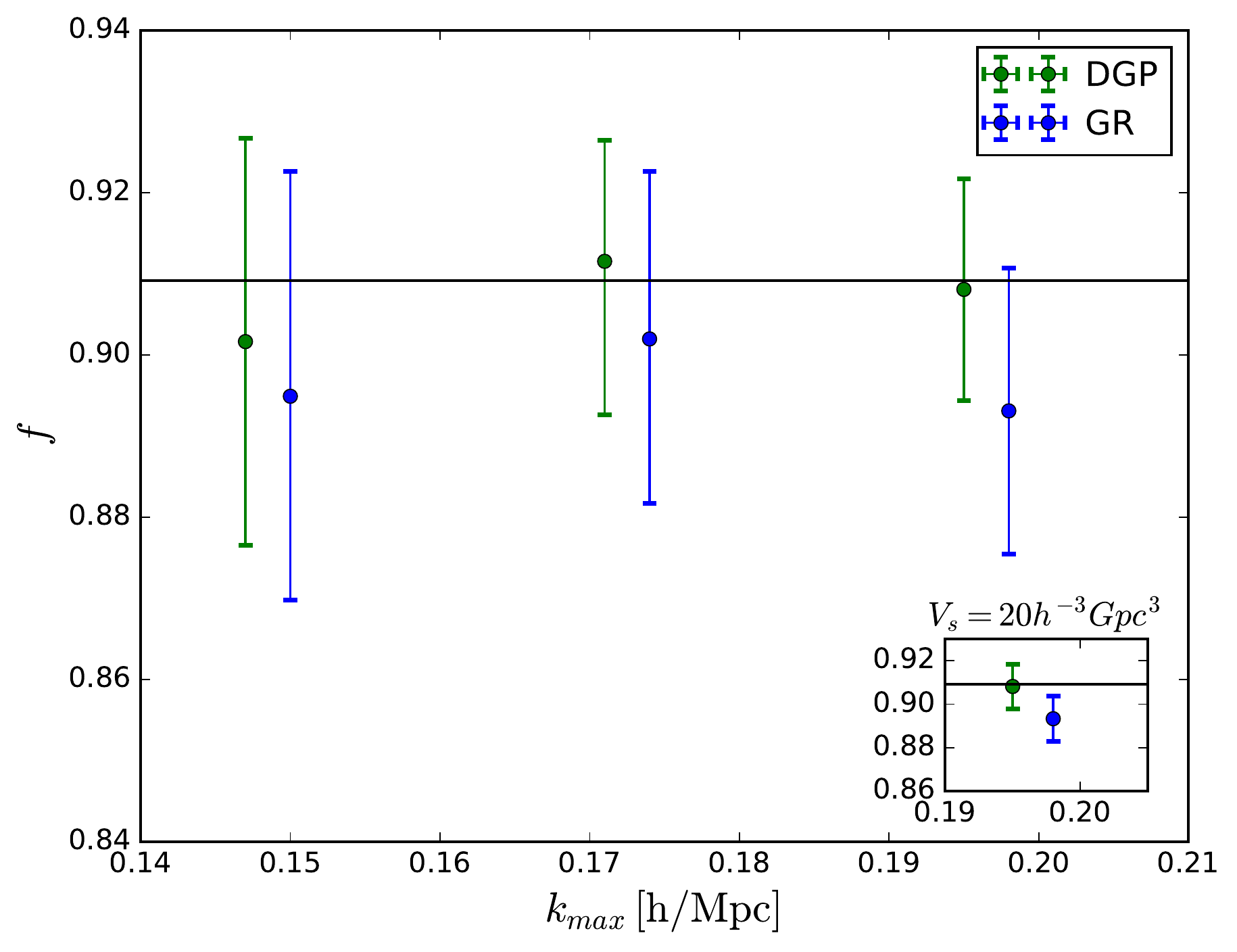} 
  \caption{The best fit value for $f$ as a function of $k_{\rm max}$ with the $2\sigma$ errors for the DGP and GR templates at $z=1$. The survey volume is taken to be $10\mbox{Gpc}^{3}/h^3$  with the annotated plot containing the prediction at $k_{\rm max}=0.195h$/Mpc for an increased survey volume of $20 \mbox{Gpc}^{3}/h^3$. The GR template values (blue) have been slightly shifted for better visualisation.}
\label{nbodyz1b}
\end{figure}

\begin{figure}[H]
  \captionsetup[subfigure]{labelformat=empty}
  \centering
  \subfloat[]{\includegraphics[width=7.5cm, height=7cm]{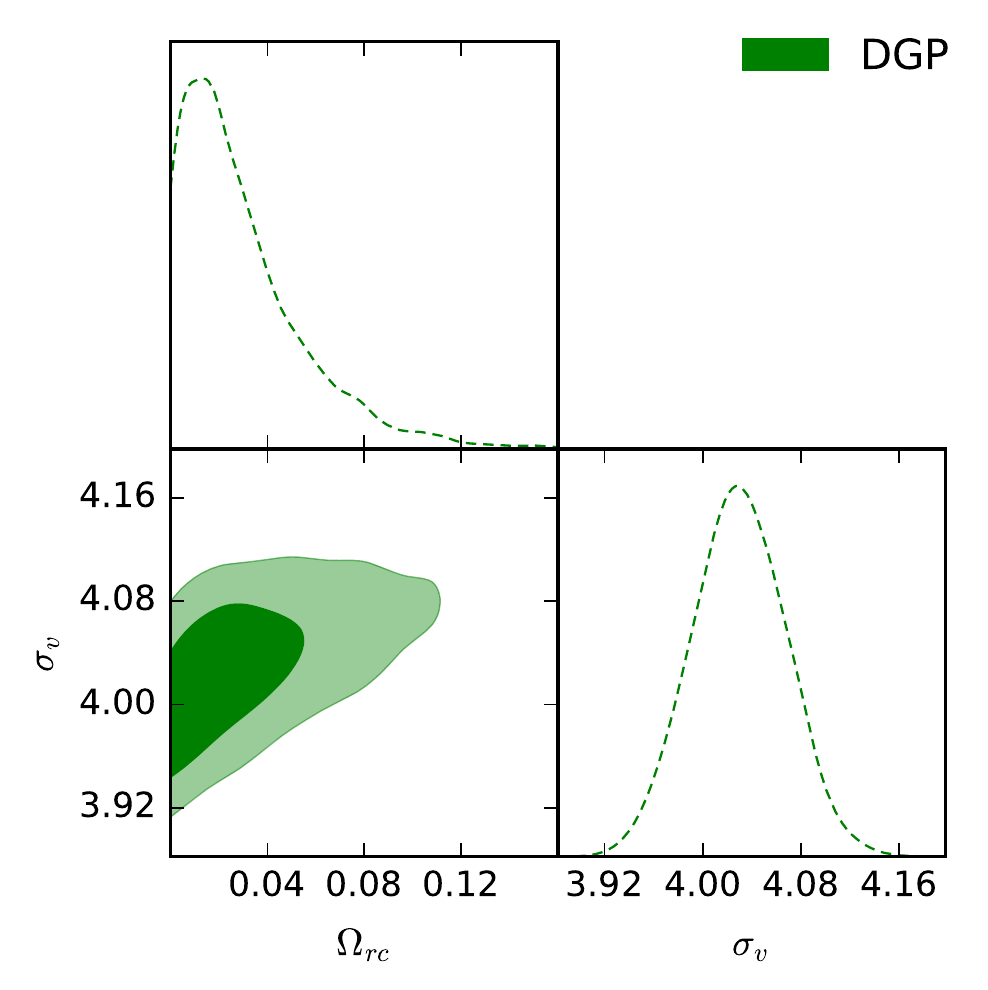}} \quad
  \subfloat[]{\includegraphics[width=7.5cm, height=7cm]{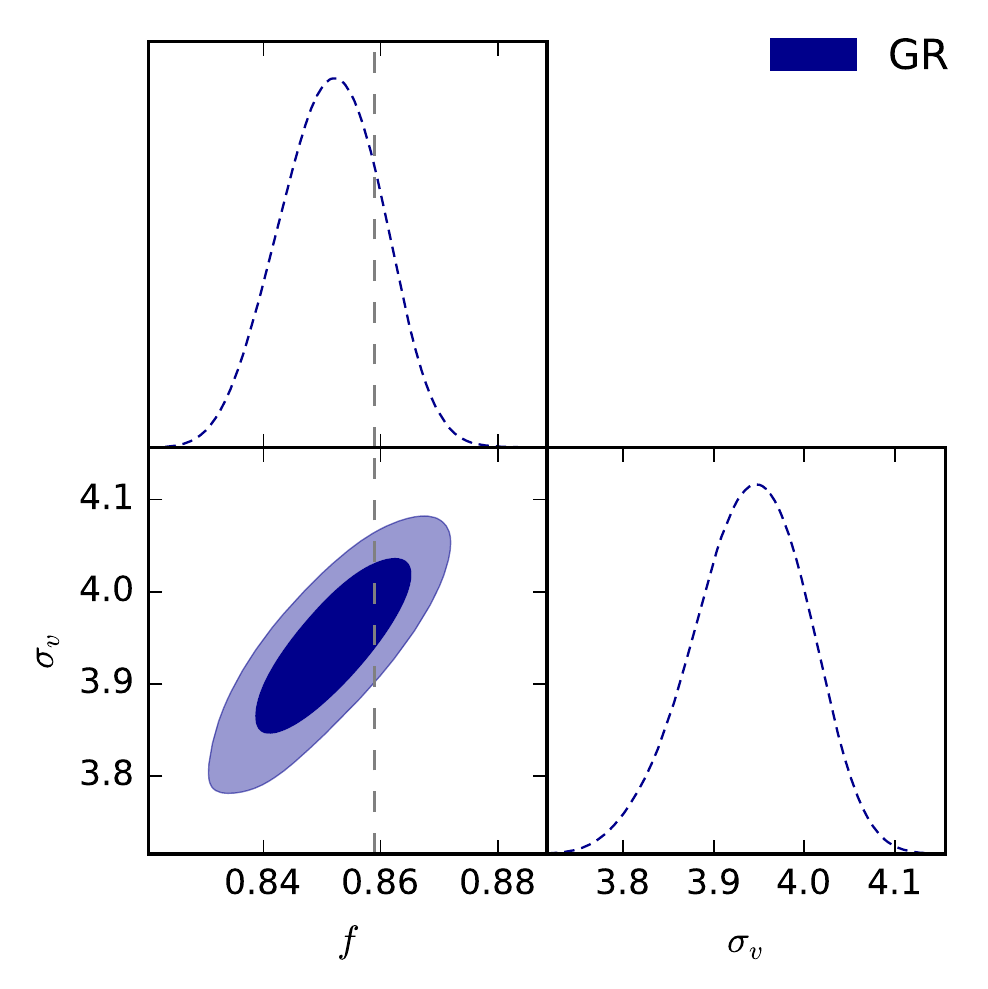}} 
  \caption{ {\bf Left:} The $1\sigma$ and $2\sigma$ confidence contours for the DGP template at $z=1$ fitting up to $k_{\rm max}=0.171h$/Mpc using 28 bins with the simulation's fiducial value for $\Omega_{rc}=0$ (the GR simulation). {\bf Right:} The $1\sigma$ and $2\sigma$ confidence contours for the GR template fitting to the same simulation as used in the left panel, using the same number of bins, but without a prior on $f$. The fiducial $f$ is marked by a dashed line. A survey of volume of $10\mbox{Gpc}^{3}/h^3$ is assumed.}
\label{nbodygrz1}
\end{figure}

\begin{figure}[H]
  \captionsetup[subfigure]{labelformat=empty}
  \centering
  \subfloat[]{\includegraphics[width=7.5cm, height=7cm]{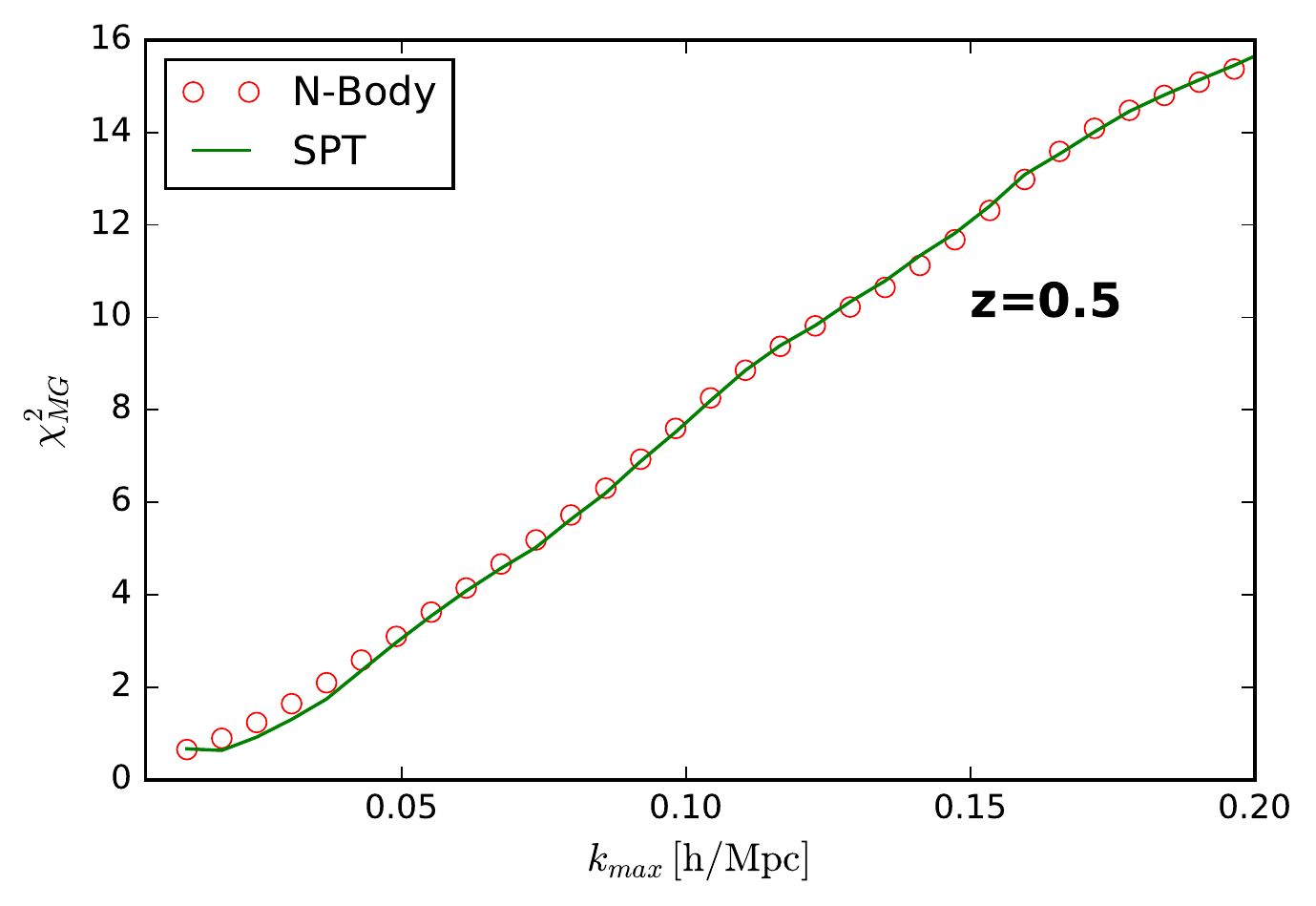}} \quad
  \subfloat[]{\includegraphics[width=7.5cm, height=7cm]{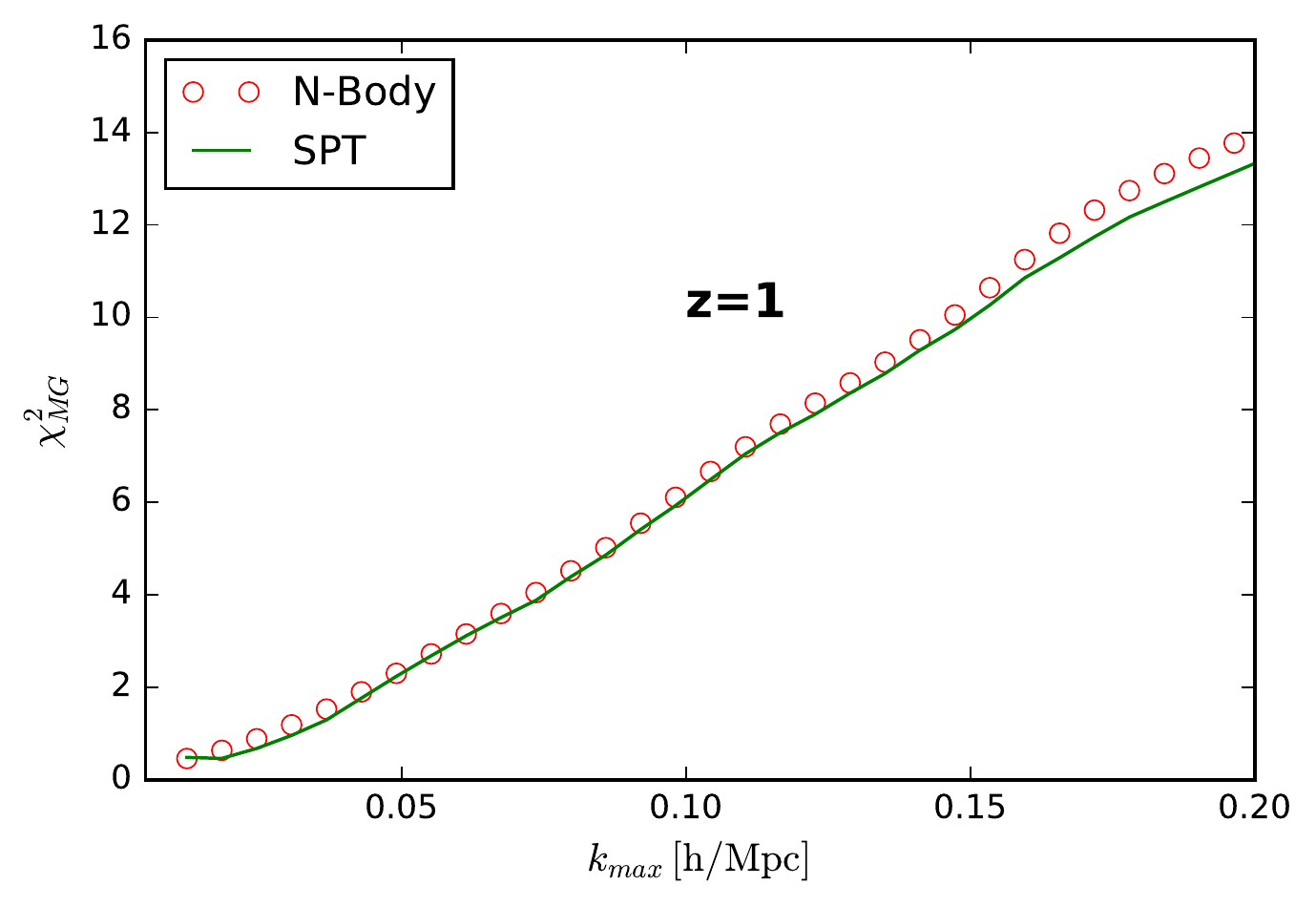}} 
  \caption{The quantity computed using Eq.\ref{chisqrt} for SPT (green line) and N-body (red circles) at $z=0.5$ (left) and $z=1$ (right).}
\label{chisqrtest}
\end{figure}

We begin by explicitly defining the theoretical templates. The GR template here means that we use the TNS model for the RSD where we set $\mu=1, \gamma_2=\gamma_3=0$ but treat $f$ as a free parameter parametrised by $\Omega_{rc}$. On the other hand, the DGP template uses $\mu, \gamma_2, \gamma_3$ as given in Eq.\ref{mudgp} - Eq.\ref{g3dgp} to model the non-linearity in the DGP model properly. We remind the reader that it is a common practice to use a GR template to estimate $f$ from survey data sets (eg.\cite{Beutler:2016arn}).
\newline
\newline
Our main intention in this work is to robustly assess whether using GR-only based RSD modelling (in other words, a model ignorant of any possible deviations from the GR picture of structure formation) on a non-GR universe is an accurate enough procedure to avoid biased estimates of $f$. Note that the DGP template encompasses a pure-GR scenario (by setting $\Omega_{rc}=0$) and so we should observe no biasing in using the DGP template for the GR data. Further, we want to determine on what scales and what amplitudes biasing becomes significant. As we fix the amplitude of density perturbations, both theoretical templates are equivalent on linear scales, but as we include increasingly non-linear scales in the analysis the templates deviate producing a bias. This bias may be masked with the freedom of the TNS damping parameter $\sigma_v$, although the model specific non-linearities act beyond pure damping of small scale power and so $\sigma_v$ cannot perfectly substitute for incorrect perturbation modelling. In other words, model bias should still be expected at some level. 
\newline
\newline
Our analysis follows the techniques described in \cite{Lewis:2002ah}, with the MCMC algorithm walking in $\{\sigma_v,\Omega_{rc}\}$ parameter space. In order to measure when non-linearities become an issue in terms of theoretical modelling we complete the analysis for varying values of $k_{\rm max}$ up to the upper bounds found in the previous subsection. By going to higher $k$-modes we suppress the statistical errors. Fig.\ref{nbodyz05} shows the results at $z=0.5$. On the left hand side we present the $1\sigma$(68$\%$ C.L) and $2\sigma$(95$\%$ C.L) contours for the DGP template (green) and GR template (blue) for a matching up to $k_{\rm max}=0.147h$/Mpc using 24 bins. The marginalised statistics are shown in the side panels and the fiducial value, $\Omega_{rc}=0.438$, is marked as a dashed line. We note a slight offset of the best fit values of $\Omega_{rc}$, although the fiducial value remains in the $1\sigma$ region for both templates. There is also an offset in the best fit $\sigma_v$ which can be interpreted as the GR template's use of this parameter to compensate for non-linear effects in the DGP-PICOLA simulations.  On the right of Fig.\ref{nbodyz05}, we see the marginalised best fit values for $f(\Omega_{rc})$ with their $2\sigma$ errors for varying scale inclusion ($k_{\rm max}=0.110,0.135,0.147h$/Mpc with GR template's values shifted slightly for better visualisation). Note that as we push to higher $k_{\max}$ the errors become smaller, as expected. We here remind the reader that there is a lower bound on $f$ imposed by $\Omega_{rc}\geq 0$.  As can be seen in both plots, the GR template comfortably accommodates the data within $68\%$ (for $k_{\rm max}=0.147h$/Mpc) within the templates' validity regime.  
\newline
\newline
Fig.\ref{nbodyz1a} shows the results at $z=1$ this time matching up to a $k_{\rm max}=0.195h$/Mpc. It is in this case that we find the GR template struggling to capture the full shape of the multipoles and in the left hand plot we see that the fiducial parameter lies outside the $1\sigma$ region and just within the $2\sigma$ region. On the other hand, the DGP template is centred around the fiducial value. On the right hand side of Fig.\ref{nbodyz1a} we show the results for an analysis similar to that of the left hand side but for a larger survey with a volume of $V_s=20\mbox{Gpc}^{3}/h^3$, which is the estimated volume of DESI \cite{Aghamousa:2016zmz}. In this case the GR template fails to capture the fiducial at the $2\sigma$ level implying that inadequate theoretical modelling for such a large survey could introduce a significant error in parameter inference.  
\newline
\newline
Fig.\ref{nbodyz1b} shows the improvement in constraints as we increase $k_{\rm max}$. We have included an annotation with the same analysis done for a larger survey with a volume of $V_s=20 \mbox{Gpc}^{3}/h^3$ fitting up to $k_{\rm max}=0.195h$/Mpc. We see that the DGP template does consistently well in reproducing the fiducial while the GR begins to fail at around $k_{\rm max}=0.171h$/Mpc. 
\newline
\newline
As a consistency check of our analysis, we make use of the GR simulations at $z=1$. Again, by assuming a survey volume of $V=10 \mbox{Gpc}^3/h^3$, we repeated the analysis and obtained constraints on $\Omega_{rc}$ using the DGP template, as well as a constraint on $f$ using the GR template without the induced prior on $f$ coming from our parametrisation using $\Omega_{rc}$. Fig.\ref{nbodygrz1} shows the results. Both contours recover the fiducial parameters within $1\sigma$. 
\newline
\newline
As seen from the left panel of Fig.\ref{nbodygrz1}, the nDGP model with $\Omega_{rc}=0.438$ can be excluded with high significance ($\gg 2 \sigma$) by a survey with a volume $V_s=10 \mbox{Gpc}^3/h^3$ fitting up to $k_{max}=0.171 h$/Mpc, if our universe is described by GR. To quantify this, we computed the following quantity using the N-body measurements
\begin{equation}
\chi^2_{MG}(k_{\rm max}) = \frac{1}{N_{dof}}\sum_{\ell} \sum^{k_{\rm max}}_k \mbox{Cov}^{-1}_{\ell,\ell}(k)  [P_{\ell}^{DGP} (k) - P_{\ell}^{LCDM}(k)]^2, 
\label{chisqrt}
\end{equation}
where $\mbox{Cov}^{-1}_{\ell,\ell} $ is the covariance matrix between the multipoles assuming an ideal survey of $V_s=10 \mbox{Gpc}^{3}/h^3$. We also computed the same quantity using SPT. Fig.\ref{chisqrtest} shows the results up to $k_{\rm max}=0.2 h$/Mpc at $z=0.5$ and $z=1$. The results clearly shows that our ability to distinguish between LCDM and nDGP increases with $k_{max}$. Also we find that the $\chi^2$ from SPT is very similar to those obtained from simulation. This indicates that the TNS model is capable of providing the same level of significance of deviations from LCDM as the full non-linear treatment, making it a good estimator of structure formation in this range of scales. 
\newline
\newline
We have compiled the template results in Table.\ref{sumres1}, Table.\ref{sumres2} and Table.\ref{sumres3}. 
\begin{table}[ht]
\caption{Summary of template performances at $z=0.5$  for nDGP simulations where fiducial $f = 0.783$.}
\centering
\begin{tabular}{c c c c c c c}
\hline \hline 
Template & $k_{max} [h/\mbox{Mpc}] $ & bins & $V_s [\mbox{Gpc}/h]^3$ & $f\pm2\sigma$ &$\sigma_v \pm 2\sigma [\mbox{Mpc}/h]$ \\ 
\hline
GR & 0.110 & 18 & 10 &$ 0.774\pm^{0.034}_{0.022}$& $5.92\pm_{0.53}^{0.43}$ \\
DGP & 0.110 & 18 & 10 &$ 0.779\pm^{0.031}_{0.029}$&$ 5.86 \pm_{0.44}^{0.42} $\\
GR & 0.135 & 22 & 10 & $0.769\pm^{0.029}_{0.016}$& $ 5.86\pm_{0.16}^{0.22}$\\
DGP & 0.135 & 22 & 10 &$ 0.779\pm^{0.022}_{0.022}$ & $5.86\pm_{0.19}^{0.19}$  \\
GR & 0.147 & 24 & 10 &$ 0.777\pm^{0.025}_{0.025}$&  $5.96\pm_{0.15}^{0.14} $\\
DGP & 0.147 & 24 & 10 &$ 0.786\pm^{0.019}_{0.018}$& $5.89\pm_{0.13}^{0.12}$ \\
\hline
\end{tabular}
\label{sumres1}
\end{table}
\begin{table}[ht]
\caption{Summary of template performances for nDGP simulations at $z=1$ where fiducial $f = 0.909$.}
\centering
\begin{tabular}{c c c c c c c}
\hline \hline 
Template & $k_{max} [h/\mbox{Mpc}] $ & bins & $V_s [\mbox{Gpc}/h]^3$ &$f \pm2\sigma$ &  $\sigma_v \pm 2\sigma [\mbox{Mpc}/h]$ \\ 
\hline
GR & 0.147 & 24 & 10 &$ 0.895\pm^{0.028}_{0.025}$ & $4.08\pm_{0.26}^{0.22}$  \\
DGP & 0.147 & 24 & 10 & $0.902\pm^{0.025}_{0.025} $&$ 4.05\pm^{0.19}_{0.20}  $\\
GR & 0.171 & 28 & 10 & $0.902\pm^{0.021}_{0.020}$ &$ 4.19\pm_{0.09}^{0.10} $\\
DGP & 0.171 & 28 & 10 &$0.912\pm^{0.015}_{0.019}$ &$ 4.16\pm^{0.10}_{0.10}  $ \\
GR & 0.195 & 32 & 10 &$ 0.893\pm^{0.018}_{0.018}$&  $4.13\pm_{0.07}^{0.06} $\\
DGP & 0.195 & 32 & 10 & $0.908\pm^{0.014}_{0.014}$& $4.10\pm^{0.05}_{0.06} $\\
GR & 0.195 & 32 & 20 &$0.893\pm^{0.012}_{0.013} $& $ 4.13\pm_{0.05}^{0.04} $\\
DGP & 0.195 & 32 & 20 &$ 0.908\pm^{0.010}_{0.010}$&$ 4.10\pm^{0.04}_{0.04}$ \\
\hline
\end{tabular}
\label{sumres2}
\end{table}
\begin{table}[ht]
\caption{Summary of template performances for GR simulations at $z=1$ where fiducial $f = 0.859$.}
\centering
\begin{tabular}{c c c c c c c}
\hline \hline 
Template & $k_{max} [h/\mbox{Mpc}] $ & bins & $V_s [\mbox{Gpc}/h]^3$ & $f\pm2\sigma$ & $\sigma_v \pm 2\sigma [\mbox{Mpc}/h]$ \\ 
\hline
GR & 0.171 & 28 & 10 &$ 0.851\pm^{0.020}_{0.020} $& $3.94\pm_{0.14}^{0.13} $ \\
DGP & 0.171 & 28 & 10 & $0.869\pm^{0.019}_{0.010}$ &$ 4.01\pm^{0.09}_{0.11}$  \\
\hline
\end{tabular}
\label{sumres3}
\end{table}

\newpage

\section{An Ideal Survey: SPT Mock Data} 
\begin{figure}[H]
  \captionsetup[subfigure]{labelformat=empty}
  \centering
  \subfloat[]{\includegraphics[width=7.5cm, height=7cm]{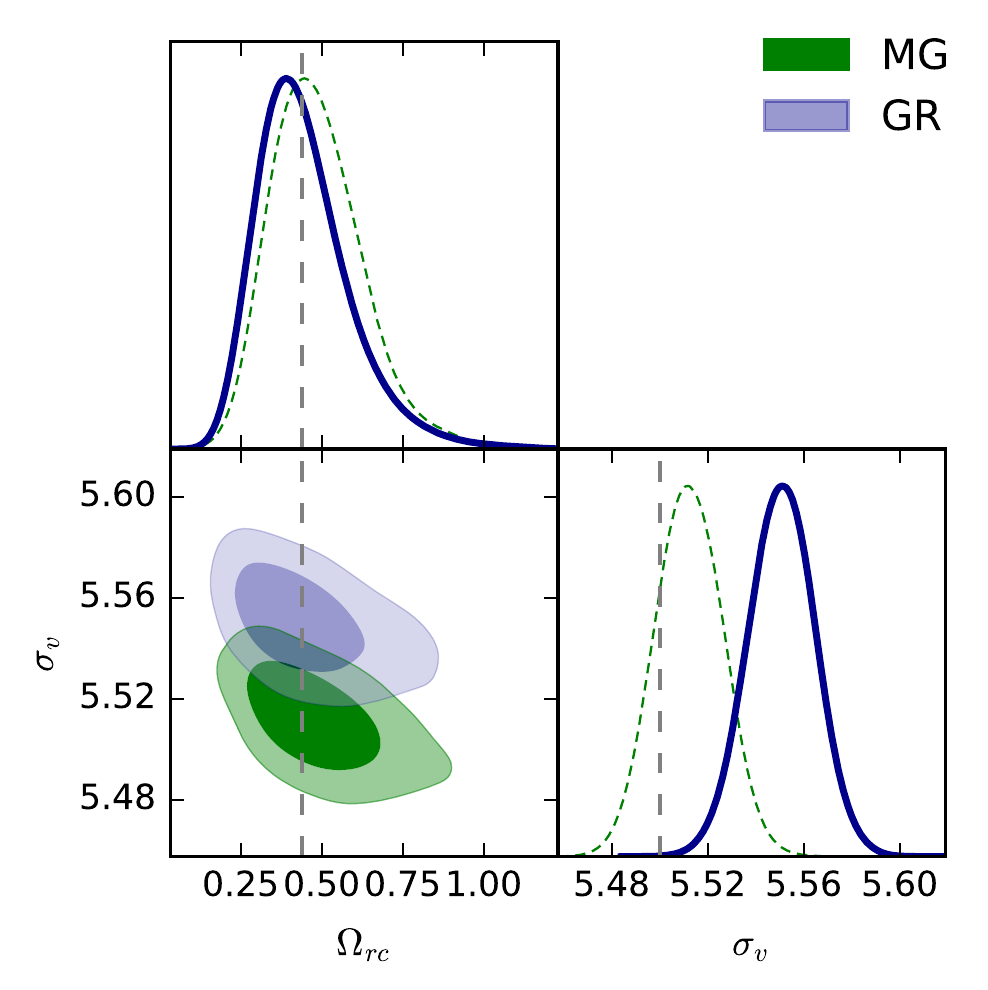}} \quad
  \subfloat[]{\includegraphics[width=7.5cm, height=7cm]{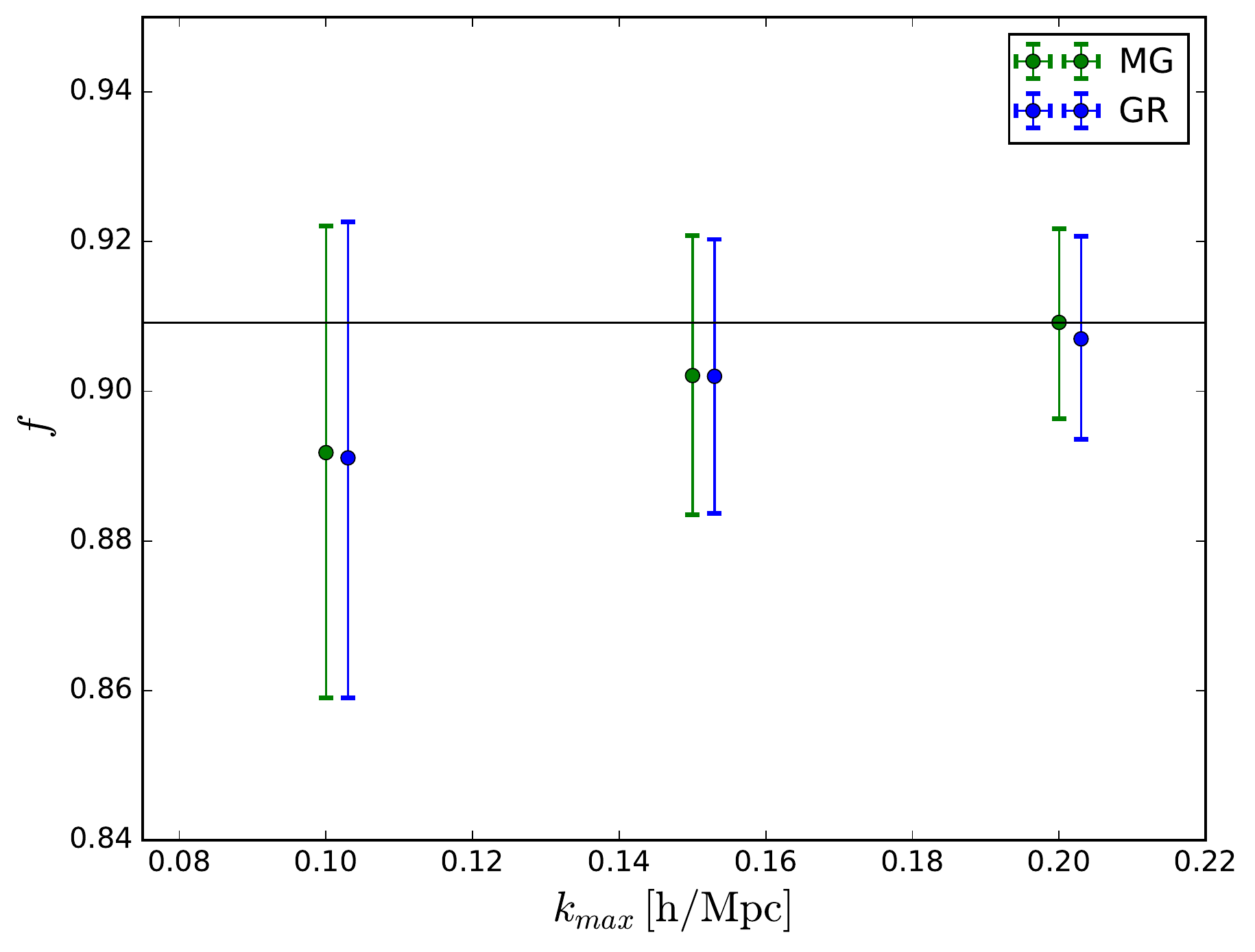}} 
  \caption{{\bf Left:}  The $1\sigma$ and $2\sigma$ confidence contours for the $\alpha$-DGP template and the GR template at $z=1$ fitting up to $k_{\rm max}=0.2h$/Mpc using 20 bins for mock data with $\alpha =1$. The mock's fiducial values for $\Omega_{rc}$ and $\sigma_v$ indicated by the dashed line. {\bf Right:} The best fit value for $f$ as a function of $k_{\rm max}$ with the $2\sigma$ errors for the $\alpha$-DGP and GR template. The GR template values (blue)  have been slightly shifted for better visualisation. A survey of volume of $10\mbox{Gpc}^{3}/h^3$ is assumed.}
\label{mocka1}
\end{figure}

\begin{figure}[H]
  \captionsetup[subfigure]{labelformat=empty}
  \centering
  \subfloat[]{\includegraphics[width=7.5cm, height=7cm]{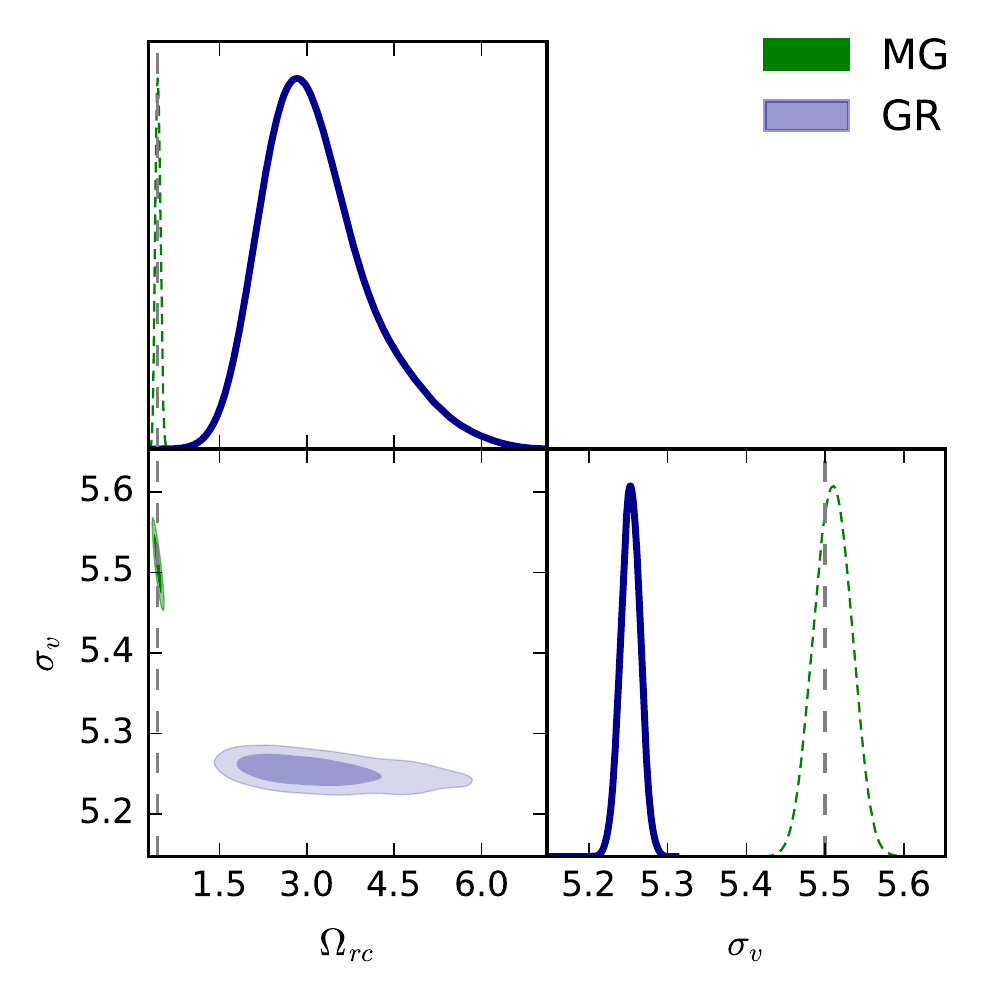}} \quad
  \subfloat[]{\includegraphics[width=7.5cm, height=7cm]{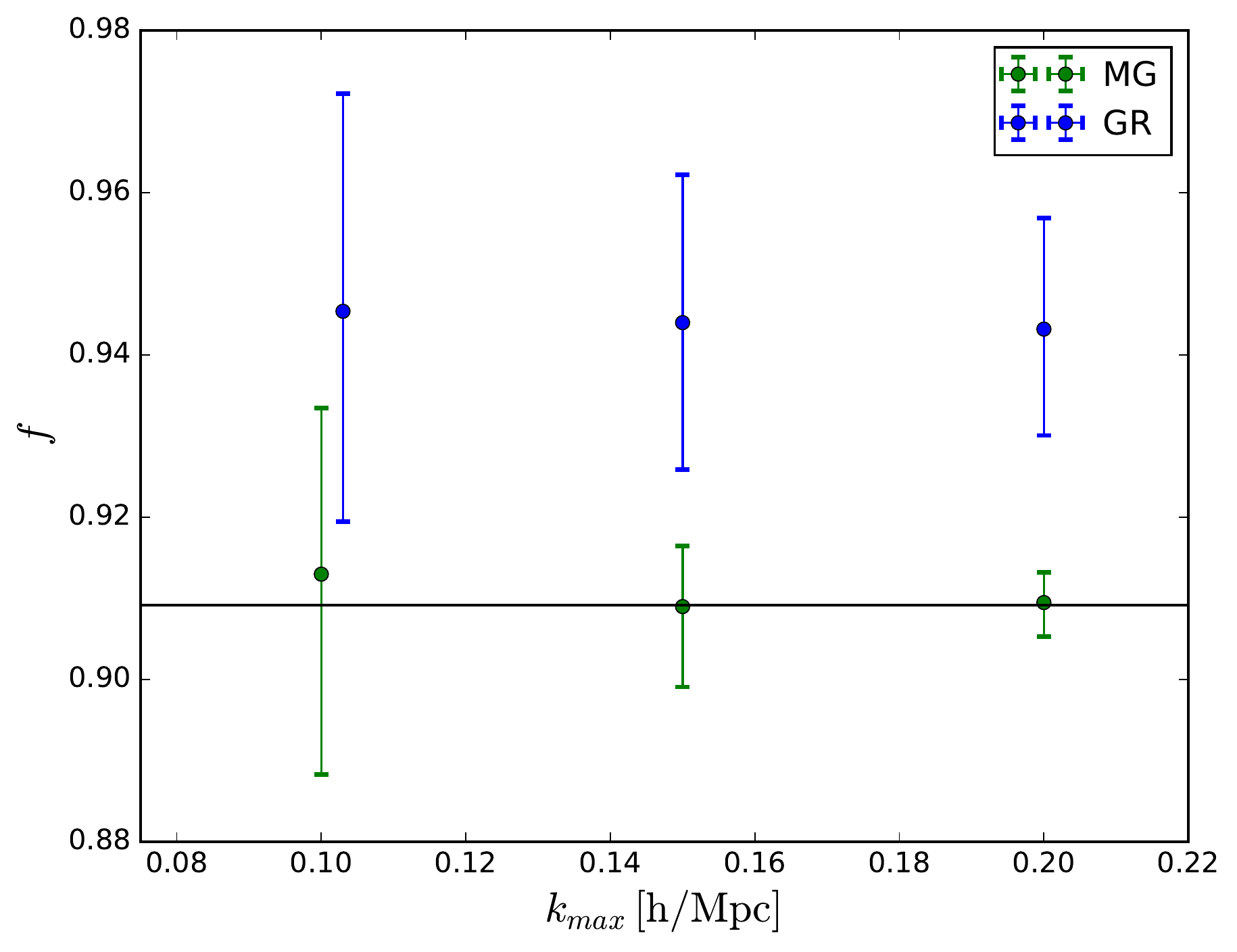}} 
  \caption{{\bf Left:}  The $1\sigma$ and $2\sigma$ confidence contours for the $\alpha$-DGP template and the GR template at $z=1$ fitting up to $k_{\rm max}=0.2h$/Mpc using 20 bins for mock data with $\alpha =15$. The mock's fiducial values for $\Omega_{rc}$ and $\sigma_v$ indicated by the dashed line. {\bf Right:} The best fit value for $f$ as a function of $k_{\rm max}$ with the $2\sigma$ errors for the $\alpha$-DGP and GR template. The GR template values (blue)  have been slightly shifted for better visualisation. A survey of volume of $10\mbox{Gpc}^{3}/h^3$ is assumed.}
\label{mocka15}
\end{figure}

\begin{figure}[H]
  \captionsetup[subfigure]{labelformat=empty}
  \centering
  \includegraphics[width=12.3cm, height=7cm]{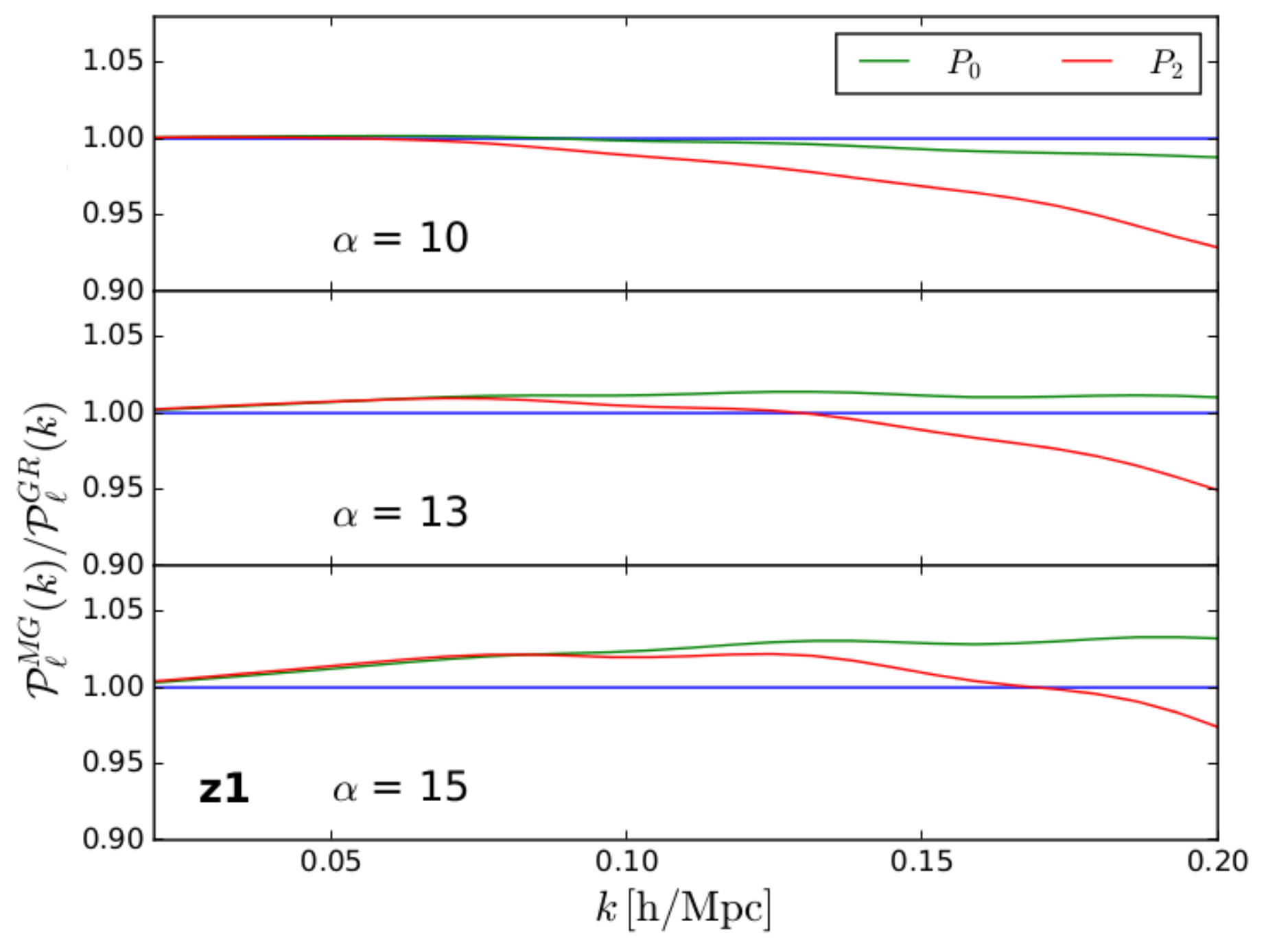}
  \caption{Ratio of the GR and  $\alpha$-DGP theoretical templates' monopole (green) and quadrupole (red) predictions for varying values of $\alpha$ at $z=1$. The blue line represents no deviation in template's predictions. }
\label{alpha}
\end{figure}

Being model dependent,  we can expect that both the scales affected as well as the magnitude of theoretical bias will in general depend on the specific phenomenology of a given gravity model. Therefore, if one wants to precisely estimate the importance of such theoretical bias for a given set of real galaxy spectroscopic data, one would need to run N-body simulations for each model under consideration and then perform a similar analysis as in the previous section. This is obviously not practical and in this section we provide a means of getting a first indication of whether or not model bias is an issue for a given model, in other words, whether or not the model can be safely encompassed by the GR template within the relevant range of scales. 
\newline
\newline
We proceed as follows. First, multipole data is produced for a given model of gravity using SPT up to some valid $k_{\rm max}$. Then the covariance matrix for this data is computed as was done for the N-body data using the parameters of an ideal survey. Finally, this data is given Gaussian errors using the covariance matrix. This provides an easily produced, idealistic, simulated mock data set which can be done for any model of gravity described by the framework discussed in the first section. A statistical analysis as done in the previous section can then be performed on this data. Here we do this for the nDGP model with the same fiducial model parameter previously used, $\Omega_{rc}=0.438$ but with $\sigma_8=0.87$. All other cosmological parameters are the same as the nDGP simulations. We choose a fiducial $\sigma_v=5.5 \mbox{Mpc}/h$ and use the ideal survey parameters $V_s=10\mbox{Gpc}^3/h^3$ and $\bar{n}= 4 \times 10^{-3} h^3 /\mbox{Mpc}^{3}$. Only $z=1$ is considered in this section and we extend our statistical analysis to $k_{\rm max}=0.2h$/Mpc. 
\newline
\newline
The question we ask here is, how much enhanced dynamics induced by MG is needed to incur a significant bias using the GR template. One could investigate this by creating mock data for larger values of $\Omega_{rc}$ say, although this becomes quite unrealistic. In fact the value of $\Omega_{rc}=0.438$ is already ruled out by BOSS LOWZ and CMASS data to within $2\sigma$ (see \cite{Barreira:2016mg}), with the authors placing an upper bound of 0.36. What we choose to do instead is rescale the non-linear mode mixing which governs the change of the scalar field's non-linear derivative interactions - the source of screening. The rescaling is done by introducing a parameter $\alpha$. We will scale $\gamma_2$ by $\alpha$ and $\gamma_3$ by $\alpha^2$ in the Euler equation's non-linear source term $S(\bfk)$ (Eq.\ref{eq:Perturb3}). By setting $\alpha =1$ we obtain the usual nDGP model but for values larger than unity the model changes to one with enhanced screening contributions. By tuning $\alpha$ we will be able to test the capabilities of the GR template to cover model non-linearities given an idealistic survey. Note this approach is not meant to represent a realistic or viable model but rather to be illustrative.
\newline
\newline
\newline
Fig.\ref{mocka1} show the results of the MCMC analysis for $\alpha=1$. The left hand plot shows the $1\sigma$ and $2\sigma$ contours for a matching done up to $k_{\rm max}=0.2h$/Mpc. We see that in this case both templates well recover the fiducial $\Omega_{rc}$. The right hand plot shows that the templates are comparable in their best fit value for $f$ as well as their $2\sigma$ constraints at all values of $k_{\rm max}$. 
\newline
\newline
Finally, Fig.\ref{mocka15} illustrates that if we set $\alpha=15$ the GR template completely fails to recover the fiducial value even at large (more linear) scales, with $\Omega_{rc}=0.438$ still lying outside the $2\sigma$ errors for $k_{\rm max}=0.1h$/Mpc (right hand side plot). This is an extreme case with the screening effects becoming important even at linear scales. Fig.\ref{alpha} shows the deviation between the templates for $\sigma_v$ and $\Omega_{rc}$ set to fiducial values, clearly indicating when the enhancement enters the linear regime and away from $\sigma_v$'s ability to suppress it. With $\alpha=13$ we note 1-2$\%$ deviations at scales as large as $k=0.05h$/Mpc. The deviation starts to become significant at larger scales, where $\sigma_v$ has minimal effect on the spectrum, at $\alpha=10$ (see Fig.\ref{nbodyplz1} for an indication of scales where $\sigma_v$ starts to have a significant impact on the spectrum). 
\newline
\newline
This concludes the chapter and all main results within this thesis. The next chapter summarises the work presented, discusses the results and their implications, and finally describes ongoing and future work. 


\chapter{Discussion and Conclusions}

\begin{chapquote}{J.D}
``So beautiful, so true, the cosmos had never lied to him. A diamond dotted crown, ordaining the world as the purifier and cultivator of thought, of emotion, conquerer of breath and death."
\end{chapquote}

Within the next decade the cosmos will be put under a microscope of unprecedented precision. Surveys covering many Gigaparsecs are about to begin and large portions of the cosmos will be explored, shedding light on effects and phenomena that previous surveys simply could not resolve. Much has changed since the times of Galileo where single phenomena could make or break theories. Cosmology has become a game of statistics, and by having such large volumes, the next generation of surveys will have minuscule statistical errors on the growth of structure and hence gravity. This leaves only observational and theoretical systematics to deal with. These are generally more relevant at smaller scales where the most information is held. This is why theoretically describing the small scales is essential in extracting the most information from data. But we must tread carefully. By having such great observational precision, one bestows more power to theoretical inaccuracies in producing biased pictures of gravity and cosmology. These biases are unacceptable if we are to ever obtain a deeper understanding of nature. 
\newline
\newline
Our goal was and is to prepare a theoretically-unbiased tool for making optimised comparisons to galaxy clustering data. The results contained in this thesis go a long way in this respect. {\tt MGCopter} is capable of the {\it consistent numerical construction} of relevant {\it quasi non-linear} LSS predictions in the perturbative framework. The consistency makes sure no modelling bias enters the prediction in constraints of gravitational theories and the numerical method ensures deviations in separability approximations do not introduce further errors. The quasi non-linear calculations enables greater extraction of information and by providing two approaches to clustering statistics, the power spectrum and correlation function, we also give a means of diminishing each's observable systematics. For selected models of gravity the pipeline was tested against measurements from fully non-linear simulations and analytic approximations, with great consistency. Parameter inference analyses can be performed against data sets using a bespoke MCMC algorithm within the code. This was used for simulation comparisons in order to understand the importance of having consistent pipelines in upcoming data analyses. 
\newline
\newline
These results are summarised and discussed below in order of presentation in the thesis. Aside from these results, many extensions and developments have been made and are being finalised for journal submission. We highlight these at the end of this chapter and present some of these preliminary results. 


\section{Summary of Results Presented}
\subsection{Numerical Results for Power Spectrum in MG}
We have introduced a code which can calculate the TNS power spectrum up to 1-loop order and its multipoles using a numerical method proposed in \cite{Taruya:2016jdt}. The calculation can easily be applied to a large class of modified gravity and dark energy models through the Poisson equation source terms. These include both Vainshtein and Chameleon screened models. An example of each of these, namely the DGP and Hu-Sawicki models, were considered.  
\newline
\newline
Our employment of the algorithm is shown to be consistent with the exact analytical expressions in the EdS cosmology. Similar sub percent matches for  the LCDM and nDGP models give information on the validity of the separability ansatz that is used in the analytic solutions. The validity of this ansatz is verified on the scales considered and at the redshifts considered. We have found that at smaller redshifts and smaller scales this ansatz begins to worsen, albeit still at the sub percent level. The biggest deviations are in the velocity velocity power spectrum, reaching about 0.5 (1.0)\% deviation at $k \sim 0.2h$/Mpc for the LCDM (nDGP) cosmology. For the EdS cosmology we find exact agreement in analytic and numerical results as expected. In \cite{Fasiello:2016qpn} the authors derive an exact treatment of the density and velocity time dependence in LCDM. Their results are consistent with Fig.\ref{convergence2}. We also successfully replicate results selected from \cite{Taruya:2013quf}, for the Hu-Sawicki $f(R)$ model where the authors use a suite of N-body simulations to compare to PT predictions. Table.\ref{work1t} gives a summary of these primary results.  
\begin{table}[ht]
\caption{Summary of Chapter 5 Results.}
\centering
\begin{tabular}{| p{13.2cm}|}
\hline \hline 
Deviation (\%) of analytic to numerical results at $k=0.2h$/Mpc at $z=0.4$  \\ 
\end{tabular}
\begin{tabular}{| p{3.cm} | p{3.cm} | p{3.cm}|  p{3.cm} |}
\hline \hline 
Model & \%$P_{\delta \delta}$& \%$P_{\theta \theta }$ & \%$P_{0}$   \\ 
\hline
EdS & $<0.1$ & $<0.1$ &  $<0.1$ \\
LCDM & $<0.3$ & $<0.7$ &  $<0.5$ \\
nDGP & $<0.5$ & $<1.0$ &  $<0.5$ \\
\hline
\end{tabular}
\begin{tabular}{| p{13.2cm}|}
\hline \hline 
Deviation (\%) of results to simulations at $k=0.12(0.15)h$/Mpc at $z=1$   \\ 
\end{tabular}
\begin{tabular}{| p{2.3cm} | p{2.3cm} | p{2.3cm}|  p{2.3cm} | p{2.3cm} |}
\hline \hline 
Model & \%$P_{\delta \delta}$& \%$P_{\theta \theta }$ & \%$P_{0}$ & \%$P_{2}$  \\ 
\hline
$f(R)$ & $<1(3)$ & $<1(3)$ &  $<1(3)$ & $<1(1)$ \\
\hline
\end{tabular}
\label{work1t}
\end{table}


\subsection{Comparison of Correlation Function Modelling}
The code was then extended to calculate the GSM non-linear redshift space correlation function as modelled in \cite{Reid:2011ar} for the same class of gravity and dark energy models. This was the focus of Chapter 6. We also extended the code to calculate the non-linear redshift space correlation function as described by the TNS model using the RegPT treatment as done in \cite{Taruya:2014faa}. To make comparisons between the two predictions the TNS power spectrum monopole and quadrupole were first compared to N-body data in order to obtain the best fit $\sigma_v$ (see Fig.\ref{pab1} and Table.II of \cite{Taruya:2014faa}). This required finding a realm of validity for the SPT predictions which was found by comparing the real space power spectra (see Appendix E). We then found fair agreement between these two treatments to within $4\%$ for GR and nDGP with $\Omega_{rc} = 0.438$ and up to a $8\%$ deviation in the treatments for the chameleon screened $f(R)$ model with $|f_{R0}|=10^{-4}$ around the BAO scale (Fig.\ref{xirs1}, Fig.\ref{xirs2} and Fig.\ref{xirs3}). We have also compared the LPT correlation function \cite{Matsubara:2008wx} in real space with that obtained using a FT of the RegPT 1-loop spectrum (Fig.\ref{xir1}). The RegPT treatment gives up to $4\%$ more damping around the BAO scale. Recently a LPT prediction for MG models has been developed \cite{Aviles:2017aor} allowing the extension of such comparisons. 
\newline
\newline
We observe large damping in the GSM and FT of TNS treatments over the linear predictions with more damping observed in the $f(R)$ and nDGP cases. This is due to enhanced gravity encoded in the additional non-linearities for these theories. The difference between the GSM and FT of TNS predictions predominantly comes from their treatment of the RSD. While the GSM is completely perturbative in making the non-linear mapping to redshift space within configuration space in a probabilistic manner, the TNS makes the mapping through the redshift-real space distance relation (see Eq.\ref{redtoreal}) and conservation of mass, and is also partly phenomenological. Although it benefits from an added degree of freedom ($\sigma_v$), the FT of TNS opens the question of how to correctly treat small scale SPT divergences and resumation is still very open (examples of such treatments include RegPT \cite{Taruya:2012ut}, renormalised perturbation theory \cite{Crocce:2005xy,Crocce:2007dt} and effective field theory prescriptions \cite{Senatore:2014via,Vlah:2015sea}). This issue has yet to be investigated thoroughly. In light of this, one cannot say with certainty which approach to the redshift space correlation function will perform better when matching simulation or observational data. This will be the focus of a future work.  
\newline
\newline
To give the GSM model extra freedom, we promote the isotropic velocity dispersion contribution to the GSM's pairwise dispersion $\sigma_{12}^2$  as a free parameter $\sigma_{\rm iso}$, which is physically equivalent to TNS's $\sigma_v$ parameter. By doing this we can enhance the PT prediction, given in eq.(\ref{sigmaiso}), and better match the TNS on small scales. We find that the PT prediction for $\sigma_{\rm iso} = 3.9$ Mpc/$h$ does well for the nDGP model and we are able to match the FT of TNS prediction at scales $s \leq 100$Mpc$/h$ to within $2\%$ (Fig.\ref{xirs2}).  For $f(R)$ and GR we find the PT prediction underestimates the small scale velocity dispersion, and we find the larger values of $\sigma_{\rm iso}^{\rm GR} = 5 (3.16)$ Mpc/$h$ and $\sigma_{\rm iso}^{f(R)} = 7.5(5.2)$ Mpc/$h$ (PT prediction in brackets) better match the FT of TNS at smaller scales, specifically in the quadrupole prediction (Fig.\ref{xirs1} and Fig.\ref{xirs3}). Around the scales $100 $ Mpc/$h \leq s \leq 180 $ Mpc/$h$ $\sigma_{\rm iso}$ has a marginal effect. The preferred values of $\sigma_{\rm iso}$  in the modified gravity theories both differ by around $30 \%$ when compared with the best fit values of $\sigma_v$ of the TNS model. The GR value of $\sigma_{\rm iso}$ is within $\sim 5\%$ of its TNS equivalent. In summary, we find that both approaches model the RSD consistently in the range $50\mbox{Mpc}/h \leq s \leq 180 \mbox{Mpc}/h$ with  the GSM requiring the promotion of $\sigma_{12}^2$ to a free parameter to be consistent with the TNS approach, particularly for the quadrupole. 
\newline
\newline
Using the best fit values for $\sigma_{\rm iso}$ we find that the differences between GR and MG-GSM predictions for the correlation function multipoles accurately follow those using the FT of TNS indicating that both approaches to modelling the RSD consistently treat modifications to gravity, with neither giving an enhanced MG signal over the other (Fig.\ref{frvgr} and Fig.\ref{dgpvgr}). The non-linear differences follow the LSM differences in all cases with the LSM generally picking up larger deviations from GR consistently in both multipoles. The survey comparisons done in \cite{Reid:2012sw} imply the GSM treatment over-damps the BAO wiggle in redshift space. This suggests a preference of the RegPT treatment to the real and redshift space correlation function although marginally. Again, we wait for the availability of simulation data to make this conclusion. In any case, the ability to compute the redshift space correlation function for generalised models should prove to be very useful when performing statistical analyses on survey data and obtaining gravitational parameter constraints. 
\newline
\newline
Lastly, by moving to smaller scales and using a fuller shape of the correlation function we expect any deviations from GR to become less able to hide in nuisance degrees of freedom such as $\sigma_v$, $\sigma_{\rm iso}$ or tracer bias. Fig.\ref{frvgr} and Fig.\ref{dgpvgr} show the difference between the MG and GR predictions for the correlation function. We see that at the BAO scale down to the scales valid for the GSM treatment, we have a significant MG signal.  By pushing into these scales we enter regions as yet unused for constraining models beyond GR \cite{Baker:2014zba}. Table.\ref{work2t} gives a summary of some of these results.  
\begin{table}[ht]
\caption{Summary of Chapter 6 Results.}
\centering
\begin{tabular}{| p{12.7cm}|}
\hline \hline 
Max deviations of various results at $z=0.5$ between $50<r,s<160$Mpc/$h$ \\ 
\end{tabular}
\begin{tabular}{| p{2.2cm} | p{2.2cm} | p{2.2cm}|  p{2.2cm} |  p{2.2cm} |}
\hline \hline 
Model & $\xi_r/\xi_L$ & $\xi_0^{TNS}/\xi_0^{GSM}$ &$\xi_2^{TNS}/\xi_2^{GSM}$ & $\sigma_v/\sigma_{iso}^{\rm best}$   \\ 
\hline
LCDM & 0.7 & 1.03 &  1.02 & 0.95\\
nDGP & 0.75 & 1.04 &  1.02 & 0.93 \\
$f(R)$ & 0.73 & 1.04 &  1.02 & 0.92 \\
\hline
\end{tabular}
\label{work2t}
\end{table}


\subsection{Relevance of MG-Modelling for Stage IV surveys} 
Chapter 7 was motivated by the question of whether constraints of the growth rate derived with a GR PT-template give an unbiased measurement of growth for the case of a universe described by a modified gravity. This question is very relevant in the context of the next stage of cosmological surveys. We provided a first level analysis of the theoretical model bias systematic and have presented a method for quick assessment of the model specific significance of this systematic.
\newline
\newline
Firstly, we compared nDGP MG-PICOLA simulation data with the GR and nDGP theoretical predictions for the redshift space power spectrum. We used the TNS model of RSD which has been validated against both GR and modified gravity simulations \cite{Taruya:2010mx,Taruya:2013quf}. This was done at the level of dark matter clustering and only the first two multipoles were considered. Idealised future survey parameters were adopted in the analysis. We found out that the small-scale velocity damping term $\sigma_v$ included in the TNS model provides a flexibility through which the template can, to some extent, accommodate the enhanced small-scale clustering of the nDGP model. This was clearly indicated by higher values of $\sigma_v$ attained by the GR-template fit.
\newline
\newline
Both templates perform well in recovering the simulation's fiducial parameter at low redshift. We point out that the real space analysis done in \cite{Barreira:2016mg} concluded that no nDGP model bias is evident at redshifts up to $z=0.57$ which is consistent with our results at $z=0.5$. That being said, a full comparison of our results is difficult as there are a number of differences between their analysis and the one done here. In particular the use of different RSD models, the modelling of survey errors and their inclusion of galaxy bias which provides more fitting freedom to the GR template. We do find that at high redshift the GR prediction becomes increasingly biased and the difference between the two templates is greater. Using $V_s=20\mbox{Gpc}^3/h^3$, which will be realisable with stage IV surveys, we find systematically biased estimates of the GR template, with it failing to recover the fiducial parameter to within $2\sigma$ at $z=1$. This apparent bias might be due to specific limitations intrinsic to the SPT approach. The current analysis is left suggestive with robustness sought in additional theoretical modelling (for example including galaxy bias) or fuller treatment of non-linear scales such as using the EFT approach \cite{Baumann:2010tm,Carrasco:2012cv}.  Nonetheless, what is clear is that for the $V_s=20 \mbox{Gpc}^3/h^3$ case the $1\sigma$ regions of both templates do not intersect which implies that the template's predictions are inconsistent at that level. Again, the inclusion of galaxy bias may relieve this. 
\newline
\newline
In our second analysis we created mock data from SPT predictions by adding Gaussian noise generated using the errors derived from an idealised survey. Two data sets were created at $z=1$ using varying levels of model dependent non-linearity. This was done to simulate modified gravity models which have an enhanced non-linear source term. We find that by increasing the magnitude of the screening terms (see Eq.\ref{eq:Perturb3}) the GR template does progressively worse in recovering the fiducial $\Omega_{rc}$, with the model bias being unimportant up to a non-linear contribution of around 10 times the base value. Above this the GR predictions become very biased and at 15 times the base non-linearity the GR template fails to recover the fiducial value even at scales of $k=0.1h$/Mpc. This exercise provides an indication on what scales and at what level of enhanced small-scale clustering a modified gravity model has to be consistently treated in RSD modelling in order to avoid significant theoretical biases that otherwise would diminish the desired accuracy of growth rate estimates. On this note, the creation of mock data can be done for any model of gravity within the framework discussed in \cite{Bose:2016qun} giving an avenue for assessing the importance of theoretical model bias in growth rate estimation from a given data set. The data quality of stage IV surveys indicates that this test will be important and is essential if we wish to put trusted constraints on modified gravity parameter space. Summaries of the main results are found in Table.\ref{sumres1}, Table.\ref{sumres2} and Table.\ref{sumres3}. 


\section{The Path Underfoot and Ahead} 
The work described here has gone a way in weeding out deadly inaccuracies in our path to a more fundamental picture of nature. There are of course still many more pitfalls and obstacles to overcome before we can extract anything meaningful from the cosmos. One main avenue of my current research has been in extending the perturbative framework discussed  in preparation for real data analyses.
\subsection{Tracer Bias} 
\begin{figure}[H]
  \captionsetup[subfigure]{labelformat=empty}
  \centering
  \subfloat[]{\includegraphics[width=12cm, height=6cm]{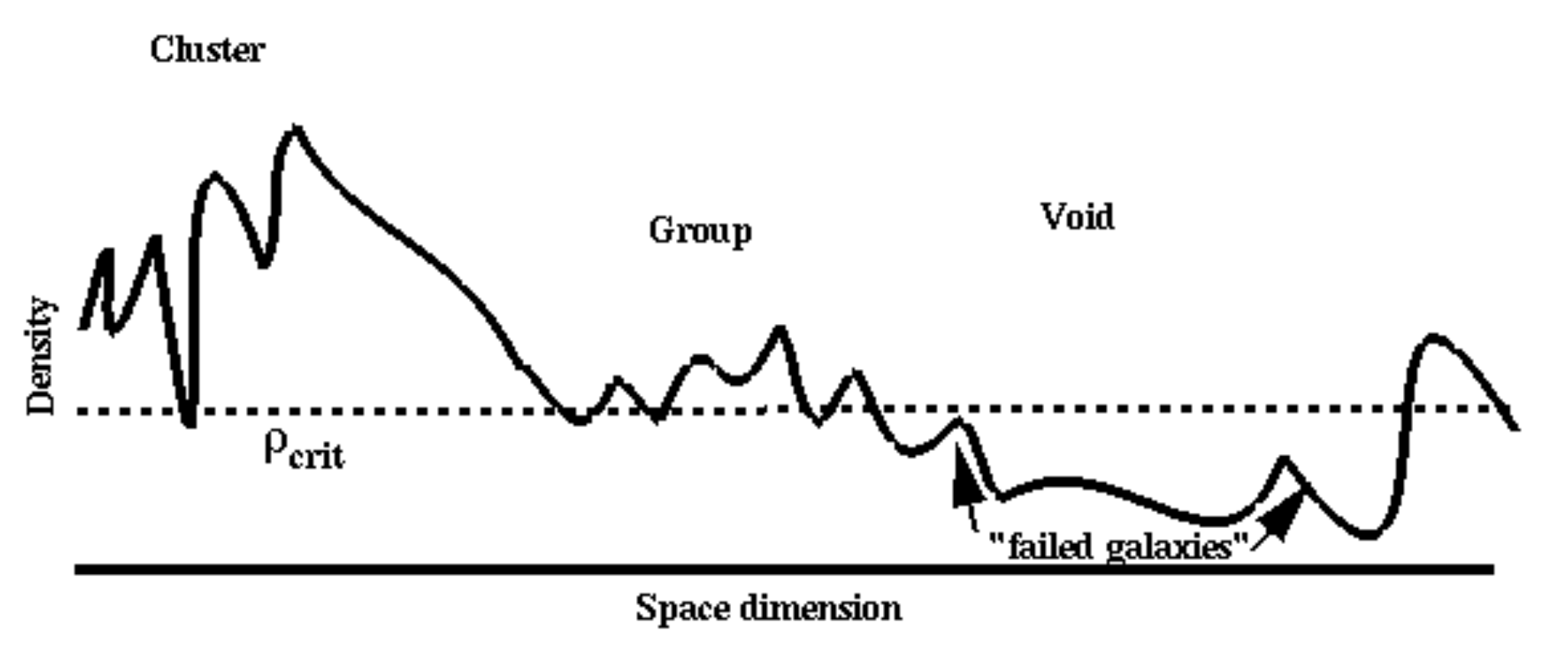}} \quad
  \subfloat[]{\includegraphics[width=7.5cm, height=7cm]{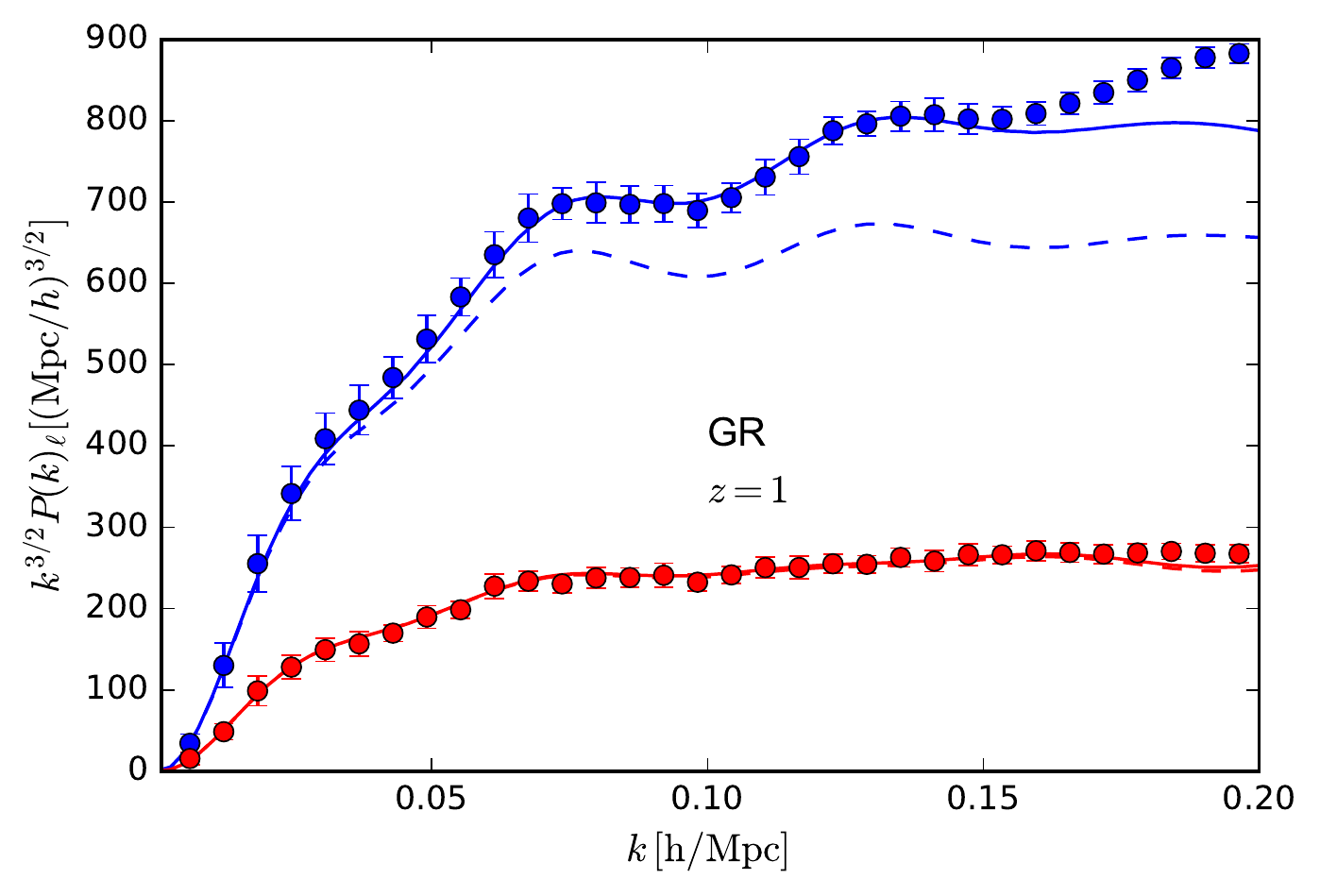}} \quad
    \subfloat[]{\includegraphics[width=7.5cm, height=7cm]{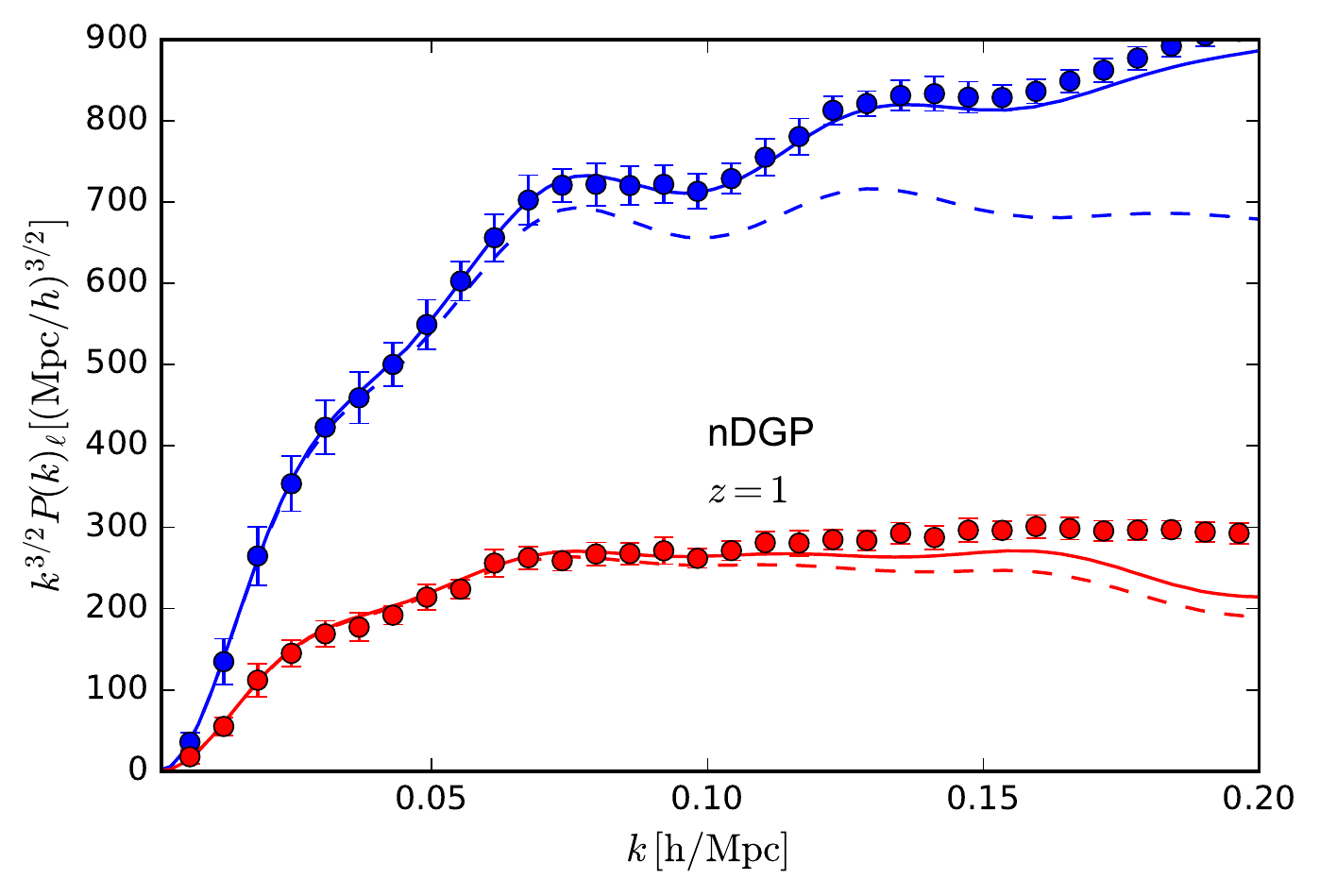}} \quad
  \caption{{\bf Top:} A picture illustrating how galaxies do not extract the full shape of the underlying density field \cite{Ostriker:1980rj}. {\bf Bottom, left:} GR-TNS monopole (blue) and quadrupole (red) with best fit bias parameters of the bias expansion (solid) and linear (dashed) bias models against PICOLA halo spectra. {\bf Bottom, right:} DGP-TNS monopole (blue) and quadrupole (red) with best fit bias parameters of the Q-bias (solid) and linear (dashed) bias models against PICOLA halo spectra. The spectra measurement from PICOLA use the number density selection $\bar{n}_h=1\times10^{-3}h^3/\mbox{Mpc}^3$.}
\label{biaspic} 
\end{figure}

One obvious extension is the issue of galaxy bias. The galaxies we observe are not perfect tracers of the underlying dark matter density field, a picture made clear on the top of Fig.\ref{biaspic}. Galaxies require a certain critical over-density to collapse, but in regions where this criteria is not met we observe none and so have no information. This implies some mapping $\delta_m \rightarrow \delta_g$, the simplest of which is a simple scaling $\delta_g = b \delta_m$ where $b$ is a constant, which works fairly well at large scales. More sophisticated treatments are required to accurately describe the mapping at smaller scales. One such phenomenological model common in past survey analyses comparisons is the Q-bias model \cite{Cole:2005sx,Song:2015oza}
\begin{equation}
b(k) = b_1\sqrt{\frac{1+Ak^2}{1+Bk}},
\end{equation}
where A,B and $b_1$ are free parameters. MCMC analyses of various bias models are currently being performed on MG halo catalogues from PICOLA simulations. In particular, I have been working on generalising the commonly used non-linear bias expansion model \cite{McDonald:2009dh} to MG theories and identifying its realm of validity. At quasi non-linear scales this model also has 3 free parameters associated with leading terms in the bias expansion. The perturbative template multipoles against data for a population of halos in a suite of 50$\times 1\mbox{Gpc}^3/h^3$ DGP-PICOLA simulations is shown at the bottom of Fig.\ref{biaspic}. From this we see that an accurate description of galaxy bias is essential in extracting correct information from LSS surveys, especially at small scales. 
\subsection{Modifying the Dark Sector} 
\begin{figure}[H]
  \captionsetup[subfigure]{labelformat=empty}
  \centering
  \subfloat[]{\includegraphics[width=7.5cm, height=6.6cm]{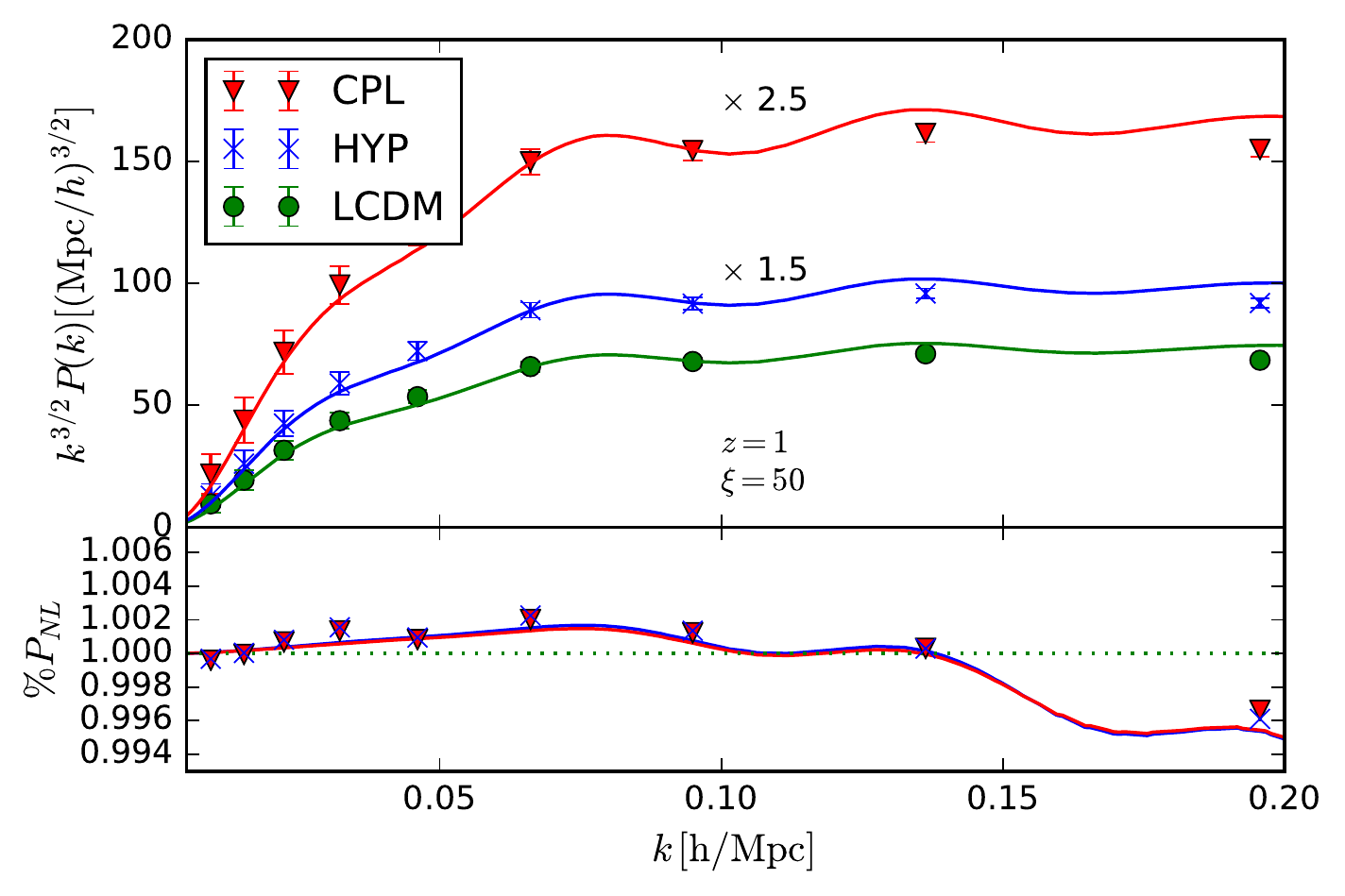}} \quad
  \subfloat[]{\includegraphics[width=7.5cm, height=6.6cm]{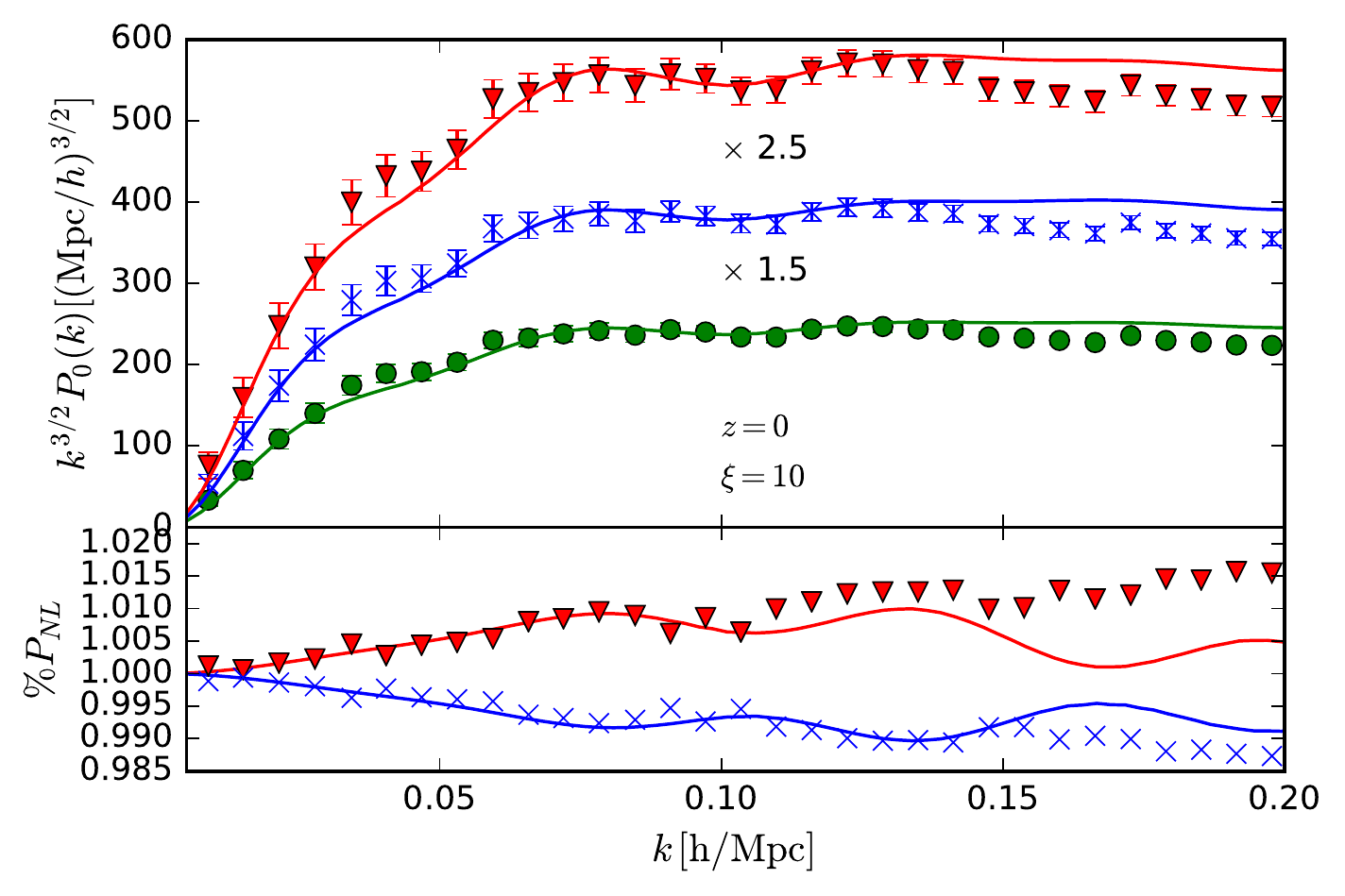}} \quad
  \caption{{\bf Left:} SPT predictions (solid) and N-body measurements (points) of the matter power spectrum in real space at $z=1$ for the HYP (blue) and CPL (red) variable $\omega$ models. The interaction strength for these models is set to $\xi =50$. {\bf Right:} The TNS monopole for $\omega =-1.1$ (blue), $\omega=-0.9$ (red) and LCDM (green) against N-body measurements. We have scaled the interacting models for better visualisation. The interaction strength for these models is set to $\xi =10$. The top panels show the power spectra scaled by $k^{3/2}$ and the bottom panels show the ratio of  $P(k)/P_L(k)$ to $P^{LCDM}(k)/P^{LCDM}_L(k)$.}
\label{interactingde} 
\end{figure}

Tentative tensions between the Planck CMB data \cite{Planck:2015xua} and late Universe data using various probes such as clusters \cite{Vikhlinin:2008ym,deHaan:2016qvy}, gravitational lensing \cite{Heymans:2013fya,Abbott:2017wau}, and RSD \cite{Blake:2011rj,Reid:2012sw,Macaulay:2013swa,Beutler:2013yhm,Gil-Marin:2015sqa,Simpson:2015yfa}, have been uncovered. These tensions consistently suggest an overestimation of structure growth by the CMB evolved model when compared to the low-$z$ measurements. This consistency seems to support a physical effect rather than systematic, although unknown systematics, such as determination of mass bias in clusters and modelling of non-linear effects in weak lensing, may still be the cause. Fifth forces from MG models considered in this work promote more structure growth. From the other side, one avenue that has been very promising in relieving these tensions is to allow for an interaction within the dark sector \cite{Simpson:2010vh,Lesgourgues:2015wza,Pourtsidou:2016ico,Baldi:2016zom,Buen-Abad:2017gxg} while keeping the theory of gravity that of general relativity. Extensions to the generality of the code have been made to include such interactions \cite{Baldi:2016zom,Pourtsidou:2013nha}. I have also extended to $\omega \neq -1$ (see Eq.\ref{rhoeqn}) as well as time evolving equations of state (see Eq.\ref{vareos}) for dark energy, specifically the Chevalier-Polarski-Linder form \cite{Chevallier:2000qy} and a hyperbolic tangent form
\begin{align}
\omega_{\rm CPL} &= \omega_0 + (1-a)\omega_a, \label{wcpl} \\  
\omega_{\rm HYP} &= \omega_0 +\frac{\omega_a}{2}[1+\tanh(\frac{1}{a}-1-a_t)]. \label{whyp}
\end{align}
Fig.\ref{interactingde} shows 1-loop dark matter power spectrum predictions of the interacting model described in \cite{Baldi:2016zom} for the Eq.\ref{wcpl} and Eq.\ref{whyp} evolving equation of states in real space (left) and the redshift space monopole (right) for fixed $\omega$ models against N-body data. The interaction between dark energy and dark matter is governed by a coupling parameter $\xi$ \cite{Simpson:2010vh,Baldi:2014ica,Baldi:2016zom}. Future LSS data sets have the possibility of uncovering clear signatures of such interactions making perturbative templates that accurately describe the interaction vital in parameter constraints. This work has recently been made public in \cite{Bose:2017jjx}.
\subsection{Extending Framework to Smaller Scales}
\begin{figure}[H]
  \captionsetup[subfigure]{labelformat=empty}
  \centering
  \subfloat[]{\includegraphics[width=7.5cm, height=7cm]{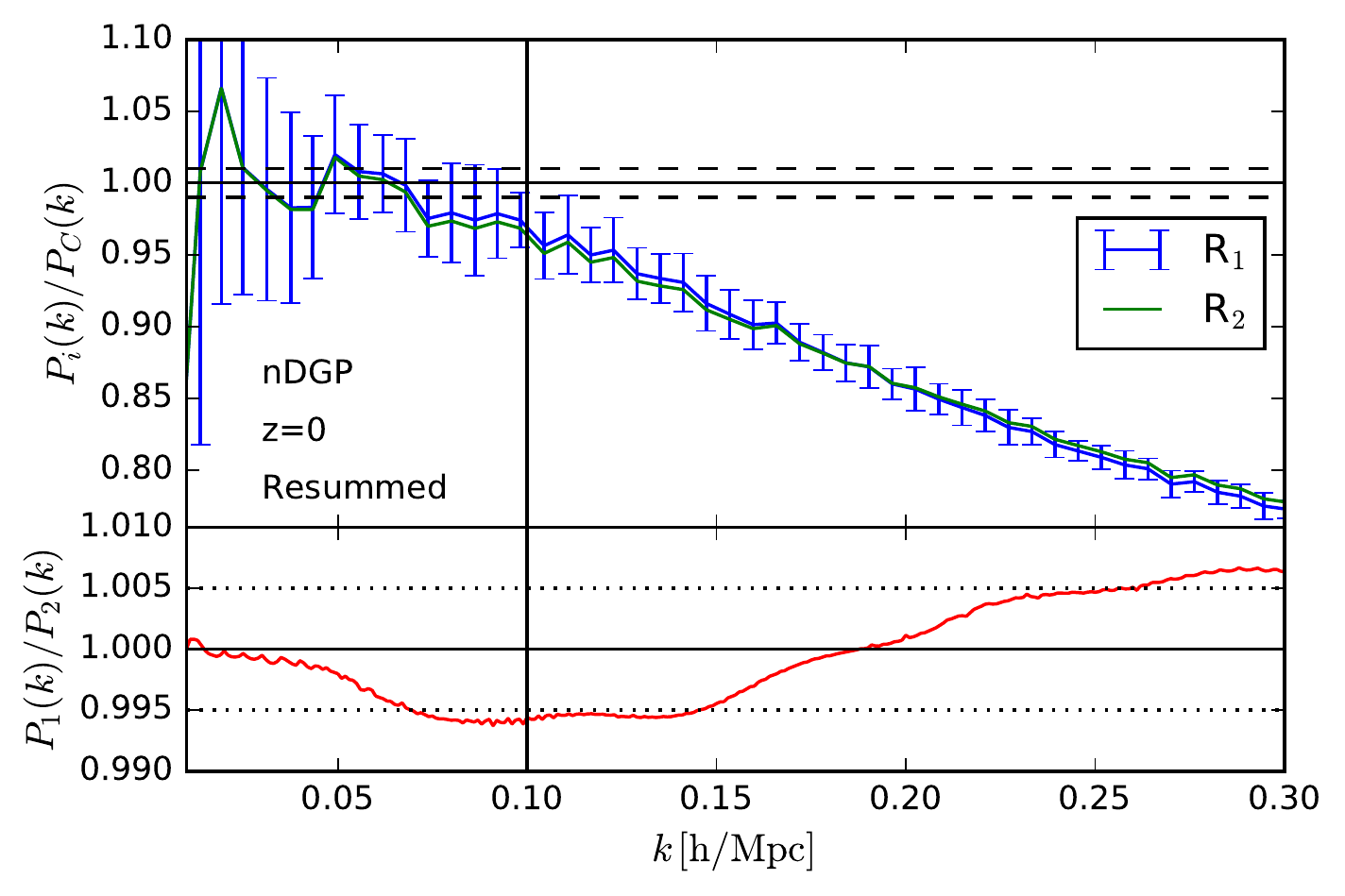}} \quad
   \subfloat[]{\includegraphics[width=7.5cm, height=7cm]{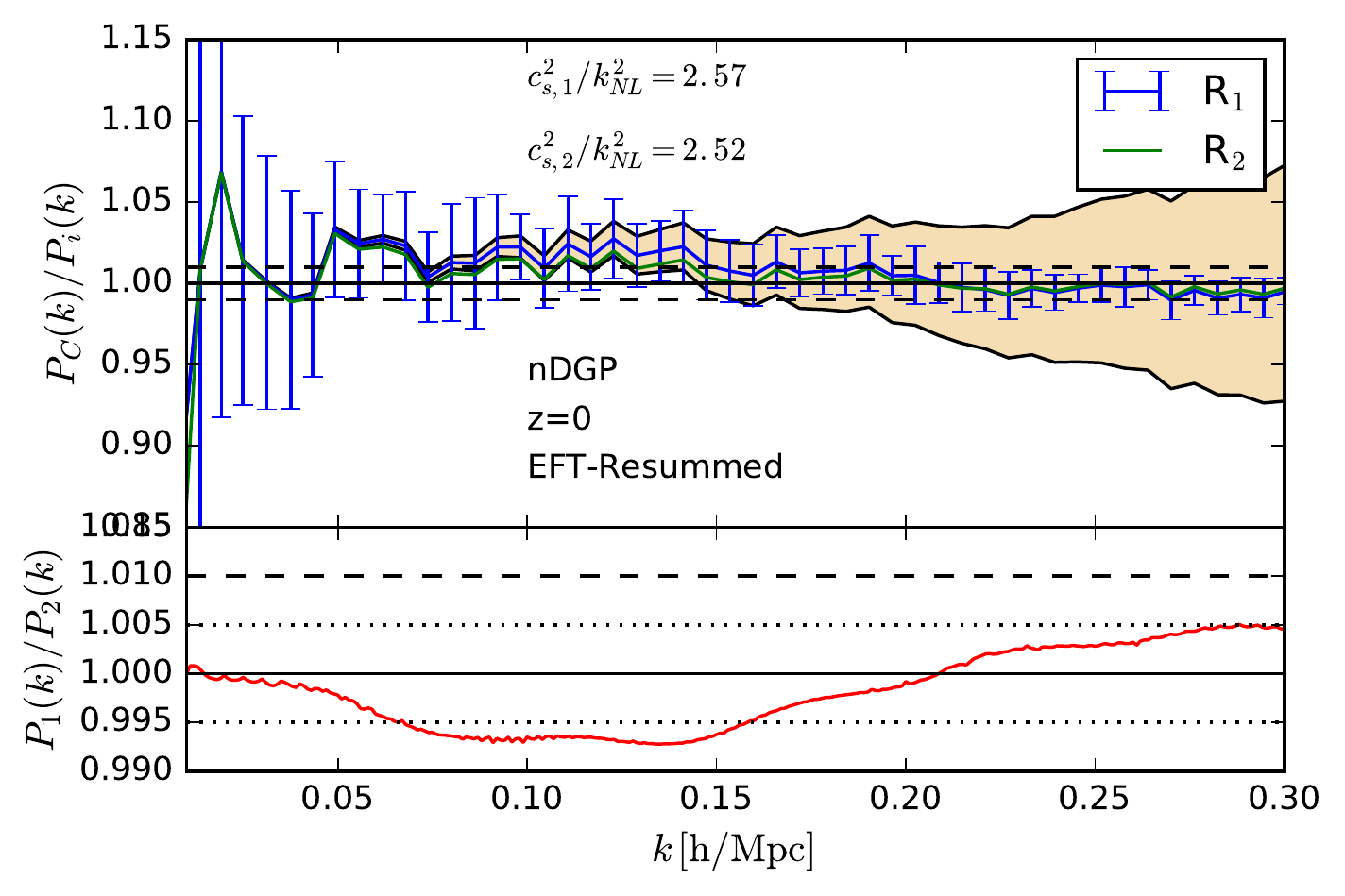}} \quad
  \caption{The 1-loop PT (left) EFToLSS (right) matter power spectrum within DGP using two different resummation methods against simulation data at $z=0$. The top panel shows the ratio of the  power spectra predications and PICOLA data. The yellow band on the right shows the theoretical error associated with the 2-loop computation. The bottom panels shows the ratio of the two predictions.}
\label{eftolss} 
\end{figure}

I have also recently been investigating extensions of the PT framework, namely using the effective field theory of LSS (EFToLSS) \cite{Baumann:2010tm,Carrasco:2012cv,Senatore:2014via,Vlah:2015sea}. This framework offers a great way of controlling small scale contaminations in the power spectrum loop expansion as well as extending the range of validity of the perturbative approach to the power spectrum. This is done by employing a parameter $c_s^2$ that quantifies uncertainty of small scale physics. This enters the 1-loop density spectrum as follows 
\begin{equation}
P^{{\rm EFT}}_{\delta \delta}(k,a) = P^{1-{\rm loop}}_{\delta \delta}(k,a) - 2 (2\pi) c_s^2 \frac{k^2}{k_{\rm NL}^2}P_{11}(k,a),
\end{equation}
where $k_{\rm NL}$ indicates a scale at which the EFToLSS breaks down which in practice can be absorbed into $c_s^2$.
\newline
\newline
 Currently, using the EFToLSS, I am looking at the effect of resummation techniques, theoretical approximations and errors incurred in a truncation of the loop expansion on the prediction for the redshift space power spectrum and quantifying these theoretical errors in the context of a percent level precision growth measurement. The right of Fig.\ref{eftolss} shows a comparison of resummation methods within 1-loop PT and EFToLSS against DGP-PICOLA data. We can see that without accounting for theoretical errors incurred in a truncation of the loop expansion, the 1-loop PT prediction exits the $1\%$ accuracy regime at around $k=0.1h$/Mpc while the EFToLSS fails at around $k=0.15h$/Mpc, where the theoretical errors indicated by the beige band become larger than the errors on the data. Comparing the EFToLSS approach with the PT-TNS approach, in relation to growth constraints using MG-PICOLA data, is also an ongoing work.  
\noindent As mentioned, the EFToLSS approach to the power spectrum offers a great way of probing small scales. Relatedly, work has also been done on constructing simulation measured source terms in the perturbative framework which would be cosmology independent \cite{Manzotti:2014loa}. The effect of screening could then imply an MG independence. This would offer a potential means of putting priors on the free parameters of EFToLSS, resulting in improved constraints on growth. In configuration space, by including non-Gaussianities in the GSM one can probe scales below $30\mbox{Mpc}/h$ \cite{Bianchi:2016qen}. To further optimise constraints, development of these extensions to the GSM in general theories of gravity is something that would be worth investigating. Combined with the EFToLSS power spectrum,  the widest range of scales currently offered by PT in both Fourier and configuration space can be used.

\subsection{Higher Order Statistics}
\begin{figure}[H]
  \captionsetup[subfigure]{labelformat=empty}
  \centering
    \subfloat[]{\includegraphics[width=7.5cm, height=7cm]{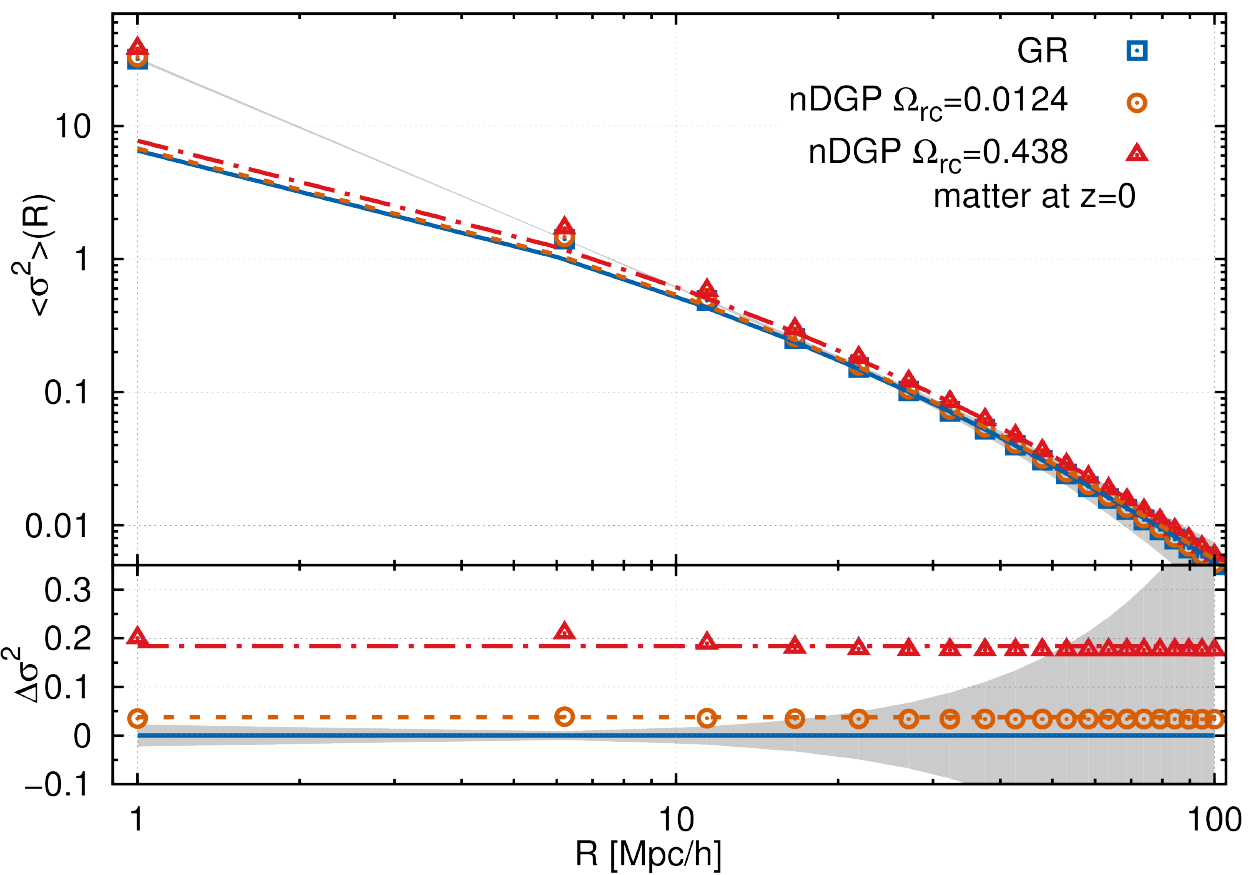}} \quad
  \subfloat[]{\includegraphics[width=7.5cm, height=7cm]{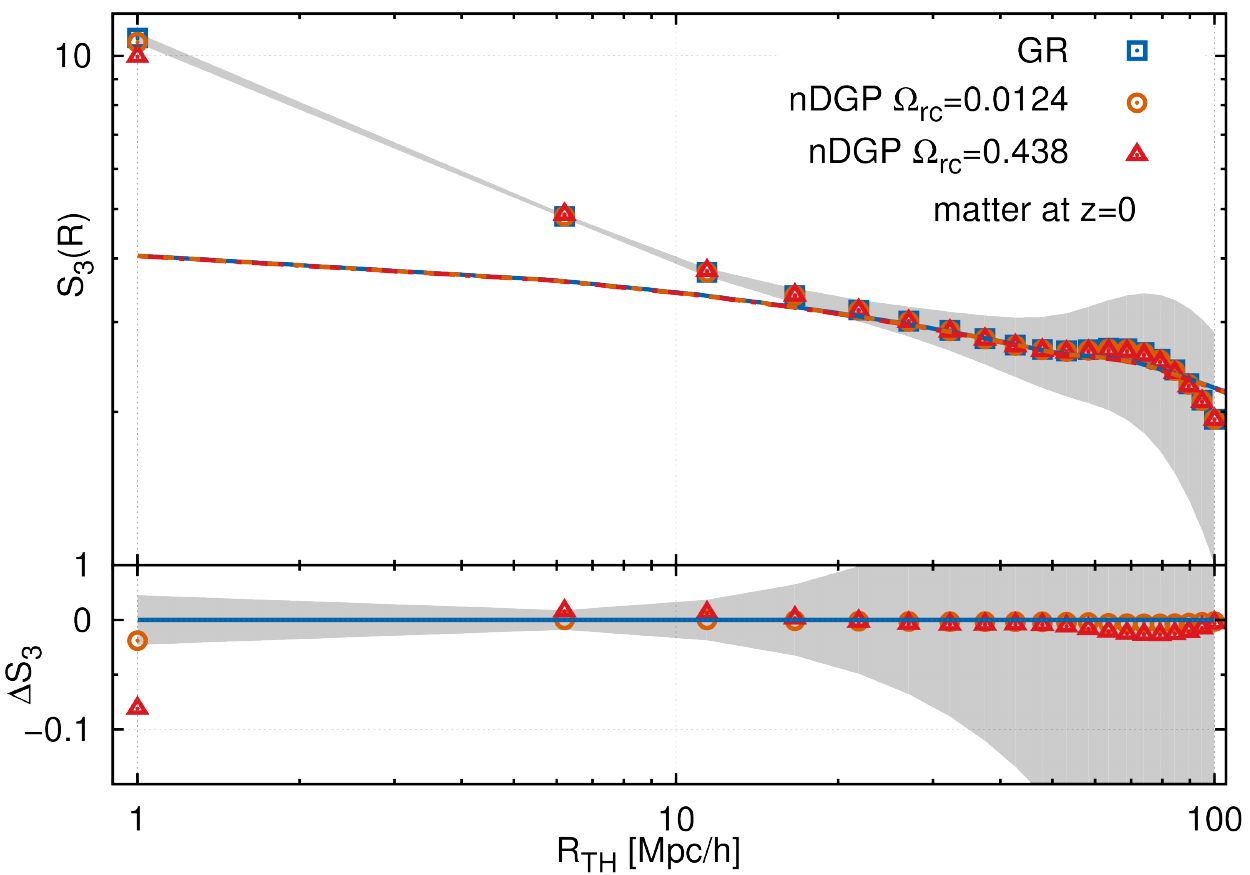}} \quad
   \subfloat[]{\includegraphics[width=15.5cm, height=7cm]{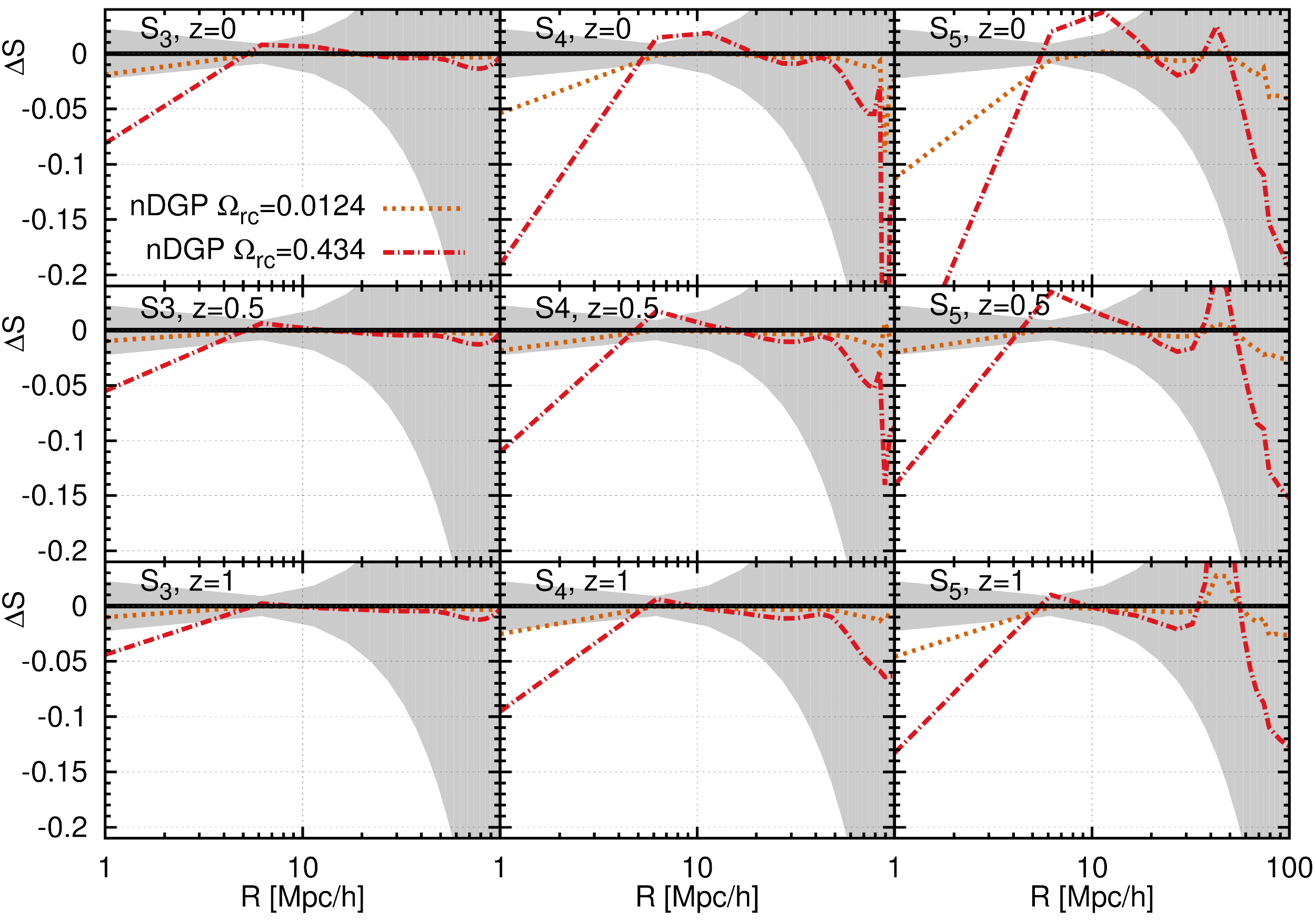}} \quad
  \caption{{\bf Top Left and Right:} Comparison of matter density variance and skewness respectively for $z=0$ estimated from N-body simulations (points) with the PT predictions as a function of smoothing scale. The lower panel shows the fractional difference of both nDGP models taken with respect to the GR case. {\bf Bottom:} Fractional differences from GR taken across three epochs $z=0, 0.5, 1$ (rows of panels from top to bottom) of the first three matter density reduced cumulants. The shaded region illustrate the cosmic variance error for the GR fiducial case.}
\label{hclus} 
\end{figure}

Finally, by having such large volumes these surveys will be able to beat down the errors in the bispectrum measurement allowing it to be used to unprecedented capacity in parameter inference. If used in conjunction with the power spectrum it can help break parameter degeneracies and studies indicate that parameter constraints can improve by 2-5 times \cite{Byun:2017fkz}. The bispectrum in MG is also poorly understood, and if it is to be used in tests of gravity that move beyond consistency tests of GR, there is much to be done. Recently a 1-loop prediction for the bispectrum was proposed with phenomenological components in the TNS style \cite{Hashimoto:2017klo}. This has shown very good match with simulation data and having this prediction within the {\tt MGCopter} framework would prove extremely powerful. 
\newline
\newline
In a similar vein as the bispectrum, the cumulants of the density and velocity perturbations can be used to break degeneracies and extract interesting information from clustering data. These statistics are simply given by
\begin{align}
\label{eqn:cumulants}
\ad{}_{\rm c} =& 0,\,\,\textrm{(the mean)}\nonumber\\
\ad{2}_{\rm c} =& \ad{2}\equiv\sigma^2,\,\,\textrm{(the variance)}\nonumber\\
\ad{3}_{\rm c} =& \ad{3},\,\,\textrm{(the skewness)}\nonumber\\
\ad{4}_{\rm c} =& \ad{4} - 3\ad{2}_{\rm c}^2.\,\,\textrm{(the kurtosis)}
\end{align}
The top two panels of Fig.\ref{hclus} show {\tt MGCopter} predictions for the variance and skewness against N-body measurements while the bottom panel shows the ratio of nDGP and GR higher order moments \cite{Hellwing:2017pmj}. We see the kurtosis offers valuable information on deviations from GR. It is an ongoing project to extend the kurtosis computation to general models of gravity as well as extend the cumulant predictions to redshift space. 
\noindent In conclusion, the proposed and ongoing projects aim at making the best use of the next generation of surveys with respect to the issue of dark energy and gravity, and more broadly, stimulate and direct future gravitational research. The ongoing goal is a comprehensive, widely encompassing and self-contained tool that can be established pre data releases, to squeeze every last drop out of nature's answer to our ever pressing question and to ultimately climb a bit further, or as at beginning of the last century, make a giant stride up the mountainside.


\appendix 
\chapter{Numerical Algorithm} 
\begin{figure}[H]
  \captionsetup[subfigure]{labelformat=empty}
  \centering
  \subfloat[]{\includegraphics[width=7.5cm, height=7cm]{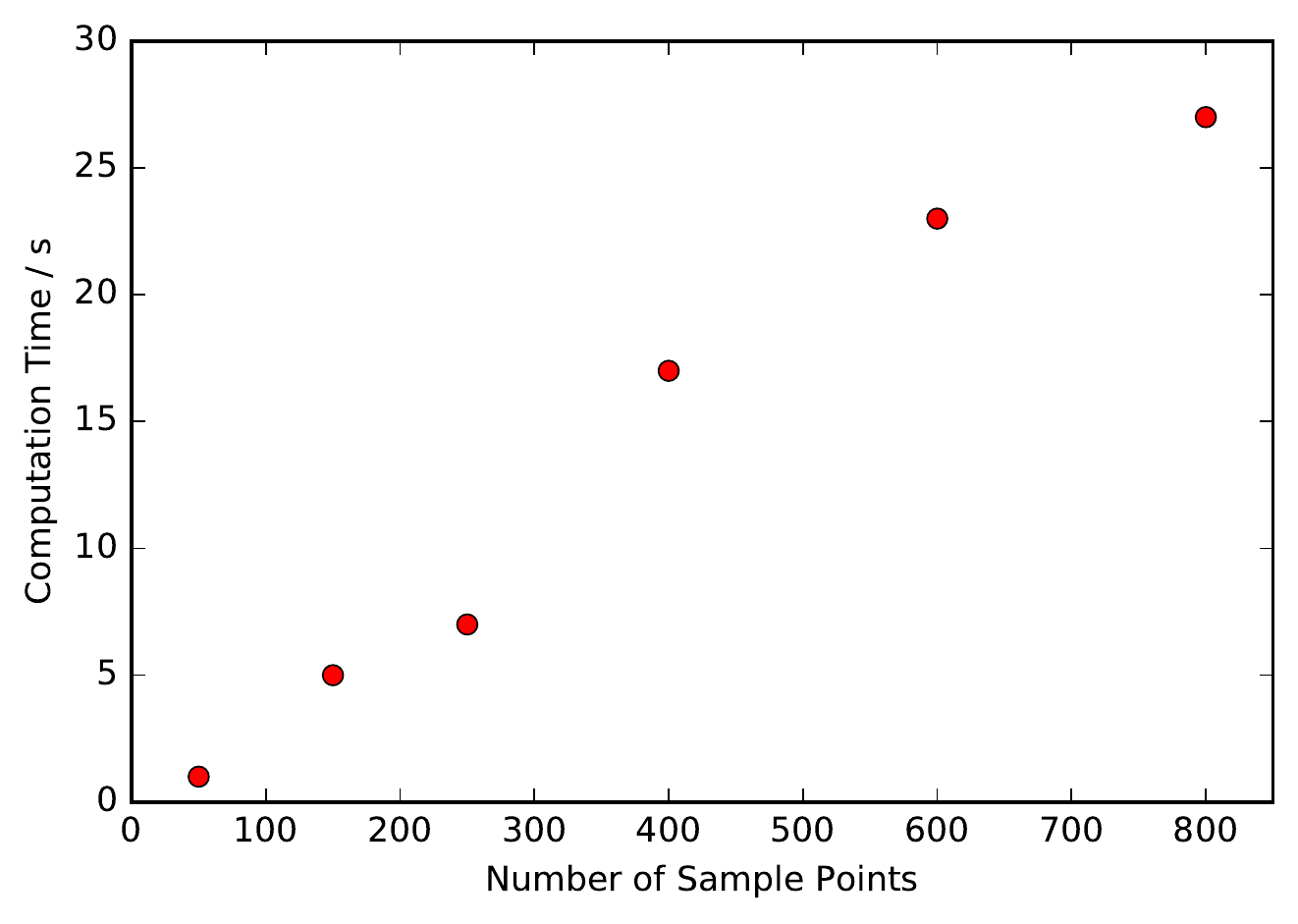}} \quad
  \subfloat[]{\includegraphics[width=7.5cm, height=7cm]{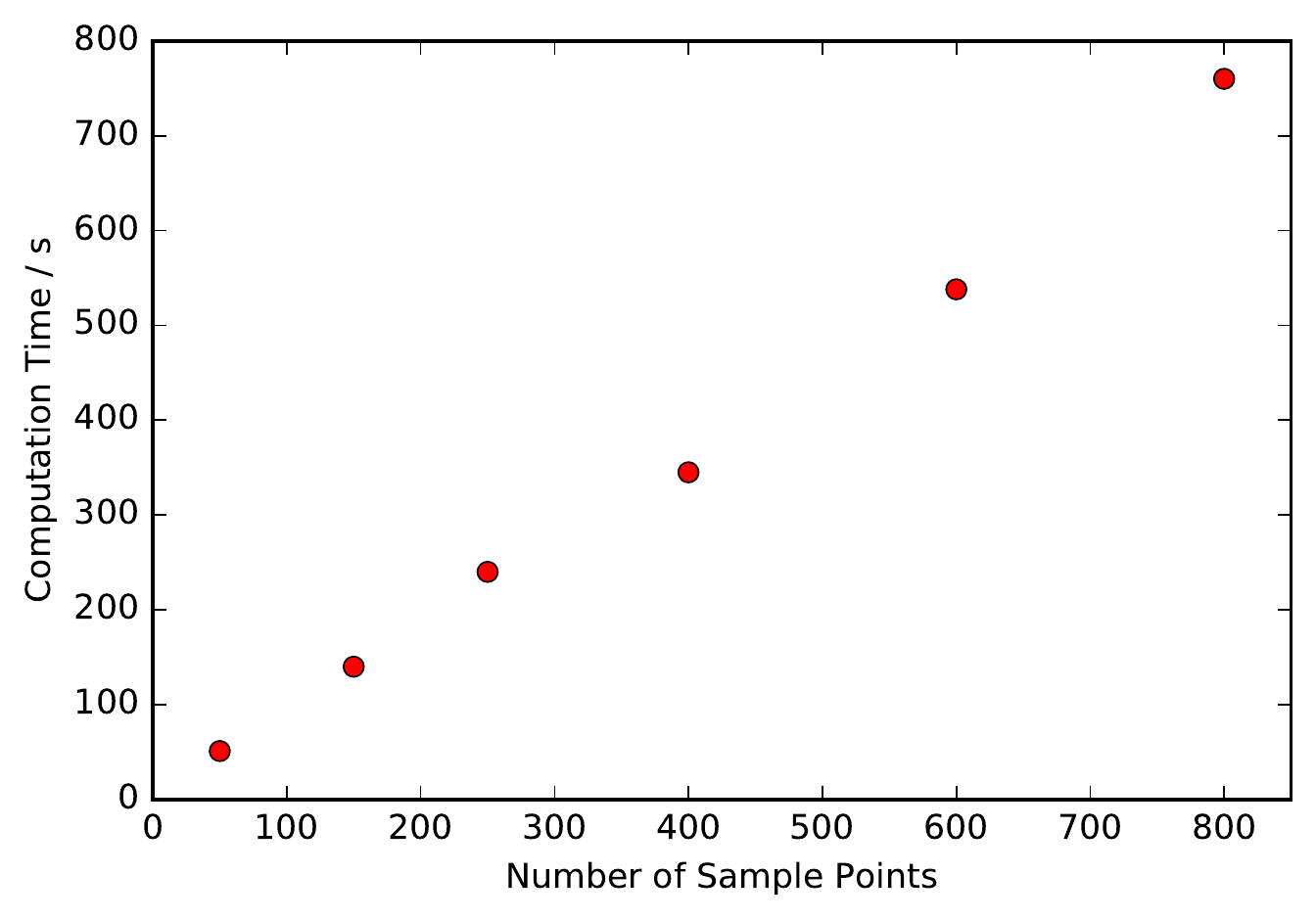}} 
  \caption{Time taken to compute 30 values of $P_{\delta \delta},P_{\theta \theta}$ and $P_{\theta \delta}$ at 1-loop for different values of $n_1$ in the scale independent case (left) and scale dependent case (right).}
\label{time1}
\end{figure}

It is well known that for the LCDM model the perturbations evolve independently of scale at linear order and so separability of the kernels into  time and  scale dependent parts becomes a good assumption allowing for an analytic solution. This is true for those models in which  $\mu(k;a)$, $\gamma_2(\bfk, \bfk_1, \bfk_2; a)$  and $\gamma_3(\bfk, \bfk_1, \bfk_2, \bfk_3;a)$ can be written as separable functions of scale and time. Analytic forms of the perturbations  for massless Horndeski's theory under the quasi-static approximation have been derived in \cite{Takushima:2015iha,Takushima:2013foa,Kimura:2011dc,DeFelice:2011hq}, going up to 3rd order as well as forms for the 1-loop power spectra in \cite{Takushima:2015iha}. In \cite{Koyama:2009me} the authors derive the analytical expressions for the perturbations up to 3rd order in DGP gravity. 
\newline
\newline
In general separability cannot be assumed, for example in the $f(R)$ class of models. In this case the Euler and continuity equations become analytically intractable and one must calculate the kernels numerically to proceed. Since our goal is to be as general as possible,  we solve the Euler and continuity equations numerically.
\newline
\newline
For our purposes we seek to calculate the density and velocity perturbations up to 3rd order which will be used in the computation of the 1-loop power spectrum as well as the TNS $A$ correction term up to consistent order. To this end, we require the, generally scale and time dependent, perturbative kernels $F_1(\bfk ; a)$, $G_1(\bfk; a)$, $F_2(\bfk_1,\bfk_2; a)$, $G_2(\bfk_1,\bfk_2; a)$, $F_3(\bfk_1,\bfk_2,\bfk_3; a)$ and $G_3(\bfk_1,\bfk_2,\bfk_3; a)$. We can write these as functions of two vector magnitudes, say $k_1$ and $k_2$ as well as the angle between them, $\mu= \hat{\bfk_1} \cdot \hat{\bfk_2}$. This is true even for the 3rd order kernels as we have a constraint coming from the integral of $\delta_D(\bfk-\bfk_{1\dots n})$. In the end we can reduce all loop integrals to 3 variables, $k_1=k$, $k_2=k_1r$ and $\mu$.
\newline
\newline
 We refer the readers to \cite{Taruya:2016jdt} and \cite{Taruya:talk} for details on the numerical algorithm. The algorithm involves finding kernel solutions for various values of integrated Fourier vector magnitude $r$ and angular parameter $\mu$. The number of $\mu$ values we sample is fixed to 256 which is the number of abscissae used in the Gauss-Legendre integration method used for the angular integral. The number of $r$ values we sample is variable and we denote this by $n_1$. This gives a total of $256 \times n_1$ solutions we need to calculate per value of $k$. Increasing $n_1$ gives a finer solution space to sample and consequently a more accurate result. 
\newline
\newline
The Euler and continuity system is solved using the $gsl$ package {\tt odeiv2}. {\tt odeiv2} is able to solve the equations using a host of methods. The default for our code is a Runge-Kutta Prince-Dormand (8, 9) method which works very well and quickly. The number of time steps used by the solver is adaptive and depends on the desired accuracy. Our default accuracy is based on matching the numerically calculated $P^{22}$ and $P^{13}$ with the EdS analytical versions to below $1\%$ within the desired range of scales. For the numerically calculated kernels we use EdS initial conditions in our system of equations, as in most dark energy and gravity models the EdS approximations holds at early times. Explicitly 
\begin{align}
F_1(\bfk;a_i) &= a_i,  \\ 
G_1(\bfk; a_i) &= -a_i, \\ 
F_2(\bfk_1,\bfk_2; a_i) &= a_i^2 F_2^{EdS}(\bfk_1,\bfk_2), \\ 
G_2(\bfk_1,\bfk_2;a_i) &= -a_i^2 G_2^{EdS}(\bfk_1,\bfk_2), \\ 
F_3(\bfk_1,\bfk_2,\bfk_3; a_i)& = a_i^3 F_3^{EdS}(\bfk_1,\bfk_2,\bfk_3), \\ 
G_3(\bfk_1,\bfk_2,\bfk_3; a_i)& = -a_i^3 G_3^{EdS}(\bfk_1,\bfk_2,\bfk_3), 
\end{align}
where $a_i = 0.0001$ is the initial scale factor and the symmetric analytic kernels on the right hand side are given by Eq.\ref{eds2ndordera}-\ref{eds2ndorderb} and Eq.\ref{eds3rdordera}-\ref{eds3rdorderb}. The complete system of Euler and continuity equations to be solved consists of 9 sets 
\begin{itemize}
\item
3 first order sets which is solved for $F_1(k_i; a)$ and $G_1(k_i;a)$, where \newline $k_i \in \{k,k_1,|k-k_1|\}$. 
\item 
5 second order sets which are solved for  $F_2(\bfk_1, \bfk_2; a)$ and $G_2(\bfk_1, \bfk_2;a)$. The 5 sets are for the following pairs of $k$-mode inputs : $(\bfk_1, \bfk-\bfk_1)$, $(\bfk-\bfk_1, -\bfk)$, $(-\bfk_1, \bfk)$, $(\bfk_1, \bfk)$ and $(\bfk_1, -\bfk_1)$.  The first of these is for $P^{22}$ while the last 4 are used to construct the 3rd order kernel and the RSD A-term. 
\item 
1 third order set which is solved for  $F_3(\bfk, \bfk_1, -\bfk_1 ; a)$ and $G_3(\bfk, \bfk_1, -\bfk_1;a)$.
\end{itemize} 
The left of Fig.\ref{time1} shows the time cost of finer sampling in the independent case, where the solution space is initialised once. This can be done when there is no $k$ dependence in the $\mu$, $\gamma_2$ and $\gamma_3$ terms as in DGP (see Eq.\ref{mudgp} - Eq.\ref{g3dgp}). The right of Fig.\ref{time1} shows the time cost of initialising the kernels for each $k$ mode as needed in the scale dependent case, for example in $f(R)$ gravity.  From Fig.\ref{time1} and Fig.\ref{frconv} we see that increasing the sampling gives a large time cost for very little improvement in accuracy when compared to the N-body predictions. The time cost for an output of 30 values of $P_{\delta \delta}, P_{\delta \theta}$ and $P_{\theta \theta}$ at 1-loop level for $n_1= 150$ is 2 minutes on the laptop described on the next page. This can only be cut down so much as the amount of time needed to solve the 9 pairs of equations puts a lower bound on the time cost. In Chapter 5 we have used $n_1=150$ unless otherwise stated. 
\newline
\newline
An additional point is that by increasing the sampling within the algorithm one gets a convergence of the numerical solution which provides a handle on the error arising from finite sampling. Increased sampling comes at a computation time cost and we find good accuracy can be achieved for relatively low sampling.  We note that this cost is easily alleviated through parallelisation methods of the independent $k$ modes. This is still not ideal for large amount of computation, say in MCMC analyses where thousands of initialisations would be necessary. For this reason a library has been developed which constructs interpolated functions for the power spectra over the MG parameters, thus requiring only a limited number of initialisations if an MCMC analysis needs to be performed on scale dependent theories. 
\newline
\newline
All timing results were obtained on a MacBook Pro laptop computer, with a 2.52 GHz Intel Core 2 Duo processor and running on Mac OS X version 10.6.8.  The code is both OpenMP and MPI enabled but  parallelisation was not used for the timing results. 

\chapter{Horndeski A,B, and C Terms} 
In this appendix we reproduce the $A_0$,$A_1$,$A_2$,$B_0$,$B_1$, $B_2$,$B_3$,$C_0$ and $C_1$ terms as listed in Appendix A of \cite{Takushima:2015iha}
\begin{align} 
A_0  =  &\frac{\dot{\Theta}}{H^2} + \frac{\Theta}{H} + \mathcal{F}_T - 2 \mathcal{G}_T-2 \frac{\dot{\mathcal{G}_T}}{H} - \frac{\mathcal{E} + \mathcal{P}}{2H^2}, \\ 
A_1  = &\frac{\dot{\mathcal{G}_T}}{H} + \mathcal{G}_T - \mathcal{F}_T, \\ 
A_2  =& \mathcal{G}_T - \frac{\Theta}{H}, \\
B_0  = &\frac{X}{H} \Big[ \dot{\phi} G_{3,X} + 3 ( \dot{X} + 2HX) G_{4,XX} + 2X\dot{X} G_{4,XXX} -3\dot{\phi}G_{4,\phi X} + 2\dot{\phi}XG_{4,\phi XX} \nonumber \\ 
& + (\dot{H}+H^2)\dot{\phi}G_{5,X} + \dot{\phi}\left[2H\dot{X}+(\dot{H}+H^2)X\right]G_{5,XX} +H\dot{\phi}X\dot{X}G_{5,XXX} \nonumber \\ 
&-2(\dot{X} +2HX)G_{5,\phi X} - \dot{\phi} X G_{5,\phi \phi X} -X (\dot{X} -2HX) G_{5,\phi XX}\Big], \\
B_1 = & 2X\left[G_{4,X} +\ddot{\phi}(G_{5,X}+XG_{5,XX})-G_{5,\phi}+XG_{5,\phi X}\right], \\ 
B_2 = & -2X\left(G_{4,X}+2XG_{4,XX}+H\dot{\phi}G_{5,X} +H\dot{\phi}XG_{5,XX}-G_{5,\phi}-XG_{5,\phi X}\right),\\
B_3 = & H\dot{\phi}XG_{5,X}, \\ 
C_0= & 2X^2G_{4,XX} + \frac{2X^2}{3}\left(2\ddot{\phi}G_{5,XX}+\ddot{\phi}XG_{5,XXX}-2G_{5,\phi X} + X G_{5,\phi XX}\right), \\ 
C_1 = & H\dot{\phi} X \left(G_{5,X} + X G_{5, XX}\right), 
\end{align}
 where 
 \begin{align}
 \mathcal{F}_T = &2\left[G_4 - X\left(\ddot{\phi}G_{5,X} + G_{5,\phi}\right) \right], \\ 
 \mathcal{G}_T = &2\left[G_4 -2XG_{4,X}-X(H\dot{\phi}G_{5,X} -G_{5,\phi})\right],\\
 \Theta = & -\dot{\phi} X G_{3,X} +2HG_4 - 8 HXG_{4,X} - 8HX^2G_{4,XX}+\dot{\phi}G_{4,\phi}+2X\dot{\phi}G_{4,\phi X} \nonumber \\ 
 &-H^2\dot{\phi}(5XG_{5,X}+2X^2G_{5,XX})+2HX(3G_{5,\phi}+2XG_{5,\phi X}), \\ 
 \mathcal{E} = &2XG_{2,X}-G_2 +6X\dot{\phi}HG_{3,X}-2XG_{3,\phi}-6H^2G_4 \nonumber \\ 
 &+24H^2X(G_{4,X} +XG_{4,XX})-12HX\dot{\phi}G_{4,\phi X}-6H\dot{\phi}G_{4,\phi}\nonumber \\ 
 &+2H^3X\dot{\phi}(5G_{5,X}+2XG_{5,XX})-6H^2X(3G_{5,\phi}+2XG_{5,\phi X}), \\ 
 \mathcal{P} = & G_2 -2X(G_{3,\phi}+\ddot{\phi}G_{3,X}) +2(3H^2+2\dot{H})G_4-12H^2XG_{4,X}-4H\dot{X}G_{4,X} \nonumber \\ 
 & -8\dot{H}XG_{4,X}-8HX\dot{X}G_{4,XX}+2(\ddot{\phi}+2H\dot{\phi})G_{4,\phi}+4XG_{4,\phi \phi} \nonumber \\
 & +4X(\ddot{\phi}-2H\dot{\phi})G_{4,\phi X} -2X(2H^3\dot{\phi}+2H\dot{H}\dot{\phi}+3H^2\ddot{\phi})G_{5,X} \nonumber \\ 
 & -4H^2X^2\ddot{\phi}G_{5,XX} + 4HX(\dot{X}-HX)G_{5,\phi X} \nonumber \\ 
 &+ 2\left[2(\dot{H}X + H\dot{X}) + 3H^2X\right] G_{5,\phi} + 4HX\dot{\phi}G_{5,\phi \phi}.
 \end{align}
Consecutive subscripts after the comma indicate consecutive partial derivatives with respect to the indicated variable. As a reminder, the overdot represents a derivative with respect to FLRW metric time $t$. 

|
\chapter{Statistical Inference and MCMC}
\begin{figure}[H]
  \captionsetup[subfigure]{labelformat=empty}
  \centering
  \subfloat[]{\includegraphics[width=15.2cm, height=7.cm]{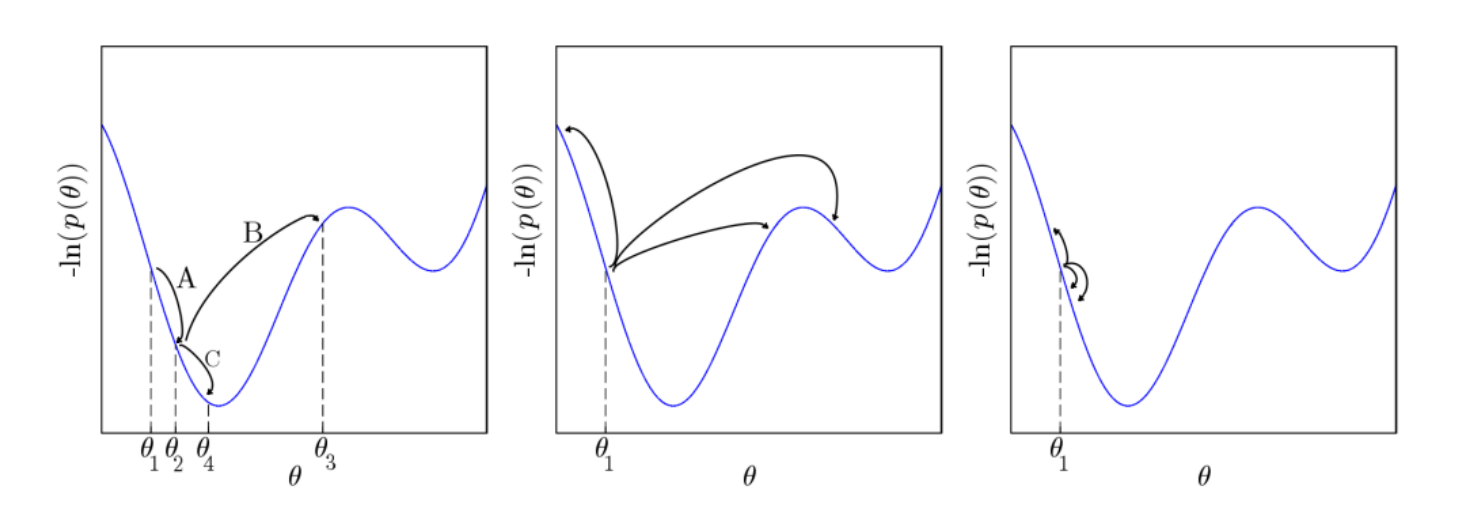}} 
  \caption{{\bf Left}: The Markov chain starts at $\theta_1$. $\theta_2$ is then proposed and accepted (A). $\theta_3$ is then proposed and refused (B). $\theta_4$ is proposed and accepted (C). The chain goes $\{ \theta_1, \theta_2, \theta_2, \theta_4, \dots \}$. {\bf Centre}: An instance of too large a step size, with the chain getting stuck because of unlikely proposed parameter sets. {\bf Right}: An instance of too small a step size, with the chain probing the posterior extremely slowly. The figure was taken from \cite{Leclercq:2014jda}.}
\label{mcmcstep}
\end{figure}

In this appendix we give a summary of statistical inference of parameters of interest and a commonly used method in doing so. To begin we quote the hinge of the relevant methods, Bayes' theorem 
\begin{equation}
{\rm P}(D) {\rm P}(H | D)  = {\rm P}(H)  {\rm P}(D | H), 
\end{equation}
where ${\rm P}(H)$ is the probability that hypothesis H is true and ${\rm P}(H|D)$ is the probability that  H is true given that D is true. Put into a cosmological context, H will usually be some gravitational theory/ cosmology and D will be the measurement we've made from simulations or the real universe. Then, the quantity we are interested in is ${\rm P} (H | D)$, commonly called the {\bf posterior} probability distribution function. 
\newline
\newline
Next we have ${\rm P}(H)$, which is called the {\bf prior} because it constitutes all the a priori information we have on our theory. For example, from mathematical consistency or physical arguments (e.g. $\Omega_b > 0$). The prior can also come from other experiments/data $\tilde{D}$. Basically, it is all the information you have on your parameters, gravitational theory, cosmology, prior to the measurement $D$. 
\newline
\newline
Once you make the measurement D the idea is to determine the posterior which then can become your prior for future measurements. To update the prior you also need a {\bf likelihood} distribution, ${\rm P}(D| H)$, which denotes the probability of observing the data given your theoretical model. ${\rm P}(D)$ is the probability of making the measurement you've made. Generally this probability is difficult to determine because we have one universe to observe. It is generally treated as a normalisation factor in analyses. In the end we are left with the following relationship for successive measurements $D_1$ and $D_2$
\begin{align}
 {\rm P}(H | D_1, D_2 ) & \propto   {\rm P}(H | D_1) {\rm P}(D_2 | H) \nonumber \\ 
  & \propto   {\rm P}(H) {\rm P}(D_2 | H) {\rm P}(D_1 | H).  
\end{align}
This can be repeated many times and in the end we obtain a better estimate of the probability that our gravitational theory and cosmology are true. In LSS analyses, the likelihood function is usually given by a $\chi^2$ distribution as shown in Eq.\ref{covarianceeqn}, where the model is compared to the data with some associated errors. The priors depend on the observers and their previous knowledge. In the cases described in this thesis the only priors on parameters come from physical considerations, for example densities should be positive. Given these two, we then have a means of constructing our posterior distribution. Markov Chain Monte Carlo (MCMC) methods are a way of doing exactly this. 
\newline
\newline
A Markov Chain is a process that jumps from state $H_n$ to state $H_m$ given some probability that only depends on $H_n$. In practice these states will be a set of our theoretical parameters. The chain will go to a new set of parameters based on our likelihood and the probability of the set $H_n$ compared to that of $H_m$. In this way the chain moves towards regions where the probability of $H_i$s is high given the data (recall the data is integrated in the likelihood). These jumps essentially map out our posterior distribution (see Fig.\ref{nbodygrz1} for example). 
\newline
\newline
The algorithm {\tt MGCopter} uses is the popular Metropolis-Hastings MCMC method. This proceeds as follows 
\begin{enumerate}
\item
Choose a starting set of parameters $H_i$. 
\item
Propose a new set of parameters $H_j$ according to some symmetric proposal distribution $q(H_j | H_i)$. 
\item 
Calculate the ratio 
\begin{equation}
a = \frac{ {\rm P}(D | H_j)}{ {\rm P}(D | H_i)},
\end{equation}
using the likelihood distribution. 
\item
If $a>1$, then we jump to $H_j$ point because it gives a higher probability according to the data. If $a<1$ then we jump to $H_j$ with probability $a$. If it is not selected because of this we remain at $H_i$. This helps us to move around parameter space and to not get stuck in a local maximum of the posterior. 
\item
Repeat steps 2 to 5 until the posterior appears to have been well mapped. This can be checked according to various convergence criterions and we direct the reader to \cite{cowles:mcmc} for more details. 
\end{enumerate}
One pitfall for such algorithms is the proposal distribution. The step sizes need to be efficient in probing parameter space as shown in Fig.\ref{mcmcstep}, taken from \cite{Leclercq:2014jda}.
\noindent In our analyses of the parameter set $\{\Omega_{rc},\sigma_v\}$ we run initial chains with a step size that gives us a fair acceptance rate $\sim 30\%$.  We then rerun with longer chains that give us a fair acceptance rate but the steps of each parameter are chosen according to the standard deviation derived from the initial chains. We run multiple chains that make thousands of steps each to get a converged result for the probability of our parameters. This is possible by running {\tt MGCopter} in parallel, done on the supercomputer {\tt Sciama}, at the University of Portsmouth.  


\chapter{N-Body Simulations and COLA}
In this appendix we give a qualitative summary of N-body simulations and the approximate COLA (COmoving Lagrangian Acceleration) approach \cite{Tassev:2013pn}. Full N-body simulations allow us to describe clustering of dark matter down to very small scales ($\sim5h/\mbox{Mpc}$), while COLA offers varying degrees of accuracy depending on parameters and chosen time steps. In chapters 5 and 7 we use measurements of the power spectra from such simulations as a benchmark of accuracy for our perturbative approach. 
\newline
\newline
We begin with a summary of the N-body approach but refer the interested reader to one of the many reviews on the topic \cite{Bagla:2004au} for a more detailed description. Generally, N-body simulations consist of following and recording the motions of particles in a box which is associated with a cosmological volume. Of course cosmological numbers of actual dark matter particles cannot be simulated and so the simulation particles generally represent a bound group of dark matter particles. These are then given masses $m$ and have an associated length $\epsilon \propto m^{1/3}$ by sphericity. This length governs the spatial resolution of the simulation, meaning smaller masses give better resolution. The N-body process then follows these main steps
\begin{enumerate} 
\item
The particles are given some initial positions. 
\item
The gravitational force acting on each particle is calculated by solving the Poisson equation (Eq.\ref{eq:poisson0}). Note that $\delta$ is the non-linear density contrast calculated from the distribution of the particles. This gives us an acceleration for each particle by then solving $a^2 \ddot{\bfx} = -\nabla \Phi$, where $\bfx$ is the current particle position in comoving coordinates. 
\item
The positions and velocities of each particle are updated according to a time step. Usually the Leapfrog  integrator is used which shifts velocities and positions out of phase by half a time step.  
\item
Repeat steps 2 and 3 until the simulation is finished. 
\end{enumerate} 
Varying methods are used in solving the Poisson and acceleration equations. One method, called the particle mesh method proceeds as follows. The density of the box is calculated by interpolating each particles mass over a grid. 
Over the same grid the Poisson equation is solved in Fourier space by using a fast Fourier transform approach, and the calculated forces are then assigned on the particles. There are of course issues with this method. At the grid cells the assigned force may become anisotropic because of the cell's geometry. Further, there is a problem of selecting the optimal sized cells constituting the mesh. Finer mesh's are good for high density regions but are more computationally expensive and vice-versa. In the case of MG, one has the added complication of having to solve the non-linear KGE as well.  Special techniques to deal with MG's non-linear KGE are needed and one cannot simply fast Fourier transform the Poisson equation. 
 \newline
 \newline
 The adaptive mesh refinement (AMR) method is a good alternative to the fast Fourier transform, uniform mesh approach. It refines the mesh only in areas with increased clustering or dynamics to improve resolution. This of course means that we run into problems in assigning the force as the mesh can now take on more complicated shapes. To overcome this and the non-linear KGE we need to solve, a technique called relaxation is used in place of the fast Fourier transform approach. This method employs a root finding algorithm for the Poisson equation
 \begin{equation}
F(\Phi_{i,j,k}) - \frac{3 a \Omega_m(a) \rho_{i,j,k}}{2} = 0
\label{rooteqn}
\end{equation}
where $\rho_{i,j,k}$ is the density assigned to cell $(i,j,k)$ and $F(\Phi_{i,j,k})$ is a function of the potential in this cell. It iteratively moves closer to the true solution with a truncation of this process governed by some convergence criteria. 
\newline
\newline
The simulations used in this thesis make use of the AMR code, ECOSMOG \cite{Li:2011vk}, which efficiently solves the dynamics of the scalar field producing accurate, high-resolution simulations. These are still time costly of course, and if we want to have converged results of our observables, that are not plagued by sample variance, we will need to run many such simulations and take the average. Further, for our purposes, those of testing perturbative techniques, we only care about accuracy in the relatively large scales ($k<0.5h$/Mpc).  This is why we have resorted to using the COLA approach for most of our comparisons.  
\newline
\newline
The COLA approach \cite{Tassev:2013pn} is an approximate method that is based off the N-body process described above. The big issue with N-body is that small time steps are required for high accuracy results. This demand can be relaxed for the scales we are concerned with. What COLA does is assumes the particles are generally at the position given by 2nd order LPT, $\bfx_{\rm LPT}$, plus some small correction $\delta \bfx$. If we then substitute $\bfx = \bfx_{\rm LPT} + \delta \bfx$ we then need solve the following equations 
\begin{align}
\nabla^2 \Phi =& \frac{a^2 \kappa \rho \delta}{2},  \\ 
 a^4 \ddot{\delta \bfx} =&   \nabla \Phi - a^4 \ddot{\bfx}_{\rm LPT}, \\ 
 a^2 \dot{\delta \bfx} = & \delta \bfv, 
 \end{align}
 where $\delta \bfv$ is the perturbation around the LPT prediction for the velocity. Since the LPT value is easily predicted and the correction to this is small at larger scales, then the simulation can afford to increase the time steps. Of course at small scales this results in worsened accuracy as $\delta \bfx$ will no longer be small (nor a correction as LPT breaks down). 
 \newline
 \newline
 This method has been parallelised \cite{Howlett:2015hfa} and extended to the class of MG used in this work in \cite{Winther:2017jof} by making use of the parametrisation described here and incorporating it into 2nd order LPT. As previously mentioned, we use simulations from \cite{Winther:2017jof} for most of Chapter 7. 


\chapter{Fourier Space Comparisons: nDGP} 
\begin{figure}[H]
  \captionsetup[subfigure]{labelformat=empty}
  \centering
  \subfloat[]{\includegraphics[width=7.5cm, height=7cm]{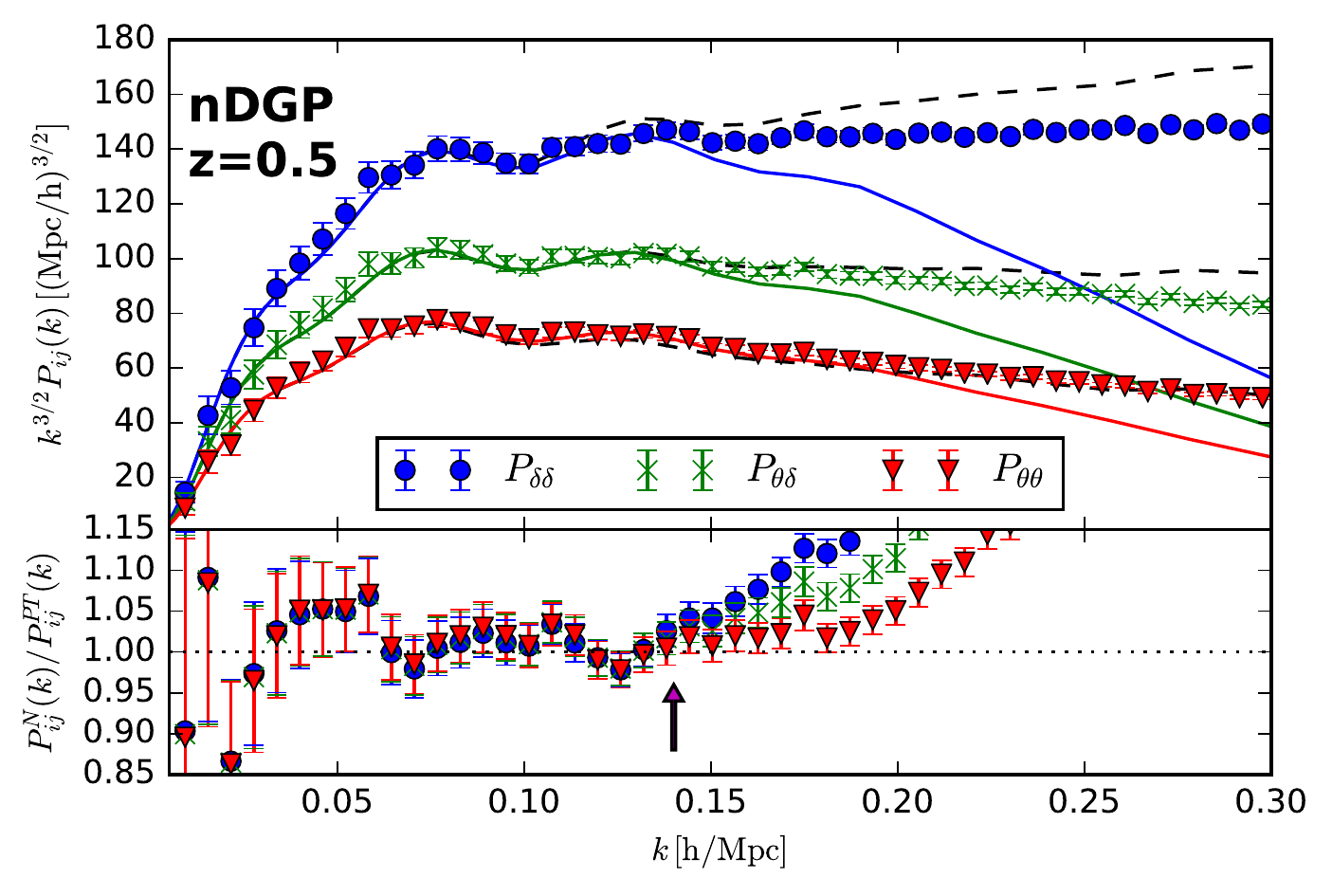}} \quad
  \subfloat[]{\includegraphics[width=7.5cm, height=7cm]{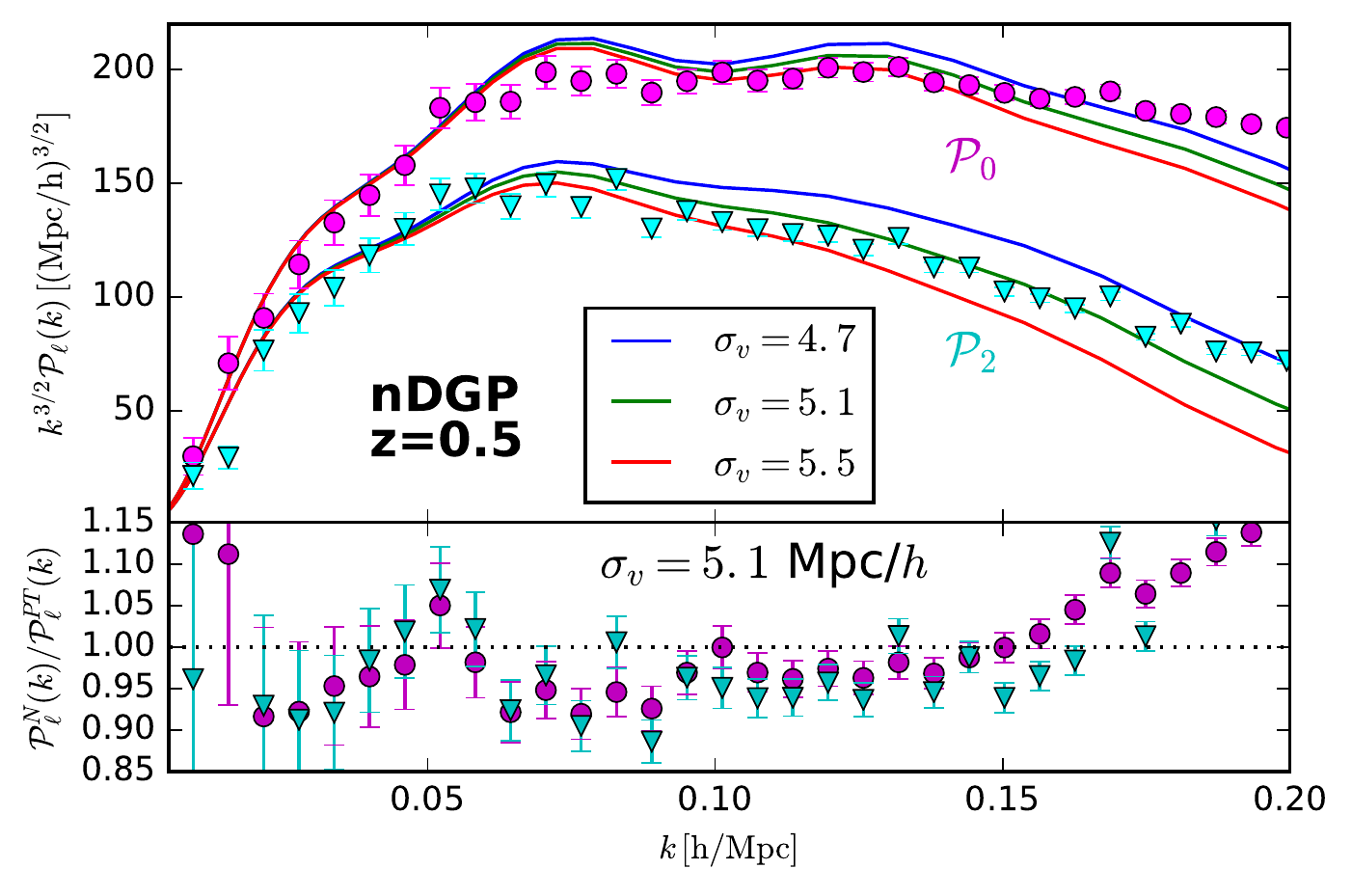}} 
  \caption{{\bf Left:} Comparison of N-body measurements of the auto matter (blue), matter-velocity divergence (green) and auto velocity divergence (red) power spectra in real space at $z=0.5$ for nDGP. The top panels show the RegPT (solid) and SPT (dashed) power spectra multiplied by $k^{3/2}$ and the bottom panels show the fractional difference between N-body and RegPT predictions with range of $1\%$ deviation indicated by a solid arrow. {\bf Right:} Comparison of N-body measurements of the redshift space monopole (magenta) and quadrupole (cyan) power spectra at $z=0.5$ for nDGP. The top panels show the multipoles multiplied by $k^{3/2}$ calculated for three values of $\sigma_v$ and the bottom panels show the fractional difference between N-body and TNS predictions for the best fit $\sigma_v$. The nDGP parameter is $\Omega_{rc}=0.438$.}
\label{pab1}
\end{figure}

In this appendix we determine the best fit $\sigma_v$ parameter of the TNS model against N-body data. This provided a good accuracy benchmark for comparison in Chapter 6 for the GSM model. We begin by finding the best fit $\sigma_v$. To do this we must first determine the range of validity of SPT. The left pane of Fig.\ref{pab1} shows the real space matter-matter (blue), matter-velocity divergence (green) and velocity divergence (red) power spectra using SPT (dashed) and RegPT (solid) against N-body data for the nDGP model of gravity.  The $k_{\rm max}h$/Mpc  we use for the fitting of $\sigma_v$ in the multipoles is given by the solid arrow which delimits the $1\%$ deviation region. We have fitted Gaussian error bars to the data assuming a survey volume of $1 \mbox{Gpc}^3/h^3$.  
\newline
\newline
With a range of validity we can now fit the TNS free parameter $\sigma_v$. We consider the multipoles of Eq.(\ref{redshiftps}) given by Eq.(\ref{multipolesl}). The monopole and quadrupole are then fit up to the $k_{\rm max}$ found previously. Higher order multipoles have a very low signal to noise ratio making them problematic to measure in practice and so we will not consider them in our results. 
\newline
\newline
The right pane of Fig.\ref{pab1} shows the monopole (magenta) and quadrupole (cyan) N-body measurements against the RegPT-TNS predictions for three different values of $\sigma_v$. The fractional difference of the best fit $\sigma_v$ with N-body is shown in the bottom panels. The best fit value for $\sigma_v$ is found to be $5.1$Mpc/$h$. The best fit value for $f(R)$ and GR were found to be $6$Mpc/$h$ and $4.75$Mpc/$h$ respecitvely in \cite{Taruya:2014faa}.


\chapter{Validating use of MG-PICOLA}
\begin{figure}[H]
  \captionsetup[subfigure]{labelformat=empty}
  \centering
  \subfloat[]{\includegraphics[width=7.5cm, height=6.6cm]{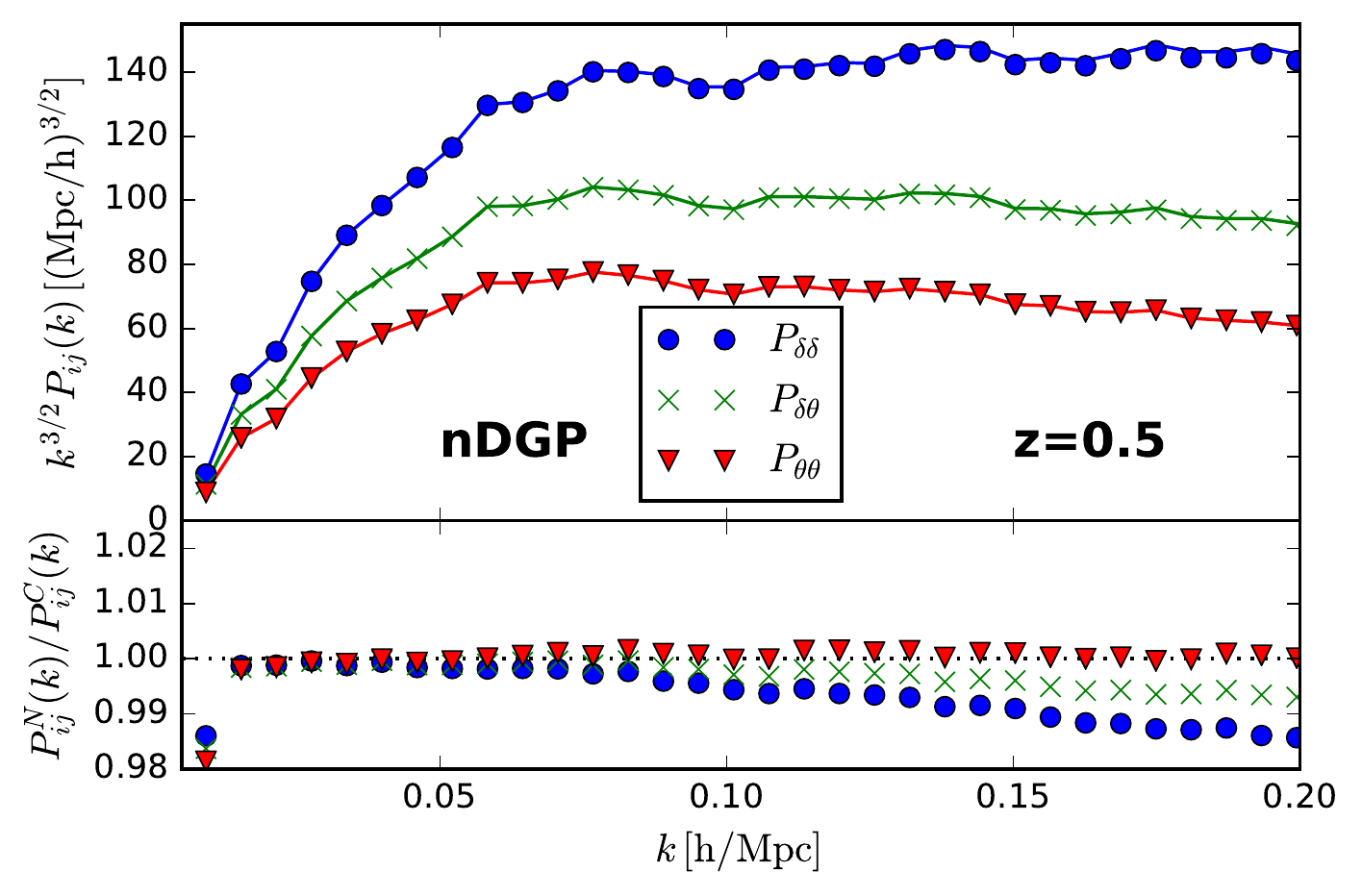}} \quad
  \subfloat[]{\includegraphics[width=7.5cm, height=6.6cm]{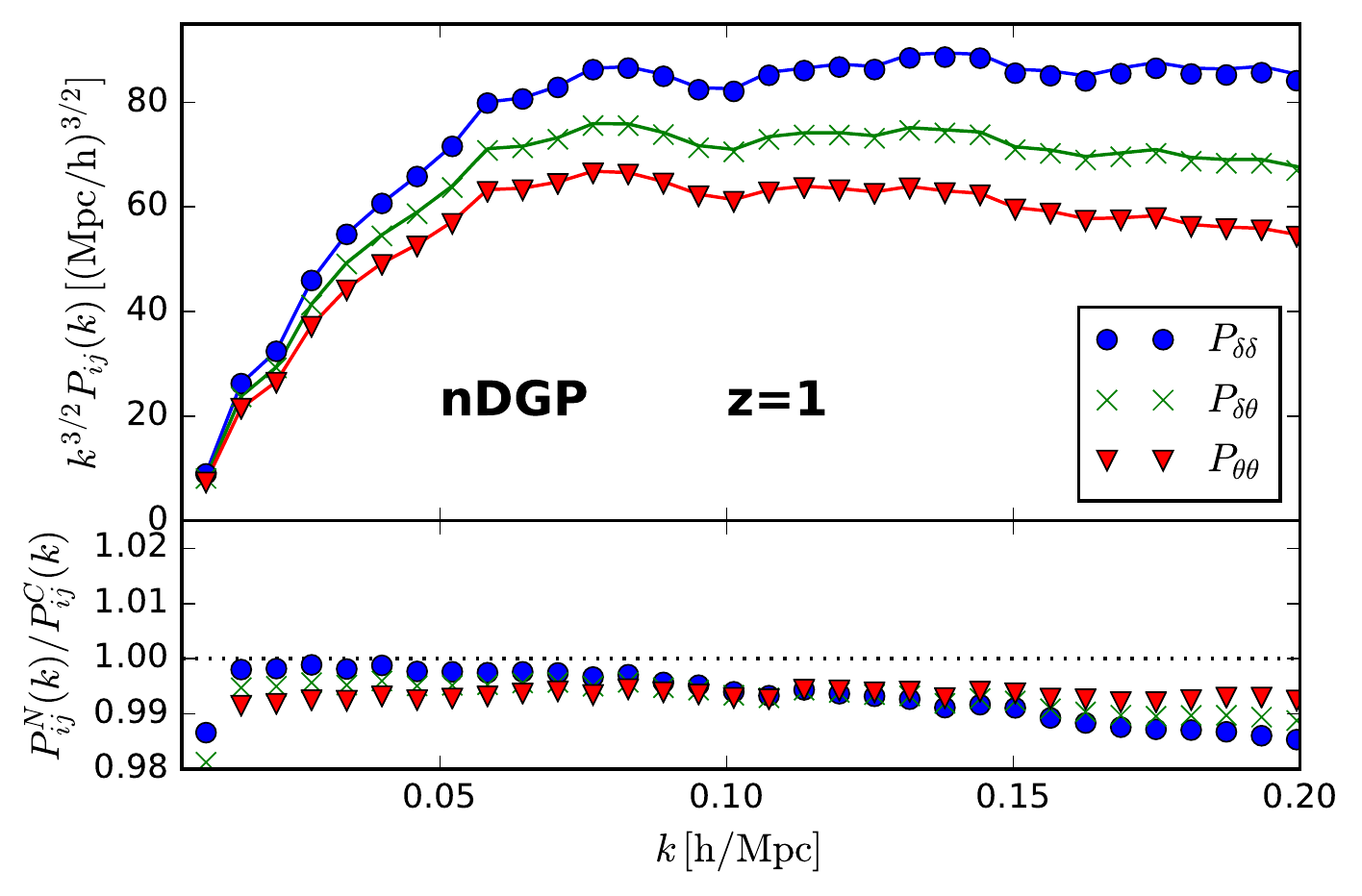}} 
  \caption{COLA1 (solid) and N-body measurements (points) of the auto and cross power spectra of density and velocity fields in real space for nDGP at $z=0.5$ (left) and $z=1$ (right).  The top panels show the power spectra scaled by $k^{3/2}$ and the bottom panels show the ratio of the two measurements.}
\label{cola1vnbody1}
\end{figure}

\begin{figure}[H]
  \captionsetup[subfigure]{labelformat=empty}
  \centering
  \subfloat[]{\includegraphics[width=7.5cm, height=6.6cm]{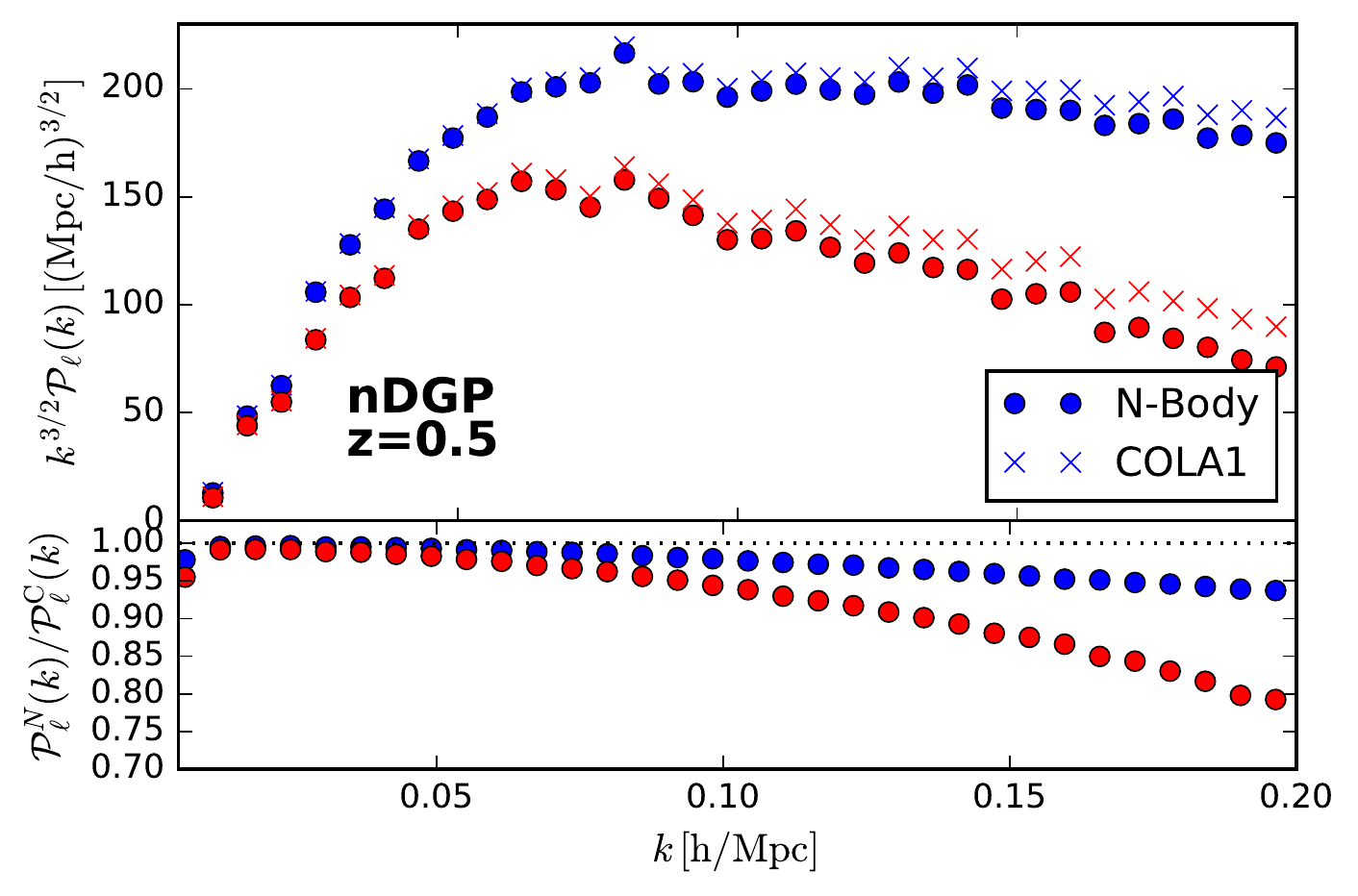}} \quad
  \subfloat[]{\includegraphics[width=7.5cm, height=6.6cm]{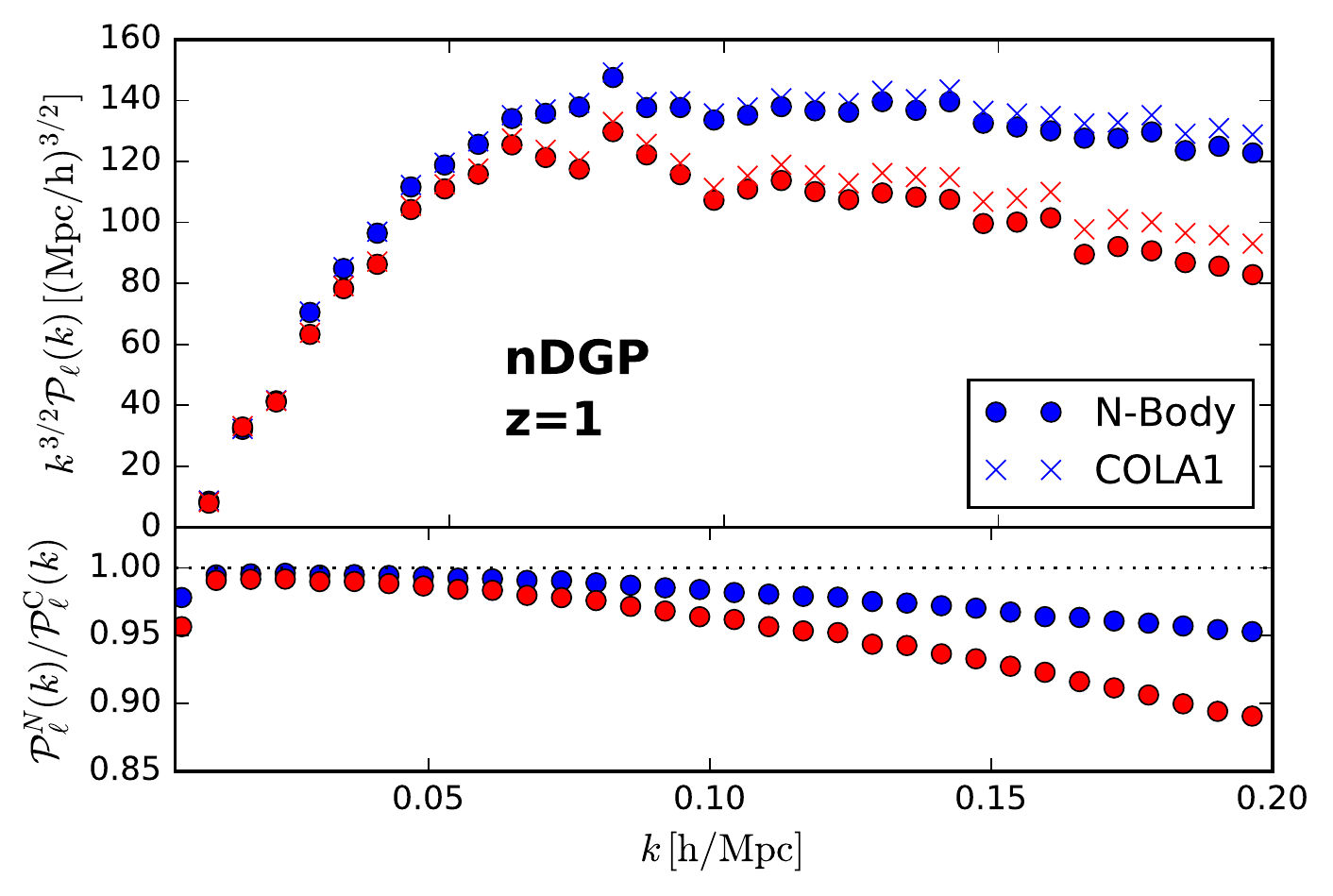}} 
  \caption{COLA1 (crosses) and N-body measurements (circles) of the redshift space monopole (blue) and quadrupole (red) for nDGP at $z=0.5$ (left) and $z=1$ (right).  The top panels show the power spectra scaled by $k^{3/2}$ and the bottom panels show the ratio of the two measurements.}
\label{cola1vnbody2}
\end{figure}

\begin{figure}[H]
  \captionsetup[subfigure]{labelformat=empty}
  \centering
  \subfloat[]{\includegraphics[width=7.5cm, height=6.6cm]{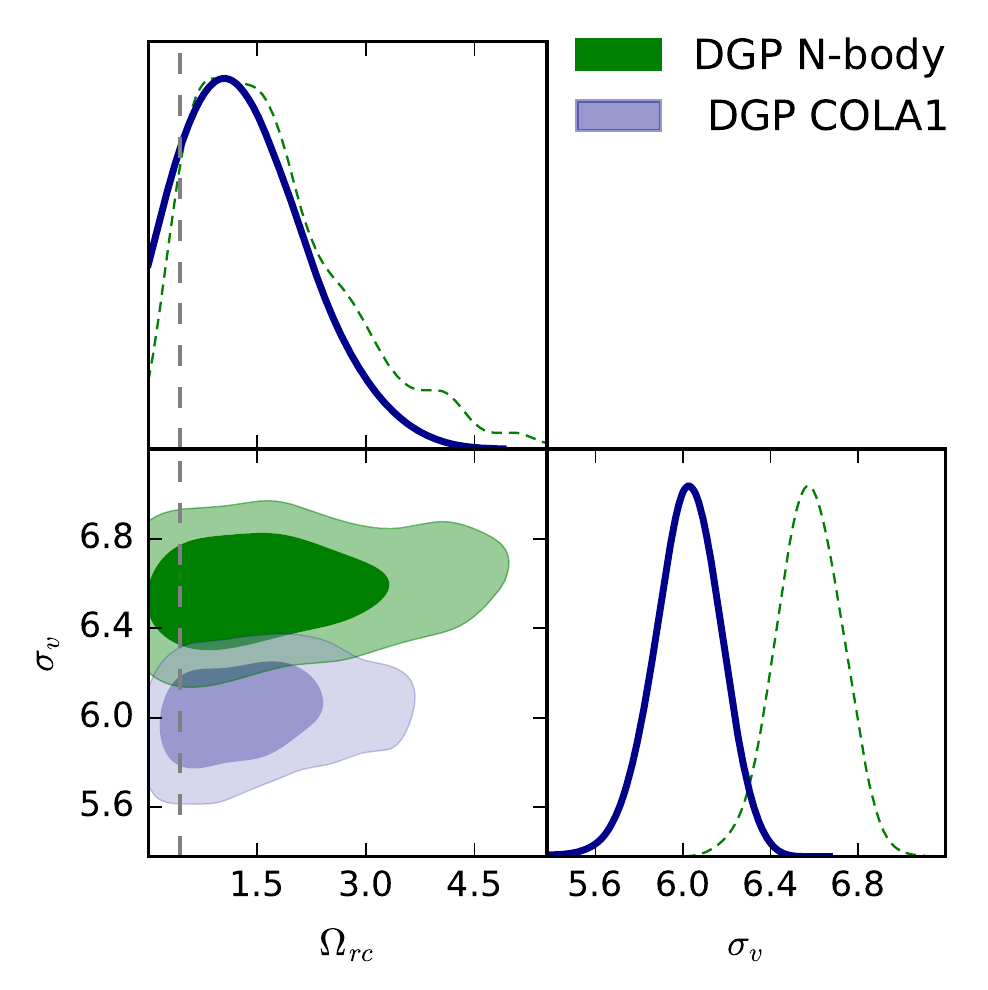}} \quad
  \subfloat[]{\includegraphics[width=7.5cm, height=6.6cm]{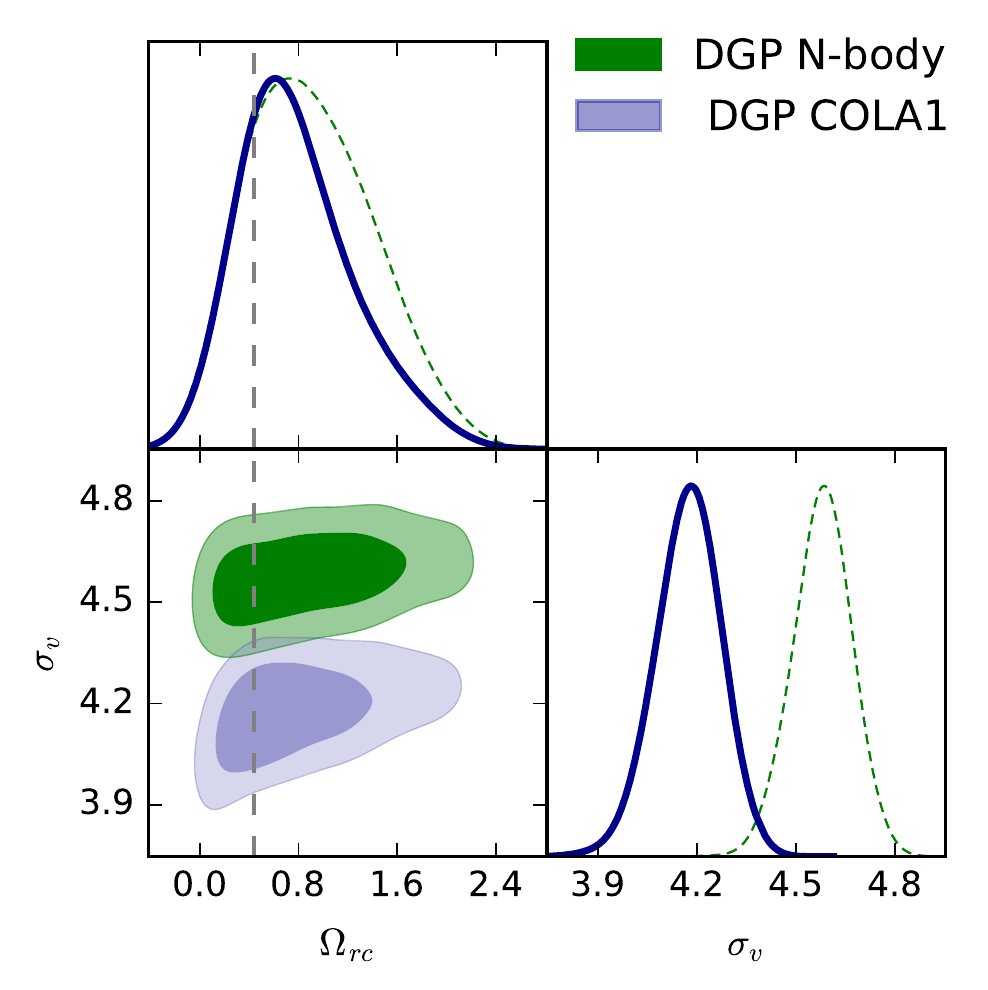}} 
  \caption{ The $1\sigma$ and $2\sigma$ confidence contours for the DGP template using the N-body (green) and COLA1(blue) data sets at $z=0.5$(left) and $z=1$(right) fitting up to $k_{\rm max}=0.147h$/Mpc and $k_{\rm max}=0.171h$/Mpc respectively with the simulation's fiducial value for $\Omega_{rc}$ indicated by the dashed line. A survey of volume of $1\mbox{Gpc}^{3}/h^3$ is assumed. }
\label{cola1vnbody3}
\end{figure}

\begin{figure}[H]
  \captionsetup[subfigure]{labelformat=empty}
  \centering
  \subfloat[]{\includegraphics[width=7.5cm, height=6.6cm]{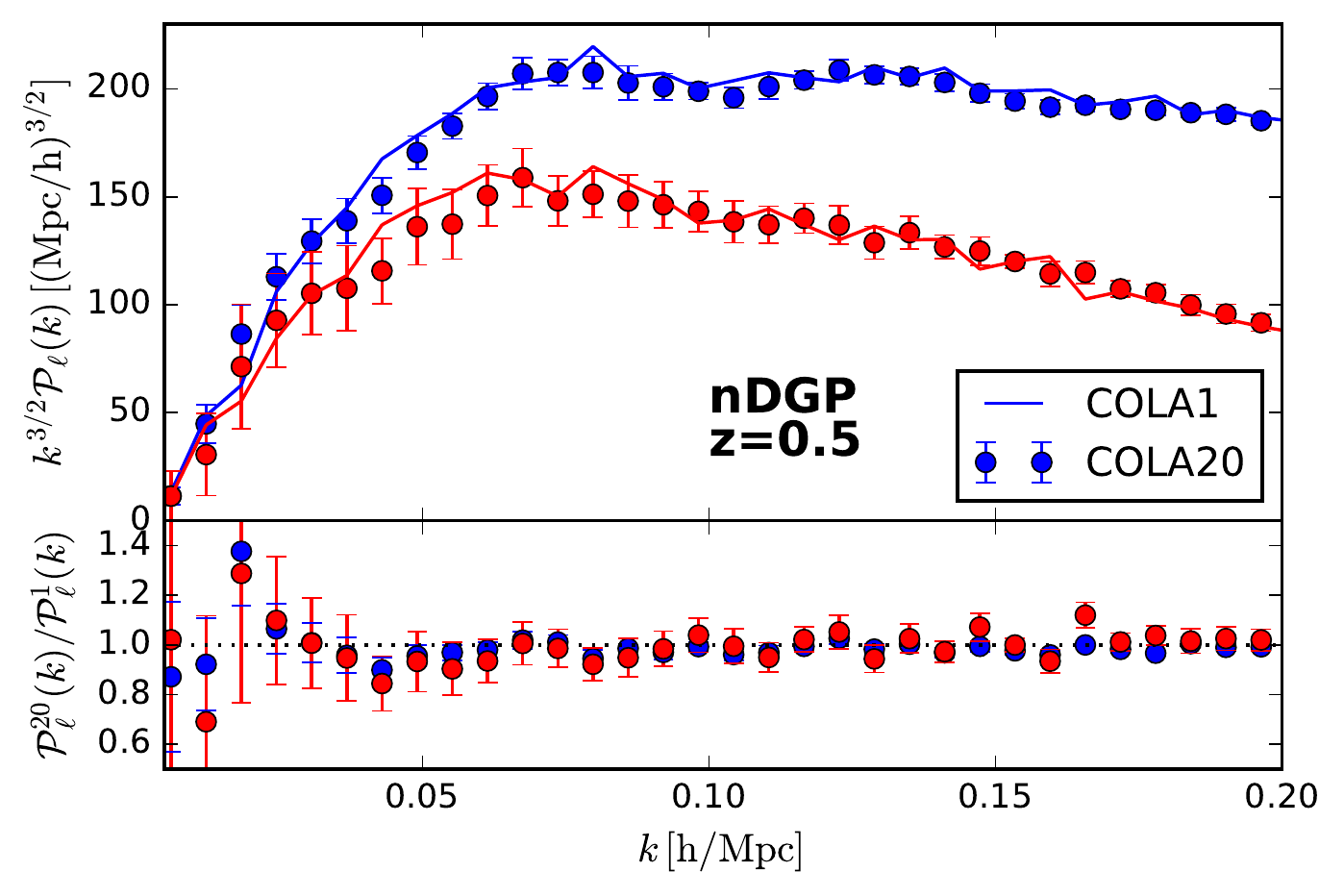}} \quad
  \subfloat[]{\includegraphics[width=7.5cm, height=6.6cm]{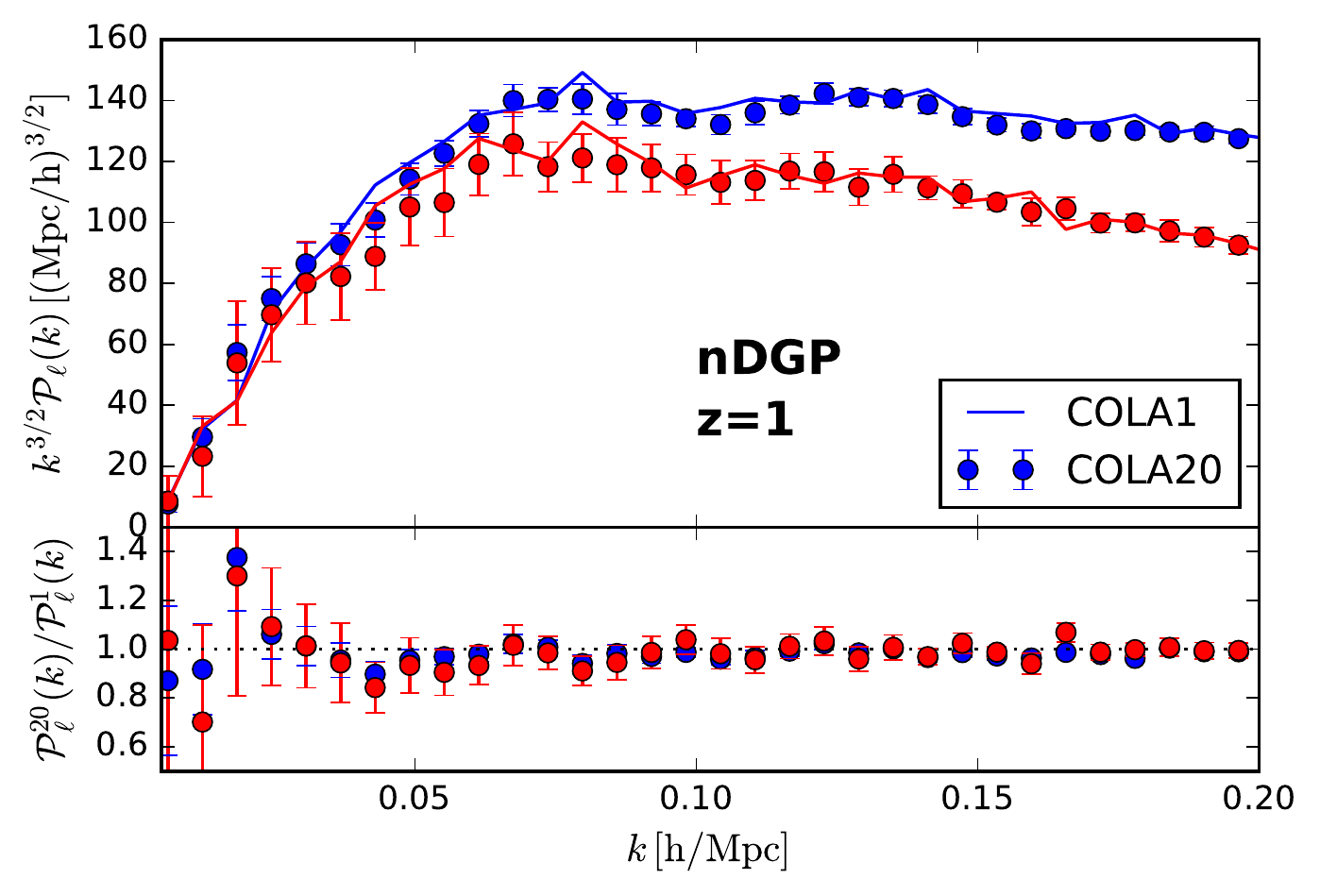}} 
  \caption{ COLA1 (solid) and COLA20 (circles) of the redshift space monopole (blue) and quadrupole (red) for nDGP at $z=0.5$ (left) and $z=1$ (right).  The top panels show the power spectra scaled by $k^{3/2}$ and the bottom panels show the ratio of the two measurements. The errors bars are the variance of the 20 runs of COLA20. }
\label{cola20vcola1}
\end{figure}

In this appendix we validate our use of MG-PICOLA in Chapter 7. We will refer to a single PICOLA simulation with the same initial conditions as N-body as COLA1 and we will refer to the averaged measurements from the 20 PICOLA runs as COLA20. We perform a number of tests listed below. 
\begin{enumerate}
\item
We compare the real space spectra from the full N-body simulation to COLA1. This serves to test the accuracy of PICOLA's evolution of structure. Fig.\ref{cola1vnbody1} shows that the COLA method reproduces the full non-linear real space spectra to within $2\%$ up to $k=0.2h$/Mpc at $z=0.5$ and $z=1$. This asserts that the full non-linear dynamics and evolution is sufficiently captured by MG-PICOLA  at the scales of interest. 
\item
We then compare the multipoles from the full N-body simulation to COLA1 and test for fiducial parameter recovery using the DGP template for both measurements. We use a direct FFTW estimation of the multipoles from the N-body data.  Fig.\ref{cola1vnbody2} shows the multipole comparisons. The redshift space multipoles show less damping in MG-PICOLA simulations compared to the full N-body measurements due to less non-linear structures in these simulations, which give less FoG effects. Since the TNS model has the free parameter $\sigma_v$, that models this non-linear effect, the reduced damping can be accounted for by a smaller value of $\sigma_v$. Fig.\ref{cola1vnbody3} shows the results from an MCMC analysis using multipole data sets from COLA1 and N-body. It shows that we get a very good match in the marginalised posterior distribution of our parameter of interest $\Omega_{rc}$ and the contours are only shifted along $\sigma_v$. We use $1\mbox{Gpc}^3/h^3$ survey errors in accordance with the size of the simulations. 
\item 
Finally, we compare the redshift space multipoles from COLA1 to COLA20. This checks to see if the initial phases used in the full N-body simulation are outside the variance of the 20 runs. Fig.\ref{cola20vcola1} shows that COLA1 is within the variance of the 20 runs and nothing is unusual about the N-body's initial seeds. We also note that the initial condition for N-body was generated by {\tt MPGrafic} which uses the Zeldovich approximation while MG-PICOLA employs 2nd order Lagrangian Perturbation theory to generate initial conditions. 
\end{enumerate}

\chapter{Covariance Between Spectra Multipoles}
In this appendix we give the expressions for the covariance between power spectra multipoles used in Chapter 7. If we omit non-Gaussian contributions only correlations between the same Fourier modes remain. The covariance is then given as \cite{Yamamoto:2005dz,Yamamoto:2002bc,Taruya:2010mx}
\begin{align}
{\rm Cov}_{\ell,\ell'}(k)= &\frac{2}{N_k}\frac{(2\ell +1)(2\ell' +1)}{2} \nonumber \\ 
&\times \int_{-1}^1d\mu \mathcal{P}_\ell(\mu)\mathcal{P}_{\ell'}(\mu)\left[P^{S}(k,\mu) + \frac{1}{\bar{n}}\right]^2.
\end{align}
$N_k$ is the number of modes within a bin at $k$, given by 
\begin{equation}
N_k = 4\pi k^2 \frac{\Delta k }{(2\pi/(V_s)^{1/3})^3},
\end{equation}
where $\Delta k$ is the bin width and $V_s$ is the survey volume. To achieve an analytic estimate for the covariance matrix we can assume linear theory given by Eq.\ref{linkais}. For dark matter the components are  given by 
\begin{align}
{\rm Cov}_{0,0}(k)=&\frac{2}{N_k}\left[\left(1+\frac{4f}{3} +\frac{6f^2}{5} + \frac{4f^3}{7}+\frac{1f^4}{9}\right)F_1^4P_0(k)^2\right. \nonumber \\ 
&\left.+\frac{2}{\bar{n}}\left(1+\frac{2f}{3}+\frac{1f^2}{5}\right)F_1^2P_0(k) +\frac{1}{\bar{n}}\right], \\ \nonumber \\ 
{\rm Cov}_{2,2}(k)=&\frac{2}{N_k}\left[\left(5+\frac{220f}{21} +\frac{90f^2}{7} + \frac{1700f^3}{231}+\frac{2075f^4}{1287}\right)F_1^4P_0(k)^2\right. \nonumber \\ 
&\left.+\frac{2}{\bar{n}}\left(5+\frac{220f}{21}+\frac{30f^2}{7}\right)F_1^2P_0(k) +\frac{5}{\bar{n}}\right], \\ \nonumber \\ 
{\rm Cov}_{0,2}(k)=&\frac{2}{N_k}\left[\left(\frac{8f}{3} +\frac{24f^2}{7} + \frac{40f^3}{21}+\frac{40f^4}{99}\right)F_1^4P_0(k)^2\right. \nonumber \\ 
&\left.+\frac{2}{\bar{n}}\left(\frac{4f}{3}+\frac{4f^2}{7}\right)F_1^2P_0(k) \right],
\end{align}
where ${\rm Cov}_{0,2} = {\rm Cov}_{2,0}$ and $f = G_1/F_1$. 

\renewcommand{\bibname}{Bibliography}
 \bibliographystyle{unsrt}
\bibliography{mybib}

\end{document}